\pdfoutput=1
\documentclass[twocolumn]{aastex61}%,linenumbers
\usepackage{amsmath}	% Advanced maths commands
\usepackage{amssymb}	% Extra maths symbols
\usepackage{placeins}
\newcommand\aastex{AAS\TeX}

\received{XXX}
\revised{YYY}
\accepted{ZZZ}
\shorttitle{\aastex\ Asteroseismology of magnetic cycles}
\shortauthors{A. R. G. Santos et al.}
\usepackage{hyperref}

\begin{document}

\title{Signatures of magnetic activity in the seismic data of solar-type stars observed by \textit{Kepler}}

\author{A. R. G. Santos}%0000-0001-7195-6542
\email{asantos@spacescience.org}
\affil{Space Science Institute, 4750 Walnut Street, Suite 205, Boulder CO 80301, USA}
\affil{Instituto de Astrof\'{i}sica e Ci\^{e}ncias do Espa\c{c}o, Universidade do Porto, CAUP, Rua das Estrelas, PT-4150-762 Porto, Portugal}
\affil{Departamento de F\'{i}sica e Astronomia, Faculdade de Ci\^{e}ncias, Universidade do Porto, Rua do Campo Alegre 687, PT-4169-007 Porto, Portugal}
\affil{School of Physics and Astronomy, University of Birmingham, Edgbaston, Birmingham B15 2TT, UK}

\author{T. L. Campante}
\affil{Instituto de Astrof\'{i}sica e Ci\^{e}ncias do Espa\c{c}o, Universidade do Porto, CAUP, Rua das Estrelas, PT-4150-762 Porto, Portugal}
\affil{Departamento de F\'{i}sica e Astronomia, Faculdade de Ci\^{e}ncias, Universidade do Porto, Rua do Campo Alegre 687, PT-4169-007 Porto, Portugal}
\affil{School of Physics and Astronomy, University of Birmingham, Edgbaston, Birmingham B15 2TT, UK}

\author{W. J. Chaplin}
\affil{School of Physics and Astronomy, University of Birmingham, Edgbaston, Birmingham B15 2TT, UK}
\affil{Stellar Astrophysics Centre, Department of Physics and Astronomy, Aarhus University, Ny Munkegade 120, DK-8000 Aarhus C, Denmark}

\author{M. S. Cunha}
\affil{Instituto de Astrof\'{i}sica e Ci\^{e}ncias do Espa\c{c}o, Universidade do Porto, CAUP, Rua das Estrelas, PT-4150-762 Porto, Portugal}
\affil{Departamento de F\'{i}sica e Astronomia, Faculdade de Ci\^{e}ncias, Universidade do Porto, Rua do Campo Alegre 687, PT-4169-007 Porto, Portugal}

\author{M. N. Lund}
\affil{School of Physics and Astronomy, University of Birmingham, Edgbaston, Birmingham B15 2TT, UK}
\affil{Stellar Astrophysics Centre, Department of Physics and Astronomy, Aarhus University, Ny Munkegade 120, DK-8000 Aarhus C, Denmark}

\author{R. Kiefer}
\affil{Kiepenheuer-Institut f\"{u}r Sonnenphysik, Sch\"{o}neckstra\ss e 6, 79104 Freiburg, Germany}

\author{D. Salabert}
\affil{IRFU, CEA, Universit\'e Paris-Saclay, F-91191 Gif-sur-Yvette, France}
\affil{Universit\'e Paris Diderot, AIM, Sorbonne Paris Cit\'e, CEA, CNRS, F-91191 Gif-sur-Yvette, France}

\author{R. A. Garc\'{i}a}
\affil{IRFU, CEA, Universit\'e Paris-Saclay, F-91191 Gif-sur-Yvette, France}
\affil{Universit\'e Paris Diderot, AIM, Sorbonne Paris Cit\'e, CEA, CNRS, F-91191 Gif-sur-Yvette, France}

\author{G. R. Davies}
\affil{School of Physics and Astronomy, University of Birmingham, Edgbaston, Birmingham B15 2TT, UK}
\affil{Stellar Astrophysics Centre, Department of Physics and Astronomy, Aarhus University, Ny Munkegade 120, DK-8000 Aarhus C, Denmark}

\author{Y. Elsworth}
\affil{School of Physics and Astronomy, University of Birmingham, Edgbaston, Birmingham B15 2TT, UK}
\affil{Stellar Astrophysics Centre, Department of Physics and Astronomy, Aarhus University, Ny Munkegade 120, DK-8000 Aarhus C, Denmark}

\author{R. Howe}
\affil{School of Physics and Astronomy, University of Birmingham, Edgbaston, Birmingham B15 2TT, UK}
\affil{Stellar Astrophysics Centre, Department of Physics and Astronomy, Aarhus University, Ny Munkegade 120, DK-8000 Aarhus C, Denmark}

\begin{abstract}

In the Sun, the frequencies of the acoustic modes are observed to vary in phase with the magnetic activity level. These frequency variations are expected to be common in solar-type stars and contain information about the activity-related changes that take place in their interiors. The unprecedented duration of {\it Kepler} photometric time-series provides a unique opportunity to detect and characterize stellar magnetic cycles through asteroseismology. In this work, we analyze a sample of 87 solar-type stars, measuring their temporal frequency shifts over segments of length 90 days. For each segment, the individual frequencies are obtained through a Bayesian peak-bagging tool. The mean frequency shifts are then computed and compared with: 1) those obtained from a cross-correlation method; 2) the variation in the mode heights; 3) a photometric activity proxy; and 4) the characteristic timescale of the granulation. For each star and 90-d sub-series, we provide mean frequency shifts, mode heights, and characteristic timescales of the granulation. Interestingly, more than $60\%$ of the stars show evidence for (quasi-)periodic variations in the frequency shifts. In the majority of the cases, these variations are accompanied by variations in other activity proxies. About $20\%$ of the stars show mode frequencies and heights varying approximately in phase, in opposition to what is observed for the Sun.
\end{abstract}


\section{Introduction} \label{sec:intro}

Solar-type pulsators, such as low-mass main-sequence stars, exhibit acoustic oscillations (p modes) which are stochastically excited by near-surface convection \citep [e.g.][]{Goldreich1977}. Furthermore, convection, together with stellar differential rotation, also plays an important role in the generation of magnetic fields and activity cycles \citep[e.g.][]{Brun2017}.

As the magnetic fields affect the medium where the acoustic waves propagate, the p modes are sensitive to changes in the magnetic activity. In the Sun, the mode frequencies and the damping rates are observed to vary in phase with the activity level, while the mode amplitudes vary in anti-phase \citep[e.g.][]{Woodard1985,Elsworth1990,Libbrecht1990a,Chaplin1998,Howe2015}. Although the frequency shifts are found to be well correlated with other activity indicators (e.g. 10.7-cm flux, sunspot number, sunspot area, magnetic plage strength index, photometric activity proxy) over the solar cycle \citep[e.g.][]{Chaplin2007,Tripathy2007,Jain2009,Broomhall2015,Santos2016,Santos2017,Salabert2017}, they also show a temporal offset (being ahead in time) in relation to those activity proxies \citep[e.g.][]{Jimenez-Reyes1998,Moreno-Insertis2000,Jain2009,Salabert2009,Salabert2015}.

In addition to the long-term variation on the timescale of the 11-yr solar cycle, the mode frequencies also vary on a quasi-biennial timescale \citep[e.g.][]{Fletcher2010,Broomhall2012,Simoniello2012,Simoniello2013,Salabert2015,Broomhall2015}. The quasi-biennial signal is present in all phases of activity, being modulated by the 11-yr cycle, i.e. having the largest amplitudes around the solar maximum. 
Quasi-biennial variations are also detected in other solar phenomena and activity indicators, such as sunspot number and area coverage, flare and coronal mass ejection rates, total and spectral solar irradiance, 10.7-cm flux, and photometric activity proxy $S_\text{ph}$ \citep[e.g.][]{Bazilevskaya2014,McIntosh2015,Salabert2017}. In fact, these mid-term variations are found to be strongly correlated with those in the solar acoustic frequencies \citep{Broomhall2012,Broomhall2015}. 

Multiple periodicities in the magnetic activity level, measured from chromospheric and photometric proxies, are also observed in several solar-type stars \citep[e.g.][]{Baliunas1995,Olah2009,Metcalfe2010,Metcalfe2013,Egeland2015,Flores2016}.
Similarly  to the solar case, for some of those stars, the longer cycle seems to modulate the shorter cycle \citep{Olah2009,Metcalfe2013}. 

The activity-related changes in the seismic properties are also expected to be common among solar-type stars. The first seismic detection of such signature in a star other than the Sun was made by \citet{Garcia2010}. The authors found evidence for an activity cycle in the photometric and seismic indicators of a solar-type star (HD 49933) observed by the CoRoT \citep[Convection, Rotation, and planetary Transits;][]{Baglin2006} space telescope. Similarly to what is observed in the Sun, the mode amplitudes and frequency shifts are anti-correlated in time and show a temporal offset in relation to the photometric indicator. The second detection of activity-related frequency shifts in a star other than the Sun was reported by \citet{Salabert2016}. The authors found temporal frequency shifts varying consistently with the photometric activity indicator, $S_\text{ph}$ \citep{Mathur2014,Garcia2014}, in the active solar-type star KIC~10644253, observed by the {\it Kepler} main mission \citep{Borucki2010}. They also mentioned the possibility that the frequency shift they found could be related to a short-term modulation, similar to the quasi-biennial modulation in the Sun. In spite of the large uncertainties, \citet{Regulo2016} found evidence for frequency shifts in the active solar-type star KIC~3733735, which seem to be ahead in time relative to the $S_\text{ph}$. \citet{Kiefer2017} analyzed 24 solar-type stars observed by {\it Kepler}, searching for variations in the mode frequencies through a cross-correlation method and in the height of the p-mode envelope. The authors reported significant frequency shifts in 23 stars and evidence for activity-related frequency shifts in six of them. \citet{Salabert2017b} studied the frequency dependence of the frequency shifts observed in {\it Kepler} solar-type stars. The results for the four best stars in their sample suggest that the main source for the observed frequency shifts in the Sun and in those stars may be different.

In this work, we analyze the short-cadence data of 87 solar-type stars observed by {\it Kepler} main mission. The main goal is to search for temporal variations of the acoustic frequencies, that may be related to stellar activity. To that end, we developed a Bayesian peak-bagging tool to estimate the mode parameters. For each star, we present and provide the temporal mean frequency shifts (including for those for the individual angular degrees), the temporal evolution of the mode heights, and characteristic timescale of the granulation. %%The structure of the article is as follows: Section~\ref{sec:meth} describes the methodology; Sections~\ref{sec:kepler} and ~\ref{sec:conclusions} present and summarize the results from the analysis of the stars in our sample.  %Section~\ref{sec:kepler} presents the results from the analysis of the stars in our sample. Finally, in Section~\ref{sec:conclusions}, we summarize the main results from our analysis.  

\section{Methodology}\label{sec:meth}

\subsection{Target sample}

\begin{figure}
\includegraphics[width=\hsize]{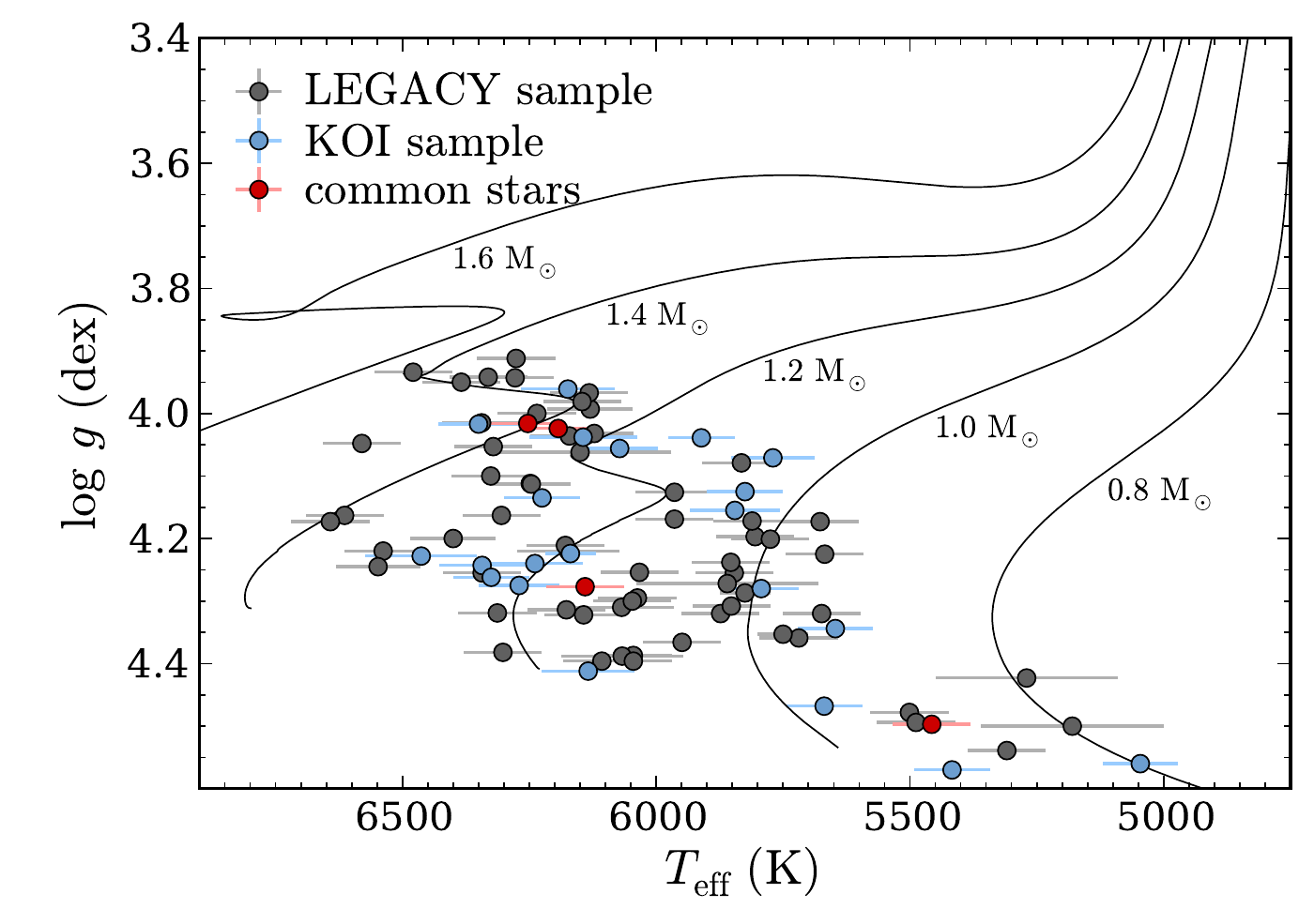}
\caption{Kiel-diagram for the stars in the target sample, which is composed of the LEGACY sample (gray dots) and a sample of 25 KOIs (blue dots). The red dots mark the stars that are common to both sub-samples. The black solid lines show the solar-calibrated evolutionary tracks obtained with the evolution code Modules for Experiments in Stellar Astrophysics \citep[MESA;][]{Paxton2011,Paxton2013}.}\label{fig:KICsample}
\end{figure}

The main goal of this work is to search for temporal frequency shifts, possibly related to magnetic activity, in a large sample of {\it Kepler} targets. In order to successfully detect and characterize solar-type oscillations, we analyze the short-cadence ($\Delta t=58.85\rm\,s$) data of 87 {\it Kepler} solar-type stars, selected from two samples previously studied in the literature (with four common stars). Most of these stars (66) are high signal-to-noise ratio solar-type pulsators that constitute the LEGACY sample \citep{Lund2017,SilvaAguirre2017}. The other group of targets is composed of 25 solar-type {\it Kepler} Objects of Interest (KOIs) that were analyzed by \citet{Campante2016} in the context of the spin-orbit alignment of the exoplanet systems. These KOIs were also part of a larger sample analyzed by \citet{SilvaAguirre2015} and \citet{Davies2016}. The stellar parameters of the 87 {\it Kepler} solar-type stars are summarized in Table~1 and
Figure~\ref{fig:KICsample} displays the target sample in a Kiel-diagram (i.e. $\log g$ versus $T_{\rm eff}$).\vspace{-0.2cm}%\ref{tab:KICtable0}

\startlongtable
\begin{deluxetable*}{ccccrrccrc}\vspace{-0.5cm}
\tablecaption{Stellar parameters of the target sample composed of 87 {\it Kepler} solar-type stars. References: Frequency of the maximum power ($\nu_\text{max}$) and large frequency separation ($\Delta\nu$) from \citet{Davies2016} or \citet{Lund2017}. Effective temperature ($T_\text{eff}$) and metallicity ($[\text{Fe/H}]$) from (1) \citet{Lund2017}; (2) \citet{SilvaAguirre2015}; (3) \citet{Chaplin2014a}; (4) \citet{Casagrande2014}; (5) \citet{Pinsonneault2012}; (6) \citet{Pinsonneault2014}; (7) \citet{Huber2013}; or (8) \citet{Ramirez2009}. Surface gravity ($\log g$) from \citet{SilvaAguirre2015,SilvaAguirre2017}. Rotation period ($P_\text{rot}$) from (9) \citet{Garcia2014}; (10) \citet{McQuillan2014}; (11) \citet{Ceillier2016}; (12) \citet{McQuillan2013a}; (13) \citet{Walkowicz2013a}; (14) \citet{Karoff2013a}; (15) \citet{Dumusque2014a}; or (16) \citet{Davies2015}. Each Quarter corresponds to one-quarter of the {\it Kepler}'s year ($\sim 372.5$ days).\vspace{-0.3cm}}\label{tab:KICtable0}
\tablewidth{\hsize}
\tabletypesize{\fontsize{9}{7.}\selectfont}
\tablehead{
\colhead{\vspace{-0.003cm}KIC\vspace{-0.015cm}}& \colhead{\vspace{-0.003cm}KOI\vspace{-0.015cm}} & \colhead{\vspace{-0.003cm}Kp\vspace{-0.015cm}} & \colhead{\vspace{-0.003cm}Quarters\vspace{-0.015cm}}& \colhead{\vspace{-0.003cm}$\nu_\text{max}$\vspace{-0.015cm}} & \colhead{\vspace{-0.003cm}$\Delta\nu$\vspace{-0.015cm}}& \colhead{\vspace{-0.003cm}$T_\text{eff}$\vspace{-0.015cm}} & \colhead{\vspace{-0.003cm}[Fe/H]\vspace{-0.015cm}}& \colhead{\vspace{-0.003cm}$\log g$\vspace{-0.015cm}}& \colhead{\vspace{-0.003cm}$P_\text{rot}$\vspace{-0.015cm}}\\
\colhead{\vspace{-0.003cm}}&\colhead{\vspace{-0.003cm}}&\colhead{\vspace{-0.003cm}}&\colhead{\vspace{-0.003cm}}&\colhead{($\mu$Hz)\vspace{-0.003cm}}&\colhead{($\mu$Hz)\vspace{-0.003cm}}&\colhead{(K)\vspace{-0.003cm}}&\colhead{(dex)\vspace{-0.003cm}}&\colhead{(dex)\vspace{-0.003cm}}&\colhead{(days)\vspace{-0.003cm}}
} 
\startdata
1435467 & -- & 8.88 & Q5.1-Q17.2 & $1406.7_{-8.4}^{+6.3}$ & $70.369_{-0.033}^{+0.034}$ & 6326$\pm$77\textsuperscript{1} & 0.01$\pm$0.1\textsuperscript{1} & 4.100$_{-0.009}^{+0.009}$ & 6.68$\pm$0.89\textsuperscript{9}\\
2837475 & -- & 8.48 & Q5.1-Q17.2 & $1557.6_{-9.2}^{+8.2}$ & $75.729_{-0.042}^{+0.041}$ & 6614$\pm$77\textsuperscript{1} & 0.01$\pm$0.1\textsuperscript{1} & 4.163$_{-0.007}^{+0.007}$ & 3.68$\pm$0.36\textsuperscript{9}\\
3425851 & 268 & 10.56 & Q6.1-Q8.3 & $2038.0_{-60.0}^{+60.0}$ & $92.600_{-2.500}^{+2.500}$ & 6343$\pm$85\textsuperscript{2} & -0.04$\pm$0.1\textsuperscript{2} & 4.243$_{-0.008}^{+0.008}$ & 7.873$\pm$0.001\textsuperscript{10}\\
3427720 & -- & 9.11 & Q5.1-Q17.2 & $2737.0_{-17.7}^{+10.7}$ & $120.068_{-0.032}^{+0.031}$ & 6045$\pm$77\textsuperscript{1} & -0.06$\pm$0.1\textsuperscript{1} & 4.387$_{-0.005}^{+0.004}$ & 13.94$\pm$2.15\textsuperscript{11}\\
3456181 & -- & 9.66 & Q5.1-Q11.3 & $970.0_{-5.9}^{+8.3}$ & $52.264_{-0.039}^{+0.041}$ & 6384$\pm$77\textsuperscript{1} & -0.15$\pm$0.1\textsuperscript{1} & 3.950$_{-0.007}^{+0.005}$ & --\\
3544595 & 69 & 9.93 & Q2.3-Q17.2 & $3366.0_{-81.0}^{+81.0}$ & $147.770_{-0.450}^{+0.450}$ & 5669$\pm$75\textsuperscript{2} & -0.18$\pm$0.1\textsuperscript{2} & 4.468$_{-0.003}^{+0.003}$ & --\\
3632418 & 975 & 8.22 & Q5.1-Q17.2 & $1166.8_{-3.8}^{+3.0}$ & $60.704_{-0.018}^{+0.019}$ & 6193$\pm$77\textsuperscript{1} & -0.12$\pm$0.1\textsuperscript{1} & 4.024$_{-0.007}^{+0.005}$ & 12.591$\pm$0.036\textsuperscript{10}\\
3656476 & -- & 9.52 & Q5.1-Q17.2 & $1925.0_{-6.3}^{+7.0}$ & $93.194_{-0.020}^{+0.018}$ & 5668$\pm$77\textsuperscript{1} & 0.25$\pm$0.1\textsuperscript{1} & 4.225$_{-0.008}^{+0.010}$ & 31.67$\pm$3.53\textsuperscript{9}\\
\enddata
\end{deluxetable*}

{\color{white}pagebreak

pagebreak

}

\subsection{Data preparation}

The pixel data of the sample stars were collected from the {\it Kepler} Asteroseismic Science Operations Center (KASOC; \url{http://kasoc.phys.au.dk}) and then corrected using the KASOC filter \citep{Handberg2014}. For further details see \citet[Section 2.1]{Lund2017}.

To search for temporal variations of the acoustic frequencies of the {\it Kepler} targets, the original time-series were divided into sub-series of 90 days overlapped by 45 days. The respective power density spectra are obtained as the periodogram of each sub-series normalized so that Parseval's theorem is satisfied.

For the photometric magnetic activity proxy, $S_\text{ph}$ (see Section \ref{sec:sph}), we use KADACS ({\it Kepler} Asteroseismic Data Analysis and Calibration Software) light curves obtained from the long-cadence data, being corrected following approach described in \citet{Garcia2011} and high-pass filtered at 20 and 55 days, allowing to measure rotation periods as long as 90 days.

\subsection{Photometric magnetic activity proxy}\label{sec:sph}

\citet{Garcia2010} showed, in the case of the CoRoT solar-type star HD\,49933, that the fluctuations associated with the presence of spots or magnetic features rotating on the surface of the star can provide a global proxy of stellar magnetic activity. However, brightness variability can have different origins with various timescales, such as convective motions, oscillations, stellar companion, or instrumental problems. Therefore, to compute such magnetic activity proxy, the stellar rotation period, $P_\text{rot}$, needs to be taken into account. The so-called photometric activity proxy, $S_\text{ph}$, is estimated as the standard deviations calculated over sub-series of length $5\times P_\text{rot}$. \citet{Mathur2014} demonstrated that $S_\text{ph}$ provides a global proxy only related to magnetism and not to other sources of variability. Furthermore, both long- (11-yr) and short-term (quasi-biennial) variations can be monitored through the $S_\text{ph}$, as shown in the case of the Sun \citep{Salabert2017}. Also, \citet{Salabert2016b} showed the complementarity between $S_\text{ph}$ and the chromospheric activity measured as the Ca K-line emission index \citep{Wilson1978}. Unlike chromospheric activity proxies, $S_\text{ph}$ can be easily estimated from space photometric observations for a large number of stars with known rotation period. Chromospheric activity proxies require a large amount of ground-based telescope time to collect enough spectroscopic data for each individual target, and this is only possible for bright targets.

\subsection{Background signal}\label{sec:backg}

The power spectrum of solar-type stars may enclose the signature of different stellar phenomena, such as active regions, granulation, faculae, and acoustic oscillations.
We describe the granular and facular components through a Harvey-like profile \citep[e.g.][]{Harvey1985,Garcia2009,Mathur2011,Handberg2011,Campante2011,Karoff2012,Davies2016}
\begin{equation}
f(\nu)=\dfrac{H}{1+(2\pi\nu\tau)^\alpha},\label{eq:harvey}
\end{equation}
where $H=4\sigma^2\tau$ is the amplitude of the granulation power, $\sigma$ and $\tau$ are, respectively, the characteristic amplitude and timescale, and $\alpha$ is the slope of the power law. 
For the activity component, we use a power law of the form
\begin{equation}
f(\nu)=\dfrac{H_\text{act}}{\nu^2},
\end{equation}
which results from considering a Harvey-like profile (Equation~\ref{eq:harvey}) in the limit $2\pi\tau_\text{act}\gg1$ with $\alpha=2$, found to be adequate in describing the exponential decay of active regions \citep[e.g][]{Garcia2009,Campante2016}.
Finally, to properly model the background signal, one needs to consider a flat component, $N$, related to the photon shot-noise.

We start by considering two competing background models. The first model considers three components ($p=5$ parameters): activity, granulation, and photon shot-noise. The second background model further includes a facular component ($p+n=8$ parameters). 
To test the statistical significance of the additional model parameters, we compute the respective likelihood ratio, $\Lambda$ \citep[e.g.][]{Appourchaux1998,Karoff2012,Karoff2013}. When $\Lambda\ll1$, we can conclude that the $n$ additional parameters are not needed to describe the background signal from a statistical viewpoint.

For a large number of sub-series, we find that the facular component is not significant enough. Therefore, in order to be consistent in the analysis, the final background model, which we apply to all {\it Kepler} targets, corresponds to the first of the tested models, i.e. 
\begin{equation}
\mathcal{B}(\nu)=\left[\dfrac{H_\text{act}}{\nu^2}+\dfrac{H_\text{gran}}{1+(2\pi\nu\tau_\text{gran})^{\alpha_\text{gran}}}\right]\eta^2(\nu)+N,
\label{eq:back}\end{equation}
where
\begin{equation}
\eta={\rm sinc}\left(\dfrac{\pi}{2}\dfrac{\nu}{\nu_{\rm Nyquist}}\right)
\end{equation}
describes the apodization resulting from the sampling of the signal \citep{Chaplin2011}, and $\nu_{\rm Nyquist}$ is the Nyquist frequency.

In this step of the analysis, we exclude the low-frequency range (with frequency cutoff defined as $200\times\nu_\text{max}/\nu_{\text{max},\odot}\,\mu\text{Hz}$), which is dominated by an activity component. Still, we model the activity component in order to prevent any contamination (due to spectral leakage) into the frequency range where the granulation component becomes important. While fitting the background, we also exclude the frequency range  of the p modes (centered at $\nu_\text{max}$ and with a width of $2/3\nu_\text{max}$).

\begin{figure}[h]
\includegraphics[width=\hsize]{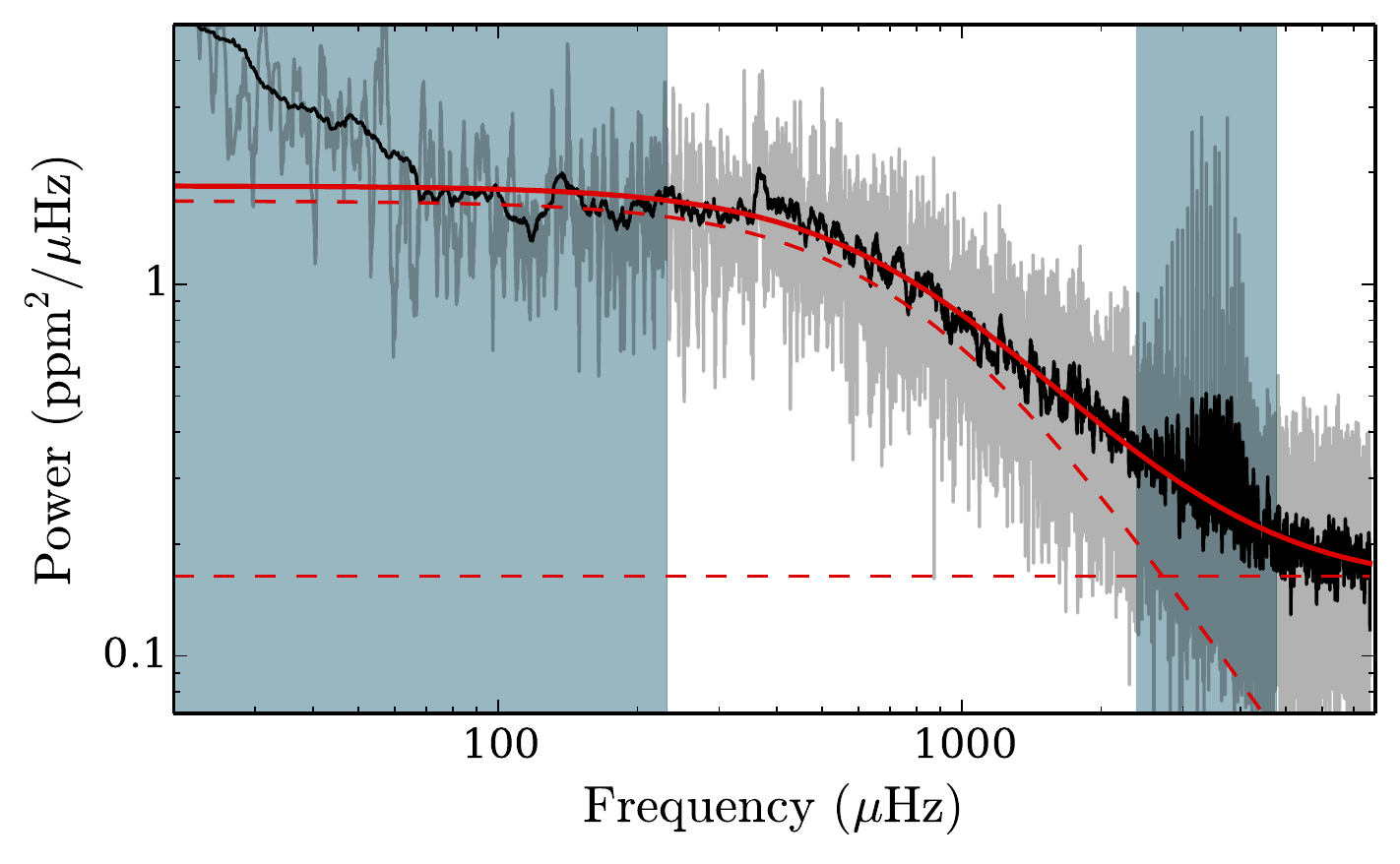}\vspace{-0.1cm}
\caption{Power density spectrum (gray and black; with different smoothing for illustrative purposes) of a 90-d sub-series for the solar-type star KIC~8006161. The red solid line shows the best background model and the dashed lines show the different background components. The blue areas mark the frequency intervals neglected from the fitting process.}\label{fig:back}
\end{figure}

The model parameters that best describe the background signal are obtained through Maximum Likelihood Estimation (MLE) and the formal errors are derived from the inverse Hessian matrix \citep[e.g.][]{Toutain1994,Campante2011}. 
The background parameters are then fixed for each sub-series in the subsequent peak-bagging analysis (Section \ref{sec:peakbag}).
As an example, Figure~\ref{fig:back} shows the power density spectrum of a 90-d sub-series for one of the stars in the sample, namely KIC~8006161. 

As the convection is the mechanism behind both granulation and acoustic oscillations, the granulation characteristic timescale, $\tau_\text{gran}$, is expected to be related to the timescale of the p-modes and, thus, to $\nu_\text{max}$. \citet{Huber2009} suggested that $\tau_\text{gran}$ scales inversely to $\nu_\text{max}$, i.e.  $\tau_\text{gran}\propto1/\nu_\text{max}$ \citep[see also][]{Kjeldsen2011}. This means that larger stars are expected to have longer granulation timescales than smaller stars. \citet[based on CoRoT red giant and main-sequence solar-type pulsators]{Kallinger2010}, \citet[based on {\it Kepler} red giants]{Mathur2011}, and \citet[based on {\it Kepler} main-sequence stars, sub-giants, and red giants]{Kallinger2014} confirmed the empirical prediction by \citet{Huber2009}, finding that the granulation timescale is approximately proportional to $\nu_\text{max}^{-1}$.

\begin{figure}[h!]
\includegraphics[width=\hsize]{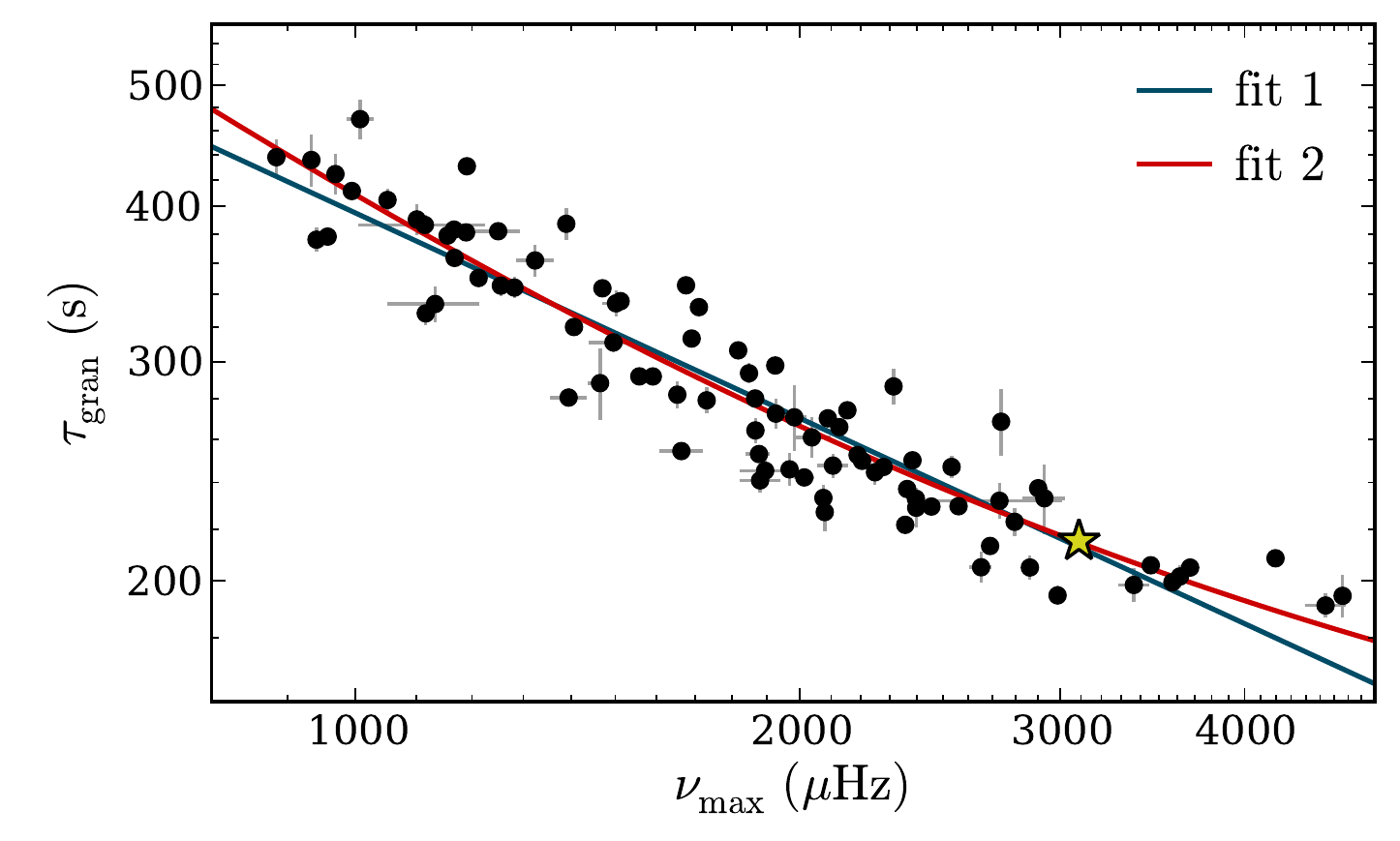}\vspace{-0.1cm}
\caption{Granulation timescale as a function of $\nu_\text{max}$ for the stars in the sample. The blue and red lines show the best fits ($\tau_\text{gran}=A\nu_\text{max}^B$ and $\tau_\text{gran}=A\nu_\text{max}^B+C$, respectively). For comparison, the yellow star marks the position of the Sun with $\nu_{\text{max},\odot}=3090\,\mu \text{Hz}$ \citep[e.g.][]{Huber2011} and $\tau_{\text{gran},\odot}$ computed in Section \ref{sec:virgo}.}\label{fig:tgnumax}
\end{figure}

Figure~\ref{fig:tgnumax} shows the granulation timescale for the stars in the sample, computed as the weighted average over the independent sub-series (i.e. every second sub-series), as a function of $\nu_\text{max}$. The yellow star marks the position of the Sun, and the red and blue lines the best fits to the data. The fit shown in blue is obtained when fitting a function of the form $\tau_\text{gran}=A\nu_\text{max}^B$ (fit~1), where $A=(1.7\pm0.3)\!\times\!10^4$ and $B\!=\!-0.55\pm0.02$. If we consider an extra parameter, a constant $C$, i.e. $\tau_\text{gran}=A\nu_\text{max}^B+C$ (fit~2),
we will find the best fit shown in red with $A=(2.1\pm2.7)\!\times\!10^5$, $B\!=\!-1.0\pm0.2$, and $C\!\!=\!110\pm40$. The value found for the exponent $B$ differs from that found in previous studies, where $B=-0.89$ \citep{Mathur2011,Kallinger2014}. This difference may arise from the type of stars used in the different studies. The results presented here are based on a sample of solar-type stars with $\nu_\text{max}$ between 880 and 4660 $\mu\text{Hz}$.

\subsection{Bayesian peak-bagging tool}\label{sec:peakbag}

In order to perform the global (simultaneous) fit of the acoustic modes, we follow a Bayesian approach \citep[e.g.][]{Benomar2009,Campante2011,Handberg2011,Davies2016,Lund2017}, through the implementation of the Affine Invariant Markov Chain Monte Carlo (MCMC) Ensemble sampler \texttt{emcee} \citep[][]{Goodman2010,Foreman-Mackey2013}.
In the Bayesian framework, the posterior probability for a set of model parameters, $\bf\Theta$, given the observational data $D$ and the available prior information $I$, i.e. $p({\bf\Theta}|D,I)$, is obtained through Bayes' theorem\vspace{-0.2cm}
\begin{equation}
p({\bf\Theta}|D,I)=\dfrac{p({\bf\Theta}|I)p(D|{\bf\Theta},I)}{p(D|I)},
\end{equation}
where $p({\bf\Theta}|I)$ is the prior probability of the parameters, $p(D|{\bf\Theta},I)$ is the likelihood function, and $p(D|I)$ is a normalization factor.

The power density spectrum of a solar-type star contains a number of oscillation modes characterized by the wave numbers $n$, $l$, and $m$ (radial order, angular degree, and azimuthal order, respectively). To model the power density spectrum of each oscillation mode, we use a standard Lorentzian profile. The final model of the acoustic spectrum is then given by
\begin{equation}
\mathcal{M}(\nu)=\sum_{n=n_0}^{n_1}\sum_{l=0}^{l_1}\sum_{m=-l}^l\,\,\dfrac{H_{nlm}}{1+\dfrac{4}{\Gamma_{nlm}^2}(\nu-\nu_{nl}-m\nu_{\rm s})^2},
\end{equation}
where $n_0$ and $n_1$ are the first and last radial orders considered in the global fit, $l_1$ is the highest angular degree visible within a radial order $n$, $\nu_{nl}$ and $\Gamma_{nlm}$ are the mode frequency and linewidth, and $\nu_{\rm s}$ represents the rotational splitting. The mode height $H_{nlm}$ is given by
\begin{equation}
H_{nlm}=\mathcal{E}_{lm}(i)\widetilde{V}_l^2S_{nl},
\end{equation}
where $\mathcal{E}_{lm}$ represents the relative mode visibility within the $(n,l)$ multiplet \citep{Gizon2003}, $i$  is the stellar inclination angle, $\widetilde{V}_l=V_l/V_{l=0}$ and $V_l$ is related to the geometrical visibility of the multiplet, and $S_{nl}$ is the multiplet's overall height. The geometrical visibilities for {\it Kepler} targets are taken from \citet{Handberg2011}.

The final model of the stellar power spectrum corresponds to
\begin{equation}
\mathcal{P}(\nu;{\bf\Theta})=\mathcal{M}(\nu)+\mathcal{B}(\nu).\label{eq:pdsmodel}
\end{equation}

In our analysis, only the linewidths and heights of the radial modes, $\Gamma_{n0}$ and $S_{n0}$, are considered as free parameters. For the quadrupole modes, we consider the linewidth and height of the closest radial mode, since their frequencies do not differ by much. For the dipolar modes, we linearly interpolate between the two closest radial orders. Thus, the final set of free parameters is
\begin{equation}
{\bf\Theta}=\{\nu_{nl},S_{n0},\Gamma_{n0},\nu_\text{s},i\}.
\end{equation}

Assuming a $\chi^2$ with two degrees of freedom statistics for the power spectrum \citep{Duvall1986,Anderson1990,Gabriel1994}, the logarithm of the likelihood function is given by\vspace{-0.1cm}
\begin{equation}
\mathcal{L}({\bf\Theta})\equiv\ln L({\bf\Theta})=-\sum_j\left\{\ln\mathcal{P}(\nu_j;{\bf\Theta})+\dfrac{P_j}{\mathcal{P}(\nu_j;{\bf\Theta})}\right\},
\end{equation}
where $\mathcal{P}(\nu_j;{\bf \Theta})$ corresponds to the mean power spectrum, which we model (Equation~\ref{eq:pdsmodel}), and $P_j$ is the observed power. We use logarithmic probabilities to ensure numerical stability.

One of the main strengths of a Bayesian approach is the possibility of using prior knowledge in the analysis. In what follows, we summarize the prior functions that we assume for the model parameters (mode frequencies, heights and linewidths of the radial modes, rotational splitting, and stellar inclination angle) and the fitting method.

\subsubsection{Prior probabilities}\label{sec:prior}

{\it Mode frequencies -} For the mode frequencies we use uniform priors, whose lower and upper limits are defined as $\nu_{nl}^0\pm4\,\mu\text{Hz}$, where the values $\nu_{nl}^0$ are taken from the literature, namely from \citet{Davies2016} and \citet{Lund2017}. We further constrain the mode frequencies by using priors on the large and small frequency separations ($\Delta\nu$ and $d\nu$, respectively), which are expected to vary smoothly from one order to the next. Following the approach of \citet{Davies2016}, we define the prior functions for the large and small frequency separations as \vspace{-0.1cm}
\begin{equation}
\ln f(\Delta\nu_l)=-0.125\sum_{n=n_0}^{n_\text{max}}\left(\dfrac{\partial^2\nu_{nl}}{\partial n^2}\right)^2,\vspace{-0.1cm}
\end{equation}
\begin{equation}
\ln f(d\nu_{l,l+2})=-0.25\sum_{n=n_0}^{n_\text{max}}\left(\dfrac{\partial d\nu_{l,l+2}}{\partial n}\right)^2.
\end{equation}
To compute the prior on the large separation, at least five modes of angular degree $l$ are needed. For this reason, the prior on the large separation is only applied when this condition is met. The prior on the small separation is only applied to $d\nu_{0,2}(n)$, which was found to be enough for a stable fit \citep{Davies2016}.\\

{\it Mode linewidths -} For the mode linewidths, we apply a uniform prior with lower and upper limits of $0$ and $12\,\mu\text{Hz}$.\\

{\it Mode heights -} We also apply a uniform prior on the height of the radial modes, whose lower limit is fixed at $0\,\text{ppm}^2/\mu\text{Hz}$ and upper limit varies from power spectrum to power spectrum, being estimated as follows:
\begin{enumerate}
\item the frequency range of interest, the p-mode envelope, is defined as $[\nu_-^0-\Delta\nu/4,\nu_+^0+\Delta\nu/4]$, where $\nu_-^0$ and $\nu_+^0$ are the minimum and maximum mode frequencies we consider (again, 0 denotes the values from the literature);
\item the contribution of the acoustic background is removed from the power spectrum;
\item the resulting power spectrum is smoothed by applying a uniform filter with size equal to the reciprocal of the resolution of the spectrum;
\item the height upper limit is defined as the maximum height of the smoothed power spectrum.\\
\end{enumerate}

{\it Stellar rotational splitting and inclination angle -} We use the posterior distributions obtained in previous studies (based on the complete light curves) by \citet{Davies2016} and \citet{Lund2017} as priors on the stellar rotational splitting and stellar inclination angle.
\subsubsection{Fitting method}\label{sec:emcee}

The global fit to the acoustic modes is performed through implementation of the algorithm \texttt{emcee} \citep[][]{Foreman-Mackey2013}, based on the Affine Invariant Markov Chain Monte Carlo (MCMC) Ensemble sampler \citep[][]{Goodman2010}. \texttt{emcee} makes use of an interacting ensemble of so-called ``walkers". Each walker has its own separate MCMC chain but the proposal distribution, i.e. the next step in the chain, depends on the positions of the remaining walkers. Furthermore, in order to ensure an efficient sampling of the parameter space, we also employ parallel tempering \citep[][]{Earl2005}, which is useful to avoid cases where a given walker gets trapped in a local maximum and to access broader regions of the parameter space.

For each power spectrum, we use $500$ walkers (initialized by sampling the prior distributions) and three temperatures defined according to \citet{Benomar2009}. Each chain runs for $10^4$ steps after a burn-in phase, long enough to ensure the convergence of the chains and a swap acceptance rate between adjacent temperatures of about $50\%$.

For each model parameter, the posterior distribution is obtained by computing the histogram of its sampled values. The final parameter estimates and uncertainties are based on the 16\textsuperscript{th}, 50\textsuperscript{th}, and 84\textsuperscript{th} percentiles, i.e. given by the median and the $68\%$ credible region of the distribution. 

\subsection{Mean frequency-shift and mean height estimation}\label{sec:fshifts}

In order to search for temporal variations of the acoustic mode frequencies, the original time-series are divided into 90-d sub-series overlapped by 45 days (in average, 22 sub-series per star).

Having the mode frequencies, $\nu_{nl}$, and the respective uncertainties (obtained with the peak-bagging tool described above) for each sub-series, one can then estimate the temporal frequency shifts. The reference mode frequencies, $\nu_{nl}^\text{ref}$, are taken as the weighted time averages of the mode frequencies. Then, for each multiplet $(n,l)$, we compute the variation in frequency with respect to the reference frequencies as 
\begin{equation}
\delta\nu_{nl}(t)=\nu_{nl}(t)-\nu_{nl}^\text{ref}.
\end{equation}
Finally, following the approach used by, e.g. \citet{Chaplin2007} and \citet{Tripathy2007}, we obtain the weighted mean frequency shifts as
\begin{equation}
\delta\nu(t)=\dfrac{\sum_{nl}\delta\nu_{nl}(t)/\sigma^2_{nl}(t)}{\sum_{nl}1/\sigma^2_{nl}(t)}.
\end{equation}
The uncertainties on the mean frequency shifts are
\begin{equation}
\sigma(t)=\left(\sum_{nl}\dfrac{1}{\sigma_{nl}^2(t)}\right)^{-1/2}.
\end{equation}
Note that for the low-degree modes, the mode inertia dependency on the angular degree is not significant \citep[e.g.][]{Chaplin2007} and, therefore, we have neglected the inertia ratio in the frequency-shift estimation.

In the Sun, the acoustic frequencies and heights show an anti-correlated behavior with time: while the frequencies increase with increasing activity level, the mode heights decrease. Evidence for such anti-correlated behavior was also found in solar-type observed by CoRoT and {\it Kepler} \citep{Garcia2014,Kiefer2017}. Therefore, we also search for temporal variations in the latter.

The mode height distributions follow a log-normal distribution. For this reason, we use the logarithm of the mode heights in the calculations for the mean height estimates. Thus, the mean logarithmic heights are computed as the weighted average of the logarithm of the mode heights, following the same approach used for the frequency shifts, i.e.
\begin{equation}
\ln S (t)=\dfrac{\sum_n\ln S_{n0}(t)/\sigma_{\ln S_{n0}}^2(t)}{\sum_{n}1/\sigma_{\ln S_{n0}}^2(t)},
\end{equation}
\begin{equation}
\sigma_\text{S}(t)=\left(\sum_n\dfrac{1}{\sigma_{\ln S_{n0}}^2(t)}\right)^{-1/2}.
\end{equation}

\section{Peak-bagging analysis}\label{sec:kepler}

Before applying the peak-bagging tool to the {\it Kepler} data, we validated the tool with real and artificial solar data. The results from the validation tests are shown in Appendix~\ref{sec:sun}. The current section presents the results for the {\it Kepler} targets.

Considering the background model, obtained in Section~\ref{sec:backg}, for each 90-d sub-series for each star, we apply the peak-bagging tool as described in Section~\ref{sec:peakbag}, obtaining the marginal posterior probability distributions for the model parameters.

\begin{figure}[h]
\includegraphics[width=\hsize]{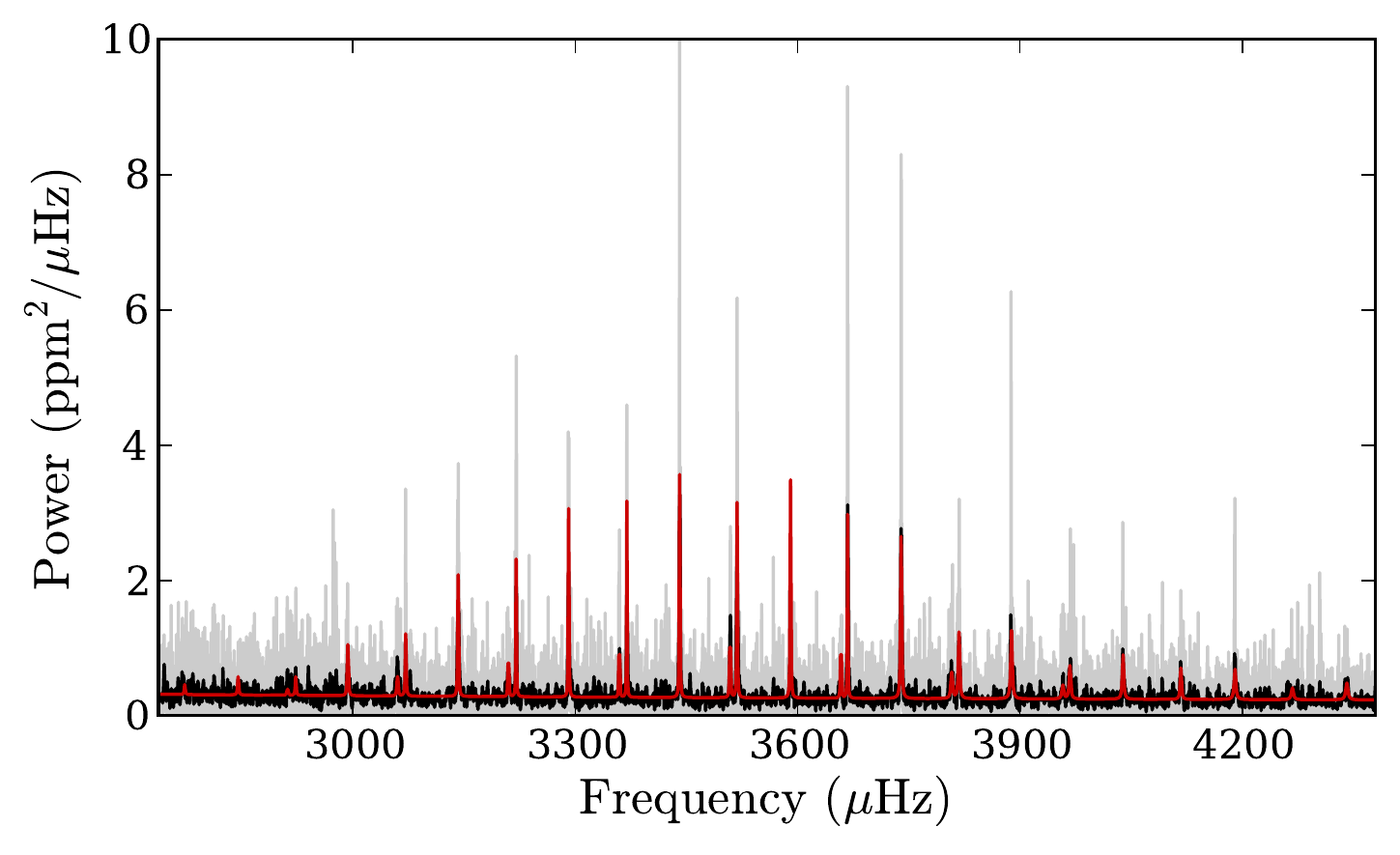}\vspace{-0.27cm}
\caption{Power density spectrum (gray and black; with different smoothing for illustrative purposes) for a given 90-d sub-series of the solar-type star KIC 8006161. The red lines show the best fit to the power spectrum obtained with the Bayesian peak-bagging tool.}\label{fig:pdskepler}\vspace{-0.45cm}
\end{figure}

The final parameter estimates are given by the median of the posterior probability distribution and the uncertainties are determined based on the $68\%$ credible region (see Section~\ref{sec:emcee}). Figure~\ref{fig:pdskepler} compares the power spectrum and the best fit obtained with the peak-bagging tool for a given sub-series of one of the stars in the sample (KIC~8006161). 

\subsection{Mean frequency shifts and mode heights}

Having the mode frequencies for each sub-series, we then compute the weighted mean frequency shifts over time for each star (see Section~\ref{sec:fshifts} for details). We compare them with frequency shifts which we obtained with a cross-correlation method which is described in detail in \citet{Kiefer2017}. With this method, estimates of the frequency shifts and the corresponding uncertainties are obtained in the following way: First, 200 realizations of the time series of each segment are generated with a resampling approach and their periodograms are calculated. Subsequently, the p-mode range of the periodograms of each segment are cross-correlated with the p-mode range of the periodograms of the reference segment. The resulting cross-correlation functions are fitted with a Lorentzian profile. The mean of the centroids of the 200 Lorentzian fits is used as the value for the frequency shift and the standard deviation of the centroids is used as the uncertainty.

\begin{figure}[h]
\includegraphics[width=\hsize]{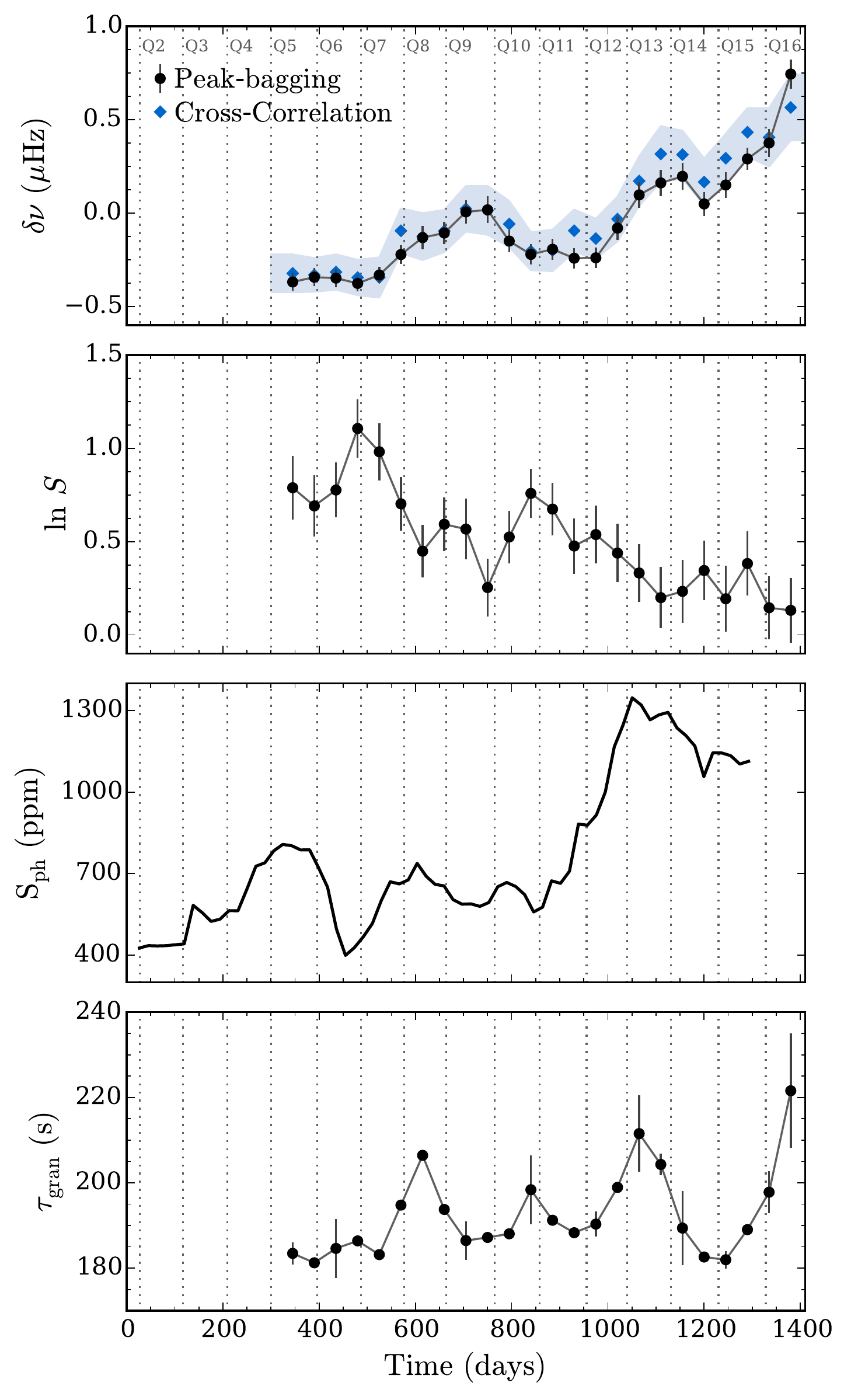}
\caption{Results for the solar-type star KIC 8006161. {\it Top panel:} Comparison between the mean frequency shifts obtained with the Bayesian peak-bagging tool (black) and the cross-correlation method \citep[blue; method described in][]{Kiefer2017}. {\it Second panel:} Logarithmic mode height obtained from the peak-bagging analysis. {\it Third panel:} Photometric magnetic activity proxy. {\it Bottom panel:} Characteristic timescale of the granulation. Vertical dotted lines mark the start/end of {\it Kepler} quarters.}\label{fig:fshKIC8006161}
\end{figure}

The top panel of Figure~\ref{fig:fshKIC8006161} compares the frequency shifts obtained with the Bayesian peak-bagging tool with those from the cross-correlation method for the solar-type star KIC~8006161. The frequency shifts presented here are obtained while considering only the five central orders (closest to $\nu_\text{max}$), which usually have the largest signal-to-noise ratio in the p modes.
Note that the frequency shifts derived from our method are relative to the average value (see Section~\ref{sec:fshifts}), while the frequency shifts from the cross-correlation method are estimated in relation to the first sub-series. Therefore, the latter are displaced by their average value in Figure~\ref{fig:fshKIC8006161}. 
First, this comparison shows that the results obtained with both methods agree very well. Second, the uncertainties on the estimated frequency shifts are smaller by a factor of a few when employing the peak-bagging method\footnote{For the common stars to both studies, the uncertainties on the frequency shifts obtained through the cross-correlation method presented here are larger than those in \citet[][]{Kiefer2017} due to the length of the sub-series, which is shorter for this work.\vspace{-0.3cm}}. This reassures us that our peak-bagging tool is able to successfully recover accurate mode frequencies and, consequently, their temporal variations. Table~\ref{tab:fshifts8006161} lists the results for KIC 8006161.

\begin{table*}\centering
\fontsize{9}{9.}\selectfont
\begin{tabular}{ccc|ccccc|c}
\multicolumn{9}{c}{KIC 8006161}\\\hline\hline
time&duty&$\tau_\text{gran}$&\multicolumn{5}{c|}{Peak-bagging}&Cross-correlation\\
(d)&cycle& (s)& $\delta\nu_{l=0}$ ($\mu$Hz) & $\delta\nu_{l=1}$ ($\mu$Hz) & $\delta\nu_{l=2}$ ($\mu$Hz) & $\delta\nu$ ($\mu$Hz)& $\ln\,S$ & $\delta\nu\,(\mu$Hz)\\\hline
345 & 0.97 & 183.4 $\pm$ 2.6 & -0.38 $\pm$ 0.06 & -0.35 $\pm$ 0.07 & -0.32 $\pm$ 0.13 & -0.37 $\pm$ 0.05 & 0.79 $\pm$ 0.17 & -0.32 $\pm$ 0.10\\
390 & 0.97 & 181.2 $\pm$ 0.3 & -0.26 $\pm$ 0.07 & -0.45 $\pm$ 0.07 & -0.44 $\pm$ 0.15 & -0.34 $\pm$ 0.05 & 0.69 $\pm$ 0.17 & -0.33 $\pm$ 0.09\\
435 & 0.96 & 184.6 $\pm$ 6.9 & -0.32 $\pm$ 0.06 & -0.39 $\pm$ 0.08 & -0.45 $\pm$ 0.14 & -0.35 $\pm$ 0.05 & 0.78 $\pm$ 0.15 & -0.32 $\pm$ 0.09\\
480 & 0.97 & 186.4 $\pm$ 0.2 & -0.35 $\pm$ 0.05 & -0.43 $\pm$ 0.07 & -0.30 $\pm$ 0.12 & -0.38 $\pm$ 0.04 & 1.11 $\pm$ 0.16 & -0.35 $\pm$ 0.09\\
525 & 0.96 & 183.2 $\pm$ 0.6 & -0.28 $\pm$ 0.05 & -0.44 $\pm$ 0.08 & -0.36 $\pm$ 0.12 & -0.33 $\pm$ 0.04 & 0.98 $\pm$ 0.15 & -0.34 $\pm$ 0.11\\
570 & 0.80 & 194.8 $\pm$ 0.6 & -0.22 $\pm$ 0.06 & -0.23 $\pm$ 0.09 & -0.17 $\pm$ 0.15 & -0.22 $\pm$ 0.05 & 0.70 $\pm$ 0.14 & -0.10 $\pm$ 0.12\\
615 & 0.78 & 206.4 $\pm$ 0.4 & -0.10 $\pm$ 0.08 & -0.17 $\pm$ 0.09 & -0.13 $\pm$ 0.18 & -0.13 $\pm$ 0.06 & 0.45 $\pm$ 0.14 & -0.13 $\pm$ 0.13\\
660 & 0.89 & 193.7 $\pm$ 0.0 & -0.03 $\pm$ 0.08 & -0.21 $\pm$ 0.09 & -0.14 $\pm$ 0.17 & -0.11 $\pm$ 0.06 & 0.59 $\pm$ 0.14 & -0.10 $\pm$ 0.11\\
705 & 0.93 & 186.5 $\pm$ 4.5 & 0.12 $\pm$ 0.09 & -0.10 $\pm$ 0.09 & -0.12 $\pm$ 0.17 & 0.01 $\pm$ 0.06 & 0.57 $\pm$ 0.16 & 0.02 $\pm$ 0.12\\
750 & 0.97 & 187.2 $\pm$ 0.2 & 0.16 $\pm$ 0.10 & -0.13 $\pm$ 0.10 & -0.18 $\pm$ 0.14 & 0.02 $\pm$ 0.07 & 0.25 $\pm$ 0.16 & 0.01 $\pm$ 0.13\\
795 & 0.97 & 188.0 $\pm$ 0.3 & -0.16 $\pm$ 0.08 & -0.14 $\pm$ 0.09 & -0.35 $\pm$ 0.15 & -0.15 $\pm$ 0.06 & 0.52 $\pm$ 0.14 & -0.06 $\pm$ 0.12\\
840 & 0.98 & 198.4 $\pm$ 8.1 & -0.12 $\pm$ 0.07 & -0.40 $\pm$ 0.09 & -0.57 $\pm$ 0.15 & -0.22 $\pm$ 0.05 & 0.76 $\pm$ 0.13 & -0.20 $\pm$ 0.10\\
885 & 0.94 & 191.2 $\pm$ 0.3 & -0.18 $\pm$ 0.07 & -0.22 $\pm$ 0.09 & -0.69 $\pm$ 0.13 & -0.19 $\pm$ 0.06 & 0.67 $\pm$ 0.14 & -0.20 $\pm$ 0.11\\
930 & 0.92 & 188.3 $\pm$ 0.8 & -0.33 $\pm$ 0.07 & -0.10 $\pm$ 0.09 & -0.17 $\pm$ 0.18 & -0.24 $\pm$ 0.06 & 0.48 $\pm$ 0.15 & -0.09 $\pm$ 0.11\\
975 & 0.90 & 190.3 $\pm$ 2.9 & -0.27 $\pm$ 0.07 & -0.20 $\pm$ 0.09 & -0.05 $\pm$ 0.14 & -0.24 $\pm$ 0.06 & 0.54 $\pm$ 0.16 & -0.14 $\pm$ 0.11\\
1020 & 0.90 & 198.9 $\pm$ 0.2 & -0.09 $\pm$ 0.09 & -0.07 $\pm$ 0.09 & 0.00 $\pm$ 0.15 & -0.08 $\pm$ 0.06 & 0.44 $\pm$ 0.16 & -0.03 $\pm$ 0.12\\
1065 & 0.95 & 211.5 $\pm$ 8.9 & 0.06 $\pm$ 0.09 & 0.15 $\pm$ 0.11 & -0.30 $\pm$ 0.17 & 0.10 $\pm$ 0.07 & 0.33 $\pm$ 0.15 & 0.17 $\pm$ 0.13\\
1110 & 0.89 & 204.3 $\pm$ 2.6 & 0.12 $\pm$ 0.09 & 0.23 $\pm$ 0.11 & -0.01 $\pm$ 0.22 & 0.16 $\pm$ 0.07 & 0.20 $\pm$ 0.16 & 0.32 $\pm$ 0.15\\
1155 & 0.89 & 189.4 $\pm$ 8.7 & 0.12 $\pm$ 0.11 & 0.25 $\pm$ 0.09 & 0.39 $\pm$ 0.18 & 0.20 $\pm$ 0.07 & 0.23 $\pm$ 0.17 & 0.31 $\pm$ 0.13\\
1200 & 0.95 & 182.6 $\pm$ 0.9 & -0.03 $\pm$ 0.09 & 0.14 $\pm$ 0.10 & 0.21 $\pm$ 0.14 & 0.05 $\pm$ 0.06 & 0.35 $\pm$ 0.16 & 0.17 $\pm$ 0.12\\
1245 & 0.90 & 182.0 $\pm$ 2.0 & 0.09 $\pm$ 0.09 & 0.22 $\pm$ 0.10 & 0.38 $\pm$ 0.17 & 0.15 $\pm$ 0.07 & 0.19 $\pm$ 0.18 & 0.29 $\pm$ 0.13\\
1290 & 0.89 & 189.0 $\pm$ 0.0 & 0.13 $\pm$ 0.08 & 0.51 $\pm$ 0.09 & 0.49 $\pm$ 0.20 & 0.29 $\pm$ 0.06 & 0.38 $\pm$ 0.17 & 0.43 $\pm$ 0.13\\
1335 & 0.84 & 197.8 $\pm$ 4.9 & 0.21 $\pm$ 0.10 & 0.59 $\pm$ 0.11 & 0.47 $\pm$ 0.23 & 0.38 $\pm$ 0.07 & 0.15 $\pm$ 0.17 & 0.41 $\pm$ 0.16\\
1380 & 0.84 & 221.6 $\pm$ 13.4 & 0.78 $\pm$ 0.09 & 0.65 $\pm$ 0.15 & 0.80 $\pm$ 0.18 & 0.74 $\pm$ 0.08 & 0.13 $\pm$ 0.17 & 0.57 $\pm$ 0.17\\
\end{tabular}
\caption{Results for the solar-type star KIC 8006161. {\it First column}: Time of the sub-series midpoint relative to the starting time of the observations. {\it Second column}: Duty-cycle for each sub-series. {\it Third column}: Characteristic timescale of the granulation component. {\it Fourth to Eighth Columns}: Mean frequency shifts (for radial ($\delta\nu_{l=0}$), dipolar ($\delta\nu_{l=1}$), and quadrupolar ($\delta\nu_{l=2}$) modes, and when combining the $l=0$ and $l=1$ modes ($\delta\nu$)) and logarithmic mode heights obtained from the Bayesian peak-bagging analysis. Note that the results presented here are based on the five central orders (see text), in particular, for KIC~8006161, we use modes of radial order ranging between 19 and 23. {\it Ninth column}: Frequency shifts obtained with the cross-correlation method described in \citet{Kiefer2017}.}\label{tab:fshifts8006161}
\end{table*}

In the Sun, the amplitude of the acoustic modes is also observed to vary over the solar cycle \citep[e.g.][]{Elsworth1993,Chaplin2000,Salabert2006,Howe2015}, decreasing with increasing magnetic activity. Thus, the acoustic frequencies and amplitudes show an anti-correlated temporal variation over the solar cycle.

Assuming that the underlying magnetically induced changes in other solar-type stars are similar to those in the Sun, evidence of an anti-correlated behavior between the acoustic frequencies and heights may be an important aspect to confirm the activity-related origin of the observed variations. With this in mind, we also searched for temporal variations in the mode heights. However, we note that this seismic indicator is not expected to be as robust as the frequency shifts.

The second panel of Figure~\ref{fig:fshKIC8006161} shows the temporal evolution of the weighted average of the logarithmic mode heights estimated with the peak-bagging tool for KIC~8006161 (see Section~\ref{sec:fshifts}), while the third panel of Figure~\ref{fig:fshKIC8006161} shows the photometric magnetic activity proxy, ${\rm S}_{\rm ph}$ (see Section~\ref{sec:sph}). These results suggest that the acoustic mode frequencies of KIC~8006161 (top panel) increase with increasing activity level, while the mode heights experience a decrease, which resembles what is observed for the Sun. Also, the behavior of the mode heights is in good agreement with that found for the height of the p-mode envelope by \citet{Kiefer2017}.

\begin{figure}[h]
\includegraphics[width=\hsize]{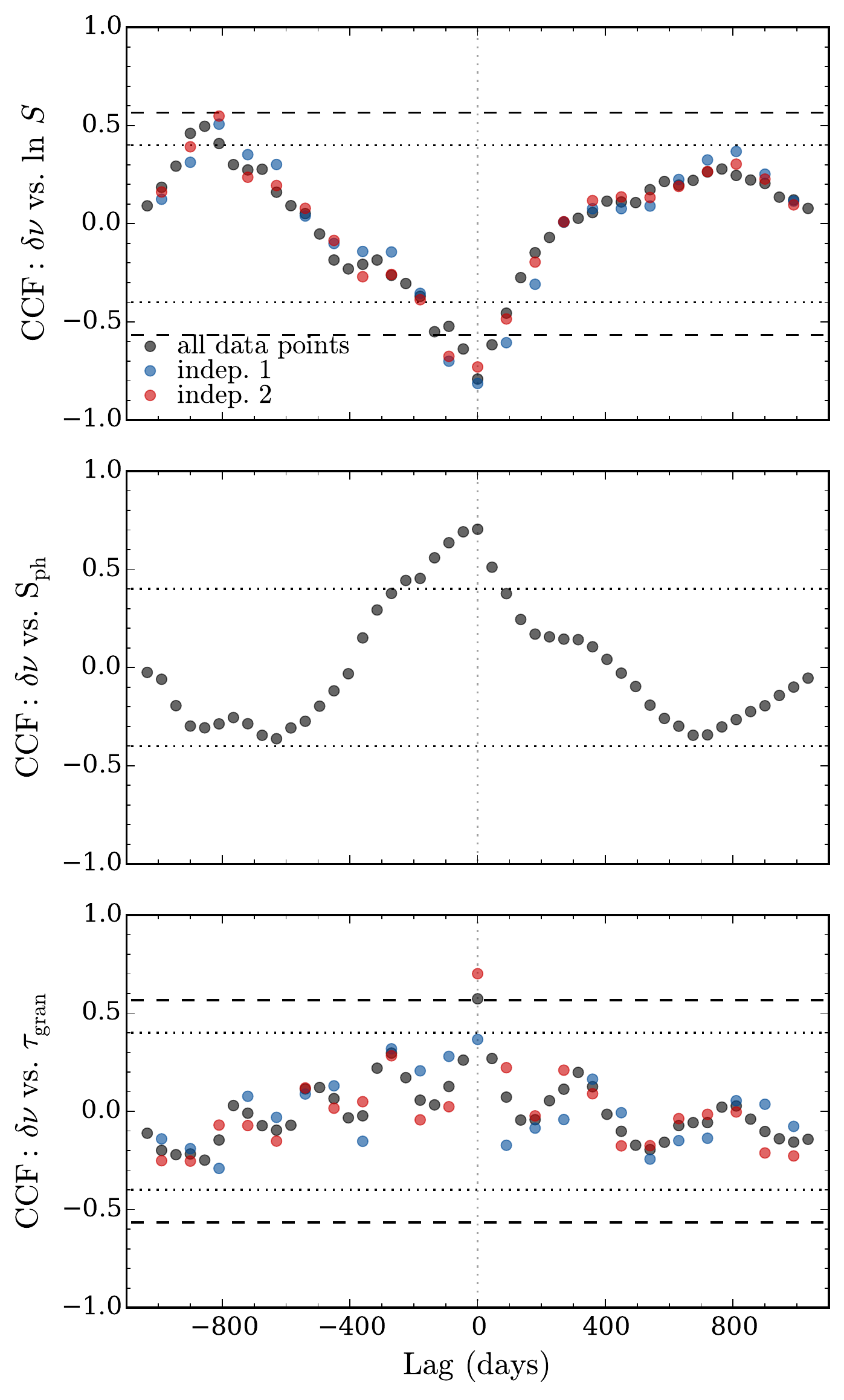}
\caption{CCF results for KIC~8006161: Cross-correlation function between frequency shifts and: logarithmic mode heights ({\it top panel}); interpolated photometric activity proxy ({\it middle panel}); characteristic granulation timescale ({\it bottom panel}). Black symbols correspond to the results for all the data points, while the blue and red symbols show the results for the independent data points. The horizontal dotted and dashed lines mark the $95\%$ significance levels for all and the independent data points, respectively. The dotted vertical lines indicate the 0-Lag.}\label{fig:ccfKIC8006161_hsph}
\end{figure}

In order to access the correlation between the different quantities and to account for possible temporal offsets, we compute the cross-correlation function (CCF) between the frequency shifts and the remainder quantities. As an example, the results for the cross-correlation function for KIC~8006161 are shown in Figure~\ref{fig:ccfKIC8006161_hsph}.
The top panel of Figure~\ref{fig:ccfKIC8006161_hsph} shows the cross-correlation function between the mean frequency shifts and the mean logarithmic mode heights. The black symbols concern the results obtained while considering all the data points, i.e. all the 90-d sub-series. The blue and red symbols concern the two sub-samples of independent data points, i.e. while considering every two sub-series of 90 d. For reference, the $95\%$ significance levels are marked. The results show that the anti-correlation between frequency shifts and mode heights is significant (absolute value above the $95\%$ significance level). The maximum anti-correlation is found at 0-Lag.
The middle panel of Figure~\ref{fig:ccfKIC8006161_hsph} shows the cross-correlation function between the frequency shifts and the interpolated photometric activity proxy (interpolated in the same times as the $\delta\nu$). The results suggest a strong correlation between the temporal frequency shifts and the photometric activity and the maximum correlation is found at 0-Lag. Table~3 summarizes the results from the cross-correlation between the frequency shifts and the remainder quantities for all stars in the sample. For simplification only the results for all the data points are listed. Also, in the table, we consider temporal lags between $\pm90$ days and we list the maximum absolute correlation within those temporal lags and the corresponding lag.%\ref{tab:CCFKIC}

\subsection{On the relation between frequency shifts and the granulation timescale}\label{sec:gran}

Although it is still a matter of debate in the literature, the properties of the solar granulation, namely amplitude and characteristic timescale, may be affected by magnetic activity. \citet{Muller2007} found evidence for a decrease in the granulation amplitude with increasing activity. However, other studies found no significant variation in this parameter \citep{Palle1995,Regulo2002,Regulo2005,Lefebvre2008,Karoff2012}. Regarding the granulation characteristic timescale, while \citet{Lefebvre2008} claim no significant variation, \citet{Regulo2002,Regulo2005} found that the granulation timescale increases with increasing activity. Moreover, although \citet{Lefebvre2008} and \citet{Karoff2012} found that the granulation properties are mostly independent of the 11-yr solar cycle, these authors found evidence for shorter quasi-periodic ($\sim$1-yr) variations, whose origin is still not understood. \citet{Lefebvre2008}, however, related those variations to instrumental effects.

For stars other than the Sun, the relation between magnetic activity and the granulation properties is also not clear. \citet{Garcia2010} found no correlation between the observed activity-related frequency shifts and the granulation timescale for the solar-type star HD~49933 observed by CoRoT. \citet{Karoff2013} analyzed the variability of the granulation component on three solar-type stars observed by {\it Kepler}. For two of these stars (KIC 6603624 and KIC 6933899, both in our sample), the authors found quasi-annual/biennial periodicities on the granulation parameters, that resemble the quasi-biennial variations in the different solar activity indicators \citep{Bazilevskaya2014,McIntosh2015,Salabert2017}, including frequency shifts \citep{Fletcher2010,Broomhall2012,Broomhall2015}. Finally, \citet{Kiefer2017} did not find evidence for a systematic correlation between the granulation timescale and the frequency shifts for their sample of 24 solar-type stars.

With the above in mind, we have been searching for correlations between the frequency shifts obtained using the Bayesian peak-bagging tool and the granulation characteristic timescale obtained in Section~\ref{sec:backg}. 

The bottom panel of Figure~\ref{fig:fshKIC8006161} shows the characteristic timescale of the granulation, $\tau_{\rm gran}$, over time for KIC~8006161. For this star the granulation timescale and the frequency shifts do not appear to have a correlated behavior. However, some of the variations seen in the granulation timescale may be related with local variations in the acoustic frequencies (top panel).

The bottom panel of Figure~\ref{fig:ccfKIC8006161_hsph} shows the cross-correlation between the mean frequency shifts and the granulation timescale as a function of the lag between the two observables. These results indicate that there is a significant correlation (above the $95\%$ significance level) between frequency shifts and granulation timescale in the case of all data points (black) and for one sample of independent data points (shown in red). The maximum correlation is found at 0-Lag.

From the 87 solar-type stars analyzed in this work, 31 stars exhibit a significant (above the $95\%$ significance level) correlation or anti-correlation between the observed frequency shifts and the granulation timescale. 
In the next subsection, we highlight the results for five additional stars (besides KIC~8006161).\vfill

\subsection{Results for the whole sample}\label{sec:extrastars} % 

The results from our analysis indicate that activity-related variations in the acoustic frequencies are likely a common phenomenon among solar-type stars. We find evidence for quasi-periodic variations in about $60\%$ of the targets. Among those, more than $70\%$ show variations in other activity indicators: in the logarithmic mode heights, in the photometric activity proxy, and/or in the granulation timescale. The detailed results for the complete target sample are presented in Appendix~B.%\ref{sec:peakapp}.

We note that for some of the analyzed targets we find (quasi-)periodic variations in the frequency shifts and/or in the background parameters with a period close to {\it Kepler}'s orbital period (i.e. close to 372.5~d). For those, one must be careful in the interpretation of the results. Similar behavior is found with different calibrated data: KASOC \citep{Handberg2014}; KADACS \citep{Garcia2011}; and PDC-MAP \citep[Presearch Data Conditioning - Maximum A Posteriori;][]{Stumpe2012,Smith2012}. Therefore, those periodicities should not result from an artifact introduced during the preparation of the light-curves. Also, they are found with both peak-bagging and cross-correlation methods. From those cases, we highlight the solar-type stars KIC~9139163 and KIC~6933899. The results for KIC~9139163 show periodic variations in the frequency shifts and characteristic granulation timescale with a period consistent with {\it Kepler}'s orbital period. For KIC~6933899, we find periodic variations with similar period in the granulation timescale, granulation amplitude, and in the frequency shifts. The frequency shifts are ahead in time in relation to the granulation parameters. The granulation component of this star was previously analyzed by \citet{Karoff2013}\footnote{\citet{Karoff2013} also found a periodic variability on the granulation component of KIC~6603624 with a period of $\sim322$~d. In this work, based on a longer time-series, we find the granulation parameters varying with a shorter periodicity of $\sim225$~d.}. Based on 13-month times-series, the authors found a temporal variation in  the granulation timescale with a period of $\sim233$~d. Here, based on a time-series three times longer, we also find a temporal variation in the granulation parameters, but with a different periodicity consistent with {\it Kepler}'s orbital period. Although less clear than the highlighted cases, other three stars (KIC~1435465, KIC~8077137, and KIC~8179536) also show evidence for variations in the acoustic frequencies with the same periodicity. If we neglect these cases, the previously indicated percentage of stars with significant frequency shifts does not change, as we used round numbers.

In what follows, we present the results for five additional solar-type stars (besides KIC~8006161) that show evidence for activity-related frequency shifts.

\subsubsection{KIC 5184732}

\begin{figure}[h]
\includegraphics[width=\hsize]{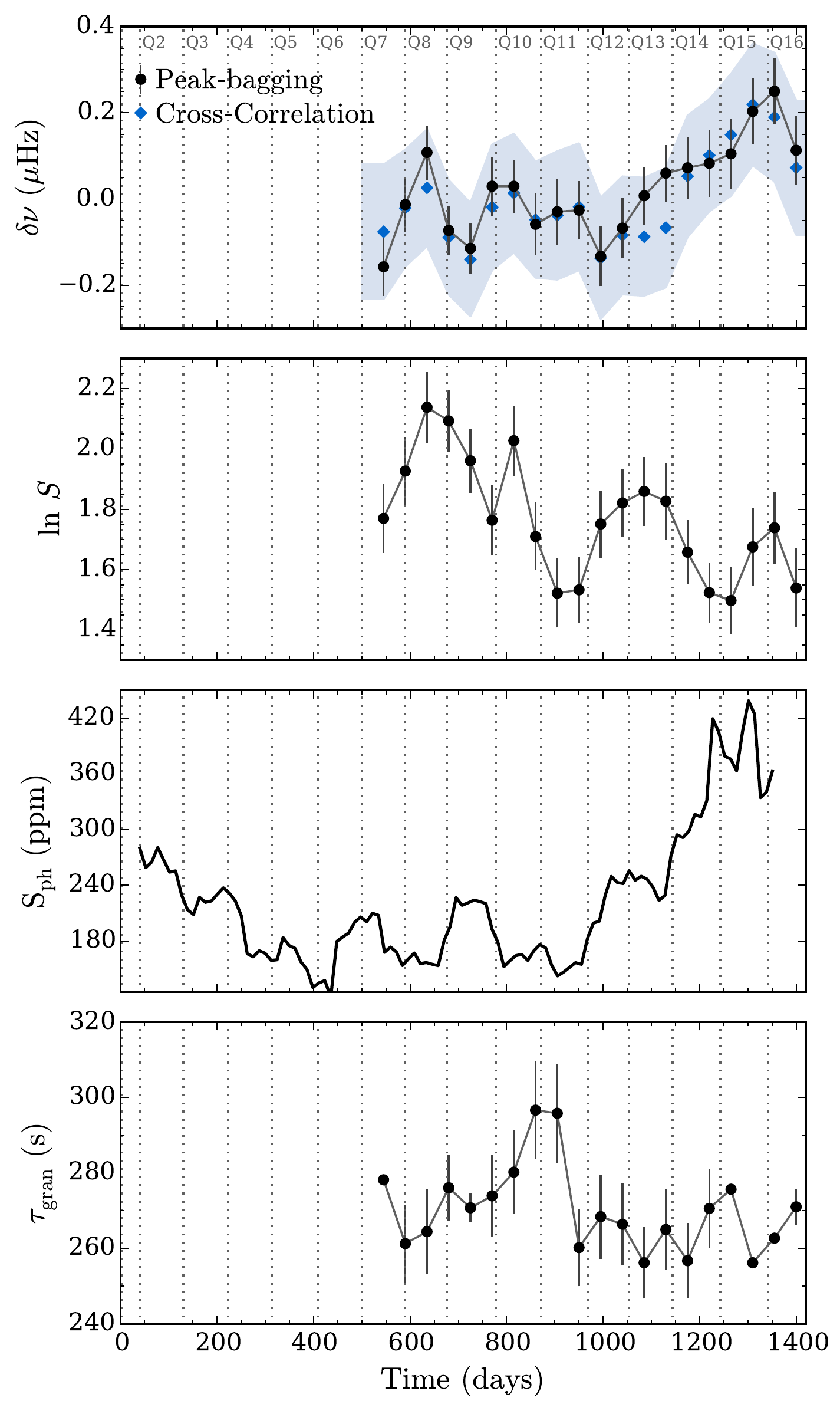}\vspace{-0.2cm}
\caption{Results for KIC 5184732 (same as in Figure~\ref{fig:fshKIC8006161}): mean frequency shifts, mode heights, photometric activity proxy, and characteristic granulation timescale. Results listed in Table~\ref{tab:5184732}.}\label{fig:KIC5184732}
\end{figure}

Figure~\ref{fig:KIC5184732} presents the results for KIC~5184732. In general, the mean frequency shifts (top panel) increase over time, while the mean mode heights (second panel) experience a decrease. This is consistent with the two quantities being anti-correlated in time, similar to the Sun and KIC~8006161. The results from the CCF (Table~3) confirm the anti-correlation between the mean frequency shifts and mode heights. Furthermore, the variation in the photometric activity proxy ($\text{S}_\text{ph}$; third panel of Figure~\ref{fig:KIC5184732}) is consistent with the variation in  the frequency shifts. In fact, these two quantities are strongly correlated (Table~3). The full $\text{S}_\text{ph}$ time-series suggests a cyclic behavior in the surface magnetism, showing what resembles the end of a cycle and the beginning of new cycle. %\ref{tab:CCFKIC)

For the characteristic granulation timescale, our results (bottom panel of Figure~\ref{fig:KIC5184732} and Table~3) suggest that there is no significant temporal variation and correlation with the frequency shifts.%\ref{tab:CCFKIC)

Finally, the results shown in the two top panels of Figure~\ref{fig:KIC5184732} are consistent with those in \citet{Kiefer2017}.

\subsubsection{KIC 9414417 (KOI 974)}

\begin{figure}[h]
\includegraphics[width=\hsize]{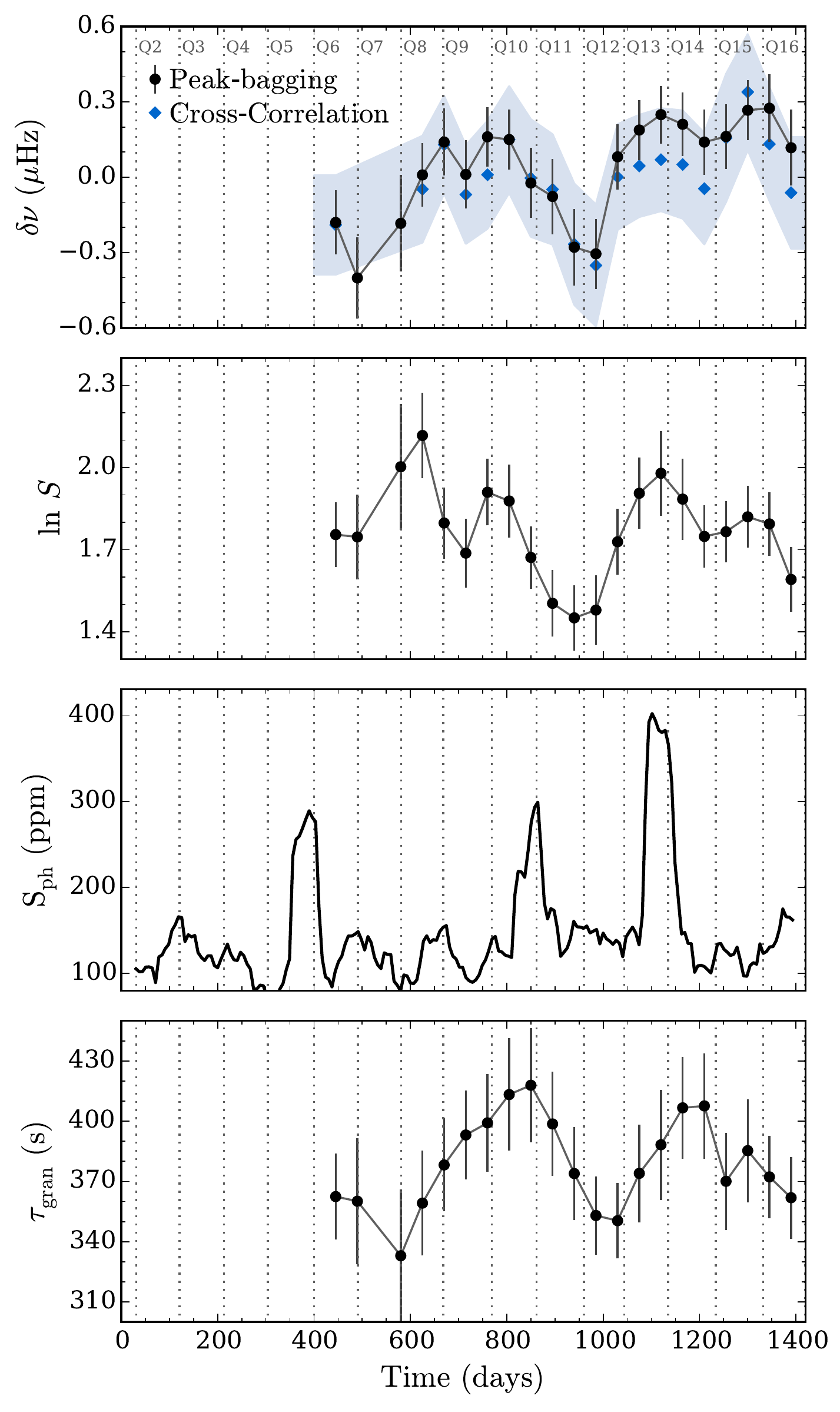}
\caption{Results for KIC 9414417 (same as in Figure~\ref{fig:fshKIC8006161}): mean frequency shifts, mode heights, photometric activity proxy, and characteristic granulation timescale. Results listed in Table~\ref{tab:9414417}.}\label{fig:KIC9414417}
\end{figure}

Figure~\ref{fig:KIC9414417} presents the results for KIC~9414417. We find quasi-periodic variations in the frequencies (top panel) and logarithmic heights (second panel) of the acoustic modes. However, for KIC~9414417, the mode frequencies and heights vary in phase, while in the Sun, these two properties vary in anti-phase over the solar activity cycle. The results from the CCF show that the correlation between frequency shift and mode heights is significant (Table~3).%\ref{tab:CCFKIC)

Interestingly, the characteristic timescale of the granulation (bottom panel of Figure~\ref{fig:KIC9414417}) also shows a quasi-periodic modulation with a temporal offset in relation to the variation in the seismic properties. The granulation timescale and frequency shifts are strongly correlated (Table~3).%\ref{tab:CCFKIC)

Knowing that the inclination angle of KIC~9414417 is close to $90^\circ$ \citep[e.g.][]{Campante2016} and assuming a spot latitudinal distribution similar to that in the Sun, the photometric activity proxy (third panel of Figure~\ref{fig:KIC9414417}) thus suggests a relatively weak photometric activity for this star. Still, there are two epochs of large $\rm S_{\rm ph}$ which coincide with epochs of large frequency shifts.\vfill

\subsubsection{KIC 10644253}

\citet{Salabert2016} found evidence for activity-related frequency shifts in KIC 10644253, which vary in agreement with the photometric activity proxy, $\rm S_{\rm ph}$. Later, \citet{Kiefer2017} confirmed the temporal variation in the acoustic frequencies of this star.

\begin{figure}
\includegraphics[width=\hsize]{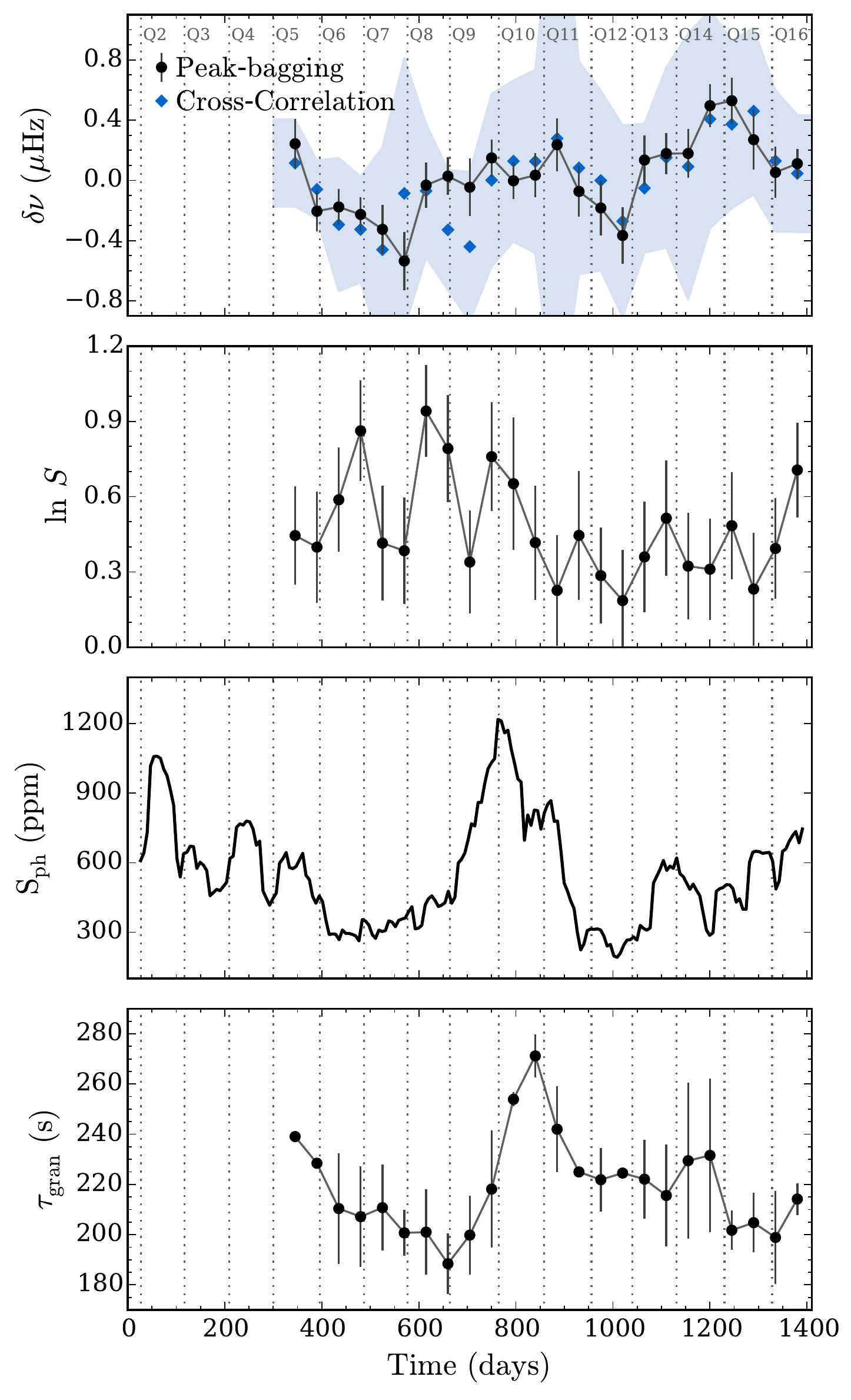}
\caption{Results for KIC 10644253 (same as in Figure~\ref{fig:fshKIC8006161}): mean frequency shifts, mode heights, photometric activity proxy, and characteristic granulation timescale. Results listed in Table~73.}\label{fig:KIC10644253}%\ref{tab:10644253}
\end{figure}

The solar-type star KIC 10644253 is also part of the target sample analyzed in this work. The black symbols in the two top panels of Figure~\ref{fig:KIC10644253} show the frequency shifts and logarithmic mode heights derived from the peak-bagging analysis (Section~\ref{sec:meth}). For comparison, the frequency shifts computed with the cross-correlation method, described in \citet{Kiefer2017}, are shown in blue in the top panel of Figure~\ref{fig:KIC10644253}. In spite of the large errors associated with the cross-correlation method, the frequency shifts obtained with the two methods are still in good agreement. Moreover, the results from the peak-bagging analysis are in good agreement with those in the literature \citep{Salabert2016,Kiefer2017}. Therefore, our analysis further confirms the temporal frequency shifts in KIC~10644253, which show a similar behavior to that of the photometric activity proxy (third panel of Figure~\ref{fig:KIC10644253}). Also, the CCF results show that frequency shifts and $S_\text{ph}$ are correlated. However, the correlation is below the $95\%$ significance level.

For the mode heights (second panel of Figure~\ref{fig:KIC10644253}) we do not find a significant variation. However, from the cross-correlation function (Table~3) there is some evidence for an anti-correlation between the frequency shifts and the mode heights.%\ref{tab:CCFKIC)

The bottom panel of Figure~\ref{fig:KIC10644253} shows the granulation timescale (estimated as described in Section~\ref{sec:backg}). For this star in particular, the variation in the acoustic frequencies and in the photometric proxy is similar to that  in the characteristic timescale of the granulation. In spite of the relatively low significance, the cross-correlation function between the frequency shifts and the granulation timescale (Table~3) suggests that the variations in these two properties are correlated.%\ref{tab:CCFKIC)

\subsubsection{KIC 7970740}

\begin{figure}[h]
\includegraphics[width=\hsize]{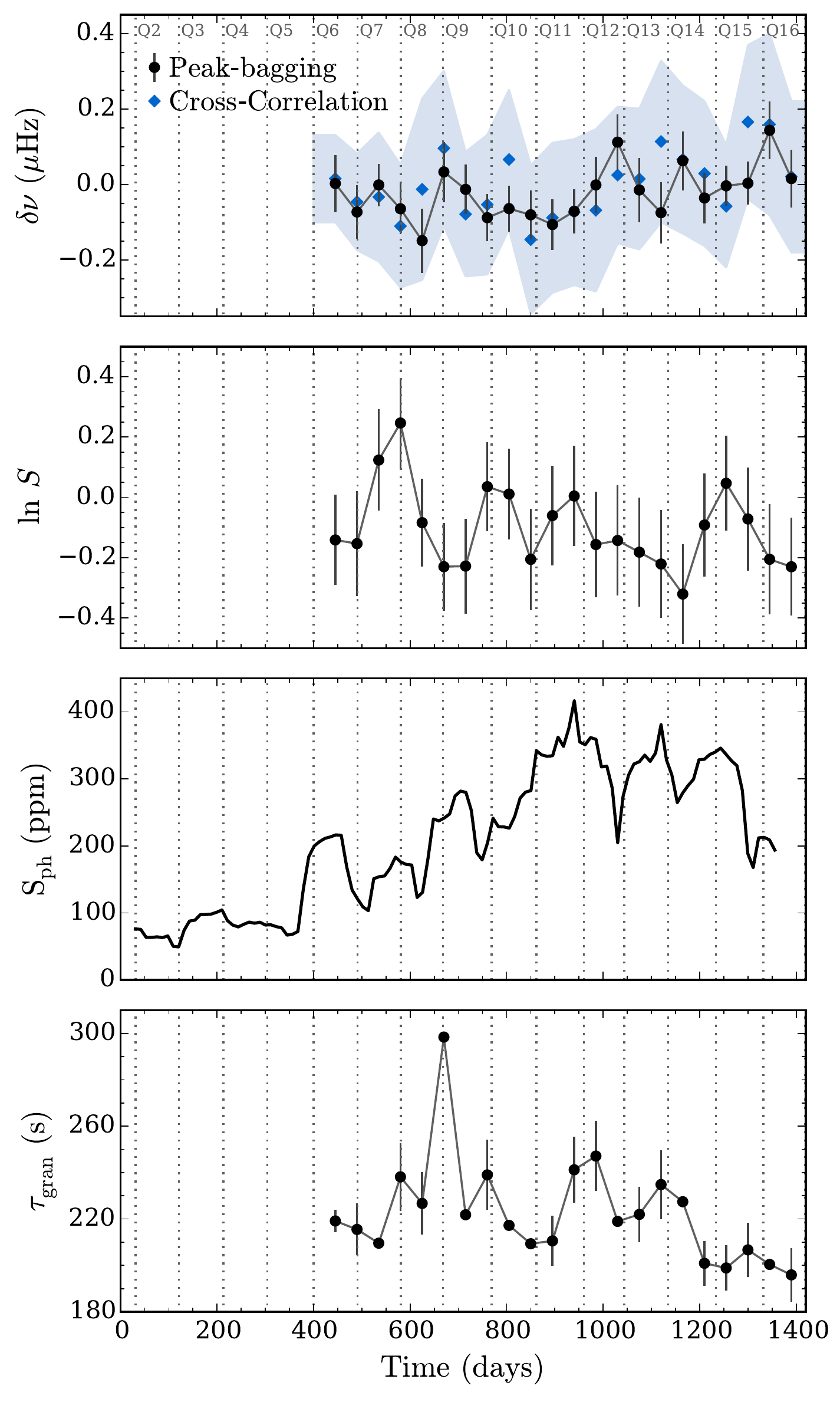}
\caption{Results for KIC~7970740 (same as in Figure~\ref{fig:fshKIC8006161}): mean frequency shifts, mode heights, photometric activity proxy, and characteristic granulation timescale. Results listed in Table~\ref{tab:7970740}.}\label{fig:KIC7970740}
\end{figure}

Figure~\ref{fig:KIC7970740} displays the results for the solar-type star KIC~7970740. The frequency shifts obtained with the peak-bagging analysis (black symbols in the top panel) increase slightly over time and agree reasonably well with those from the cross-correlation method (blue symbols). The mode heights (second panel) do not show a significant variation over time, but their overall behavior suggests a decrease. This is confirmed by the results from the CCF (Table~3), which, in spite the low significance, suggest an anti-correlation between frequency shifts and mode heights. The photometric activity proxy (third panel of Figure~\ref{fig:KIC7970740}) shows an increase on the magnetic activity, consistent with the rising phase of a cycle and consistent with the variation in the frequency shifts. The CCF results (Table~3) indicates that the frequency shifts and $\text{S}_\text{ph}$ are correlated.

The variation in the characteristic timescale of the granulation is not consistent with the variation in the frequency shifts (bottom panel of Figure~\ref{fig:KIC7970740}).

\subsubsection{KIC 8379927}

\begin{figure}[h]
\includegraphics[width=\hsize]{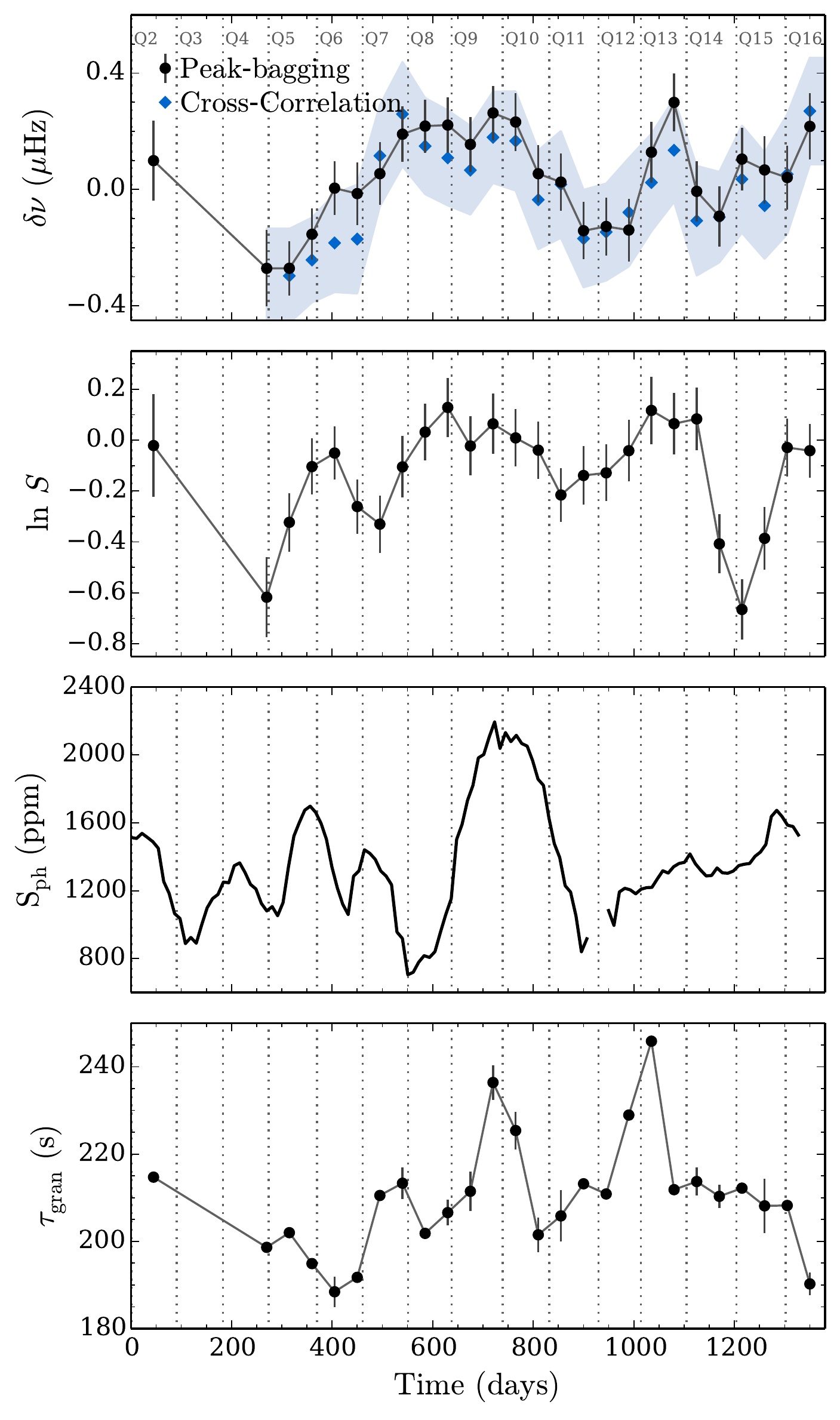}
\caption{Results for KIC 8379927 (same as in Figure~\ref{fig:fshKIC8006161}): mean frequency shifts, mode heights, photometric activity proxy, and characteristic granulation timescale. Results listed in Table~\ref{tab:8379927}.}\label{fig:KIC8379927}
\end{figure}

The black symbols in the top panel of Figure~\ref{fig:KIC8379927} show the mean frequency shifts for KIC~8379927 obtained from the peak-bagging analysis, which are in good agreement with those from the cross-correlation method (blue). The logarithmic mode heights are shown in the second panel of Figure~\ref{fig:KIC8379927} and vary approximately in phase with the frequency shifts (opposite behavior to the solar case). This is confirmed by the cross-correlation function (Table~3), which shows that the frequency shifts and mode heights are correlated. The photometric activity proxy (third panel of Figure~\ref{fig:KIC8379927}) suggests a relatively strong photometric activity for this star. The variation in the $\rm S_{\rm ph}$ index is consistent with that in the frequency shifts. 
However, the correlation between the frequency shifts and $S_\text{ph}$ is below the $95\%$ significance level (Table~3). The granulation timescale for KIC~8379927 is shown in the bottom panel of Figure~\ref{fig:KIC8379927}. The CCF results suggest a correlation between frequency shifts and granulation timescale.

KIC~8379927 is a known spectroscopic binary and, thus, one must be cautious before drawing further conclusions. Nevertheless, both frequency shifts and photometric activity proxy are observed to vary over time, and it would be interesting to understand if there is any connection between those variations and the properties of the binary.

\section{Summary and Discussion}\label{sec:conclusions}

In this work, we analyzed a large sample of solar-type stars observed by {\it Kepler} in order to search for temporal variations in the frequencies of the acoustic modes, which could possibly be related to magnetic activity.

The original data-sets were split into 90-d sub-series (with $50\%$ overlap) for which we modelled the power density spectra. In order to obtain accurate individual mode parameters, we developed a Bayesian peak-bagging tool. With the individual mode frequencies in hand, we computed the mean frequency shifts for each star.

The method was first validated with observational and artificial solar data and then applied to the {\it Kepler} targets. To further confirm the efficiency of our peak-bagging tool, we compared the mean frequency shifts with those obtained with the cross-correlation method described in \citet{Kiefer2017}. The results from the peak-bagging analysis and from the cross-correlation method are generally in good agreement. In addition, the error bars associated with the Bayesian peak-bagging analysis are smaller than those associated with the cross-correlation method. 

In order to connect the variation in the acoustic frequencies to an activity-related origin, the mean frequency shifts were compared with the mean logarithmic mode height, with the granulation timescale, and with the photometric activity proxy, $\text{S}_\text{ph}$. 

More than half of the target sample show evidences for periodic variations in the acoustic frequencies. For some of those stars, we also found consistent variations in the other parameters. In particular, we highlighted the results for five stars: KIC~8006161, KIC~5184732, KIC~9414417, KIC~10644253, KIC~7970740, and KIC~8379927. 

For KIC~8006161, KIC~5184732, and KIC~7970740, the variations in the acoustic frequencies are consistent with the $\text{S}_\text{ph}$ index and anti-correlated with the variations in the mode heights. These results resemble what is observed for the Sun: the frequency shifts vary in phase with the activity level, while the mode heights vary in anti-phase. Although the seismic properties for KIC~8006161 show a similar behavior to the solar parameters, its magnetic activity cycle is significantly stronger than the solar cycle \citep[for a detailed study of KIC~8006161, see][]{Karoff2018}.

For KIC~10644253, the temporal frequency shifts are accompanied by variations in the $\text{S}_\text{ph}$ index and in the granulation characteristic timescale.

For KIC~9414417 and KIC~8379927 (known spectroscopic binary), the frequency shifts are consistent, being approximately in phase, with the variation in the photometric proxy and in the mode heights. The relation between the frequency shifts and mode heights is opposite to the expected behavior, based on solar observations. This behavior is found in $\sim20\%$ of the analyzed targets.

In Appendix B, we present the results for the remainder of the {\it Kepler} stars in the sample.%\ref{sec:peakapp}

Taking advantage of high-quality long-term photometric time-series as those obtained by {\it Kepler}, the analysis and results of this large-scale work confirm the possibility of using asteroseismology to study stellar magnetism. A follow-up work (Santos et al. in prep) is being prepared, where, for the ensemble of solar-type stars (analyzed in the present work), we study the relation between the observed frequency shifts and the stellar fundamental properties, such as age, effective temperature, and surface rotation.

\acknowledgments

The authors thank D. Bossini for providing the evolutionary tracks shown in this paper.
This work was supported by Funda\c{c}\~{a}o para a Ci\^{e}ncia e a Tecnologia (FCT) through national funds (UID/FIS/04434/2013) and by FEDER through COMPETE2020 (POCI-01-0145-FEDER-007672). ARGS acknowledges the support by the fellowship SFRH/BD/88032/2012 funded by FCT (Portugal) and POPH/FSE (EC), from the University of Birmingham, and from NASA grant NNX17AF27G. TLC acknowledges support from grant CIAAUP-12/2018-BPD. TLC, WJC, GRD, EY, and RH acknowledge the support of the UK Science and Technology Facilities Council (STFC). MSC acknowledges the support from FCT through the Investigador FCT Contract No. IF/00894/2012. MNL acknowledges the support of The Danish Council for Independent Research | Natural Science (Grant DFF-4181-00415). 
Funding for the Stellar Astrophysics Centre (SAC) is provided by The Danish National Research Foundation (Grant agreement no.: DNRF106). RK acknowledges that the research leading to these results received funding from the European Research Council under the European Unions Seventh Framework Program (FP/2007-2013)/ERC Grant Agreement no. 307117. DS and RAG acknowledge the support from the CNES GOLF grant.  The research leading to these results has received funding from EC, under FP7, through the grant agreement FP7-SPACE-2012-312844 (SPACEINN) and PIRSES-GA-2010-269194 (ASK). The peak-bagging described in this paper was performed using the University of Birmingham's BlueBEAR HPC service, which provides a High Performance Computing service to the University's research community. See http://www.birmingham.ac.uk/bear for more details.

\bibliographystyle{aasjournal}
\bibliography{peak-bagging,pb2} 

\begin{thebibliography}{}
\expandafter\ifx\csname natexlab\endcsname\relax\def\natexlab#1{#1}\fi
\providecommand{\url}[1]{\href{#1}{#1}}

\bibitem[{Anderson {et~al.}(1990)Anderson, Duvall, \& Jefferies}]{Anderson1990}
Anderson, E.~R., Duvall, Jr., T.~L., \& Jefferies, S.~M. 1990, ApJ, 364, 699

\bibitem[{Appourchaux(2005)}]{Appourchaux2005}
Appourchaux, T. 2005, in Payload and {{Mission Definition}} in {{Space
  Sciences}}, 185

\bibitem[{Appourchaux {et~al.}(1998)Appourchaux, Gizon, \&
  Rabello-Soares}]{Appourchaux1998}
Appourchaux, T., Gizon, L., \& Rabello-Soares, M.-C. 1998, A\&AS, 132, 107

\bibitem[{Baglin {et~al.}(2006)Baglin, Auvergne, Barge, Deleuil, Catala,
  Michel, Weiss, \& {COROT Team}}]{Baglin2006}
Baglin, A., Auvergne, M., Barge, P., {et~al.} 2006, in ed. Fridlund, M.,
  Baglin, A., Lochard, J. and Conroy, L., Vol. 1306, ESA Special Publication,
  33

\bibitem[{Baliunas {et~al.}(1995)Baliunas, Donahue, Soon, Horne, Frazer,
  Woodard-Eklund, Bradford, Rao, Wilson, Zhang, Bennett, Briggs, Carroll,
  Duncan, Figueroa, Lanning, Misch, Mueller, Noyes, Poppe, Porter, Robinson,
  Russell, Shelton, Soyumer, Vaughan, \& Whitney}]{Baliunas1995}
Baliunas, S.~L., Donahue, R.~A., Soon, W.~H., {et~al.} 1995, ApJ, 438, 269

\bibitem[{Bazilevskaya {et~al.}(2014)Bazilevskaya, Broomhall, Elsworth, \&
  Nakariakov}]{Bazilevskaya2014}
Bazilevskaya, G., Broomhall, A.-M., Elsworth, Y., \& Nakariakov, V.~M. 2014,
  Space Sci. Rev., 186, 359

\bibitem[{Benomar {et~al.}(2009)Benomar, Appourchaux, \& Baudin}]{Benomar2009}
Benomar, O., Appourchaux, T., \& Baudin, F. 2009, A\&A, 506, 15

\bibitem[{Borucki {et~al.}(2010)Borucki, Koch, Basri, Batalha, Brown, Caldwell,
  Caldwell, Christensen-Dalsgaard, Cochran, DeVore, Dunham, Dupree, Gautier,
  Geary, Gilliland, Gould, Howell, Jenkins, Kondo, Latham, Marcy, Meibom,
  Kjeldsen, Lissauer, Monet, Morrison, Sasselov, Tarter, Boss, Brownlee, Owen,
  Buzasi, Charbonneau, Doyle, Fortney, Ford, Holman, Seager, Steffen, Welsh,
  Rowe, Anderson, Buchhave, Ciardi, Walkowicz, Sherry, Horch, Isaacson,
  Everett, Fischer, Torres, Johnson, Endl, MacQueen, Bryson, Dotson, Haas,
  Kolodziejczak, Van~Cleve, Chandrasekaran, Twicken, Quintana, Clarke, Allen,
  Li, Wu, Tenenbaum, Verner, Bruhweiler, Barnes, \& Prsa}]{Borucki2010}
Borucki, W.~J., Koch, D., Basri, G., {et~al.} 2010, Science, 327, 977

\bibitem[{Broomhall {et~al.}(2009)Broomhall, Chaplin, Davies, Elsworth,
  Fletcher, Hale, Miller, \& New}]{Broomhall2009}
Broomhall, A.-M., Chaplin, W.~J., Davies, G.~R., {et~al.} 2009, MNRAS, 396,
  L100

\bibitem[{Broomhall {et~al.}(2012)Broomhall, Chaplin, Elsworth, \&
  Simoniello}]{Broomhall2012}
Broomhall, A.-M., Chaplin, W.~J., Elsworth, Y., \& Simoniello, R. 2012, MNRAS,
  420, 1405

\bibitem[{Broomhall \& Nakariakov(2015)}]{Broomhall2015}
Broomhall, A.-M., \& Nakariakov, V.~M. 2015, Sol. Phys., 290, 3095,
  broomhall2015

\bibitem[{Brun \& Browning(2017)}]{Brun2017}
Brun, A.~S., \& Browning, M.~K. 2017, Living Rev. Solar Phys., 14, 4

\bibitem[{Campante {et~al.}(2011)Campante, Handberg, Mathur, Appourchaux,
  Bedding, Chaplin, Garc{\'\i}a, Mosser, Benomar, Bonanno, Corsaro, Fletcher,
  Gaulme, Hekker, Karoff, R{\'e}gulo, Salabert, Verner, White, Houdek,
  Brand{\~a}o, Creevey, Doǧan, Bazot, Christensen-Dalsgaard, Cunha, Elsworth,
  Huber, Kjeldsen, Lundkvist, Molenda-{\.Z}akowicz, Monteiro, Stello, Clarke,
  Girouard, \& Hall}]{Campante2011}
Campante, T.~L., Handberg, R., Mathur, S., {et~al.} 2011, A\&A, 534, A6

\bibitem[{Campante {et~al.}(2016)Campante, Lund, Kuszlewicz, Davies, Chaplin,
  Albrecht, Winn, Bedding, Benomar, Bossini, Handberg, Santos, Van~Eylen, Basu,
  Christensen-Dalsgaard, Elsworth, Hekker, Hirano, Huber, Karoff, Kjeldsen,
  Lundkvist, North, Silva~Aguirre, Stello, \& White}]{Campante2016}
Campante, T.~L., Lund, M.~N., Kuszlewicz, J.~S., {et~al.} 2016, ApJ, 819, 85

\bibitem[{Casagrande {et~al.}(2014)Casagrande, Silva~Aguirre, Stello, Huber,
  Serenelli, Cassisi, Dotter, Milone, Hodgkin, Marino, Lund, Pietrinferni,
  Asplund, Feltzing, Flynn, Grundahl, Nissen, Sch{\"o}nrich, Schlesinger, \&
  Wang}]{Casagrande2014}
Casagrande, L., Silva~Aguirre, V., Stello, D., {et~al.} 2014, ApJ, 787, 110

\bibitem[{Ceillier {et~al.}(2016)Ceillier, {van Saders}, Garc{\'\i}a, Metcalfe,
  Creevey, Mathis, Mathur, Pinsonneault, Salabert, \& Tayar}]{Ceillier2016}
Ceillier, T., {van Saders}, J., Garc{\'\i}a, R.~A., {et~al.} 2016, MNRAS, 456,
  119

\bibitem[{Chaplin {et~al.}(1998)Chaplin, Elsworth, Isaak, Lines, McLeod,
  Miller, \& New}]{Chaplin1998}
Chaplin, W.~J., Elsworth, Y., Isaak, G.~R., {et~al.} 1998, MNRAS, 300, 1077

\bibitem[{Chaplin {et~al.}(2000)Chaplin, Elsworth, Isaak, Miller, \&
  New}]{Chaplin2000}
Chaplin, W.~J., Elsworth, Y., Isaak, G.~R., Miller, B.~A., \& New, R. 2000,
  MNRAS, 313, 32

\bibitem[{Chaplin {et~al.}(2004)Chaplin, Elsworth, Isaak, Miller, \&
  New}]{Chaplin2004}
---. 2004, MNRAS, 352, 1102

\bibitem[{Chaplin {et~al.}(2007)Chaplin, Elsworth, Miller, Verner, \&
  New}]{Chaplin2007}
Chaplin, W.~J., Elsworth, Y., Miller, B.~A., Verner, G.~A., \& New, R. 2007,
  ApJ, 659, 1749

\bibitem[{Chaplin {et~al.}(2011)Chaplin, Kjeldsen, Bedding,
  Christensen-Dalsgaard, Gilliland, Kawaler, Appourchaux, Elsworth,
  Garc{\'\i}a, Houdek, Karoff, Metcalfe, Molenda-{\.Z}akowicz, Monteiro,
  Thompson, Verner, Batalha, Borucki, Brown, Bryson, Christiansen, Clarke,
  Jenkins, Klaus, Koch, An, Ballot, Basu, Benomar, Bonanno, Broomhall,
  Campante, Corsaro, Creevey, Esch, Gai, Gaulme, Hale, Handberg, Hekker, Huber,
  Mathur, Mosser, New, Pinsonneault, Pricopi, Quirion, R{\'e}gulo, Roxburgh,
  Salabert, Stello, \& Suran}]{Chaplin2011}
Chaplin, W.~J., Kjeldsen, H., Bedding, T.~R., {et~al.} 2011, ApJ, 732, 54

\bibitem[{Chaplin {et~al.}(2014)Chaplin, Basu, Huber, Serenelli, Casagrande,
  Silva~Aguirre, Ball, Creevey, Gizon, Handberg, Karoff, Lutz, Marques, Miglio,
  Stello, Suran, Pricopi, Metcalfe, Monteiro, Molenda-{\.Z}akowicz,
  Appourchaux, Christensen-Dalsgaard, Elsworth, Garc{\'\i}a, Houdek, Kjeldsen,
  Bonanno, Campante, Corsaro, Gaulme, Hekker, Mathur, Mosser, R{\'e}gulo, \&
  Salabert}]{Chaplin2014a}
Chaplin, W.~J., Basu, S., Huber, D., {et~al.} 2014, ApJS, 210, 1

\bibitem[{Christensen-Dalsgaard \&
  Berthomieu(1991)}]{Christensen-Dalsgaard1991}
Christensen-Dalsgaard, J., \& Berthomieu, G. 1991, in Solar {{Interior}} and
  {{Atmosphere}} (University of Arizona Press, Tucson, AZ: {Cox, A. N.,
  Livingston, W. C., Mathews M.,}), 401--478

\bibitem[{Davies {et~al.}(2015)Davies, Chaplin, Farr, Garc\'{i}a, Lund, Mathis,
  Metcalfe, Appourchaux, Basu, Benomar, Campante, Ceillier, Elsworth, Handberg,
  Salabert, \& Stello}]{Davies2015}
Davies, G.~R., Chaplin, W.~J., Farr, W.~M., {et~al.} 2015, MNRAS, 446, 2959

\bibitem[{Davies {et~al.}(2016)Davies, Silva~Aguirre, Bedding, Handberg, Lund,
  Chaplin, Huber, White, Benomar, Hekker, Basu, Campante,
  Christensen-Dalsgaard, Elsworth, Karoff, Kjeldsen, Lundkvist, Metcalfe, \&
  Stello}]{Davies2016}
Davies, G.~R., Silva~Aguirre, V., Bedding, T.~R., {et~al.} 2016, MNRAS, 456,
  2183

\bibitem[{Dumusque {et~al.}(2014)Dumusque, Bonomo, Haywood, Malavolta,
  S{\'e}gransan, Buchhave, Collier~Cameron, Latham, Molinari, Pepe, Udry,
  Charbonneau, Cosentino, Dressing, Figueira, Fiorenzano, Gettel, Harutyunyan,
  Horne, Lopez-Morales, Lovis, Mayor, Micela, Motalebi, Nascimbeni, Phillips,
  Piotto, Pollacco, Queloz, Rice, Sasselov, Sozzetti, Szentgyorgyi, \&
  Watson}]{Dumusque2014a}
Dumusque, X., Bonomo, A.~S., Haywood, R.~D., {et~al.} 2014, ApJ, 789, 154

\bibitem[{Duvall \& Harvey(1986)}]{Duvall1986}
Duvall, Jr., T.~L., \& Harvey, J.~W. 1986, in Proceedings of the {{NATO
  Advanced Research Workshop}}, Vol. 169 (Cambridge, England: {Dordrecht, D.
  Reidel Publishing Co.}), 105--116

\bibitem[{Earl \& Deem(2005)}]{Earl2005}
Earl, D.~J., \& Deem, M.~W. 2005, PCCP, 7, 3910

\bibitem[{Egeland {et~al.}(2015)Egeland, Metcalfe, Hall, \&
  Henry}]{Egeland2015}
Egeland, R., Metcalfe, T.~S., Hall, J.~C., \& Henry, G.~W. 2015, ApJ, 812, 12

\bibitem[{Elsworth {et~al.}(1993)Elsworth, Howe, Isaak, McLeod, Miller, Speake,
  Wheeler, \& New}]{Elsworth1993}
Elsworth, Y., Howe, R., Isaak, G.~R., {et~al.} 1993, MNRAS, 265, 888

\bibitem[{Elsworth {et~al.}(1990)Elsworth, Howe, Isaak, McLeod, \&
  New}]{Elsworth1990}
Elsworth, Y., Howe, R., Isaak, G.~R., McLeod, C.~P., \& New, R. 1990, Nat.,
  345, 322

\bibitem[{Fletcher {et~al.}(2010)Fletcher, Broomhall, Salabert, Basu, Chaplin,
  Elsworth, Garcia, \& New}]{Fletcher2010}
Fletcher, S.~T., Broomhall, A.-M., Salabert, D., {et~al.} 2010, ApJ, 718, L19

\bibitem[{Flores {et~al.}(2016)Flores, Buccino, Saffe, \& Mauas}]{Flores2016}
Flores, M.~G., Buccino, A.~P., Saffe, C.~E., \& Mauas, P. J.~D. 2016, ArXiv
  e-prints, 1610, arXiv:1610.03731

\bibitem[{Foreman-Mackey {et~al.}(2013)Foreman-Mackey, Hogg, Lang, \&
  Goodman}]{Foreman-Mackey2013}
Foreman-Mackey, D., Hogg, D.~W., Lang, D., \& Goodman, J. 2013, PASP, 125, 306

\bibitem[{Fr{\"o}hlich {et~al.}(1995)Fr{\"o}hlich, Romero, Roth, Wehrli,
  Andersen, Appourchaux, Domingo, Telljohann, Berthomieu, Delache, Provost,
  Toutain, Crommelynck, Chevalier, Fichot, D{\"a}ppen, Gough, Hoeksema,
  Jim{\'e}nez, G{\'o}mez, Herreros, Cort{\'e}s, Jones, Pap, \&
  Willson}]{Frohlich1995}
Fr{\"o}hlich, C., Romero, J., Roth, H., {et~al.} 1995, Sol. Phys., 162, 101

\bibitem[{Fr{\"o}hlich {et~al.}(1997)Fr{\"o}hlich, Andersen, Appourchaux,
  Berthomieu, Crommelynck, Domingo, Fichot, Finsterle, Gomez, Gough, Jimenez,
  Leifsen, Lombaerts, Pap, Provost, Cortes, Romero, Roth, Sekii, Telljohann,
  Toutain, \& Wehrli}]{Frohlich1997}
Fr{\"o}hlich, C., Andersen, B.~N., Appourchaux, T., {et~al.} 1997, Sol. Phys.,
  170, 1

\bibitem[{Gabriel(1994)}]{Gabriel1994}
Gabriel, M. 1994, A\&A, 292, 281

\bibitem[{Garc{\'\i}a {et~al.}(2010)Garc{\'\i}a, Mathur, Salabert, Ballot,
  Regulo, Metcalfe, \& Baglin}]{Garcia2010}
Garc{\'\i}a, R.~A., Mathur, S., Salabert, D., {et~al.} 2010, Science, 329, 1032

\bibitem[{Garc{\'\i}a {et~al.}(2009)Garc{\'\i}a, R{\'e}gulo, Samadi, Ballot,
  Barban, Benomar, Chaplin, Gaulme, Appourchaux, Mathur, Mosser, Toutain,
  Verner, Auvergne, Baglin, Baudin, Boumier, Bruntt, Catala, Deheuvels,
  Elsworth, Jim{\'e}nez-Reyes, Michel, P{\'e}rez~Hern{\'a}ndez, Roxburgh, \&
  Salabert}]{Garcia2009}
Garc{\'\i}a, R.~A., R{\'e}gulo, C., Samadi, R., {et~al.} 2009, A\&A, 506, 41

\bibitem[{Garc{\'\i}a {et~al.}(2011)Garc{\'\i}a, Hekker, Stello,
  Guti{\'e}rrez-Soto, Handberg, Huber, Karoff, Uytterhoeven, Appourchaux,
  Chaplin, Elsworth, Mathur, Ballot, Christensen-Dalsgaard, Gilliland, Houdek,
  Jenkins, Kjeldsen, McCauliff, Metcalfe, Middour, Molenda-Zakowicz, Monteiro,
  Smith, \& Thompson}]{Garcia2011}
Garc{\'\i}a, R.~A., Hekker, S., Stello, D., {et~al.} 2011, MNRAS, 414, L6

\bibitem[{Garc{\'\i}a {et~al.}(2014)Garc{\'\i}a, Ceillier, Salabert, Mathur,
  {van Saders}, Pinsonneault, Ballot, Beck, Bloemen, Campante, Davies, {do
  Nascimento}, Mathis, Metcalfe, Nielsen, Su{\'a}rez, Chaplin, Jim{\'e}nez, \&
  Karoff}]{Garcia2014}
Garc{\'\i}a, R.~A., Ceillier, T., Salabert, D., {et~al.} 2014, A\&A, 572, A34

\bibitem[{Gizon \& Solanki(2003)}]{Gizon2003}
Gizon, L., \& Solanki, S.~K. 2003, ApJ, 589, 1009

\bibitem[{Goldreich \& Keeley(1977)}]{Goldreich1977}
Goldreich, P., \& Keeley, D.~A. 1977, ApJ, 212, 243

\bibitem[{Goodman \& Weare(2010)}]{Goodman2010}
Goodman, J., \& Weare, J. 2010, Comm. App. Math. and Comp. Sci., 5, 65

\bibitem[{Handberg \& Campante(2011)}]{Handberg2011}
Handberg, R., \& Campante, T.~L. 2011, A\&A, 527, A56

\bibitem[{Handberg \& Lund(2014)}]{Handberg2014}
Handberg, R., \& Lund, M.~N. 2014, MNRAS, 445, 2698

\bibitem[{Harvey(1985)}]{Harvey1985}
Harvey, J. 1985, in Future {{Missions}} in {{Solar}}, {{Heliospheric}} \&
  {{Space Plasma Physics}}, Vol. 235 (ESA, Noordwijk: {Rolfe E. \& Battrick
  B.}), 199

\bibitem[{Howe {et~al.}(2015)Howe, Davies, Chaplin, Elsworth, \&
  Hale}]{Howe2015}
Howe, R., Davies, G.~R., Chaplin, W.~J., Elsworth, Y.~P., \& Hale, S.~J. 2015,
  MNRAS, 454, 4120

\bibitem[{Huber {et~al.}(2009)Huber, Stello, Bedding, Chaplin, Arentoft,
  Quirion, \& Kjeldsen}]{Huber2009}
Huber, D., Stello, D., Bedding, T.~R., {et~al.} 2009, Commun. Asteroseismol.,
  160, 74

\bibitem[{Huber {et~al.}(2011)Huber, Bedding, Stello, Hekker, Mathur, Mosser,
  Verner, Bonanno, Buzasi, Campante, Elsworth, Hale, Kallinger, Silva~Aguirre,
  Chaplin, De~Ridder, Garc{\'\i}a, Appourchaux, Frandsen, Houdek,
  Molenda-{\.Z}akowicz, Monteiro, Christensen-Dalsgaard, Gilliland, Kawaler,
  Kjeldsen, Broomhall, Corsaro, Salabert, Sanderfer, Seader, \&
  Smith}]{Huber2011}
Huber, D., Bedding, T.~R., Stello, D., {et~al.} 2011, ApJ, 743, 143

\bibitem[{Huber {et~al.}(2013)Huber, Chaplin, Christensen-Dalsgaard, Gilliland,
  Kjeldsen, Buchhave, Fischer, Lissauer, Rowe, Sanchis-Ojeda, Basu, Handberg,
  Hekker, Howard, Isaacson, Karoff, Latham, Lund, Lundkvist, Marcy, Miglio,
  Silva~Aguirre, Stello, Arentoft, Barclay, Bedding, Burke, Christiansen,
  Elsworth, Haas, Kawaler, Metcalfe, Mullally, \& Thompson}]{Huber2013}
Huber, D., Chaplin, W.~J., Christensen-Dalsgaard, J., {et~al.} 2013, ApJ, 767,
  127

\bibitem[{Jain {et~al.}(2009)Jain, Tripathy, \& Hill}]{Jain2009}
Jain, K., Tripathy, S.~C., \& Hill, F. 2009, ApJ, 695, 1567

\bibitem[{Jim{\'e}nez {et~al.}(2002)Jim{\'e}nez, Roca~Cort{\'e}s, \&
  Jim{\'e}nez-Reyes}]{Jimenez2002}
Jim{\'e}nez, A., Roca~Cort{\'e}s, T., \& Jim{\'e}nez-Reyes, S.~J. 2002, Sol.
  Phys., 209, 247

\bibitem[{Jiménez-Reyes {et~al.}(1998)Jiménez-Reyes, Régulo, Pallé, \&
  Roca~Cortes}]{Jimenez-Reyes1998}
Jiménez-Reyes, S.~J., Régulo, C., Pallé, P.~L., \& Roca~Cortes, T. 1998,
  A\&A, 329, 1119

\bibitem[{Kallinger \& Matthews(2010)}]{Kallinger2010}
Kallinger, T., \& Matthews, J.~M. 2010, ApJL, 711, L35

\bibitem[{Kallinger {et~al.}(2014)Kallinger, De~Ridder, Hekker, Mathur, Mosser,
  Gruberbauer, Garc{\'\i}a, Karoff, \& Ballot}]{Kallinger2014}
Kallinger, T., De~Ridder, J., Hekker, S., {et~al.} 2014, A\&A, 570, A41

\bibitem[{Karoff(2012)}]{Karoff2012}
Karoff, C. 2012, MNRAS, 421, 3170

\bibitem[{Karoff {et~al.}(2013{\natexlab{a}})Karoff, Metcalfe, Chaplin,
  Frandsen, Grundahl, Kjeldsen, Christensen-Dalsgaard, Nielsen, Frimann,
  Thygesen, Arentoft, Amby, Sousa, \& Buzasi}]{Karoff2013a}
Karoff, C., Metcalfe, T.~S., Chaplin, W.~J., {et~al.} 2013{\natexlab{a}},
  MNRAS, 433, 3227

\bibitem[{Karoff {et~al.}(2013{\natexlab{b}})Karoff, Campante, Ballot,
  Kallinger, Gruberbauer, Garc{\'\i}a, Caldwell, Christiansen, \&
  Kinemuchi}]{Karoff2013}
Karoff, C., Campante, T.~L., Ballot, J., {et~al.} 2013{\natexlab{b}}, ApJ, 767,
  34

\bibitem[{Karoff {et~al.}(2018)Karoff, Metcalfe, Santos, Montet, Isaacson,
  Witzke, Shapiro, Mathur, Davies, Lund, Garcia, Brun, Salabert, Avelino, van
  Saders, Egeland, Cunha, Campante, Chaplin, Krivova, Solanki, Stritzinger, \&
  Knudsen}]{Karoff2018}
Karoff, C., Metcalfe, T.~S., Santos, A. R.~G., {et~al.} 2018, ApJ, 852, 46

\bibitem[{Kiefer {et~al.}(2017)Kiefer, Schad, Davies, \& Roth}]{Kiefer2017}
Kiefer, R., Schad, A., Davies, G., \& Roth, M. 2017, A\&A, 598, A77

\bibitem[{Kjeldsen \& Bedding(2011)}]{Kjeldsen2011}
Kjeldsen, H., \& Bedding, T.~R. 2011, A\&A, 529, L8

\bibitem[{Lefebvre {et~al.}(2008)Lefebvre, Garc{\'\i}a, Jim{\'e}nez-Reyes,
  Turck-Chi{\`e}ze, \& Mathur}]{Lefebvre2008}
Lefebvre, S., Garc{\'\i}a, R.~A., Jim{\'e}nez-Reyes, S.~J., Turck-Chi{\`e}ze,
  S., \& Mathur, S. 2008, A\&A, 490, 1143

\bibitem[{Libbrecht \& Woodard(1990)}]{Libbrecht1990a}
Libbrecht, K.~G., \& Woodard, M.~F. 1990, Nat., 345, 779

\bibitem[{Lund {et~al.}(2014)Lund, Lundkvist, Silva~Aguirre, Houdek,
  Casagrande, Van~Eylen, Campante, Karoff, Kjeldsen, Albrecht, Chaplin,
  Nielsen, Degroote, Davies, \& Handberg}]{Lund2014}
Lund, M.~N., Lundkvist, M., Silva~Aguirre, V., {et~al.} 2014, A\&A, 570, A54

\bibitem[{Lund {et~al.}(2017)Lund, Silva~Aguirre, Davies, Chaplin,
  Christensen-Dalsgaard, Houdek, White, Bedding, Ball, Huber, Antia, Lebreton,
  Latham, Handberg, Verma, Basu, Casagrande, Justesen, Kjeldsen, \&
  Mosumgaard}]{Lund2017}
Lund, M.~N., Silva~Aguirre, V., Davies, G.~R., {et~al.} 2017, ApJ, 835, 172

\bibitem[{Mathur {et~al.}(2011)Mathur, Hekker, Trampedach, Ballot, Kallinger,
  Buzasi, Garc{\'\i}a, Huber, Jim{\'e}nez, Mosser, Bedding, Elsworth,
  R{\'e}gulo, Stello, Chaplin, De~Ridder, Hale, Kinemuchi, Kjeldsen, Mullally,
  \& Thompson}]{Mathur2011}
Mathur, S., Hekker, S., Trampedach, R., {et~al.} 2011, ApJ, 741, 119

\bibitem[{Mathur {et~al.}(2014)Mathur, Garc{\'\i}a, Ballot, Ceillier, Salabert,
  Metcalfe, R{\'e}gulo, Jim{\'e}nez, \& Bloemen}]{Mathur2014}
Mathur, S., Garc{\'\i}a, R.~A., Ballot, J., {et~al.} 2014, A\&A, 562, A124

\bibitem[{McIntosh {et~al.}(2015)McIntosh, Leamon, Krista, Title, Hudson,
  Riley, Harder, Kopp, Snow, Woods, Kasper, Stevens, \& Ulrich}]{McIntosh2015}
McIntosh, S.~W., Leamon, R.~J., Krista, L.~D., {et~al.} 2015, Nat. Commun., 6,
  6491

\bibitem[{McQuillan {et~al.}(2013)McQuillan, Mazeh, \&
  Aigrain}]{McQuillan2013a}
McQuillan, A., Mazeh, T., \& Aigrain, S. 2013, ApJ, 775, L11

\bibitem[{McQuillan {et~al.}(2014)McQuillan, Mazeh, \& Aigrain}]{McQuillan2014}
---. 2014, ApJS, 211, 24

\bibitem[{Metcalfe {et~al.}(2010)Metcalfe, Basu, Henry, Soderblom, Judge,
  Kn{\"o}lker, Mathur, \& Rempel}]{Metcalfe2010}
Metcalfe, T.~S., Basu, S., Henry, T.~J., {et~al.} 2010, ApJL, 723, L213

\bibitem[{Metcalfe {et~al.}(2013)Metcalfe, Buccino, Brown, Mathur, Soderblom,
  Henry, Mauas, Petrucci, Hall, \& Basu}]{Metcalfe2013}
Metcalfe, T.~S., Buccino, A.~P., Brown, B.~P., {et~al.} 2013, ApJL, 763, L26

\bibitem[{Moreno-Insertis \& Solanki(2000)}]{Moreno-Insertis2000}
Moreno-Insertis, F., \& Solanki, S.~K. 2000, MNRAS, 313, 411

\bibitem[{Muller {et~al.}(2007)Muller, Hanslmeier, \&
  Salda{\~n}a-Mu{\~n}oz}]{Muller2007}
Muller, R., Hanslmeier, A., \& Salda{\~n}a-Mu{\~n}oz, M. 2007, A\&A, 475, 717

\bibitem[{Ol{\'a}h {et~al.}(2009)Ol{\'a}h, Koll{\'a}th, Granzer, Strassmeier,
  Lanza, J{\"a}rvinen, Korhonen, Baliunas, Soon, Messina, \&
  Cutispoto}]{Olah2009}
Ol{\'a}h, K., Koll{\'a}th, Z., Granzer, T., {et~al.} 2009, A\&A, 501, 703

\bibitem[{Pall{\'e} {et~al.}(1995)Pall{\'e}, Jim{\'e}nez, Perez~Hernandez,
  R{\'e}gulo, Roca~Cort{\'e}s, \& Sanchez}]{Palle1995}
Pall{\'e}, P.~L., Jim{\'e}nez, A., Perez~Hernandez, F., {et~al.} 1995, ApJ,
  441, 952

\bibitem[{Paxton {et~al.}(2011)Paxton, Bildsten, Dotter, Herwig, Lesaffre, \&
  Timmes}]{Paxton2011}
Paxton, B., Bildsten, L., Dotter, A., {et~al.} 2011, ApJS, 192, 3

\bibitem[{Paxton {et~al.}(2013)Paxton, Cantiello, Arras, Bildsten, Brown,
  Dotter, Mankovich, Montgomery, Stello, Timmes, \& Townsend}]{Paxton2013}
Paxton, B., Cantiello, M., Arras, P., {et~al.} 2013, ApJS, 208, 4

\bibitem[{Pinsonneault {et~al.}(2012)Pinsonneault, An, Molenda-{\.Z}akowicz,
  Chaplin, Metcalfe, \& Bruntt}]{Pinsonneault2012}
Pinsonneault, M.~H., An, D., Molenda-{\.Z}akowicz, J., {et~al.} 2012, ApJS,
  199, 30

\bibitem[{Pinsonneault {et~al.}(2014)Pinsonneault, Elsworth, Epstein, Hekker,
  M{\'e}sz{\'a}ros, Chaplin, Johnson, Garc{\'\i}a, Holtzman, Mathur,
  Garc{\'\i}a~P{\'e}rez, Silva~Aguirre, Girardi, Basu, Shetrone, Stello,
  Allende~Prieto, An, Beck, Beers, Bizyaev, Bloemen, Bovy, Cunha, De~Ridder,
  Frinchaboy, Garc{\'\i}a-Hern{\'a}ndez, Gilliland, Harding, Hearty, Huber,
  Ivans, Kallinger, Majewski, Metcalfe, Miglio, Mosser, Muna, Nidever,
  Schneider, Serenelli, Smith, Tayar, Zamora, \& Zasowski}]{Pinsonneault2014}
Pinsonneault, M.~H., Elsworth, Y., Epstein, C., {et~al.} 2014, ApJS, 215, 19

\bibitem[{Ram{\'\i}rez {et~al.}(2009)Ram{\'\i}rez, Mel{\'e}ndez, \&
  Asplund}]{Ramirez2009}
Ram{\'\i}rez, I., Mel{\'e}ndez, J., \& Asplund, M. 2009, A\&A, 508, L17

\bibitem[{R{\'e}gulo {et~al.}(2016)R{\'e}gulo, Garc{\'\i}a, \&
  Ballot}]{Regulo2016}
R{\'e}gulo, C., Garc{\'\i}a, R.~A., \& Ballot, J. 2016, A\&A, 589, A103

\bibitem[{R{\'e}gulo {et~al.}(2002)R{\'e}gulo, Roca~Cort{\'e}s, \&
  V{\'a}zquez~Rami{\'o}}]{Regulo2002}
R{\'e}gulo, C., Roca~Cort{\'e}s, T., \& V{\'a}zquez~Rami{\'o}, H. 2002, in
  (ESA, Noordwijk: {Wilson, A.}), 506, 889--892

\bibitem[{R{\'e}gulo {et~al.}(2005)R{\'e}gulo, V{\'a}zquez~Rami{\'o}, \&
  Roca~Cort{\'e}s}]{Regulo2005}
R{\'e}gulo, C., V{\'a}zquez~Rami{\'o}, H., \& Roca~Cort{\'e}s, T. 2005, A\&A,
  443, 1013

\bibitem[{Salabert {et~al.}(2017{\natexlab{a}})Salabert, Garc\'{i}a,
  Jim\'{e}nez, Bertello, Corsaro, \& Pall\'{e}}]{Salabert2017}
Salabert, D., Garc\'{i}a, R.~A., Jim\'{e}nez, A., {et~al.} 2017{\natexlab{a}},
  A\&A, 608, A87

\bibitem[{Salabert {et~al.}(2009)Salabert, Garc{\'\i}a, Pall{\'e}, \&
  Jim{\'e}nez-Reyes}]{Salabert2009}
Salabert, D., Garc{\'\i}a, R.~A., Pall{\'e}, P.~L., \& Jim{\'e}nez-Reyes, S.~J.
  2009, A\&A, 504, L1

\bibitem[{Salabert {et~al.}(2015)Salabert, Garc{\'\i}a, \&
  Turck-Chi{\`e}ze}]{Salabert2015}
Salabert, D., Garc{\'\i}a, R.~A., \& Turck-Chi{\`e}ze, S. 2015, A\&A, 578, A137

\bibitem[{Salabert \& Jim{\'e}nez-Reyes(2006)}]{Salabert2006}
Salabert, D., \& Jim{\'e}nez-Reyes, S.~J. 2006, ApJ, 650, 451

\bibitem[{Salabert {et~al.}(2017{\natexlab{b}})Salabert, R\'{e}gulo,
  P\'{e}rez~Hern\'{a}ndez, \& Garc\'{i}a}]{Salabert2017b}
Salabert, D., R\'{e}gulo, C., P\'{e}rez~Hern\'{a}ndez, F., \& Garc\'{i}a, R.~A.
  2017{\natexlab{b}}, ArXiv e-prints, 1712, arXiv:1712.05691

\bibitem[{Salabert {et~al.}(2016{\natexlab{a}})Salabert, R{\'e}gulo,
  Garc{\'\i}a, Beck, Ballot, Creevey, P{\'e}rez~Hern{\'a}ndez, {do Nascimento},
  Corsaro, Egeland, Mathur, Metcalfe, Bigot, Ceillier, \&
  Pall{\'e}}]{Salabert2016}
Salabert, D., R{\'e}gulo, C., Garc{\'\i}a, R.~A., {et~al.} 2016{\natexlab{a}},
  A\&A, 589, A118

\bibitem[{Salabert {et~al.}(2016{\natexlab{b}})Salabert, Garc\'{i}a, Beck,
  Egeland, Pall\'{e}, Mathur, Metcalfe, do~Nascimento, Ceillier, Andersen, \&
  Trivi\~{n}o Hage}]{Salabert2016b}
Salabert, D., Garc\'{i}a, R.~A., Beck, P.~G., {et~al.} 2016{\natexlab{b}},
  A\&A, 596, A31

\bibitem[{Santos {et~al.}(2016)Santos, Cunha, Avelino, Chaplin, \&
  Campante}]{Santos2016}
Santos, A. R.~G., Cunha, M.~S., Avelino, P.~P., Chaplin, W.~J., \& Campante,
  T.~L. 2016, MNRAS, 461, 224

\bibitem[{Santos {et~al.}(2017)Santos, Cunha, Avelino, Chaplin, \&
  Campante}]{Santos2017}
---. 2017, MNRAS, 464, 4408

\bibitem[{Silva~Aguirre {et~al.}(2015)Silva~Aguirre, Davies, Basu,
  Christensen-Dalsgaard, Creevey, Metcalfe, Bedding, Casagrande, Handberg,
  Lund, Nissen, Chaplin, Huber, Serenelli, Stello, Van~Eylen, Campante,
  Elsworth, Gilliland, Hekker, Karoff, Kawaler, Kjeldsen, \&
  Lundkvist}]{SilvaAguirre2015}
Silva~Aguirre, V., Davies, G.~R., Basu, S., {et~al.} 2015, MNRAS, 452, 2127

\bibitem[{Silva~Aguirre {et~al.}(2017)Silva~Aguirre, Lund, Antia, Ball, Basu,
  Christensen-Dalsgaard, Lebreton, Reese, Verma, Casagrande, Justesen,
  Mosumgaard, Chaplin, Bedding, Davies, Handberg, Houdek, Huber, Kjeldsen,
  Latham, White, Coelho, Miglio, \& Rendle}]{SilvaAguirre2017}
Silva~Aguirre, V., Lund, M.~N., Antia, H.~M., {et~al.} 2017, ApJ, 835, 173

\bibitem[{Simoniello {et~al.}(2012)Simoniello, Finsterle, Salabert,
  Garc{\'\i}a, Turck-Chi{\`e}ze, Jim{\'e}nez, \& Roth}]{Simoniello2012}
Simoniello, R., Finsterle, W., Salabert, D., {et~al.} 2012, A\&A, 539, A135

\bibitem[{Simoniello {et~al.}(2013)Simoniello, Jain, Tripathy,
  Turck-Chi{\`e}ze, Baldner, Finsterle, Hill, \& Roth}]{Simoniello2013}
Simoniello, R., Jain, K., Tripathy, S.~C., {et~al.} 2013, ApJ, 765, 100

\bibitem[{Smith {et~al.}(2012)Smith, Stumpe, Van~Cleve, Jenkins, Barclay,
  Fanelli, Girouard, Kolodziejczak, McCauliff, Morris, \& Twicken}]{Smith2012}
Smith, J.~C., Stumpe, M.~C., Van~Cleve, J.~E., {et~al.} 2012, PASP, 124, 1000

\bibitem[{Stahn(2010)}]{Stahn2010}
Stahn, T. 2010, Ph.D. Thesis, Mathematisch-Naturwissenschaftlichen
  Fakult{\"a}ten der Georg-August-Universit{\"a}t zu G{\"o}ttingen

\bibitem[{Stumpe {et~al.}(2012)Stumpe, Smith, Van~Cleve, Twicken, Barclay,
  Fanelli, Girouard, Jenkins, Kolodziejczak, McCauliff, \& Morris}]{Stumpe2012}
Stumpe, M.~C., Smith, J.~C., Van~Cleve, J.~E., {et~al.} 2012, PASP, 124, 985

\bibitem[{Toutain \& Appourchaux(1994)}]{Toutain1994}
Toutain, T., \& Appourchaux, T. 1994, A\&A, 289, 649

\bibitem[{Tripathy {et~al.}(2007)Tripathy, Hill, Jain, \&
  Leibacher}]{Tripathy2007}
Tripathy, S.~C., Hill, F., Jain, K., \& Leibacher, J.~W. 2007, Sol. Phys., 243,
  105

\bibitem[{Tripathy {et~al.}(2011)Tripathy, Jain, Salabert, Garc{\'\i}a, Hill,
  \& Leibacher}]{Tripathy2011}
Tripathy, S.~C., Jain, K., Salabert, D., {et~al.} 2011, J. Phys. Conf. Ser.,
  271, 012055

\bibitem[{Walkowicz \& Basri(2013)}]{Walkowicz2013a}
Walkowicz, L.~M., \& Basri, G.~S. 2013, MNRAS, 436, 1883

\bibitem[{Wilson(1978)}]{Wilson1978}
Wilson, O.~C. 1978, ApJ, 226, 379

\bibitem[{Woodard \& Noyes(1985)}]{Woodard1985}
Woodard, M.~F., \& Noyes, R.~W. 1985, Nat., 318, 449

\end{thebibliography}

\appendix
\section{Validation of the peak-bagging tool}\label{sec:sun}

Before using the peak-bagging tool to search for temporal variations in the acoustic frequencies of {\it Kepler} targets, we performed validation tests with solar data.
The following sections present the results obtained from the analysis of VIRGO/SPM \citep[Variability of solar IRradiance and Gravity Oscillations on board SOlar and Heliospheric Observatory (SOHO), where SPM stands for sunphotometers;][]{Frohlich1995,Frohlich1997,Jimenez2002} data and artificial BiSON (Birmingham Solar-Oscillations Network) data.

\subsection{VIRGO/SPM data}\label{sec:virgo}

The first validation test is performed with $\sim12$-yr VIRGO/SPM time-series from the green channel. To estimate the temporal variations in the solar acoustic frequencies, the original light curve is divided into 90-d segments overlapped by 45 days (as performed for the analysis of {\it Kepler} targets).

\begin{figure}[h]\centering
\includegraphics[width=0.52\hsize]{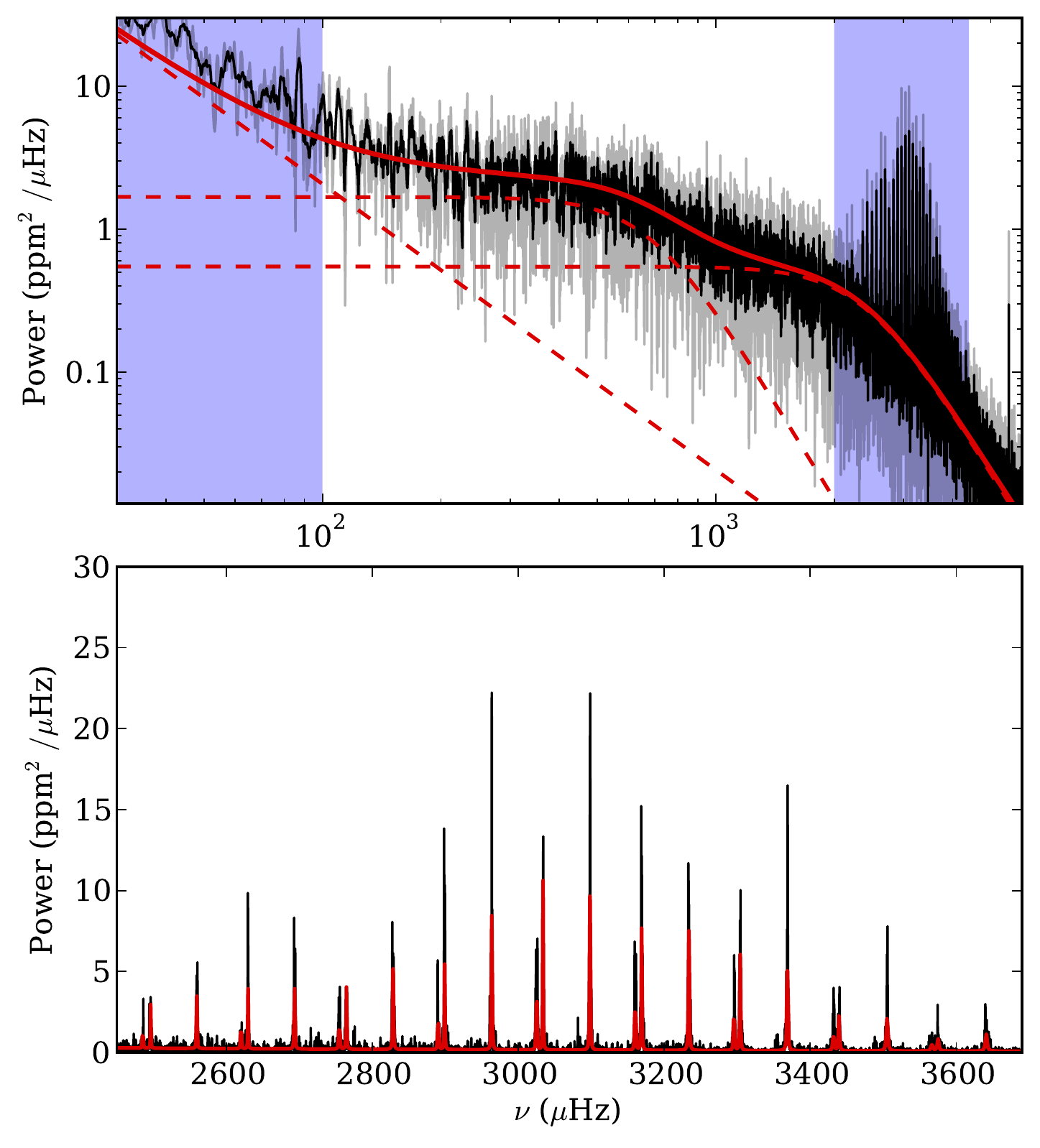}\vspace{-0.3cm }
\caption{Solar power density spectrum (black and gray) for a VIRGO/SPM sub-series of 90 days. The red solid lines show the best fits to the background signal (top panel) and to the p-modes (bottom panel; obtained with the peak-bagging tool). The dashed lines correspond to the different contributions to the brightness variability. The blue regions mark the frequency ranges neglected from the fitting process.}\label{fig:powersvirgo}\vspace{-.5cm}
\end{figure}

In the case of VIRGO/SPM data, the background model also includes a facular component with characteristic timescale fixed at the value 65.8 s \citep{Karoff2012}. The top panel in Figure~\ref{fig:powersvirgo} compares the solar power spectrum, for a given segment, with the background model that best fits the data.

Having the background model for each sub-series, we apply the peak-bagging tool (Section~\ref{sec:peakbag}) to perform the global fit to the acoustic modes. The prior functions were defined as described in Section~\ref{sec:prior}, with exception of those for the rotational splitting and inclination angle:
\begin{itemize}
\item {\it Rotational splitting} - We use a uniform prior between $0$ and $5\,\mu\text{Hz}$.
\item {\it Inclination angle} - In order to avoid boundary effects in the sampling \citep[e.g.][]{Lund2014,Campante2016,Lund2017}, the inclination is sampled from a uniform prior within $-90^\circ$ to $180^\circ$ and then folded onto the range $[0^\circ,90^\circ]$. 
\end{itemize}
The values $\nu_{nl}^0$ are taken from \citet[for VIRGO/SPM data]{Stahn2010} and \citet[for BiSON data]{Broomhall2009}.

The bottom panel of Figure~\ref{fig:powersvirgo} shows the best fit to the solar acoustic modes. The corresponding parameters are estimated from the corresponding posterior distributions as described in Section~\ref{sec:emcee}.

\begin{figure}[h]\centering
\includegraphics[width=0.62\hsize]{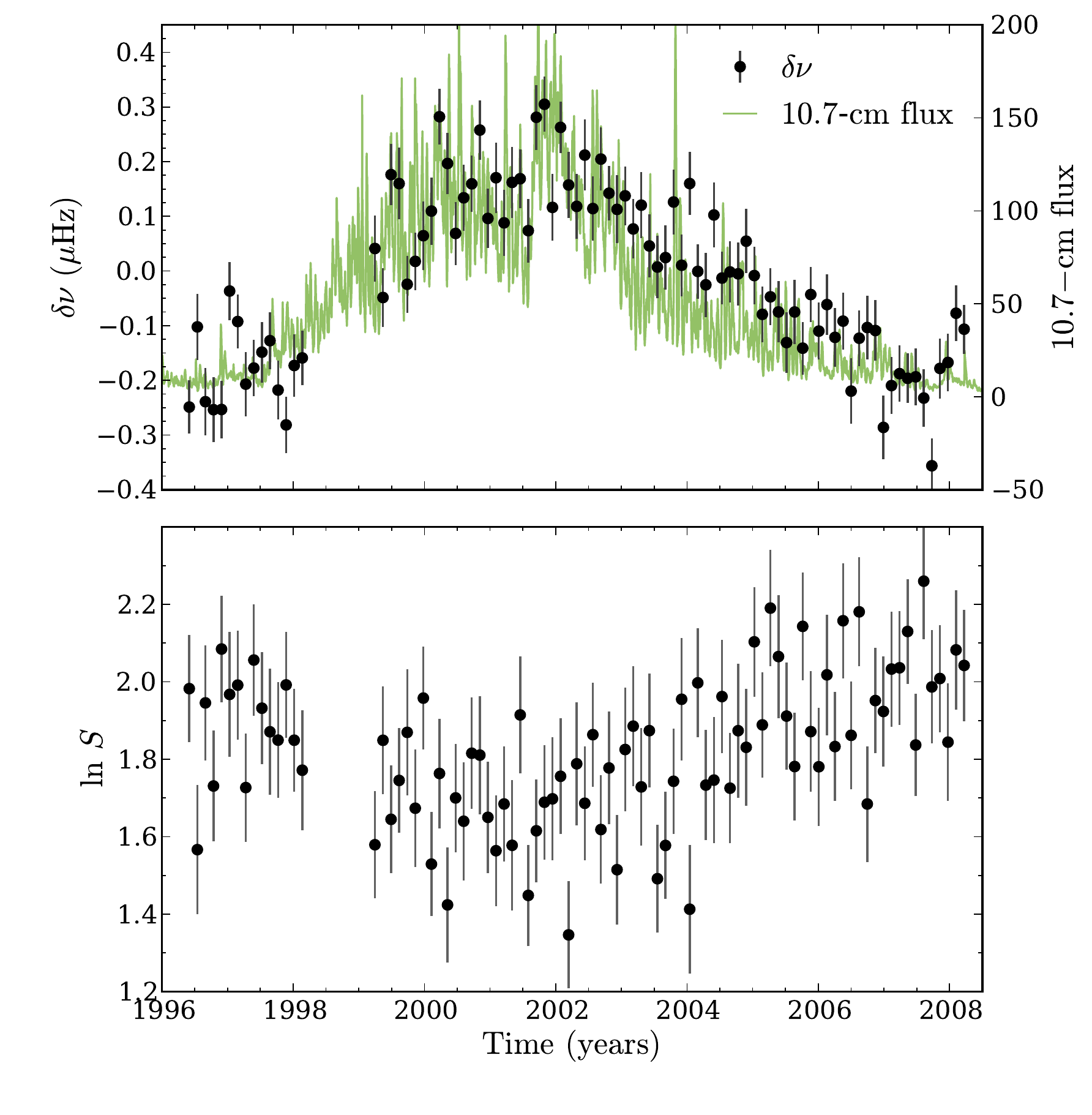}\vspace{-0.7cm}
\caption{Mean frequency shifts (black; top) and logarithmic mode heights (bottom) obtained from the peak-bagging analysis of the VIRGO/SPM data. For comparison, the daily values of the 10.7-cm flux are shown in top panel in light green. The time interval between 1998 and 1999 corresponds to the so-called SOHO vacations \citep[][]{Appourchaux2005}.}\label{fig:fshiftsvirgo}\vspace{-1.5cm}
\end{figure}

With the frequencies of the acoustic modes in hand, we then compute the observed frequency shifts. We follow the procedure described in Section~\ref{sec:fshifts} with the only difference being the the inertia ratio \citep[$Q_{nl}$;][]{Christensen-Dalsgaard1991}, which is not neglected in the solar case, i.e. we estimate the mean frequency shifts and the respective uncertainties as \citep[][]{Chaplin2007,Tripathy2007}\vspace{-0.1cm}
\begin{equation}
\delta\nu(t)=\dfrac{\sum_{nl}Q_{nl}\delta\nu_{nl}(t)/\sigma^2_{nl}(t)}{\sum_{nl}Q_{nl}/\sigma^2_{nl}(t)},\vspace{-0.1cm}
\end{equation}
\begin{equation}
\sigma(t)=\left(\sum_{nl}\dfrac{Q_{nl}}{\sigma_{nl}^2(t)}\right)^{-1/2}.
\end{equation}
The top panel of Figure~\ref{fig:fshiftsvirgo} shows the good agreement between the behavior of the frequency shifts over solar cycle~23 and the 10.7-cm flux (from NOAA/NGDC\footnote{http://www.ngdc.noaa.gov} - National Geophysical Data Center, part of the National Oceanic and Atmospheric Administration). These results are also consistent with those in the literature \citep[e.g.][]{Chaplin2007,Jain2009,Tripathy2007,Tripathy2011}. This shows that the peak-bagging tool is able to successfully recover the time behavior of the solar acoustic frequencies. 

The bottom panel of Figure~\ref{fig:fshiftsvirgo} shows the evolution of the weighted average of the logarithmic mode heights obtained with the peak-bagging tool. As expected, the mode heights show a behavior opposite to that of the frequencies and decrease towards the solar maximum.

\begin{figure}[h!]\centering
\includegraphics[width=\hsize]{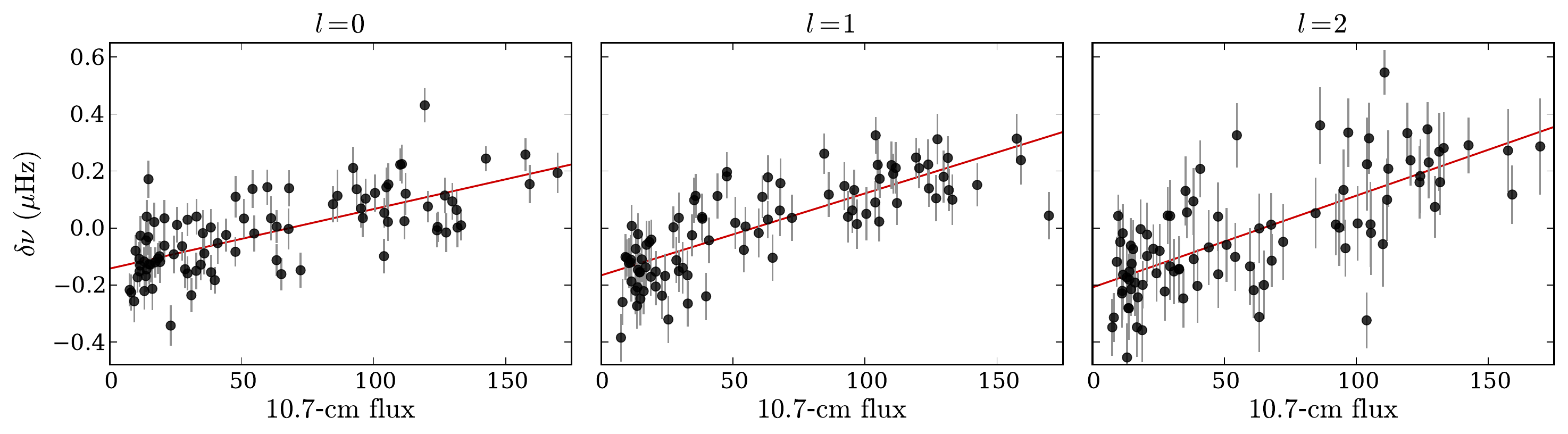}
\caption{Frequency shifts for the VIRGO/SPM data as a function of the 10.7-cm flux for the individual angular degrees: $l=0$, $l=1$, and $l=2$ (from left to right). The black solid line, the red dotted line, and the blue dashed line show the respective best linear fits.}\label{fig:fluxvirgo}
\end{figure}

\begin{figure}[h!]\centering
\includegraphics[width=0.48\hsize]{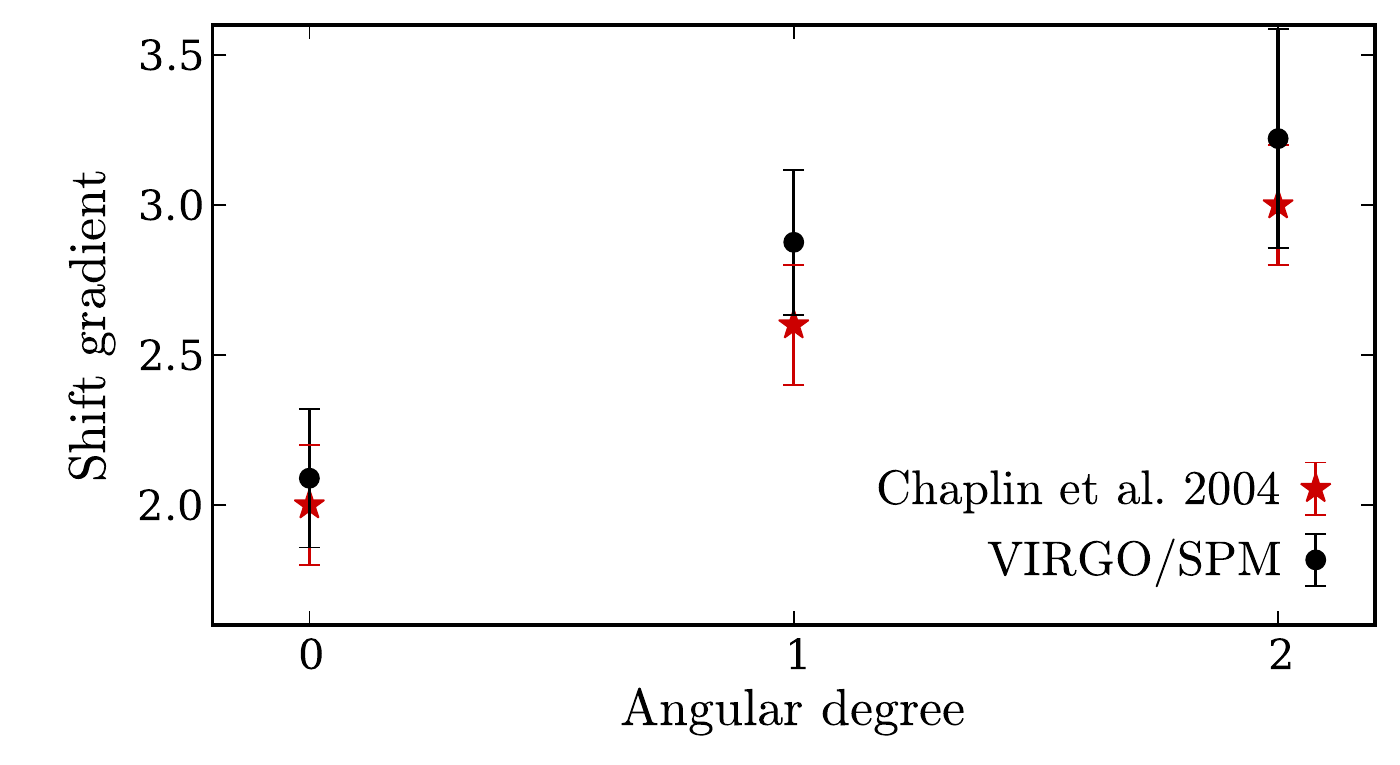}
\caption{Frequency shift per unit change in activity, shift gradient (black), as a function of the angular degree for VIRGO/SPM data. For comparison, the red stars indicate the values found by \citet{Chaplin2004}.}\label{fig:gradvirgo}
\end{figure}

Figure~\ref{fig:fluxvirgo} shows the frequency shifts for the individual angular degrees as a function of the observed 10.7-cm flux and the respective best linear fit (note that, as a first approximation, we neglect the magnetic hysteresis). The slope or shift gradient corresponds to the frequency shift per unit change in activity and is shown in Figure~\ref{fig:gradvirgo}. Our results are consistent with those from \citet{Chaplin2004}. For low-degree modes ($l\leq3$), the most import contribution for the degree dependence of the frequency shifts is the latitudinal distribution of the surface magnetic field. Thus, modes that are more sensitive to the low latitudes show larger frequency shifts \citep[e.g.][]{Chaplin2004,Broomhall2012,Salabert2015}.

\subsection{Artificial BiSON data}

The efficiency of the peak-bagging tool was also tested with an artificial BiSON data set that is used as control test. The artificial data simulate the granulation component, the acoustic oscillations (affected by the solar activity), and the photon shot-noise.

\begin{figure}[h]\centering
\includegraphics[width=0.51\hsize]{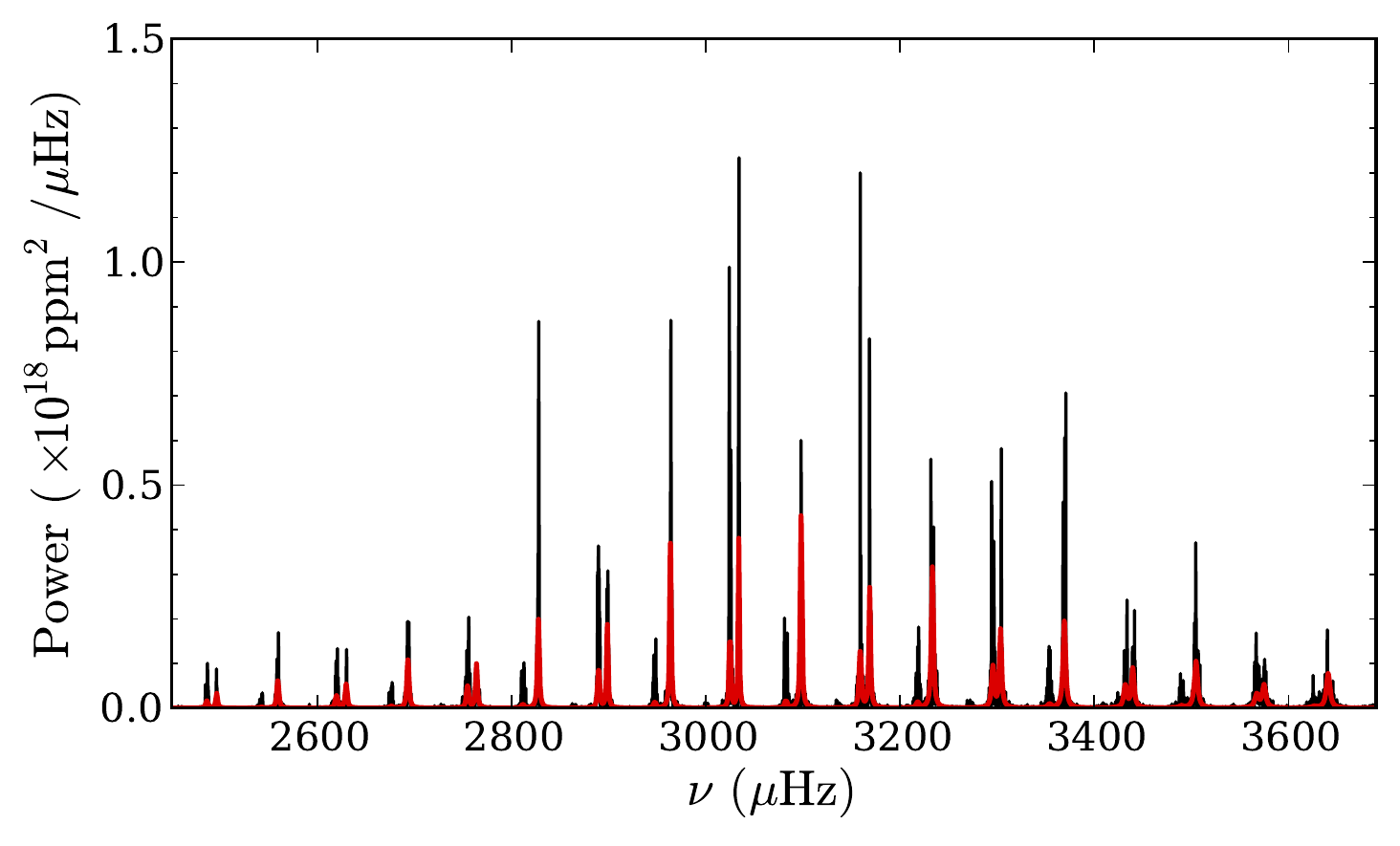}\vspace{-0.2cm}
\caption{Power density spectrum (black) for a 90-d sub-series for the artificial BiSON time-series. The red line shows the best fit to the p-modes obtained with the peak-bagging tool.}\label{fig:powersbison}
\end{figure}

\begin{figure}[h]\centering
\includegraphics[width=0.62\hsize]{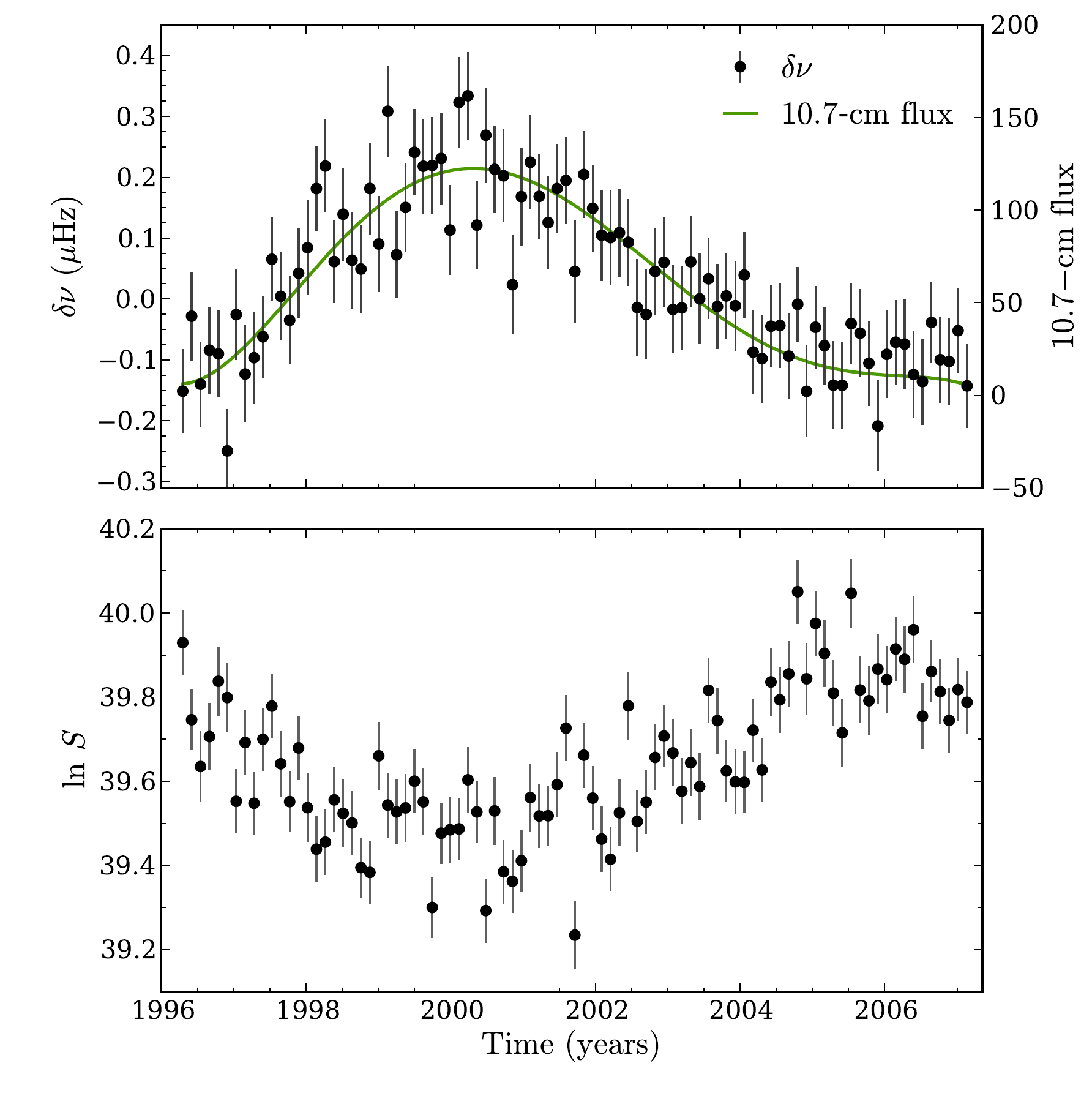}\vspace{-0.2cm}
\caption{Average frequency shifts (black; top) and logarithmic mode heights (bottom) obtained from the artificial BiSON time-series using our peak-bagging tool. For comparison, the input 10.7-cm flux is shown in green in the top panel.}\label{fig:fshiftsbison}\vspace{2cm}
\end{figure}

\begin{figure}\centering
\includegraphics[width=\hsize]{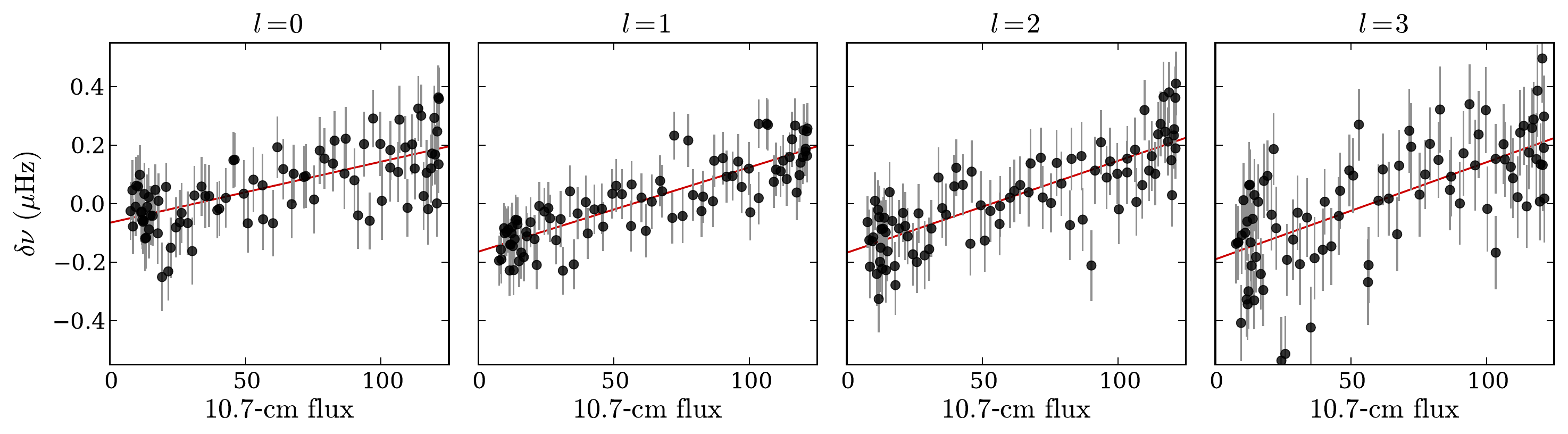}
\caption{Frequency shifts for artificial BiSON data (black) as a function of the input 10.7-cm flux for the individual angular degrees: $l=0$, $l=1$, $l=2$, and $l=3$ (from left to right). The red lines show the best fits to the data.}\label{fig:fluxbison}
\end{figure}

\begin{figure}\centering
\includegraphics[width=.48\hsize]{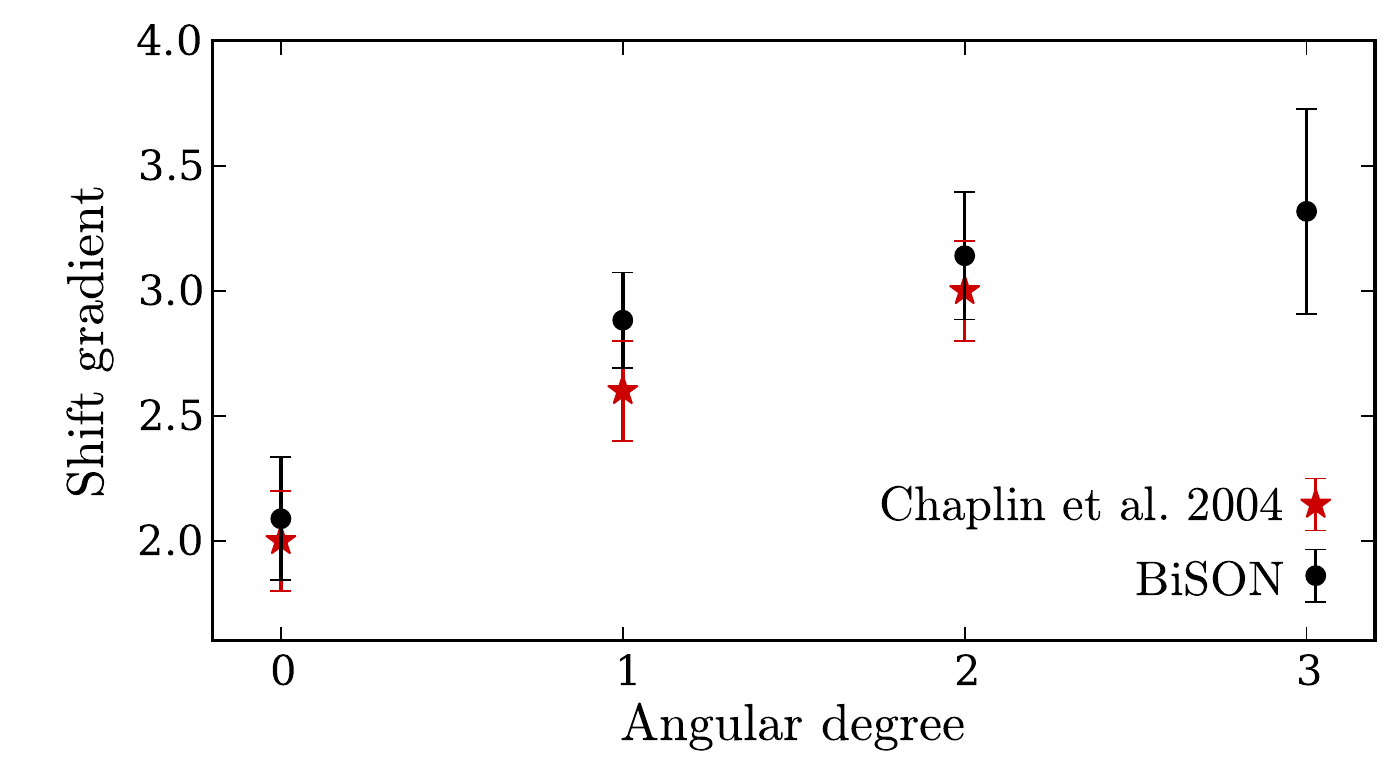}
\caption{Frequency shift per unit change in activity, shift gradient, as a function of the angular degree for artificial BiSON data (black). The red stars show the shift gradients found by \citet{Chaplin2004}.}\label{fig:gradbison}\vspace{8cm}
\end{figure}

Similarly to the previous section, the original time-series is split in 90-d segments overlapped by 45 days. Following the same procedure as in Section~\ref{sec:virgo} and assuming the respective background model (that accounts only for the granulation and noise components), we peak-bag the power density spectra obtained from each sub-series.
Figure~\ref{fig:powersbison} shows the best fit to the p-modes for a given segment of the synthetic data set. Then, we compute the temporal variations in the acoustic frequencies and mode heights and compare them with the input activity level (Figures~\ref{fig:fshiftsbison}-\ref{fig:gradbison}). These results further confirm that our peak-bagging tool is able to successfully recover the magnetic signature on the seismic data. 

\clearpage

%!TEX root = signatures-magnetic-activity.tex
\section{Peak-bagging analysis}\label{sec:peakapp}

This appendix presents the results for the {\it Kepler} stars in the sample. The results (both figures and table) for KIC~8006161 are shown in the main text, as well as the figures for KIC~5184732, KIC~7970740, KIC~8379927, KIC~9414417, and KIC~10644253. Note that for a number of stars it is not possible to measure the photometric activity proxy $S_\text{ph}$. Due to the large error bars and for representation purpose, for some stars we show the frequency shifts from the cross-correlation method were obtained with 180-d sub-series. Those cases are indicated.

In order to measure the correlation between the different quantities, we compute the cross-correlation function (CCF) between frequency shifts and logarithmic mode heights, frequency shifts and interpolated photometric activity proxy, and frequency shifts and characteristic granulation timescale. In the table 3, we consider up to a temporal lag of $\pm 90$ days. We list the maximum absolute cross-correlation value and the corresponding lag. For reference, the $95\%$ significance level is shown in the last column.%\ref{tab:CCFKIC}

\startlongtable
\begin{deluxetable*}{c|cc|cc|cc|c}\vspace{-0.5cm}
\tablecaption{Summary of the results from the cross-correlation function between frequency shifts and logarithmic mode heights (columns 2 and 3), frequency shifts and interpolated photometric activity proxy (columns 4 and 5), and frequency shifts and granulation timescale (columns 6 and 7). Column 1 lists the KIC numbers of the targets. The CCF value (columns 2, 4, and 6) correspond to the maximum absolute CCF value within the temporal lags $\pm 90$ days, and the respective lag is shown in columns 3, 5, and 7. For reference, the $95\%$ significance levels are in column 8.}\label{tab:CCFKIC}
\tablewidth{\hsize}
\tabletypesize{\fontsize{9}{6.8}\selectfont}
\tablehead{
\multicolumn1c{KIC}& \multicolumn2c{$\delta\nu$ vs. $\ln S$} & \multicolumn2c{$\delta\nu$ vs. $\text{S}_\text{ph}$}	 & \multicolumn2c{$\delta\nu$ vs. $\tau_\text{gran}$}& \colhead{$95\%$}\\
\multicolumn1c{}&\colhead{CCF value}&\multicolumn1c{lag}&\colhead{CCF value}&\multicolumn1c{lag}&\colhead{CCF value}&\multicolumn1c{lag}& \colhead{significance level}
} 
\startdata
1435467 & -0.603 & 90 & -0.266 & -90 & 0.380 & 0 & $\pm$0.400\\ 
2837475 & -0.149 & 45 & -0.217 & 0 & 0.344 & 90 & $\pm$0.400\\ 
3425851 & -0.557 & 90 & -- & -- & 0.899 & 0 & $\pm$0.980\\ 
3427720 & -0.473 & 0 & 0.262 & 0 & 0.424 & 0 & $\pm$0.400\\ 
3456181 & 0.468 & 0 & -- & -- & -0.526 & 45 & $\pm$0.524\\ 
3544595 & 0.153 & 90 & -- & -- & -0.207 & -90 & $\pm$0.358\\ 
3632418 & 0.270 & -90 & -0.457 & -45 & 0.356 & -90 & $\pm$0.400\\ 
3656476 & 0.435 & 90 & 0.365 & -90 & -0.190 & 45 & $\pm$0.400\\ 
3735871 & -0.352 & -90 & 0.347 & 0 & -0.501 & 45 & $\pm$0.400\\ 
4141376 & -0.374 & 0 & -- & -- & 0.159 & -45 & $\pm$0.418\\ 
4349452 & 0.333 & 0 & -- & -- & -0.566 & -90 & $\pm$0.400\\ 
4914423 & 0.371 & 90 & -- & -- & 0.272 & -90 & $\pm$0.438\\ 
4914923 & 0.281 & 45 & 0.322 & 45 & -0.527 & 45 & $\pm$0.400\\ 
5184732 & -0.482 & 90 & 0.689 & 45 & -0.379 & 0 & $\pm$0.438\\ 
5773345 & 0.586 & -90 & 0.371 & -45 & 0.470 & -90 & $\pm$0.591\\ 
5866724 & 0.279 & 45 & -- & -- & -0.467 & 45 & $\pm$0.370\\ 
5950854 & 0.662 & 45 & -- & -- & -0.514 & -90 & $\pm$0.566\\ 
6106415 & -0.508 & 90 & -- & -- & -0.442 & 45 & $\pm$0.428\\ 
6116048 & 0.312 & -90 & 0.137 & 45 & -0.546 & 90 & $\pm$0.400\\ 
6225718 & 0.496 & 90 & -- & -- & 0.444 & -45 & $\pm$0.418\\ 
6278762 & 0.987 & 0 & -- & -- & -0.961 & 45 & $\pm$0.980\\ 
6508366 & 0.663 & 0 & -0.305 & 45 & 0.368 & 0 & $\pm$0.400\\ 
6521045 & 0.491 & 90 & -- & -- & -0.318 & -90 & $\pm$0.370\\ 
6603624 & 0.532 & 90 & -- & -- & -0.432 & -90 & $\pm$0.400\\ 
6679371 & 0.417 & 0 & -0.502 & 0 & -0.154 & -90 & $\pm$0.400\\ 
6933899 & -0.637 & -45 & -- & -- & 0.442 & 45 & $\pm$0.400\\ 
7103006 & -0.468 & 0 & 0.325 & -45 & 0.574 & 45 & $\pm$0.400\\ 
7106245 & -0.329 & 45 & -- & -- & 0.413 & -90 & $\pm$0.418\\ 
7206837 & 0.278 & 45 & -0.192 & 90 & 0.170 & 90 & $\pm$0.400\\ 
7296438 & 0.617 & 45 & -0.534 & -90 & 0.625 & 0 & $\pm$0.620\\ 
7510397 & 0.664 & 0 & -- & -- & 0.257 & 90 & $\pm$0.438\\ 
7670943 & 0.526 & 0 & -- & -- & 0.334 & 45 & $\pm$0.418\\ 
7680114 & 0.442 & 45 & -0.438 & 0 & -0.308 & -90 & $\pm$0.400\\ 
7771282 & 0.728 & 45 & 0.438 & 0 & -0.805 & 0 & $\pm$0.524\\ 
7871531 & 0.303 & -90 & 0.325 & 0 & 0.152 & 90 & $\pm$0.400\\ 
7940546 & -0.285 & -90 & -0.642 & -45 & 0.236 & 90 & $\pm$0.438\\ 
7970740 & -0.369 & 0 & 0.481 & -90 & -0.524 & 45 & $\pm$0.418\\ 
8006161 & -0.791 & 0 & 0.704 & 0 & 0.573 & 0 & $\pm$0.400\\ 
8077137 & -0.398 & 90 & -- & -- & -0.317 & 90 & $\pm$0.418\\ 
8150065 & -0.407 & 45 & -- & -- & -0.512 & 45 & $\pm$0.566\\ 
8179536 & 0.623 & 0 & -0.254 & 90 & -0.691 & 0 & $\pm$0.524\\ 
8228742 & 0.424 & 0 & 0.643 & 45 & 0.618 & -45 & $\pm$0.400\\ 
8292840 & -0.561 & -45 & -- & -- & -0.575 & 90 & $\pm$0.400\\ 
8379927 & 0.508 & 0 & 0.299 & 0 & 0.400 & -45 & $\pm$0.392\\ 
8394589 & -0.583 & 0 & -- & -- & -0.463 & -45 & $\pm$0.400\\ 
8424992 & -0.535 & 90 & -- & -- & 0.483 & 90 & $\pm$0.693\\ 
8478994 & -0.700 & -90 & -- & -- & 0.345 & 90 & $\pm$0.400\\ 
8494142 & -0.228 & -45 & -- & -- & -0.394 & -90 & $\pm$0.438\\ 
8694723 & -0.365 & 90 & 0.436 & -45 & -0.442 & 0 & $\pm$0.400\\ 
8760414 & -0.475 & 45 & -- & -- & -0.112 & -45 & $\pm$0.400\\ 
8866102 & -0.421 & 90 & -- & -- & 0.354 & 0 & $\pm$0.370\\ 
8938364 & -0.360 & 0 & -- & -- & -0.274 & 45 & $\pm$0.418\\ 
9025370 & 0.383 & 0 & 0.347 & 0 & -0.569 & 45 & $\pm$0.400\\ 
9098294 & -0.440 & 45 & 0.196 & -45 & -0.379 & 45 & $\pm$0.400\\ 
9139151 & 0.510 & 0 & 0.355 & 0 & 0.386 & 45 & $\pm$0.400\\ 
9139163 & -0.396 & 0 & 0.401 & 90 & 0.543 & 0 & $\pm$0.400\\ 
9206432 & 0.653 & 45 & 0.418 & 0 & -0.362 & -90 & $\pm$0.490\\ 
9353712 & 0.445 & -45 & 0.258 & 0 & 0.220 & -45 & $\pm$0.490\\ 
9410862 & 0.466 & 45 & 0.488 & 0 & 0.444 & 0 & $\pm$0.418\\ 
9414417 & 0.452 & -45 & 0.221 & 45 & 0.675 & 45 & $\pm$0.428\\ 
9592705 & 0.330 & 45 & -- & -- & 0.409 & -90 & $\pm$0.418\\ 
9812850 & -0.215 & -90 & 0.235 & -90 & 0.336 & 90 & $\pm$0.400\\ 
9955598 & -0.317 & -90 & 0.141 & -90 & 0.171 & 0 & $\pm$0.400\\ 
9965715 & -0.298 & 45 & -- & -- & 0.538 & 0 & $\pm$0.462\\ 
10068307 & -0.719 & 90 & -0.282 & 90 & 0.306 & -90 & $\pm$0.438\\ 
10079226 & -0.357 & 90 & 0.467 & 0 & -0.250 & 90 & $\pm$0.693\\ 
10162436 & 0.560 & 0 & 0.255 & -90 & -0.742 & 0 & $\pm$0.418\\ 
10454113 & -0.374 & 0 & -0.270 & -45 & 0.210 & 90 & $\pm$0.400\\ 
10516096 & -0.251 & -45 & -- & -- & -0.612 & 0 & $\pm$0.400\\ 
10586004 & 0.893 & 0 & -- & -- & 0.896 & 0 & $\pm$0.980\\ 
10644253 & -0.178 & 45 & 0.347 & 0 & 0.159 & 90 & $\pm$0.400\\ 
10666592 & 0.378 & 0 & -- & -- & -0.243 & 90 & $\pm$0.346\\ 
10730618 & -0.516 & 45 & -- & -- & -0.329 & 45 & $\pm$0.438\\ 
10963065 & -0.394 & 0 & 0.264 & 0 & -0.049 & 0 & $\pm$0.392\\ 
11081729 & -0.257 & 0 & 0.477 & -45 & -0.187 & 0 & $\pm$0.400\\ 
11253226 & 0.401 & -45 & -0.076 & -90 & -0.252 & -45 & $\pm$0.400\\ 
11295426 & -0.425 & 90 & -- & -- & -0.449 & 90 & $\pm$0.400\\ 
11401755 & 0.692 & 0 & -- & -- & -0.479 & -45 & $\pm$0.418\\ 
11772920 & 0.270 & -90 & -- & -- & 0.294 & 45 & $\pm$0.400\\ 
11807274 & 0.426 & 0 & -- & -- & 0.364 & -45 & $\pm$0.418\\ 
11904151 & 0.217 & -45 & -- & -- & 0.262 & -45 & $\pm$0.358\\ 
12009504 & 0.297 & 0 & 0.169 & 90 & 0.369 & 90 & $\pm$0.400\\ 
12069127 & -0.181 & -45 & 0.365 & 45 & -0.384 & 45 & $\pm$0.524\\ 
12069424 & -0.197 & -90 & -- & -- & 0.132 & 45 & $\pm$0.438\\ 
12069449 & 0.785 & 0 & -- & -- & -0.341 & 90 & $\pm$0.438\\ 
12258514 & -0.531 & -90 & 0.427 & 0 & 0.306 & -90 & $\pm$0.400\\ 
12317678 & -0.197 & -45 & -- & -- & 0.119 & 90 & $\pm$0.400\\
\enddata
\end{deluxetable*}
%\twocolumn

\begin{figure}[ht]
\includegraphics[width=\hsize]{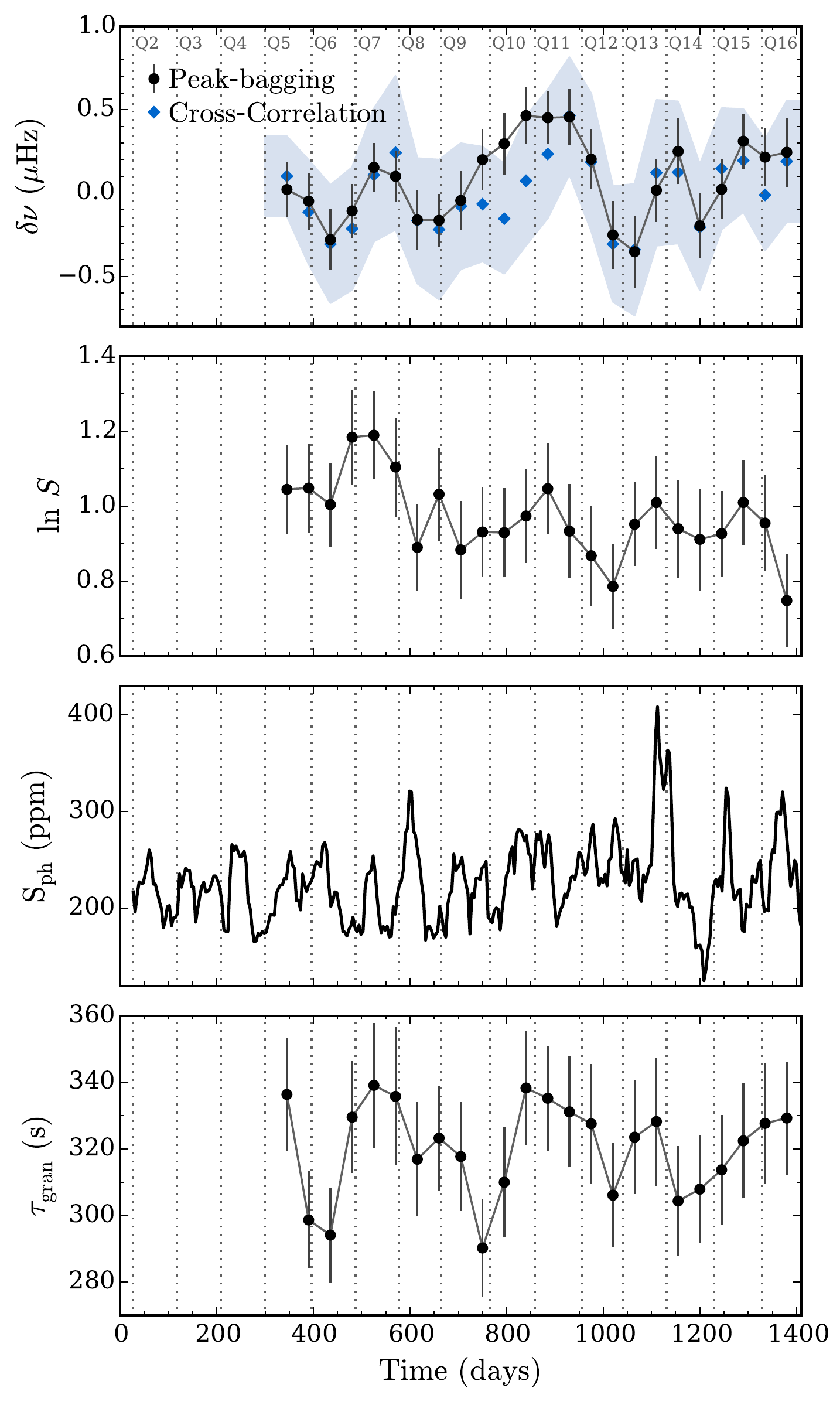}
\caption{Results for KIC 1435467 (same as in Figures~\ref{fig:fshKIC8006161}). From top to bottom, time-dependent mean frequency shifts, mode heights, photometric activity proxy, and characteristic granulation timescale. Results in Table~\ref{tab:1435467}.}\label{fig:1435467}
\end{figure}

\begin{figure}[ht]
\includegraphics[width=\hsize]{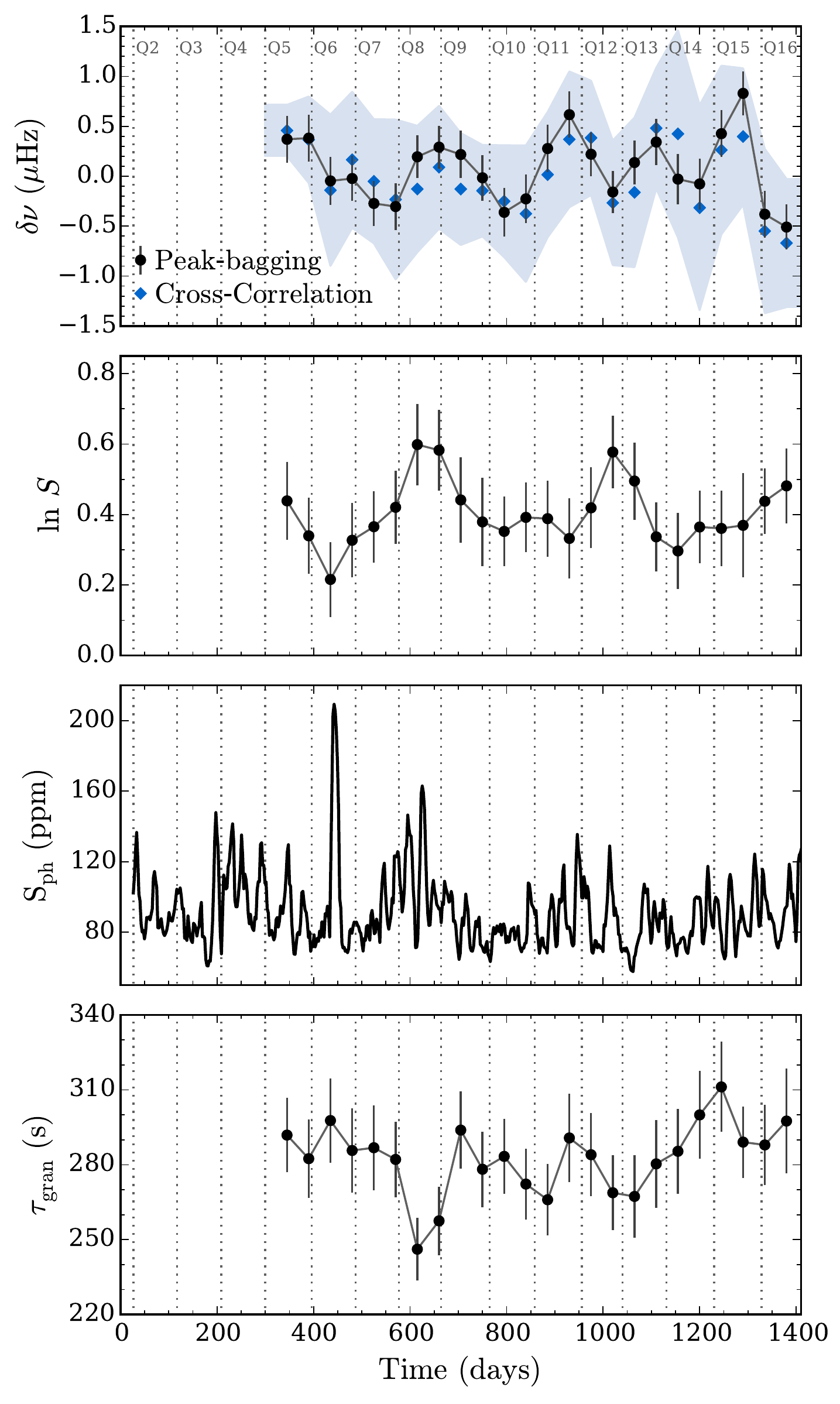}
\caption{Same as in Figure 26, but for KIC 2837475. Results in Table~\ref{tab:2837475}.}\label{fig:2837475}
\end{figure}

\begin{figure}[ht]
\includegraphics[width=\hsize]{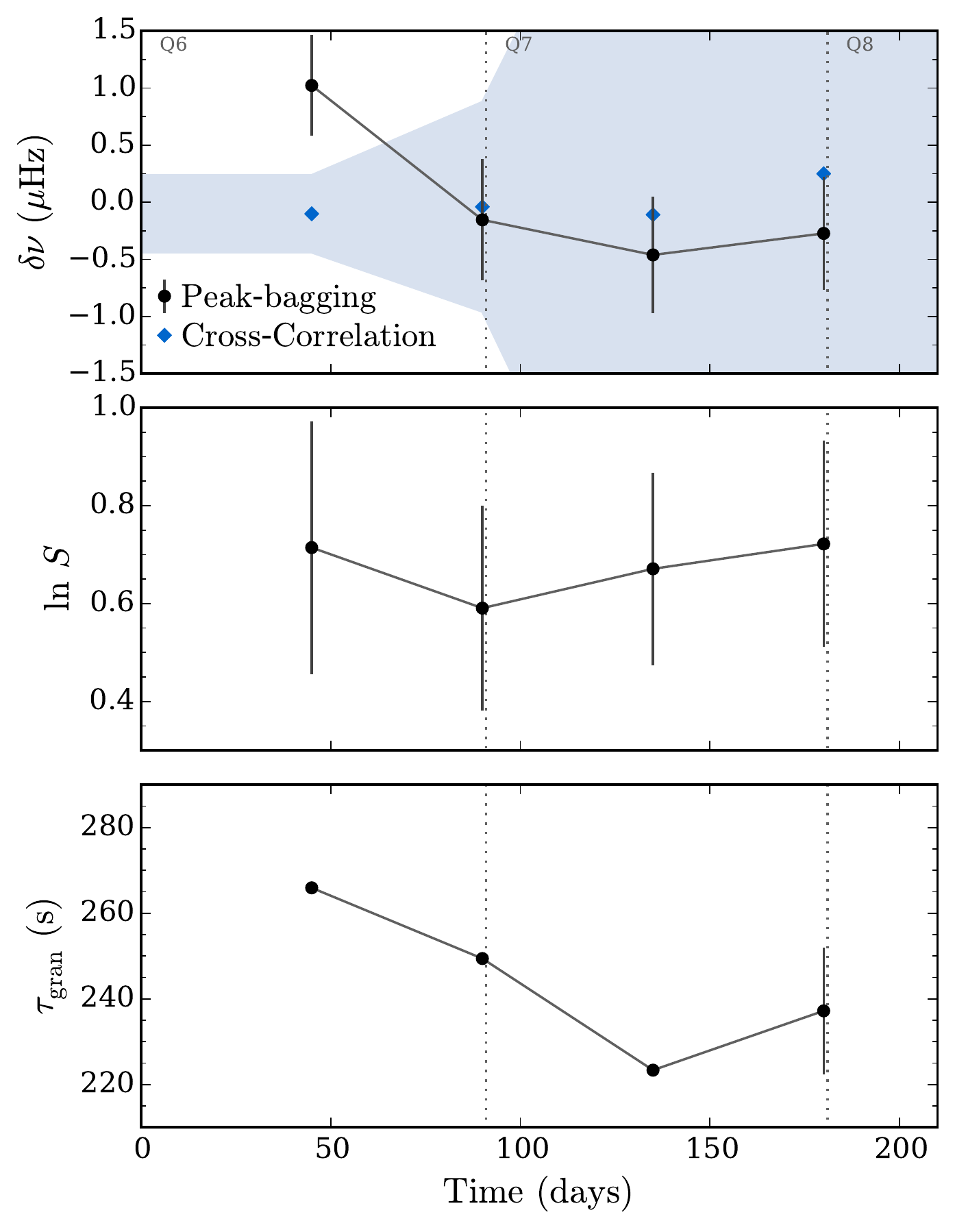}
\caption{Same as in Figure 26, but for KIC 3425851. Results in Table~\ref{tab:3425851}.}\label{fig:3425851}
\end{figure}

\begin{figure}[ht]
\includegraphics[width=\hsize]{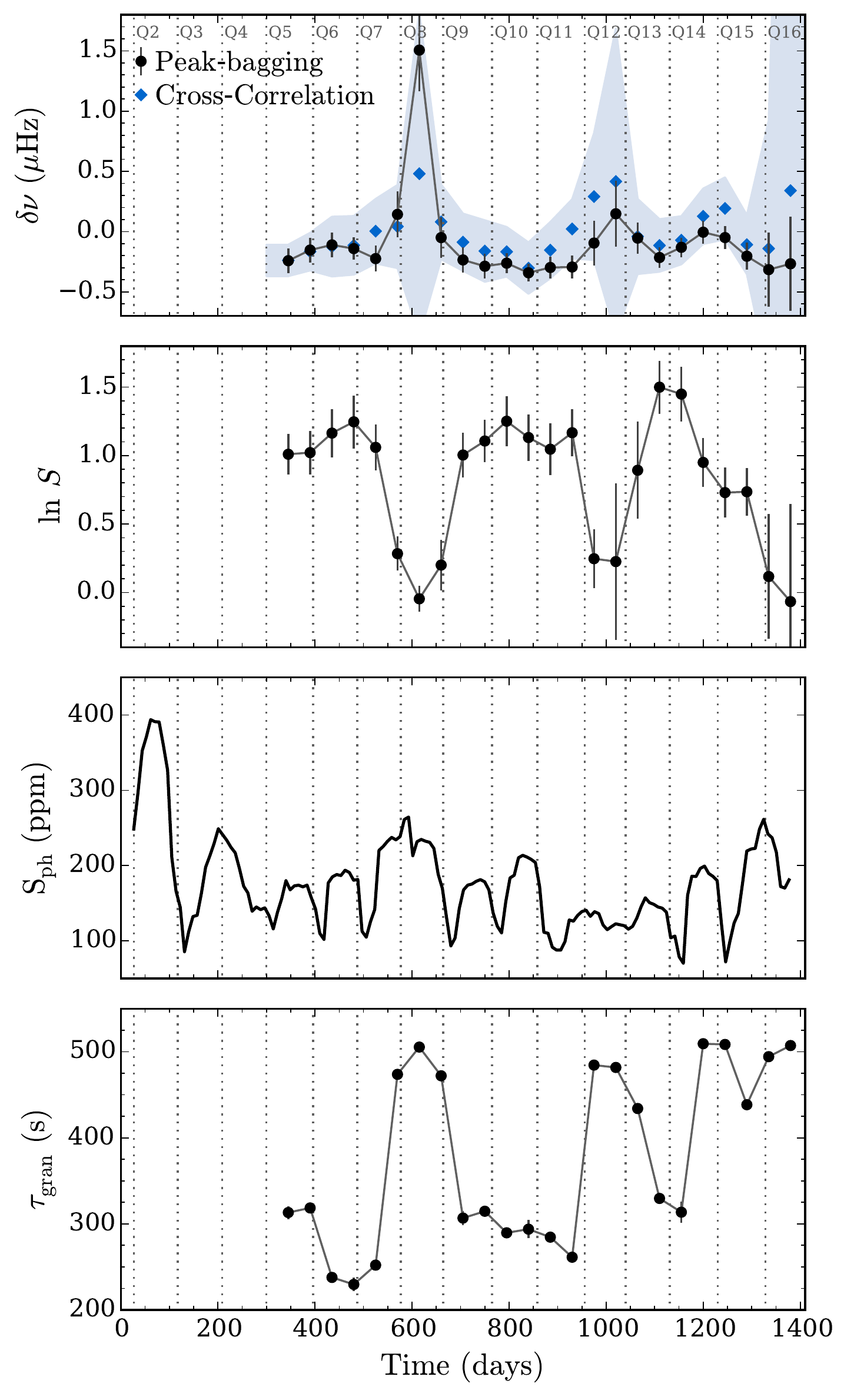}
\caption{Same as in Figure 26, but for KIC 3427720. Results in Table~\ref{tab:3427720}.}\label{fig:3427720}
\end{figure}

\begin{figure}[ht]
\includegraphics[width=\hsize]{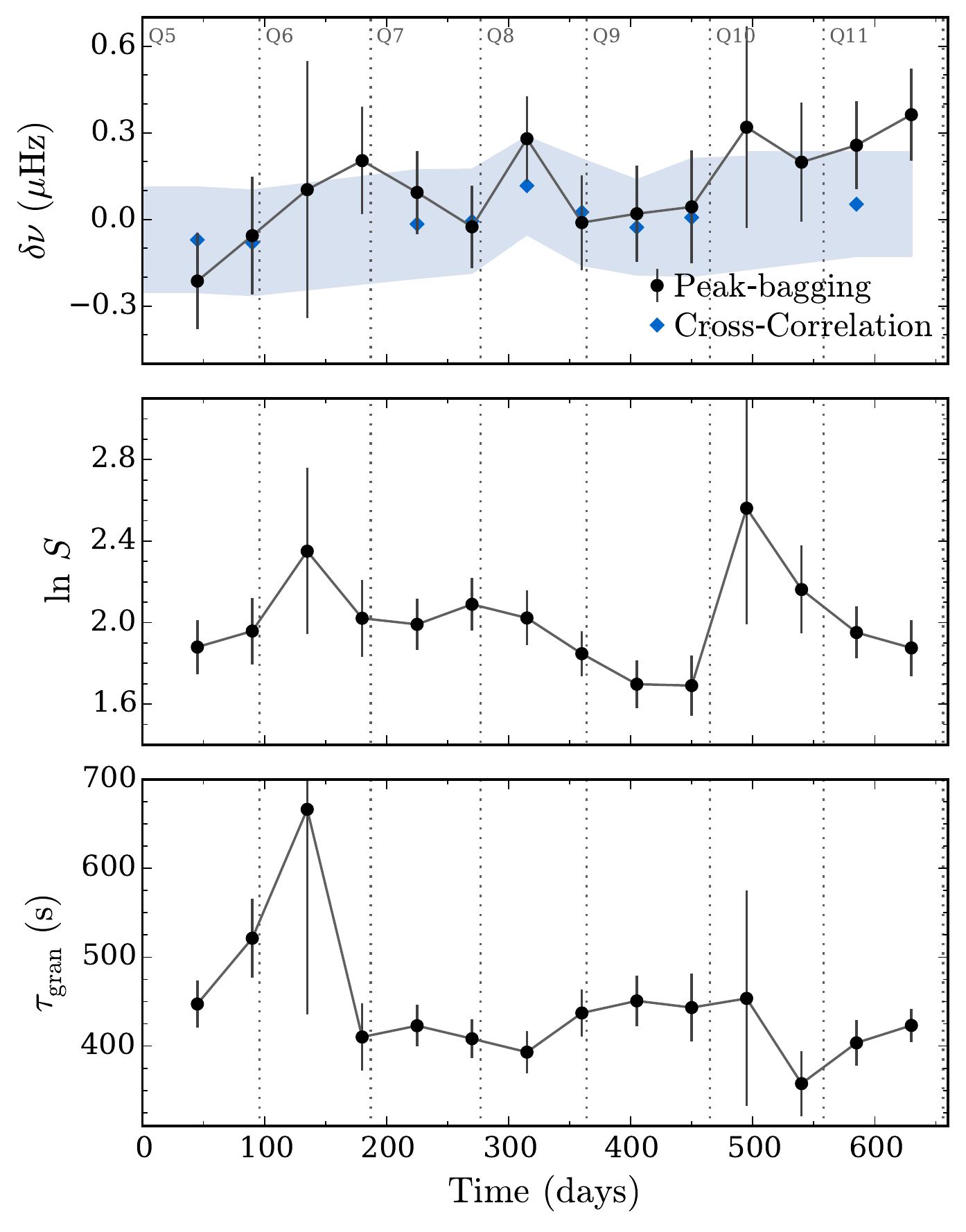}\vspace{-0.3cm}
\caption{Same as in Figure 26, but for KIC 3456181. Results in Table~\ref{tab:3456181}.}\label{fig:3456181}\vspace{-0.2cm}
\end{figure}

\begin{figure}[ht]
\includegraphics[width=\hsize]{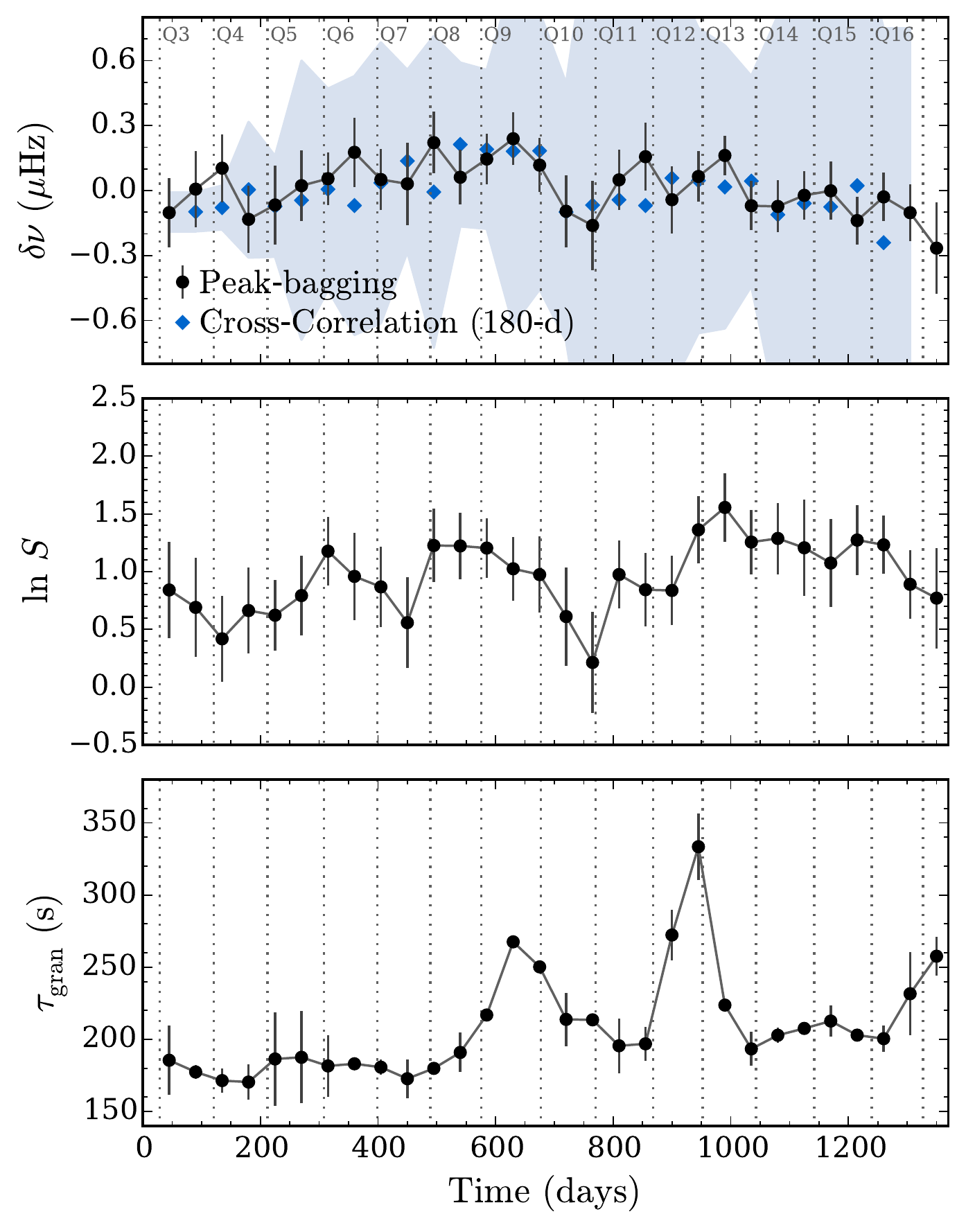}\vspace{-0.3cm}
\caption{Same as in Figure 26, but for KIC 3544595. The frequency shifts from the cross-correlation method were obtained with 180-d sub-series. Results in Table~\ref{tab:3544595}.}\label{fig:3544595}\vspace{-1.2cm}
\end{figure}

\begin{figure}[ht]
\includegraphics[width=\hsize]{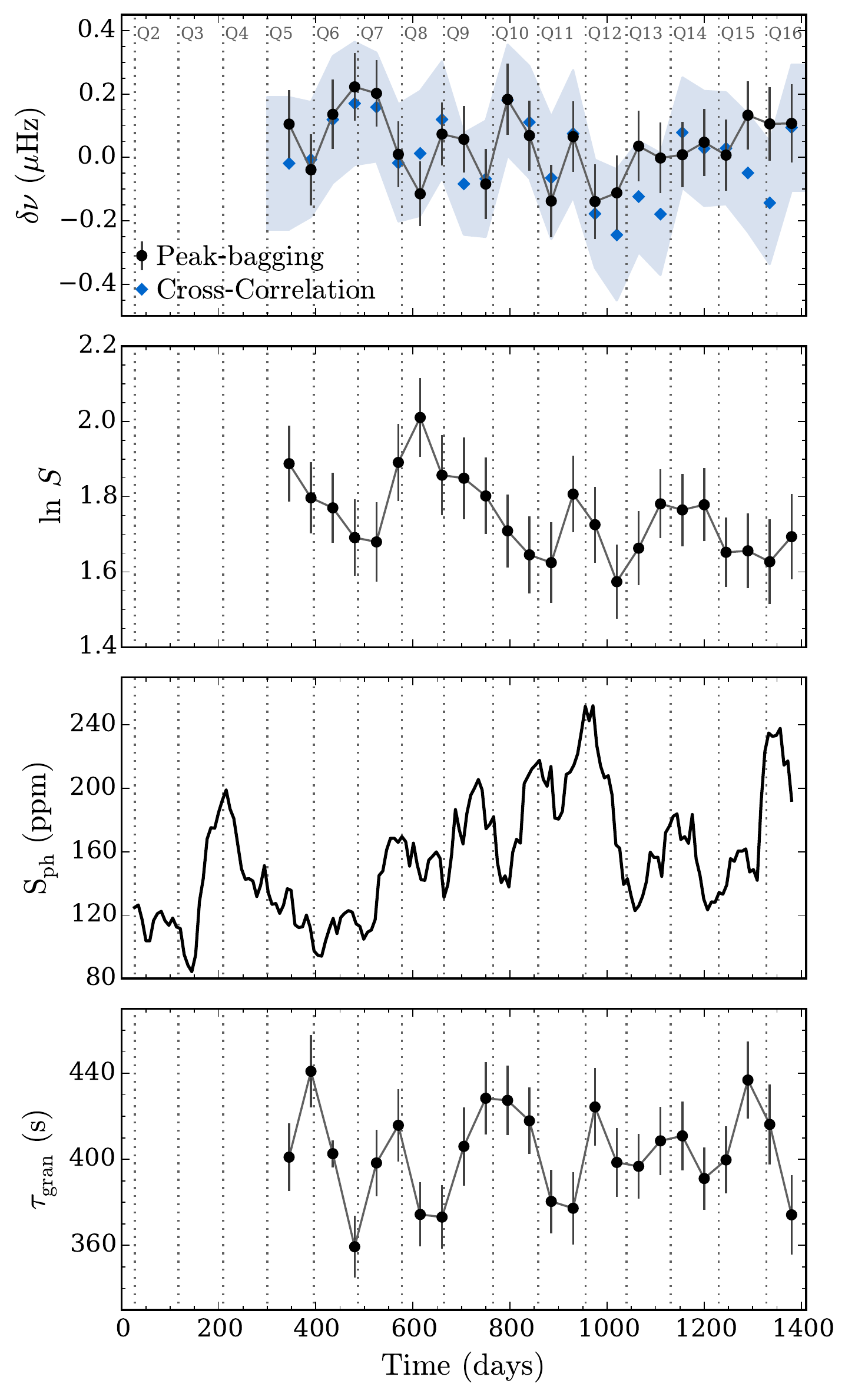}
\caption{Same as in Figure 26, but for KIC 3632418. Results in Table~\ref{tab:3632418}.}\label{fig:3632418}
\end{figure}

\begin{figure}[ht]
\includegraphics[width=\hsize]{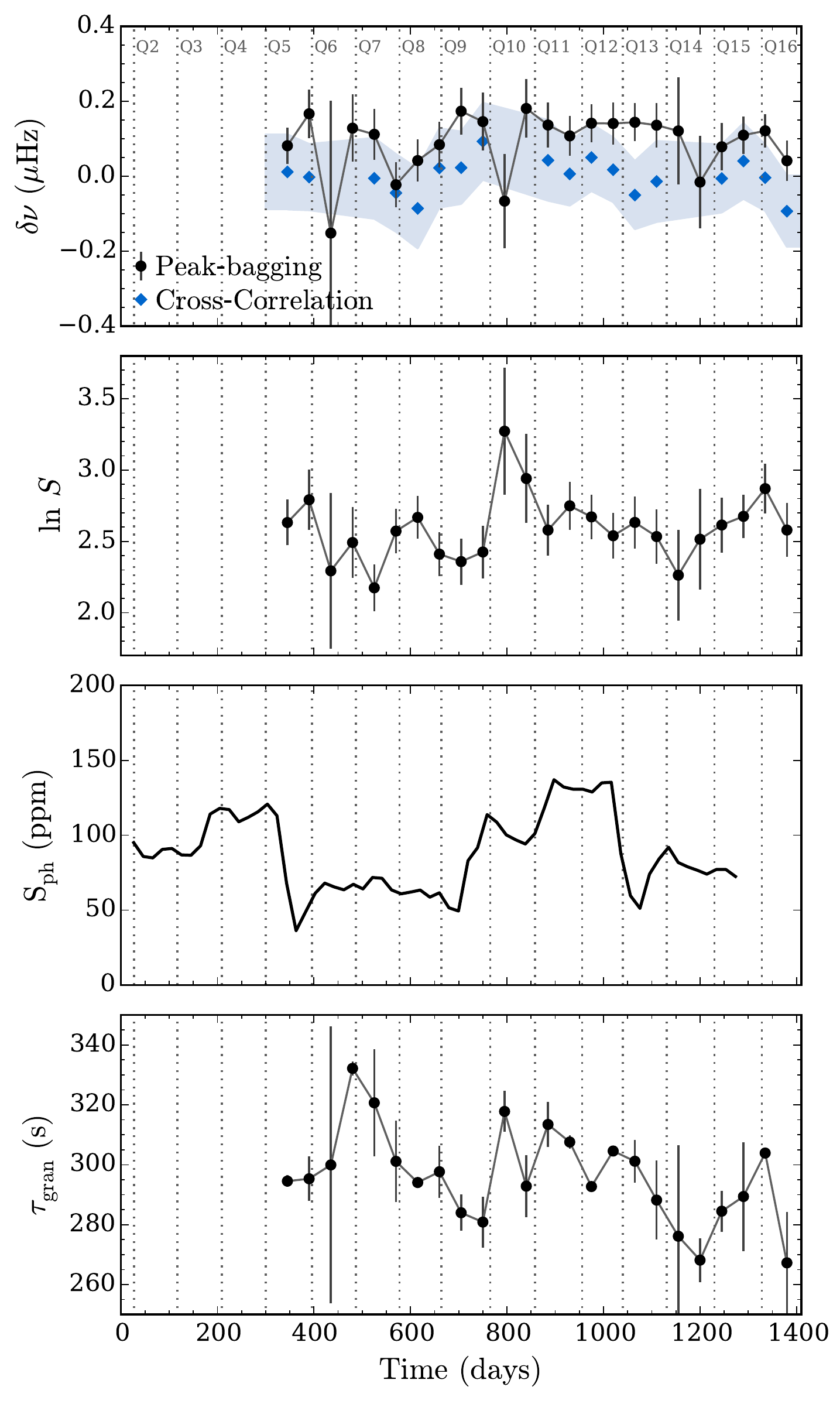}
\caption{Same as in Figure 26, but for KIC 3656476. Results in Table~\ref{tab:3656476}.}\label{fig:3656476}
\end{figure}

\begin{figure}[ht]
\includegraphics[width=\hsize]{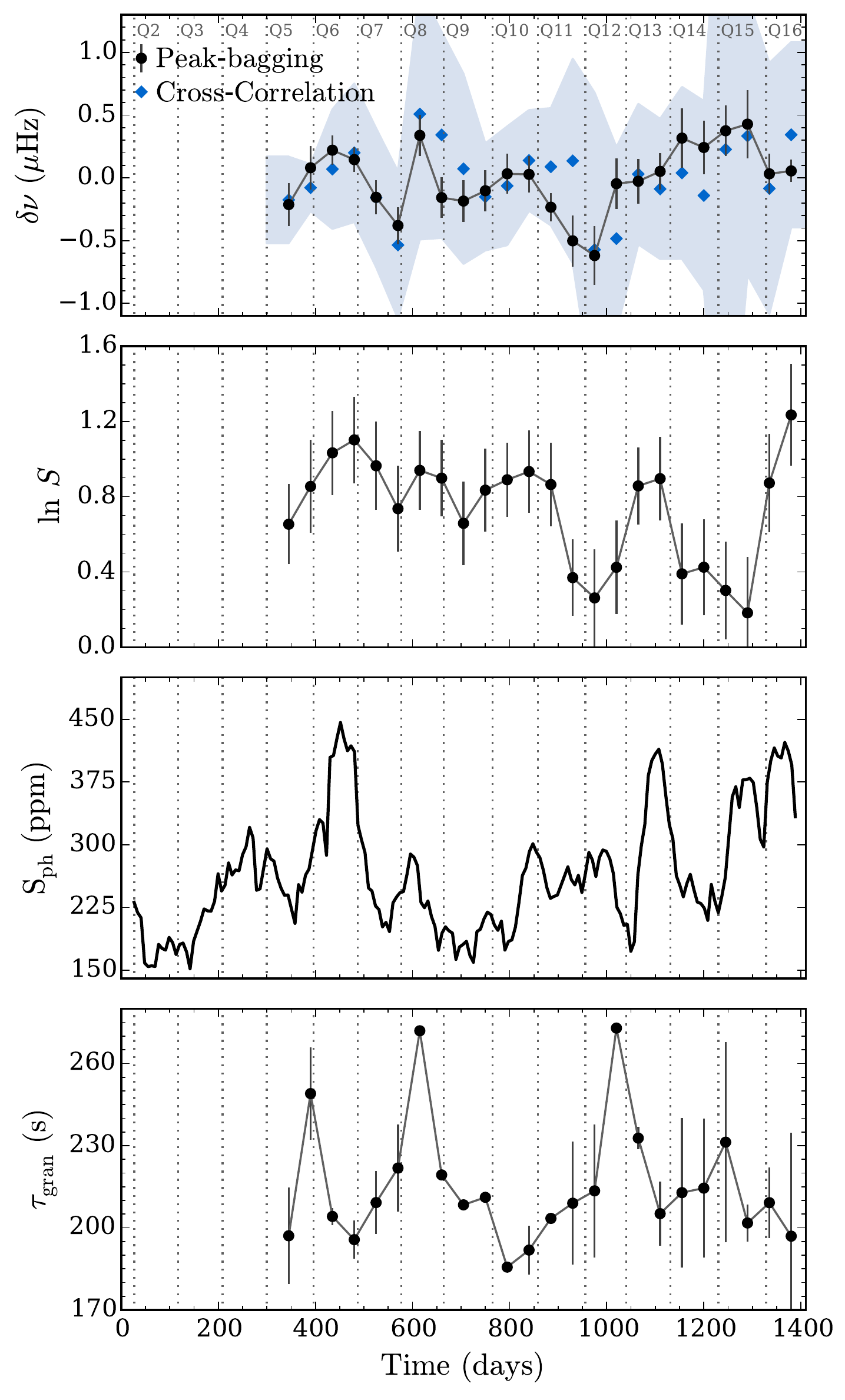}
\caption{Same as in Figure 26, but for KIC 3735871. Results in Table~\ref{tab:3735871}.}\label{fig:3735871}
\end{figure}

\begin{figure}[ht]
\includegraphics[width=\hsize]{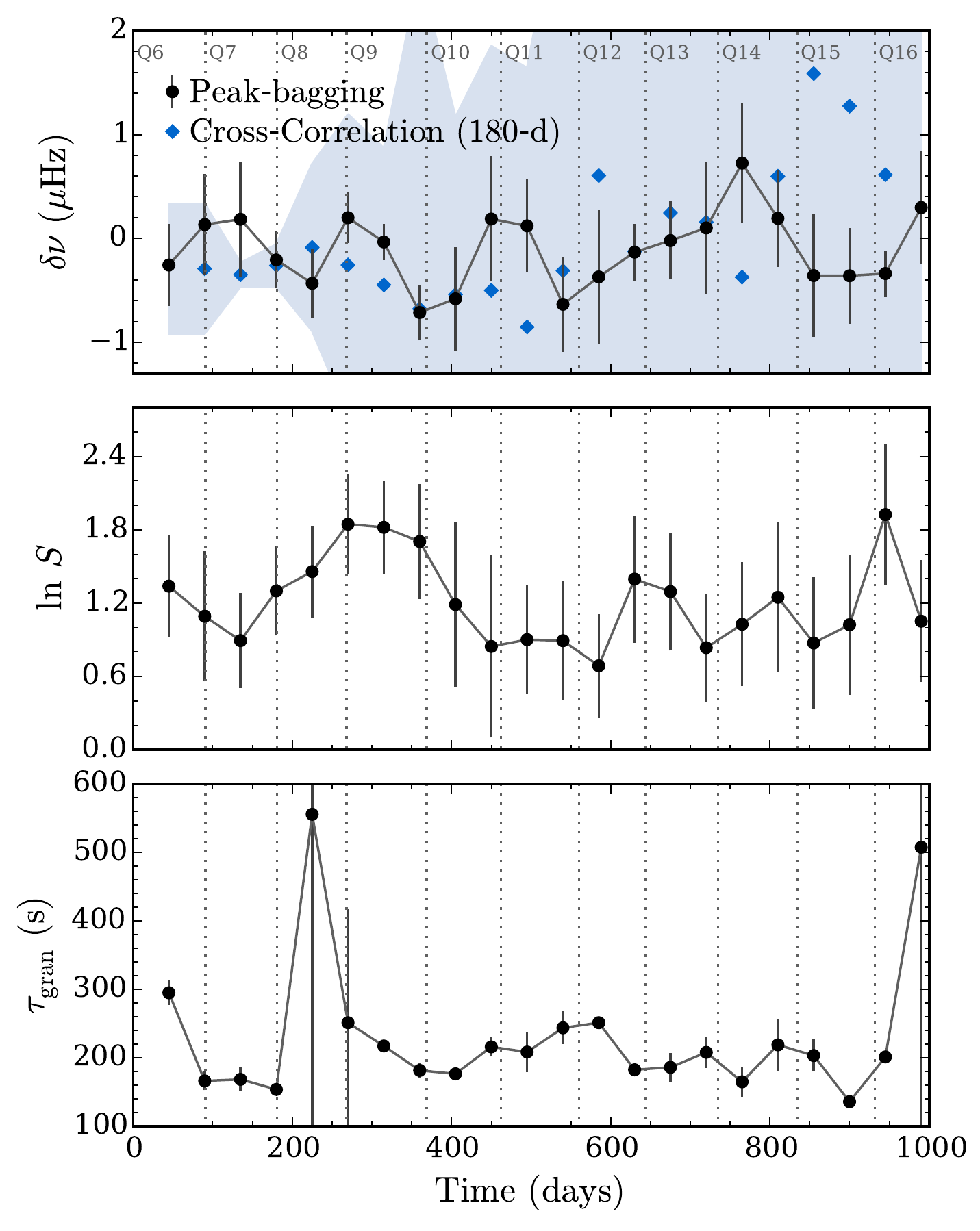}\vspace{-0.2cm}
\caption{Same as in Figure 26, but for KIC 4141376.The frequency shifts from the cross-correlation method were obtained with 180-d sub-series. Results in Table~\ref{tab:4141376}.}\label{fig:4141376}
\end{figure}

\begin{figure}[ht]
\includegraphics[width=\hsize]{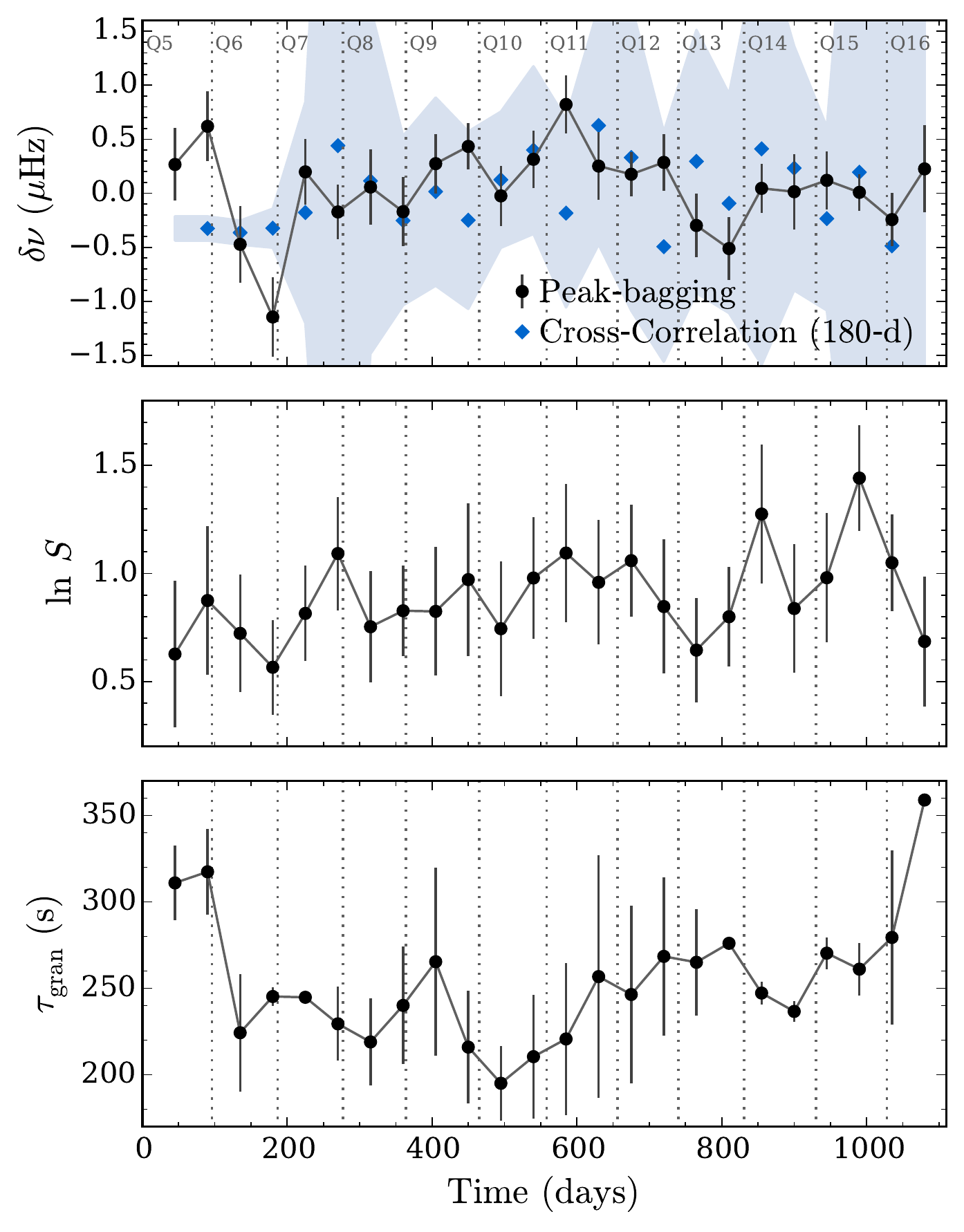}\vspace{-0.2cm}
\caption{Same as in Figure 26, but for KIC 4349452. The frequency shifts from the cross-correlation method were obtained with 180-d sub-series. Results in Table~\ref{tab:4349452}.}\label{fig:4349452}\vspace{-1.5cm}
\end{figure}

\begin{figure}[ht]
\includegraphics[width=\hsize]{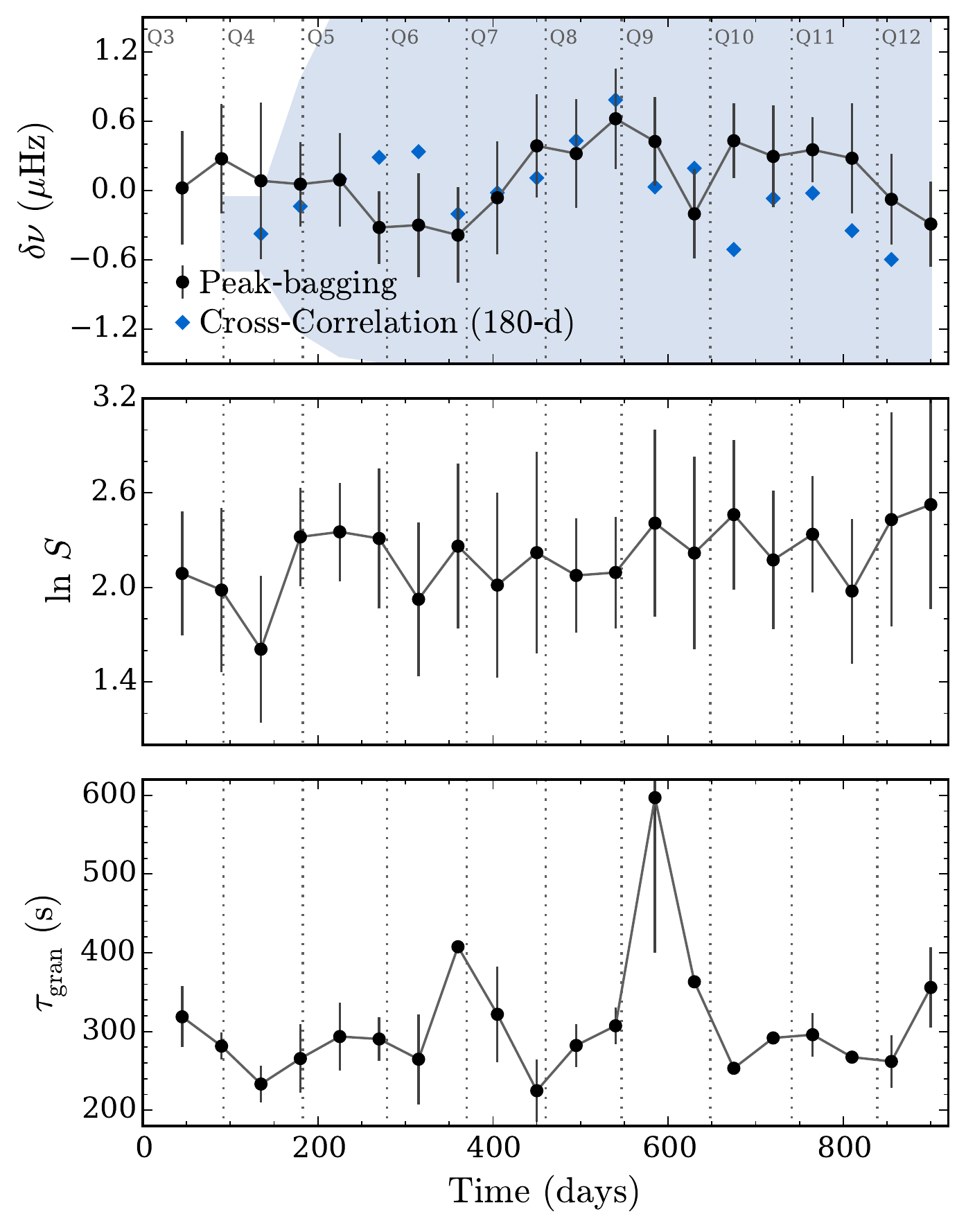}
\caption{Same as in Figure 26, but for KIC 4914423. The frequency shifts from the cross-correlation method were obtained with 180-d sub-series. Results in Table~\ref{tab:4914423}.}\label{fig:4914423}
\end{figure}

\begin{figure}[ht]
\includegraphics[width=\hsize]{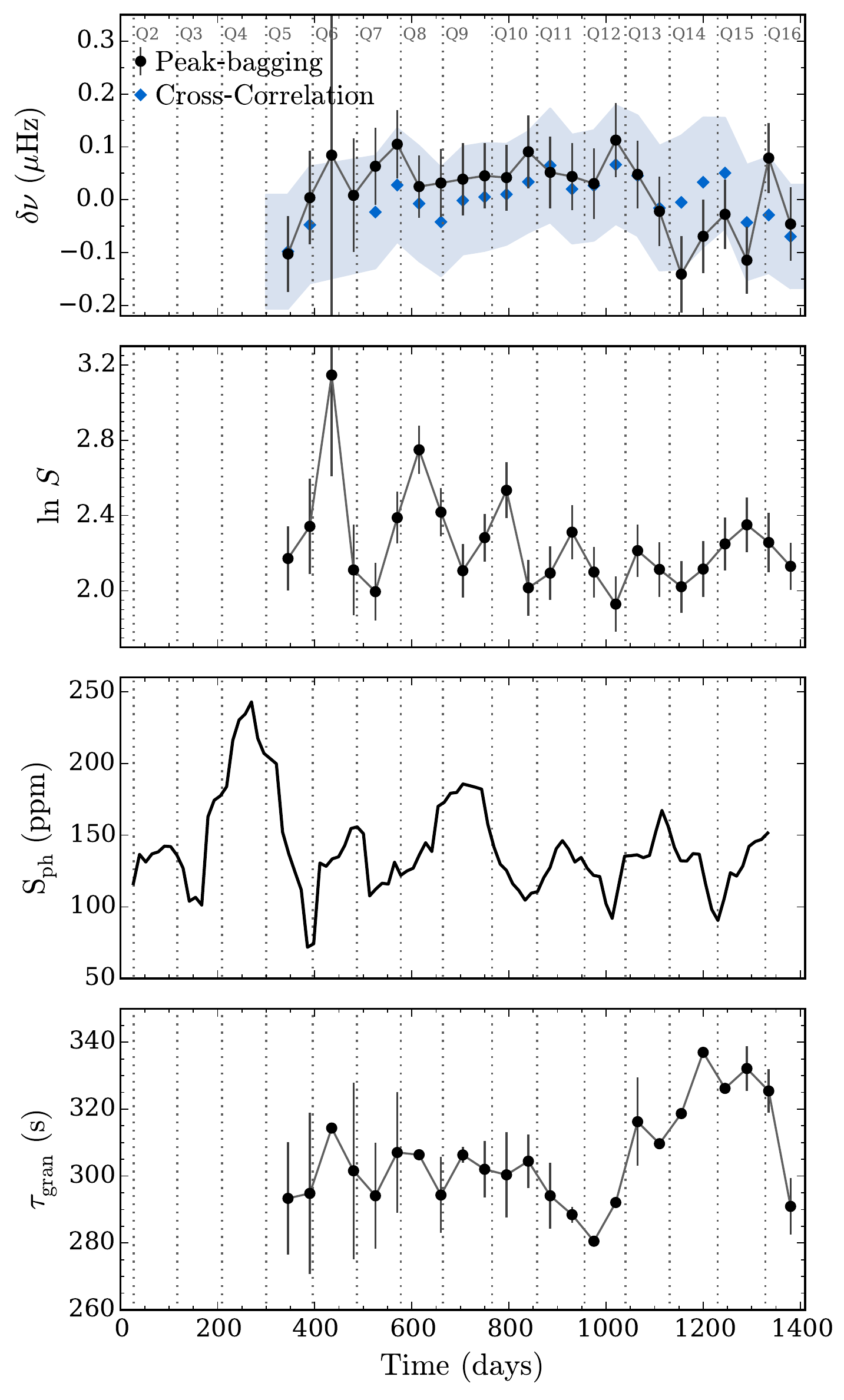}
\caption{Same as in Figure 26, but for KIC 4914923. Results in Table~\ref{tab:4914923}.}\label{fig:4914923}
\end{figure}

\begin{figure}[ht]
\includegraphics[width=\hsize]{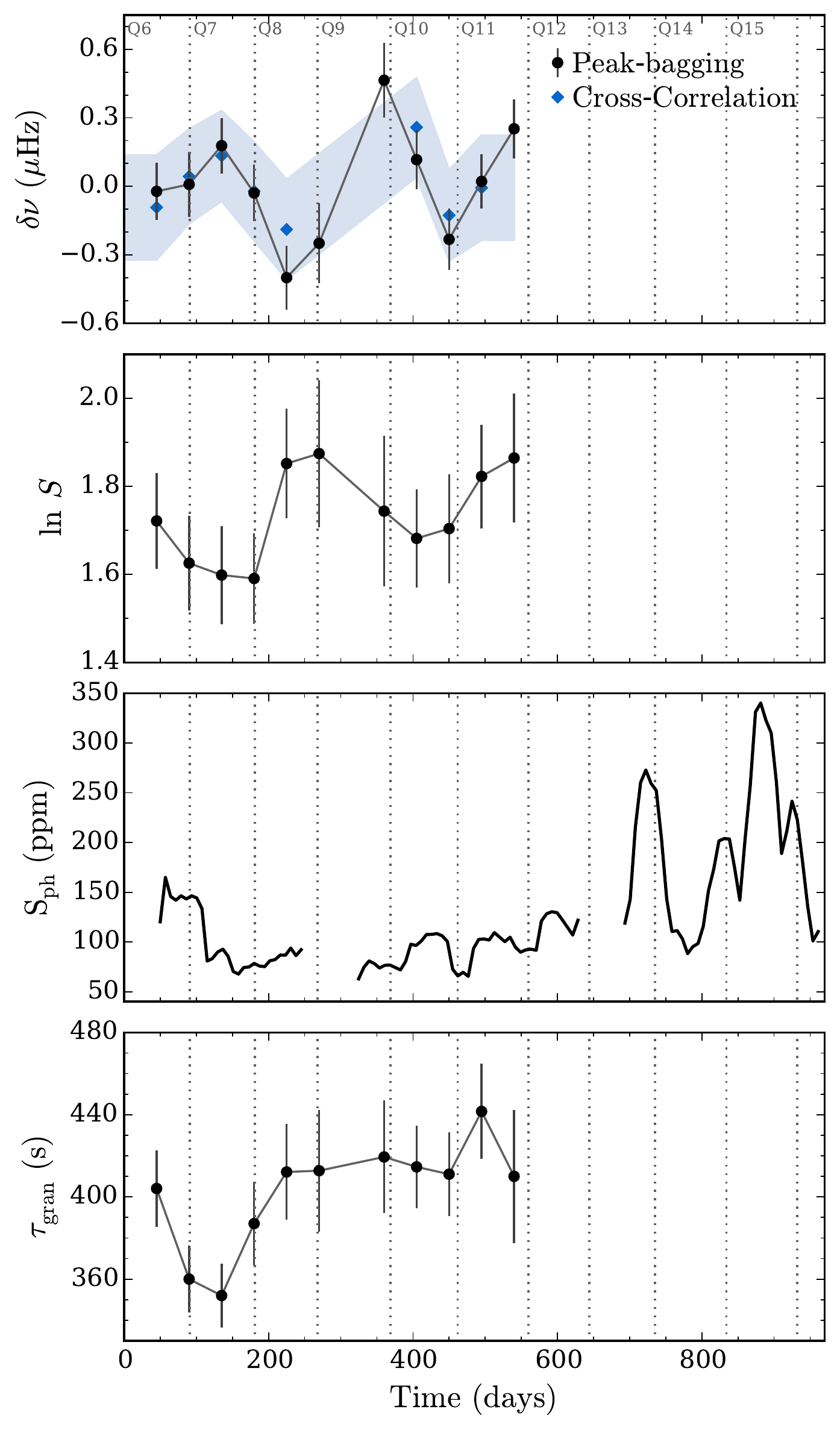}
\caption{Same as in Figure 26, but for KIC 5773345. Results in Table~\ref{tab:5773345}.}\label{fig:5773345}
\end{figure}

\FloatBarrier
\nopagebreak
%!TEX root = peakbagging.tex
\begin{figure}[ht]
\includegraphics[width=\hsize]{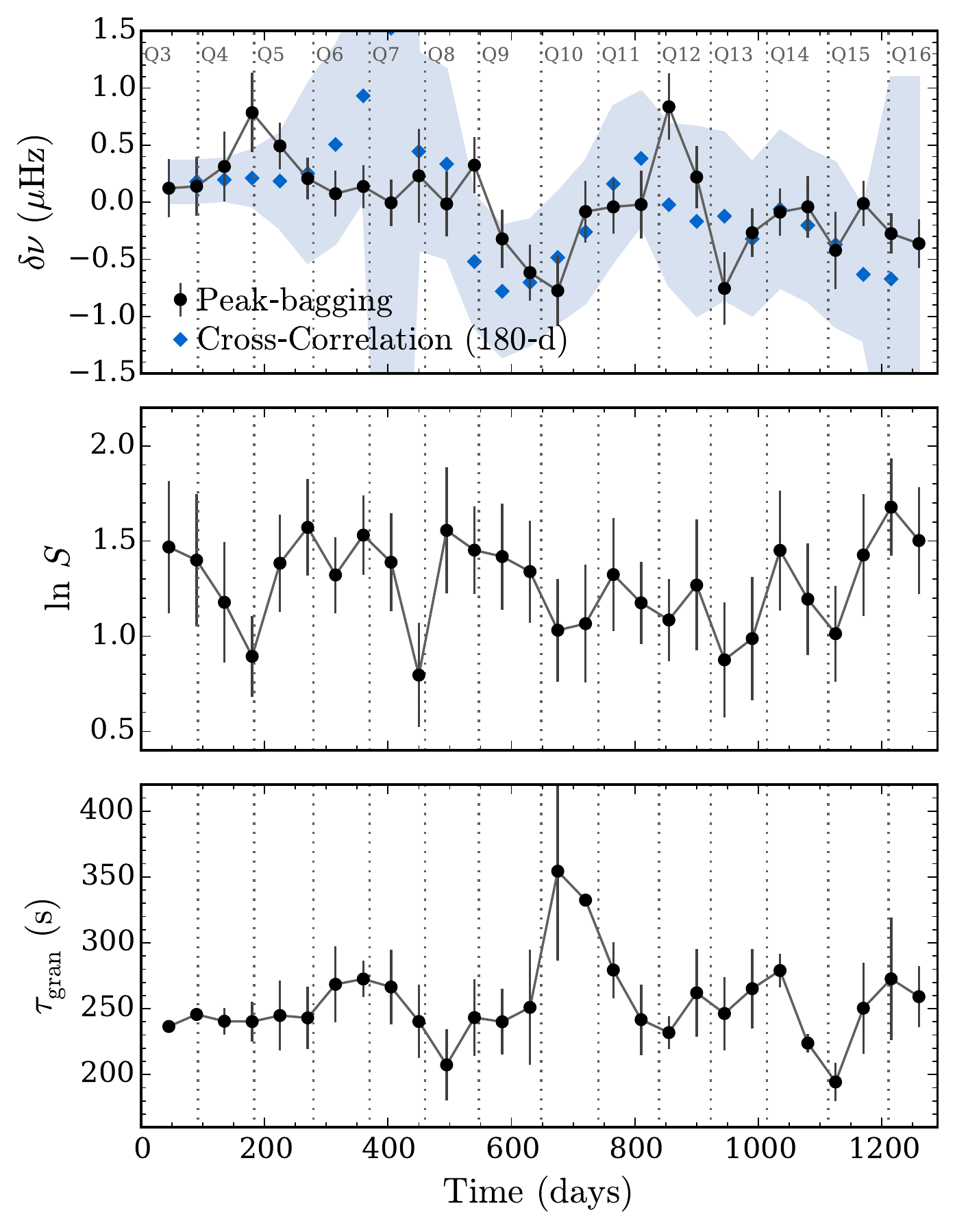}\vspace{-0.3cm}
\caption{Same as in Figure 26, but for KIC 5866724. The frequency shifts from the cross-correlation method were obtained with 180-d sub-series. Results in Table~\ref{tab:5866724}.}\label{fig:5866724}\vspace{-0.3cm}
\end{figure}

\begin{figure}[ht]
\includegraphics[width=\hsize]{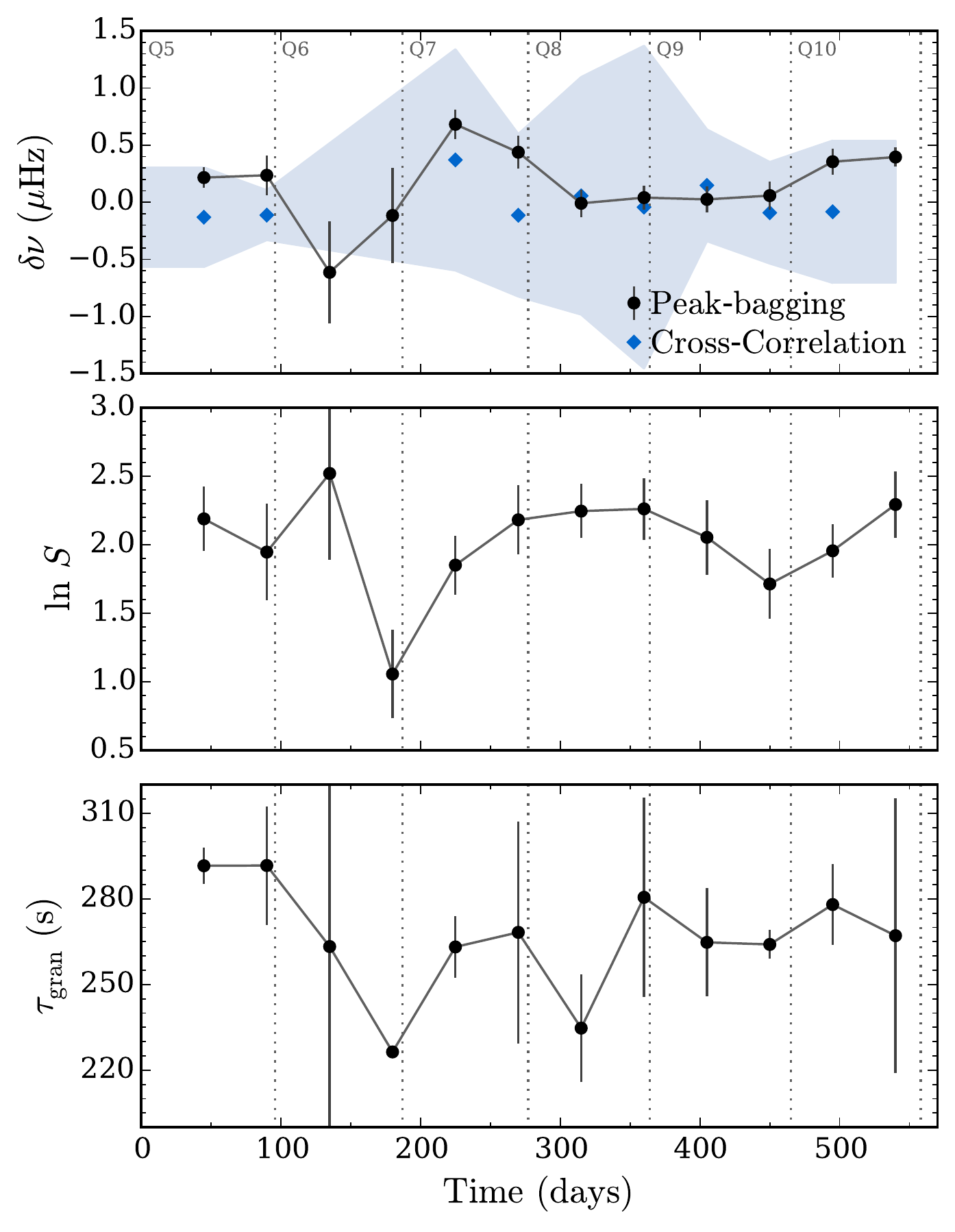}\vspace{-0.3cm}
\caption{Same as in Figure 26, but for KIC 5950854. Results in Table~\ref{tab:5950854}.}\label{fig:5950854}\vspace{-1.8cm}
\end{figure}

\begin{figure}[ht]
\includegraphics[width=\hsize]{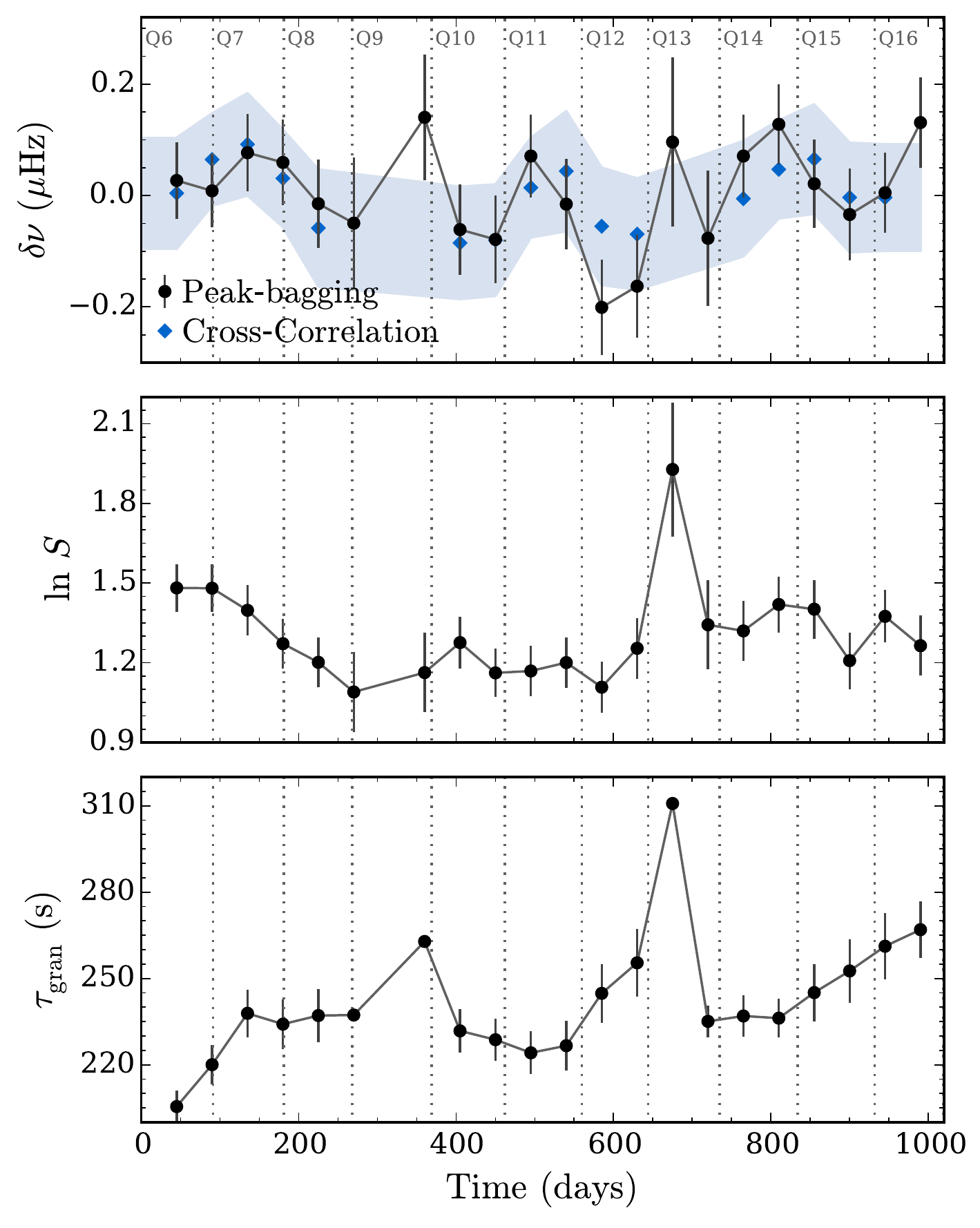}
\caption{Same as in Figure 26, but for KIC 6106415. Results in Table~\ref{tab:6106415}.}\label{fig:6106415}
\end{figure}

\begin{figure}[ht]
\includegraphics[width=\hsize]{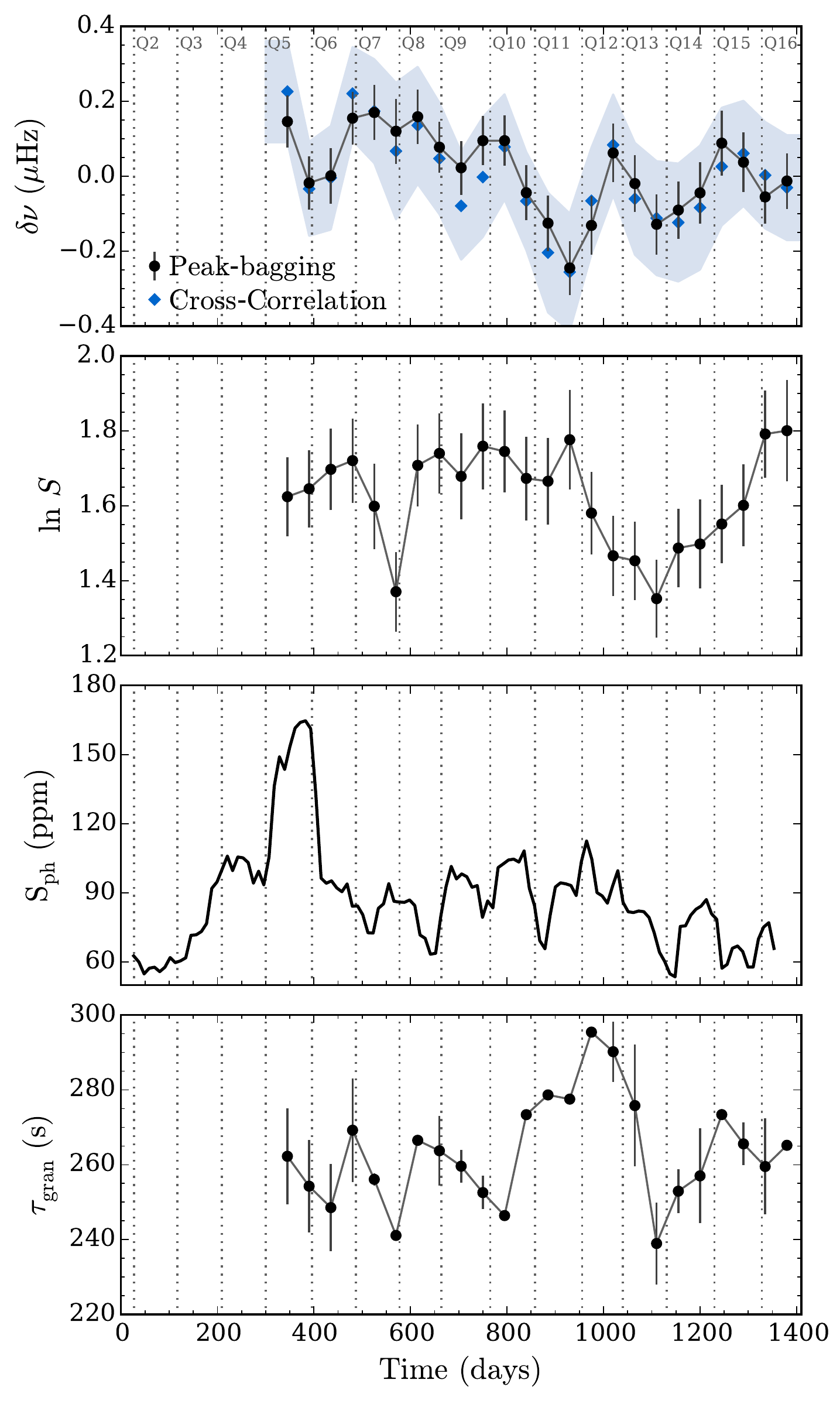}
\caption{Same as in Figure 26, but for KIC 6116048. Results in Table~\ref{tab:6116048}.}\label{fig:6116048}
\end{figure}

\begin{figure}[ht]
\includegraphics[width=\hsize]{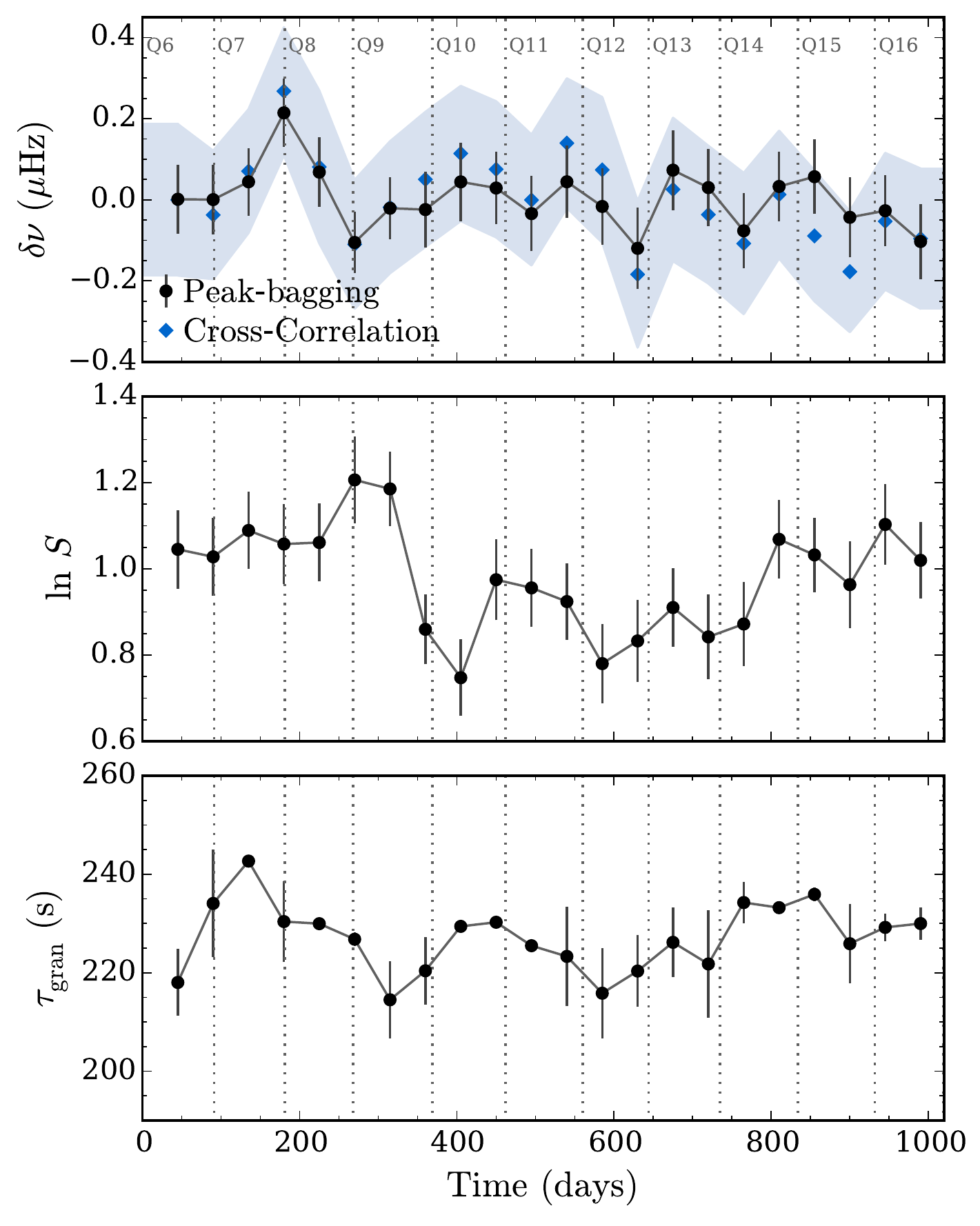}\vspace{-0.2cm}
\caption{Same as in Figure 26, but for KIC 6225718. Results in Table~\ref{tab:6225718}.}\label{fig:6225718}
\end{figure}

\begin{figure}[ht]
\includegraphics[width=\hsize]{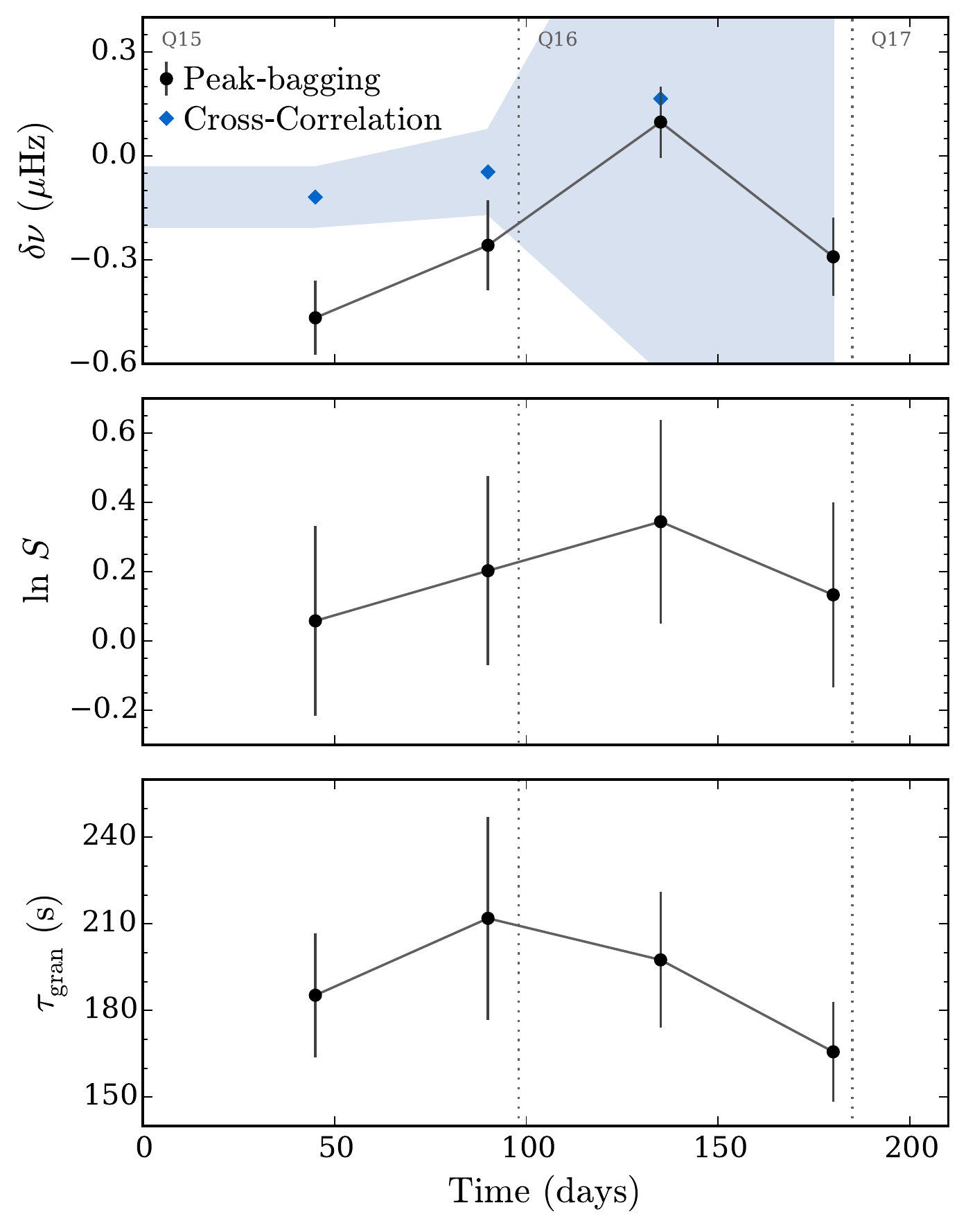}\vspace{-0.2cm}
\caption{Same as in Figure 26, but for KIC 6278762. Results in Table~\ref{tab:6278762}.}\label{fig:6278762}\vspace{-1.5cm}
\end{figure}

\begin{figure}[ht]
\includegraphics[width=\hsize]{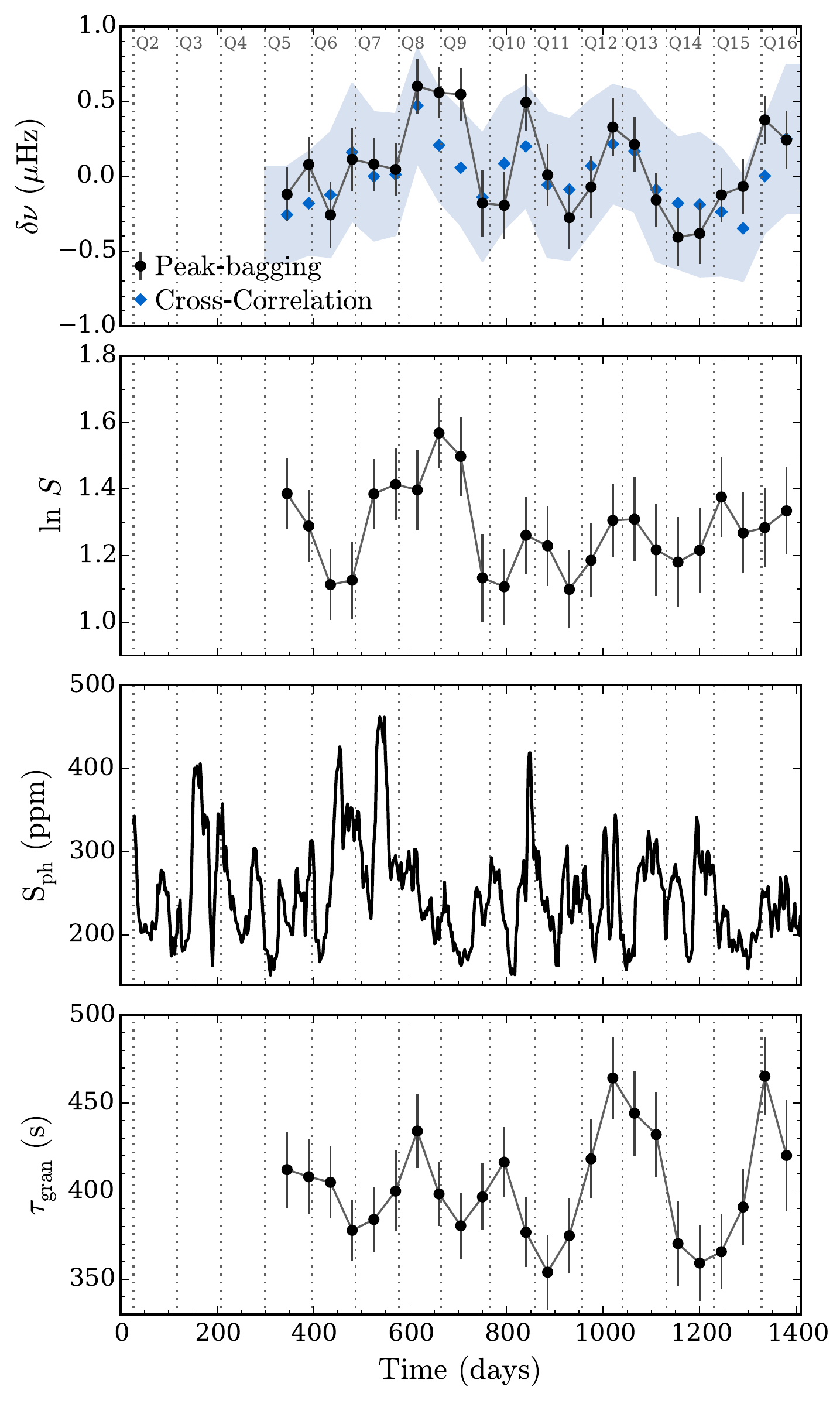}
\caption{Same as in Figure 26, but for KIC 6508366. Results in Table~\ref{tab:6508366}.}\label{fig:6508366}
\end{figure}

\begin{figure}[ht]
\includegraphics[width=\hsize]{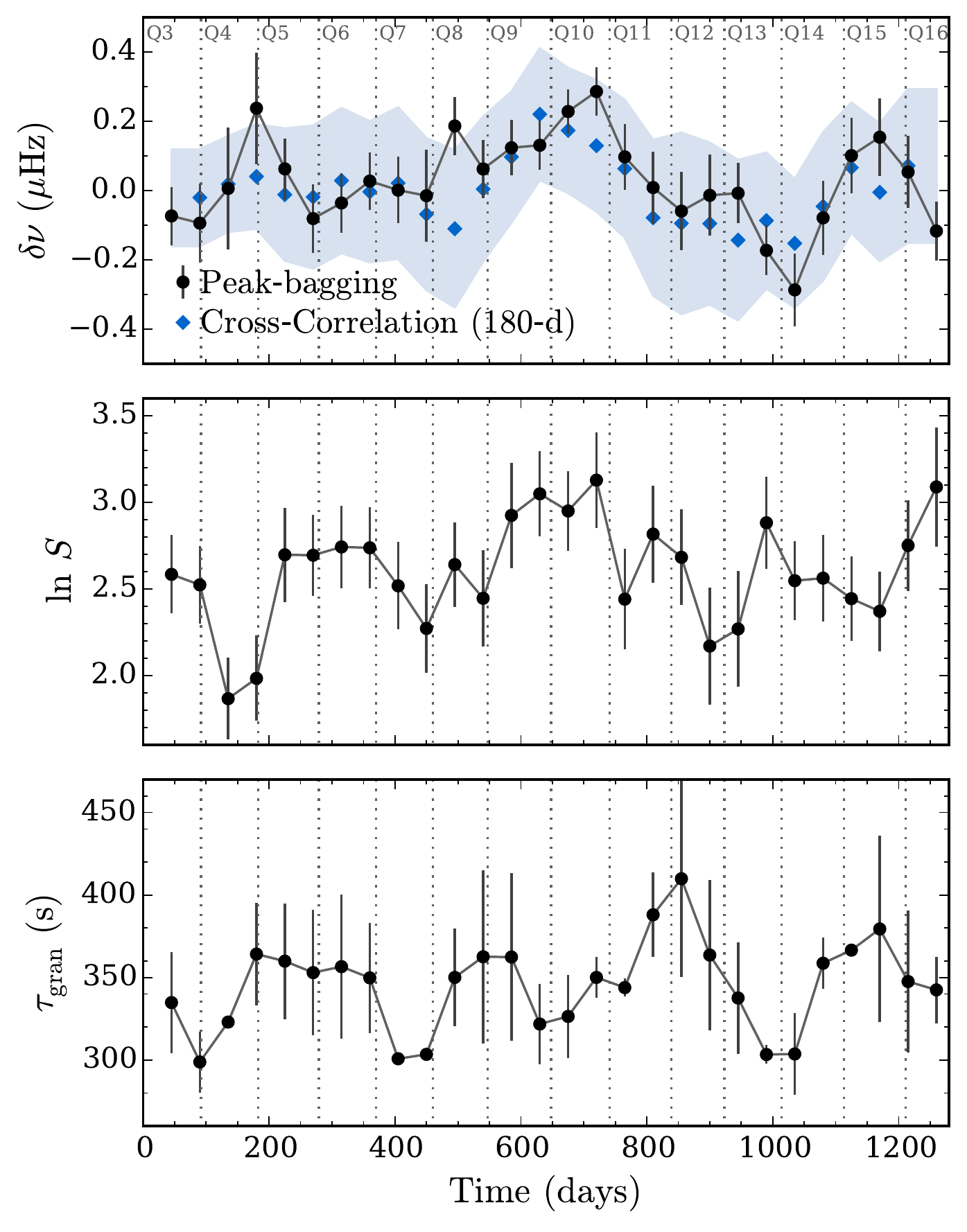}\vspace{-0.2cm}
\caption{Same as in Figure 26, but for KIC 6521045. The frequency shifts from the cross-correlation method were obtained with 180-d sub-series. Results in Table~\ref{tab:6521045}.}\label{fig:6521045}\vspace{-0.3cm}
\end{figure}

\begin{figure}[ht]
\includegraphics[width=\hsize]{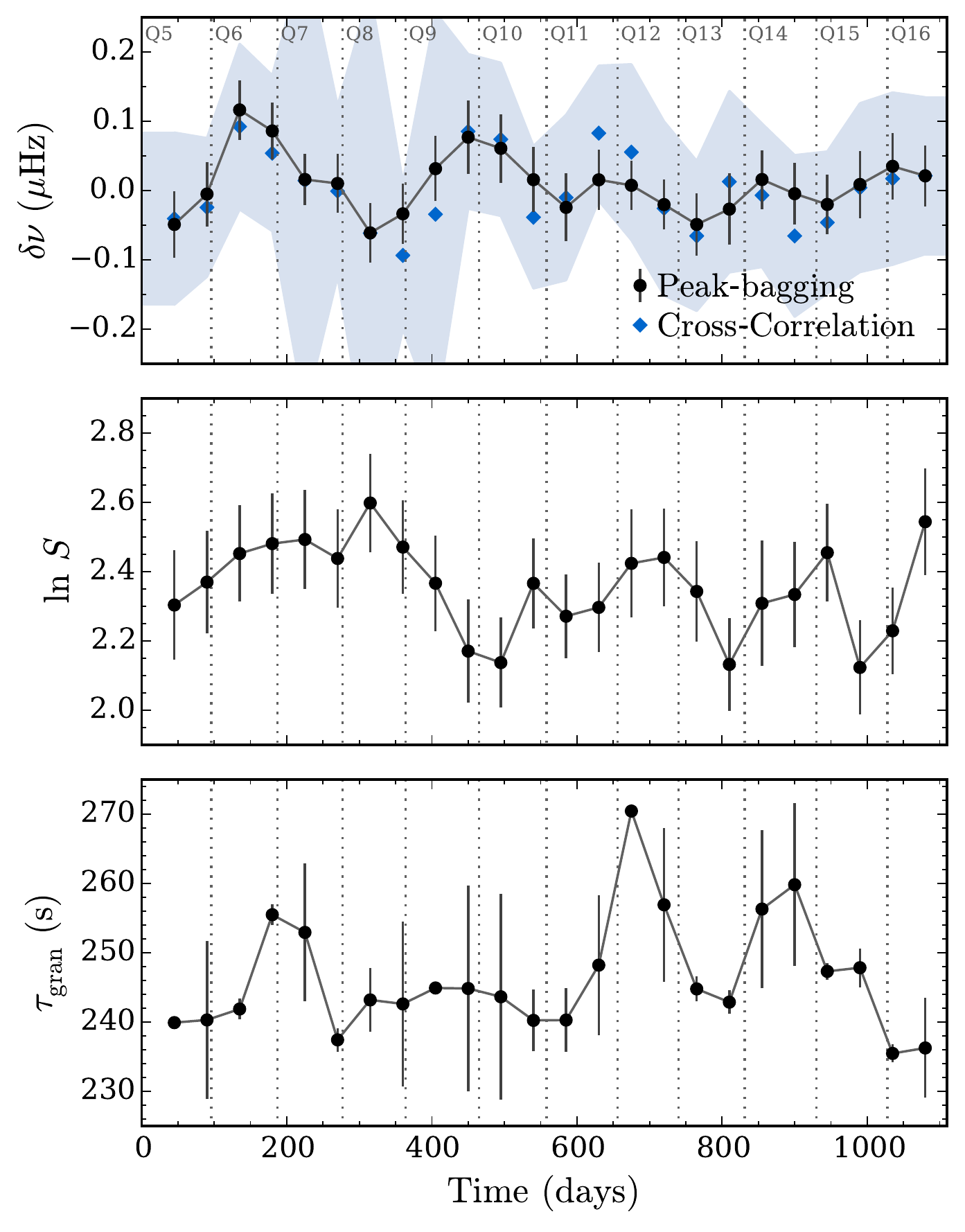}\vspace{-0.2cm}
\caption{Same as in Figure 26, but for KIC 6603624. Results in Table~\ref{tab:6603624}.}\label{fig:6603624}\vspace{-1cm}
\end{figure}

\begin{figure}[ht]
\includegraphics[width=\hsize]{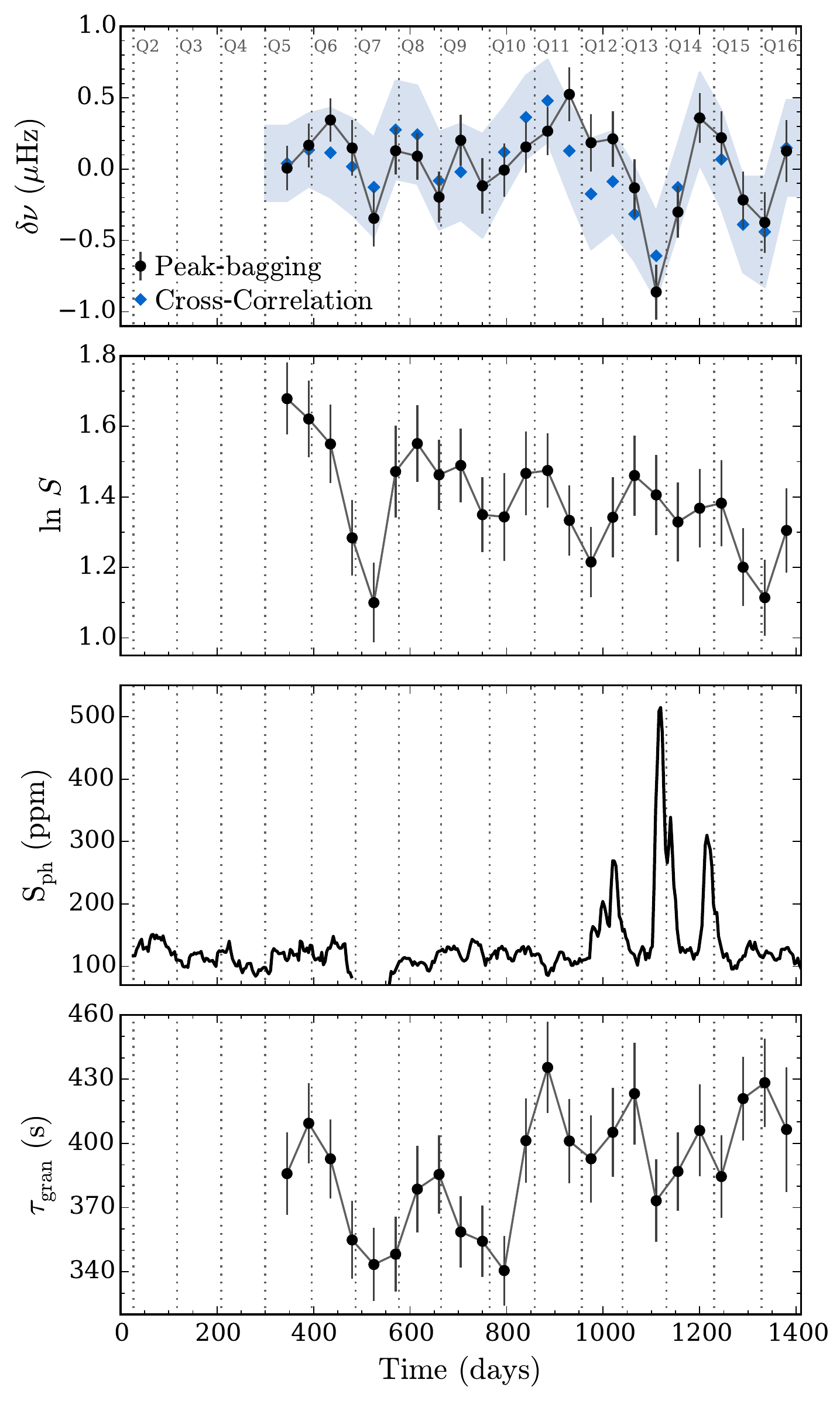}
\caption{Same as in Figure 26, but for KIC 6679371. Results in Table~\ref{tab:6679371}.}\label{fig:6679371}
\end{figure}

\begin{figure}[ht]
\includegraphics[width=\hsize]{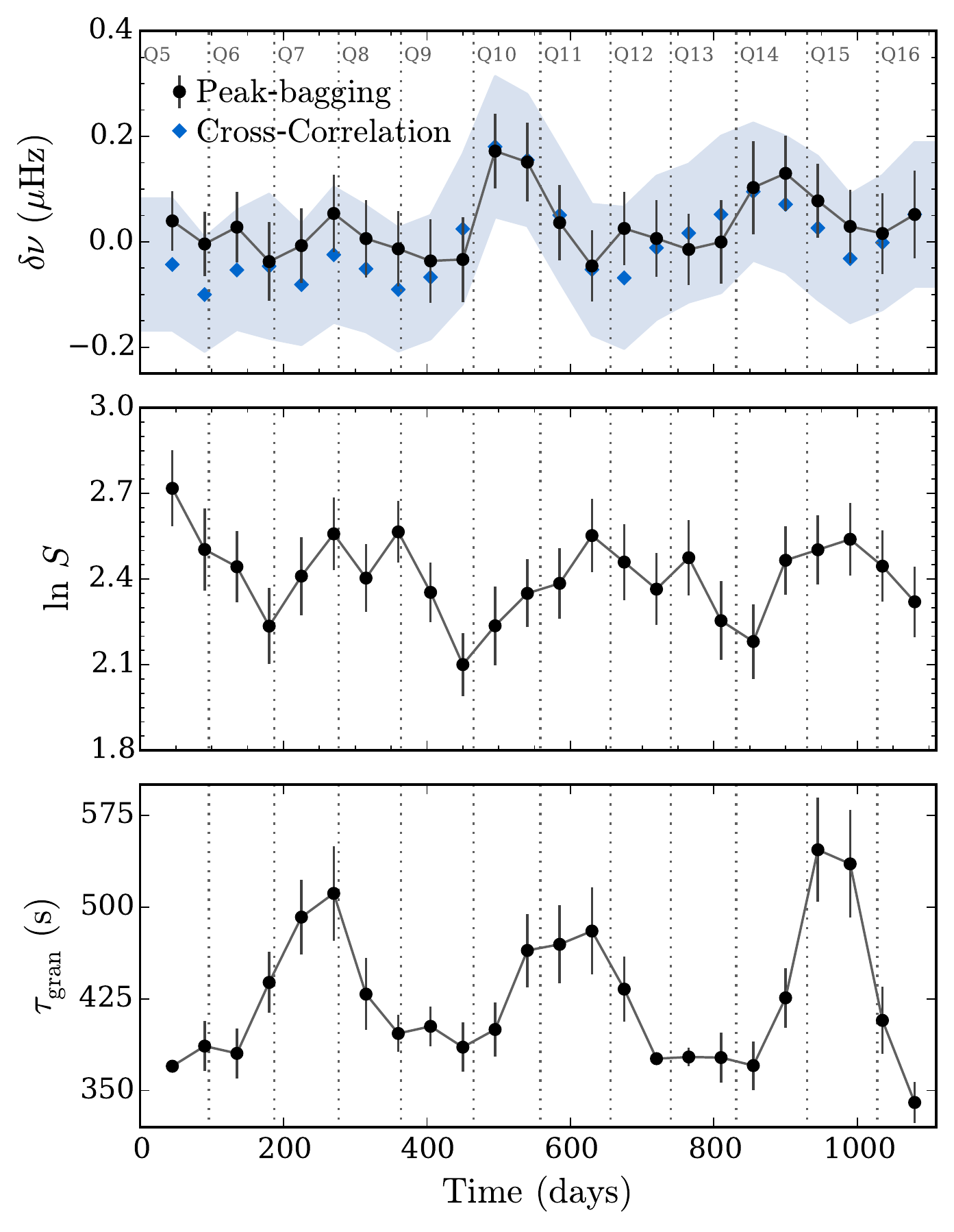}
\caption{Same as in Figure 26, but for KIC 6933899. Results in Table~\ref{tab:6933899}.}\label{fig:6933899}
\end{figure}

\begin{figure}[ht]
\includegraphics[width=\hsize]{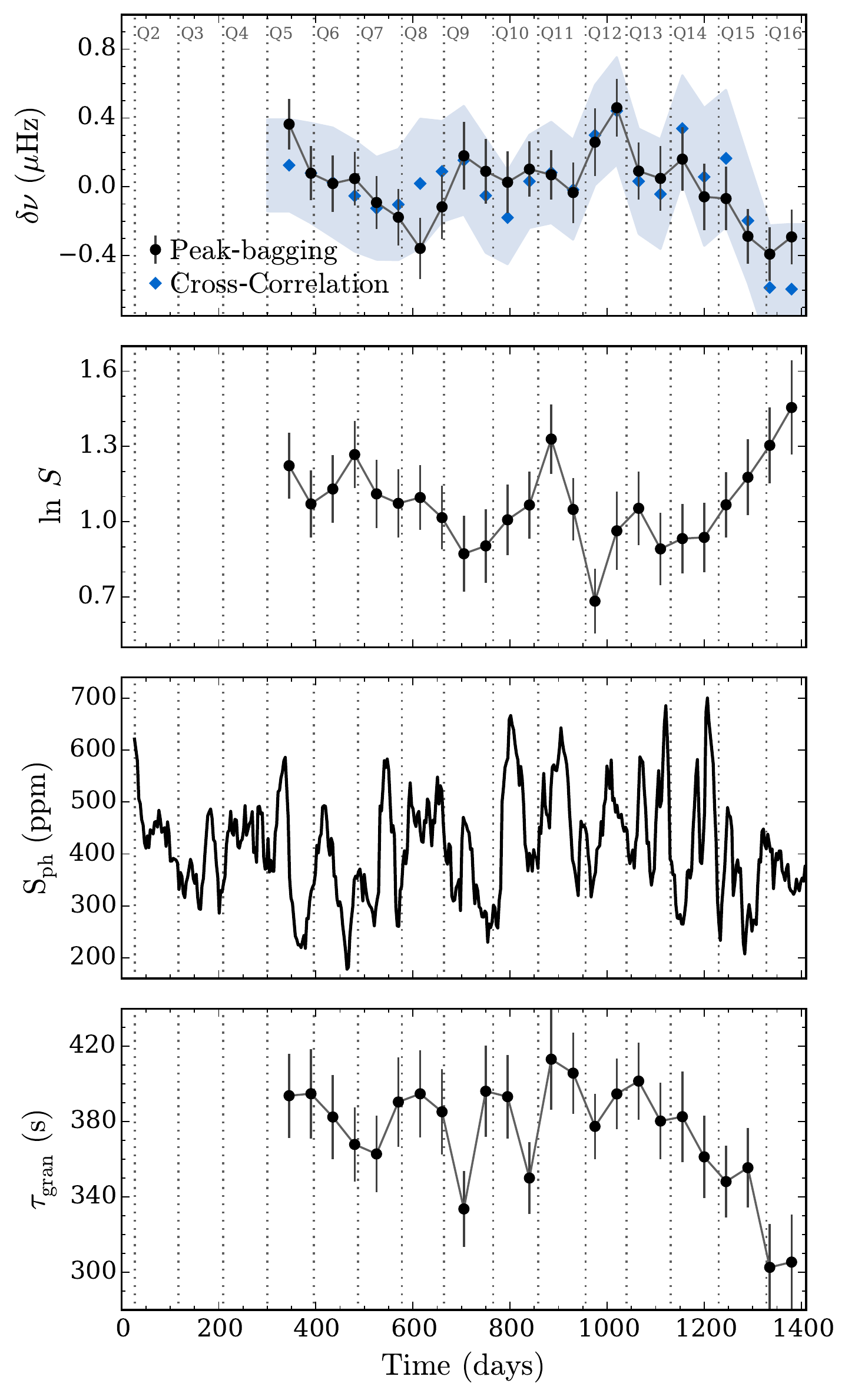}
\caption{Same as in Figure 26, but for KIC 7103006. Results in Table~\ref{tab:7103006}.}\label{fig:7103006}
\end{figure}

\begin{figure}[ht]
\includegraphics[width=\hsize]{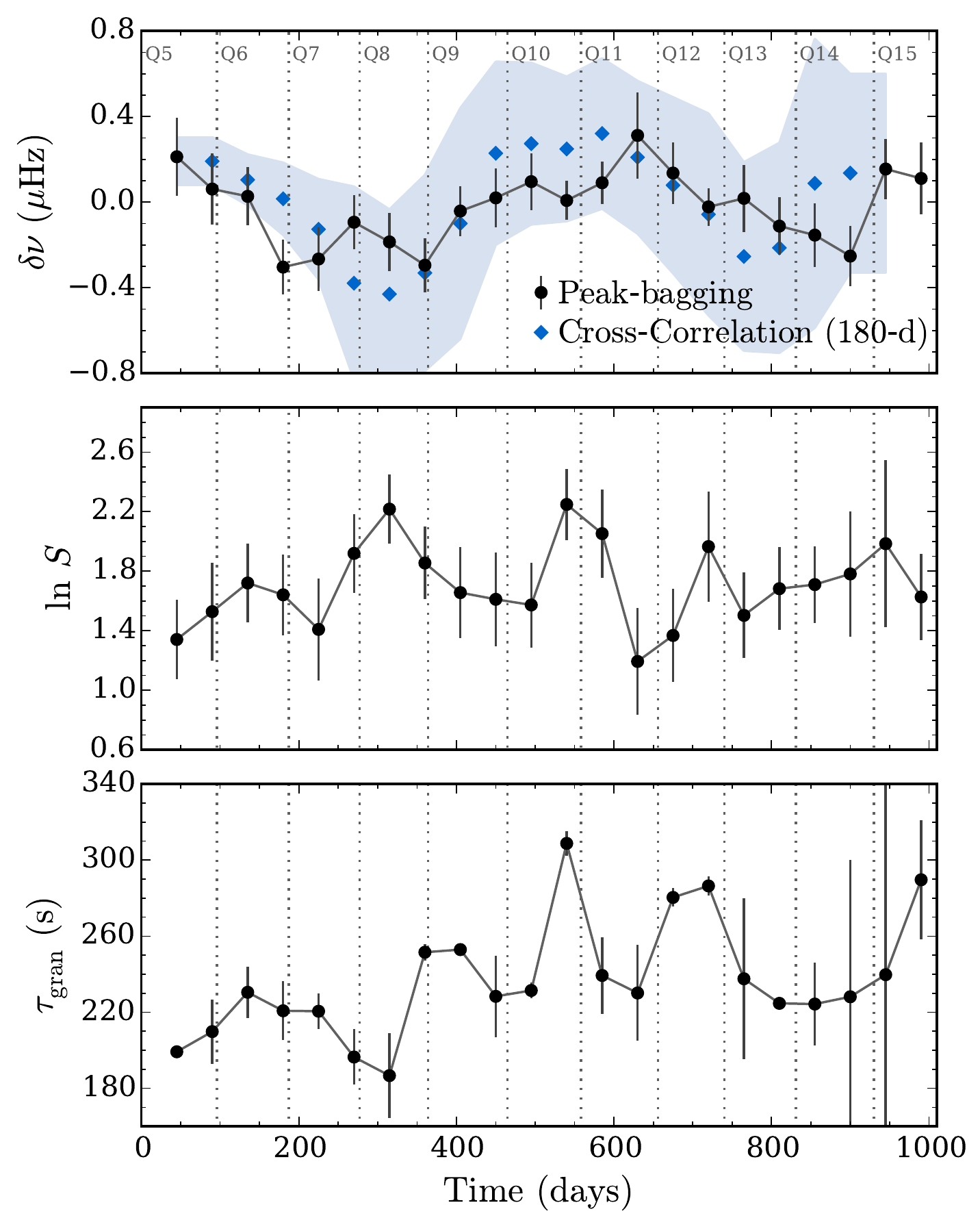}
\caption{Same as in Figure 26, but for KIC 7106245. The frequency shifts from the cross-correlation method were obtained with 180-d sub-series. Results in Table~\ref{tab:7106245}.}\label{fig:7106245}
\end{figure}

\begin{figure}[ht]
\includegraphics[width=\hsize]{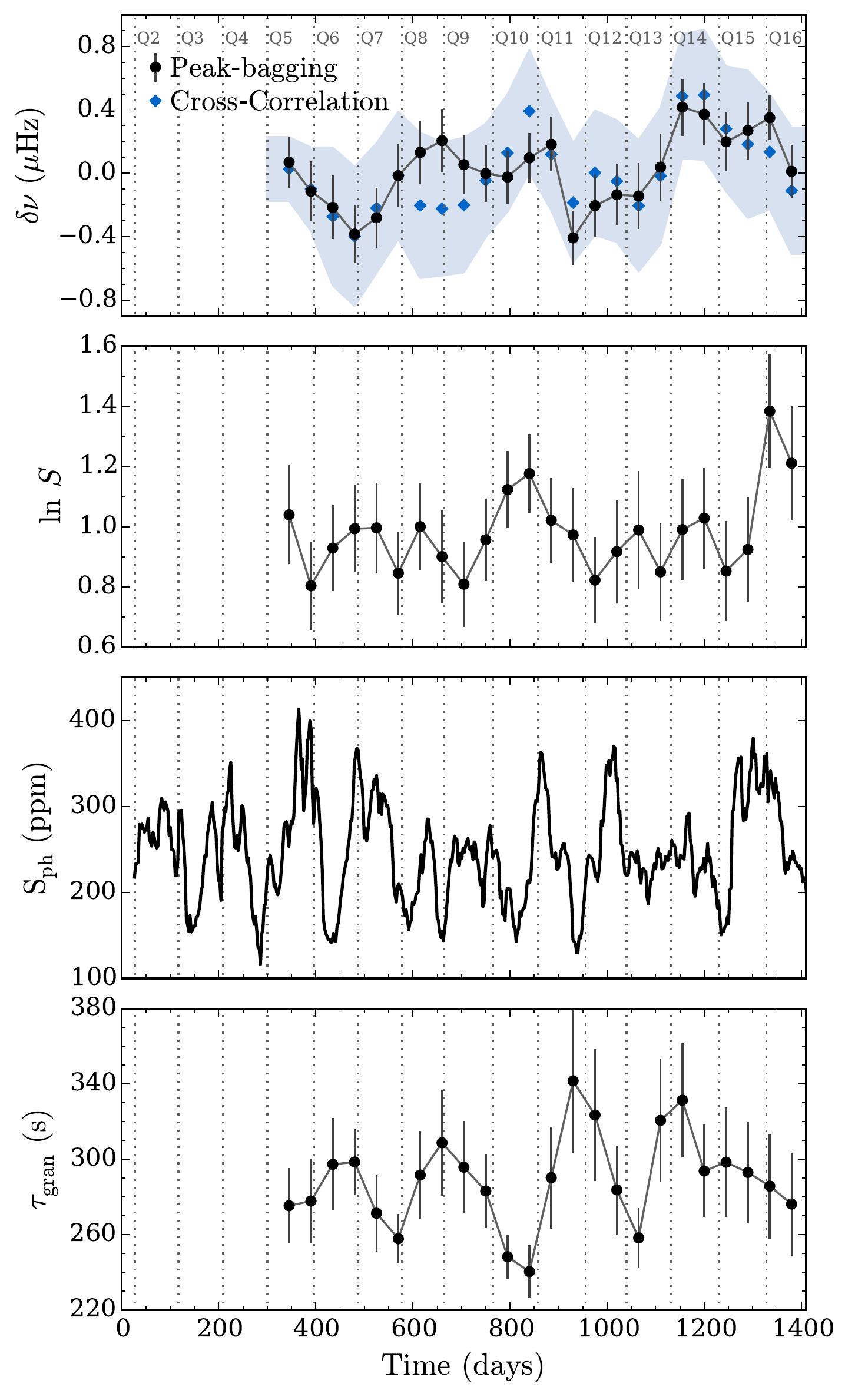}
\caption{Same as in Figure 26, but for KIC 7206837. Results in Table~\ref{tab:7206837}.}\label{fig:7206837}
\end{figure}

\begin{figure}[ht]
\includegraphics[width=\hsize]{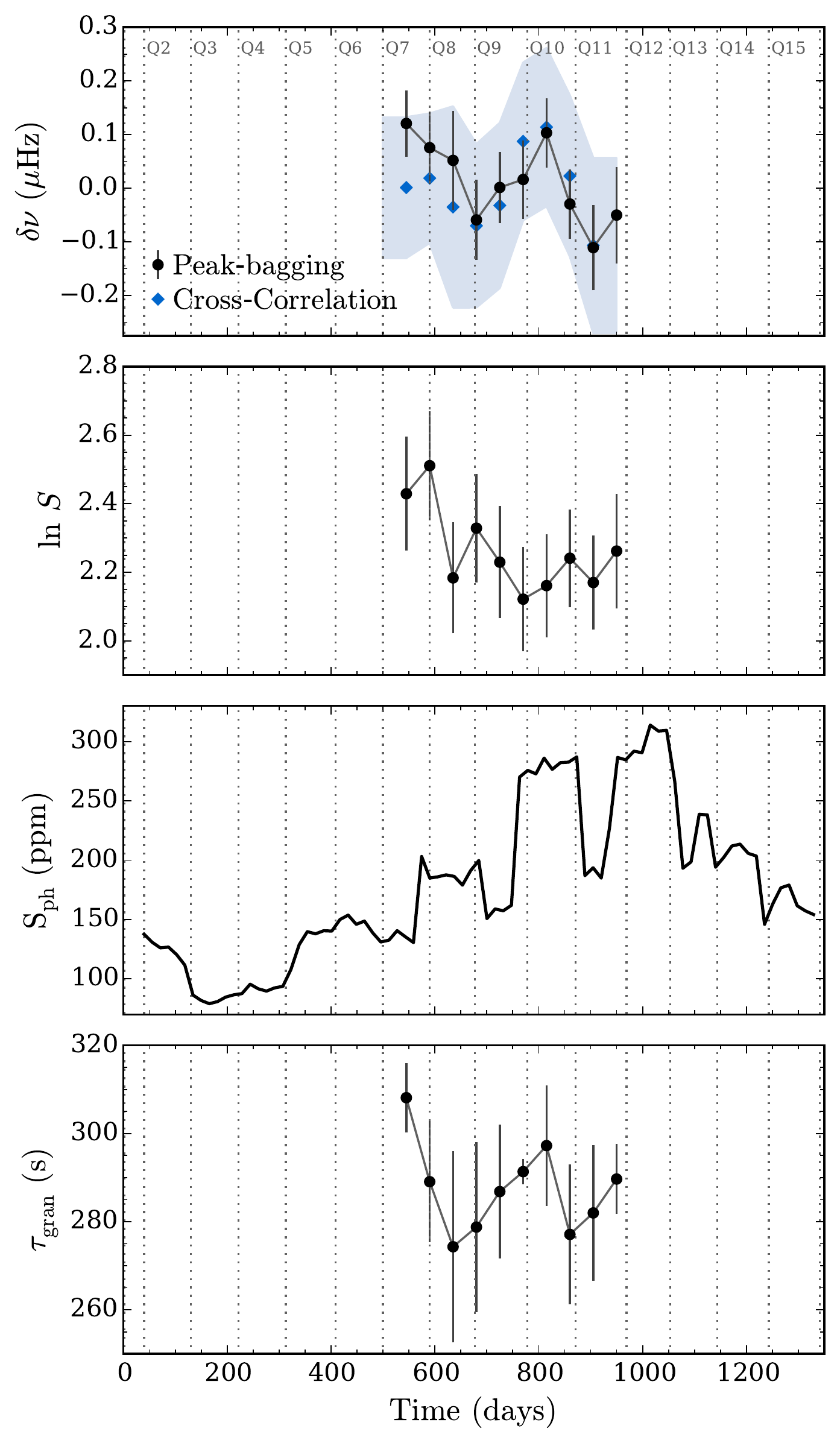}
\caption{Same as in Figure 26, but for KIC 7296438. Results in Table~\ref{tab:7296438}.}\label{fig:7296438}
\end{figure}

\begin{figure}[ht]
\includegraphics[width=\hsize]{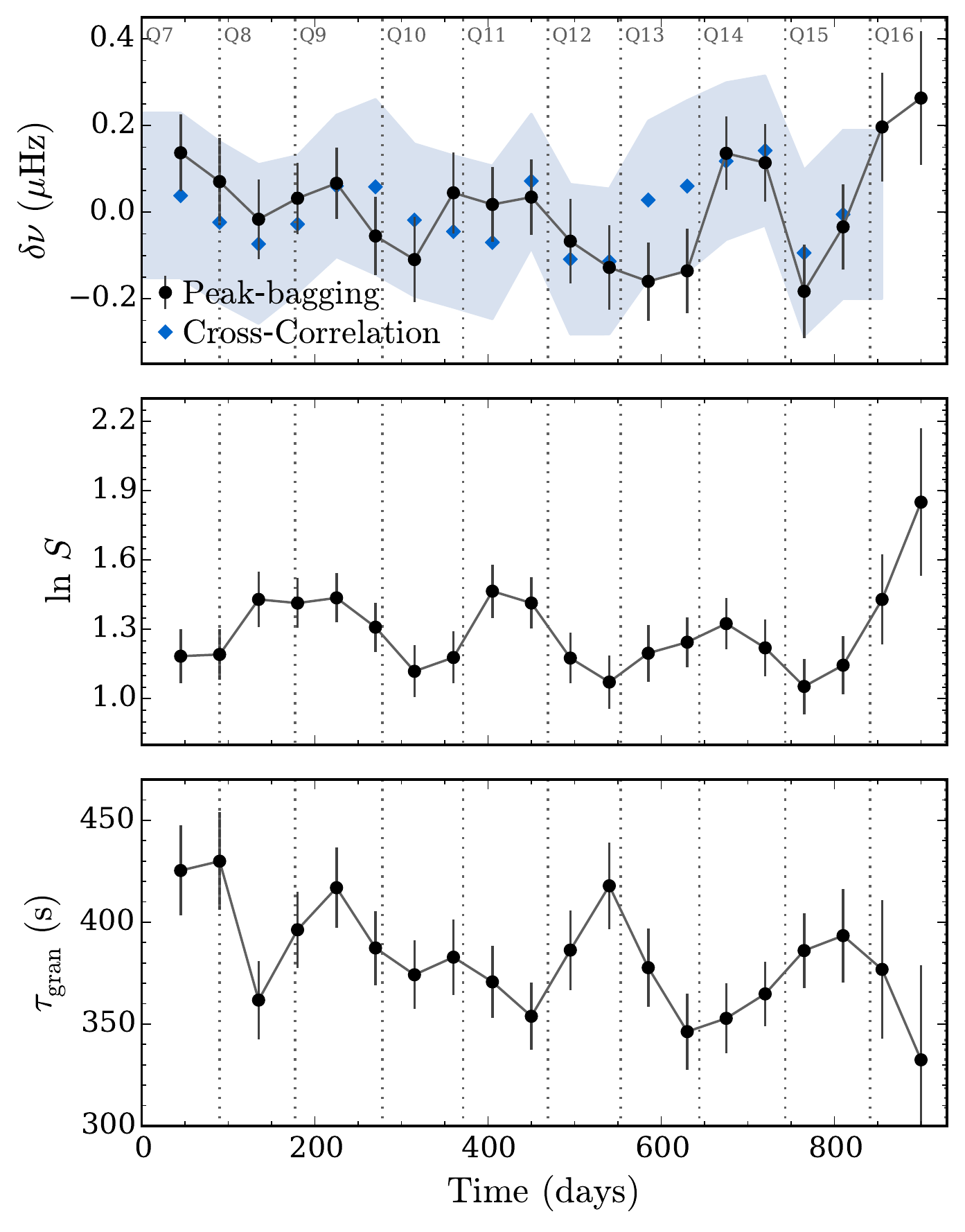}\vspace{-0.2cm}
\caption{Same as in Figure 26, but for KIC 7510397. Results in Table~\ref{tab:7510397}.}\label{fig:7510397}
\end{figure}

\begin{figure}[ht]
\includegraphics[width=\hsize]{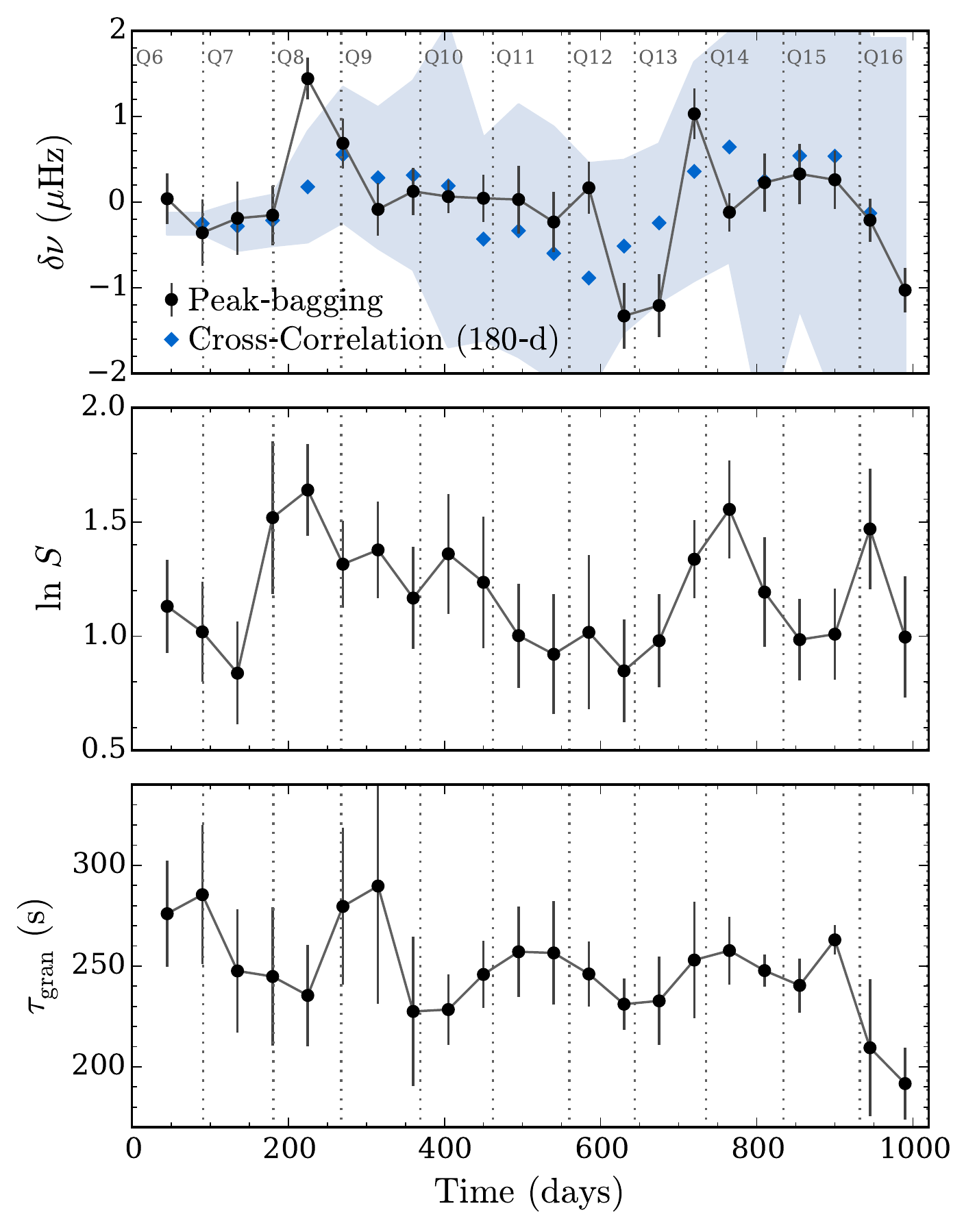}\vspace{-0.2cm}
\caption{Same as in Figure 26, but for KIC 7670943. The frequency shifts from the cross-correlation method were obtained with 180-d sub-series. Results in Table~\ref{tab:7670943}.}\label{fig:7670943}\vspace{-1.5cm}
\end{figure}

\begin{figure}[ht]
\includegraphics[width=\hsize]{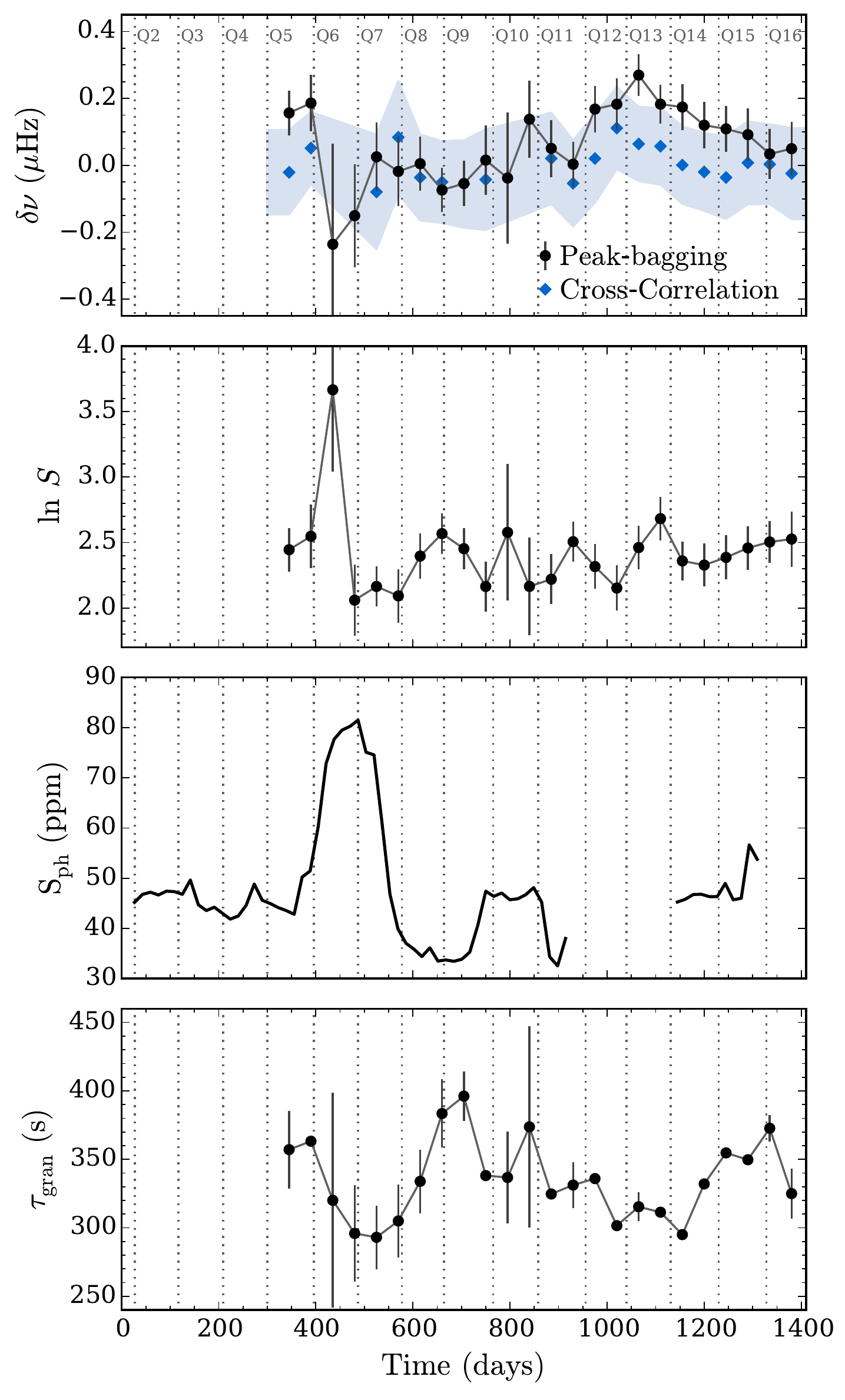}
\caption{Same as in Figure 26, but for KIC 7680114. Results in Table~\ref{tab:7680114}.}\label{fig:7680114}
\end{figure}

\FloatBarrier
\nopagebreak
%!TEX root = peakbagging.tex
\begin{figure}[ht]
\includegraphics[width=\hsize]{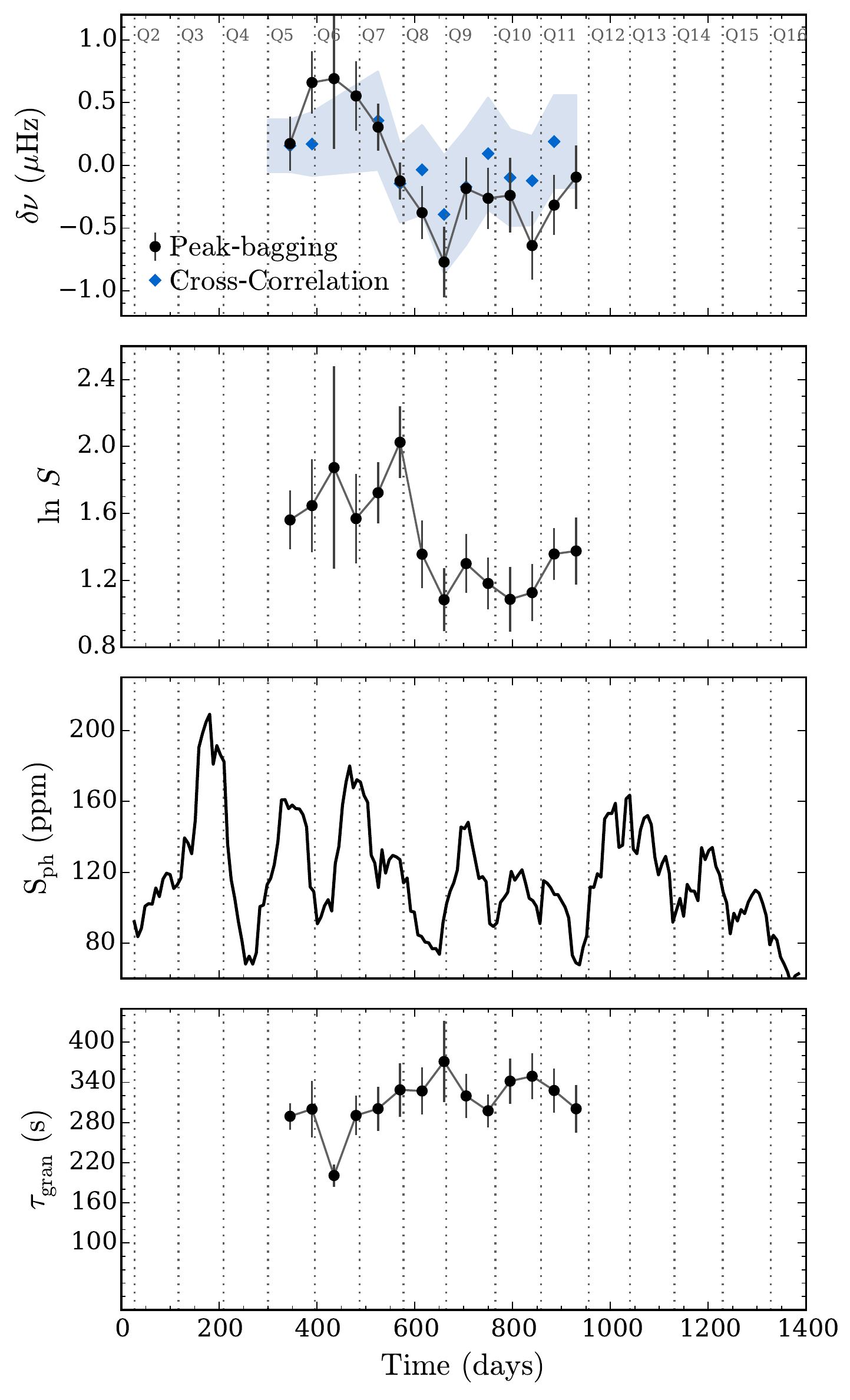}
\caption{Same as in Figure 26, but for KIC 7771282. Results in Table~\ref{tab:7771282}.}\label{fig:7771282}
\end{figure}

\begin{figure}[ht]
\includegraphics[width=\hsize]{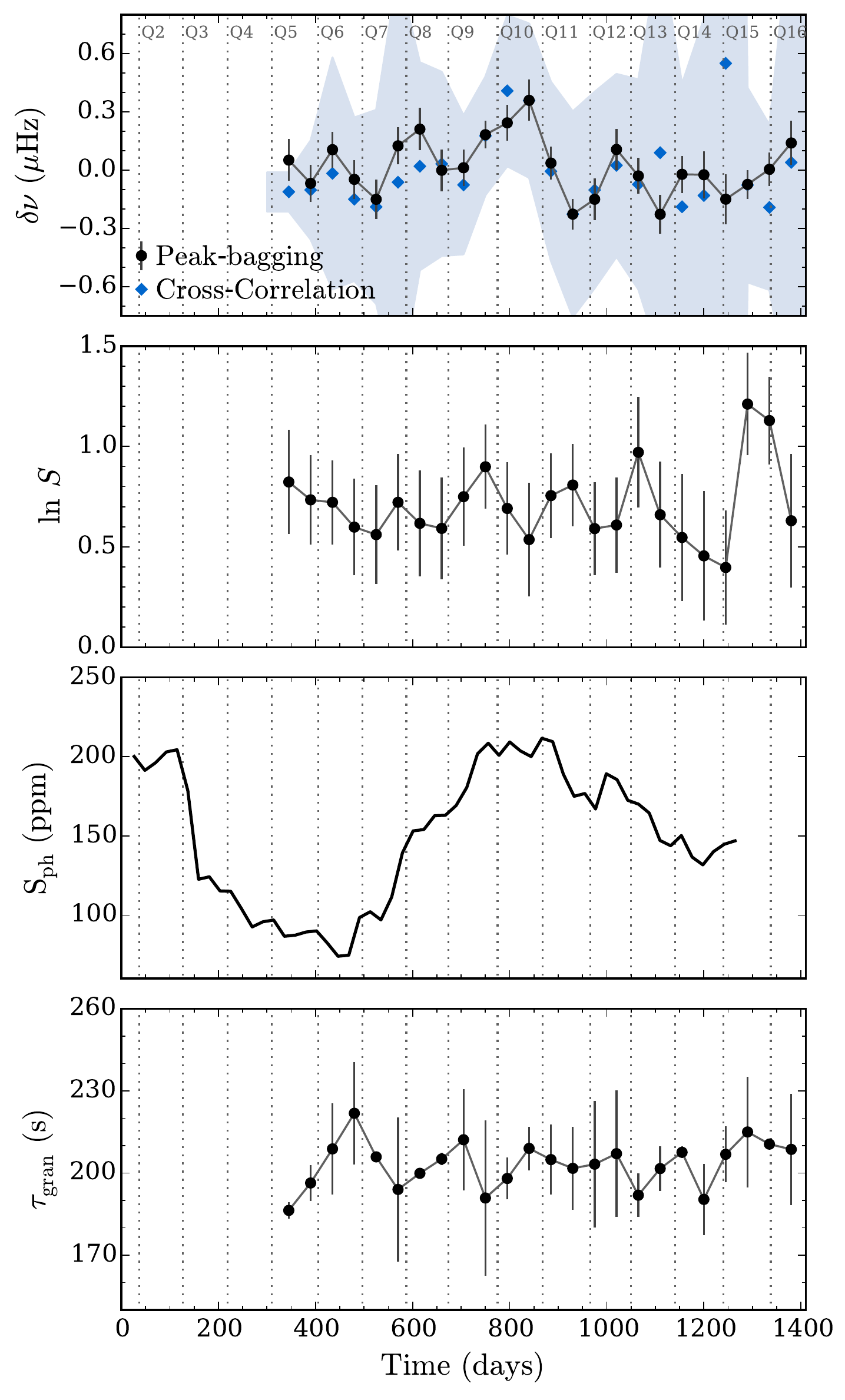}
\caption{Same as in Figure 26, but for KIC 7871531. Results in Table~\ref{tab:7871531}.}\label{fig:7871531}
\end{figure}

\begin{figure}[ht]
\includegraphics[width=\hsize]{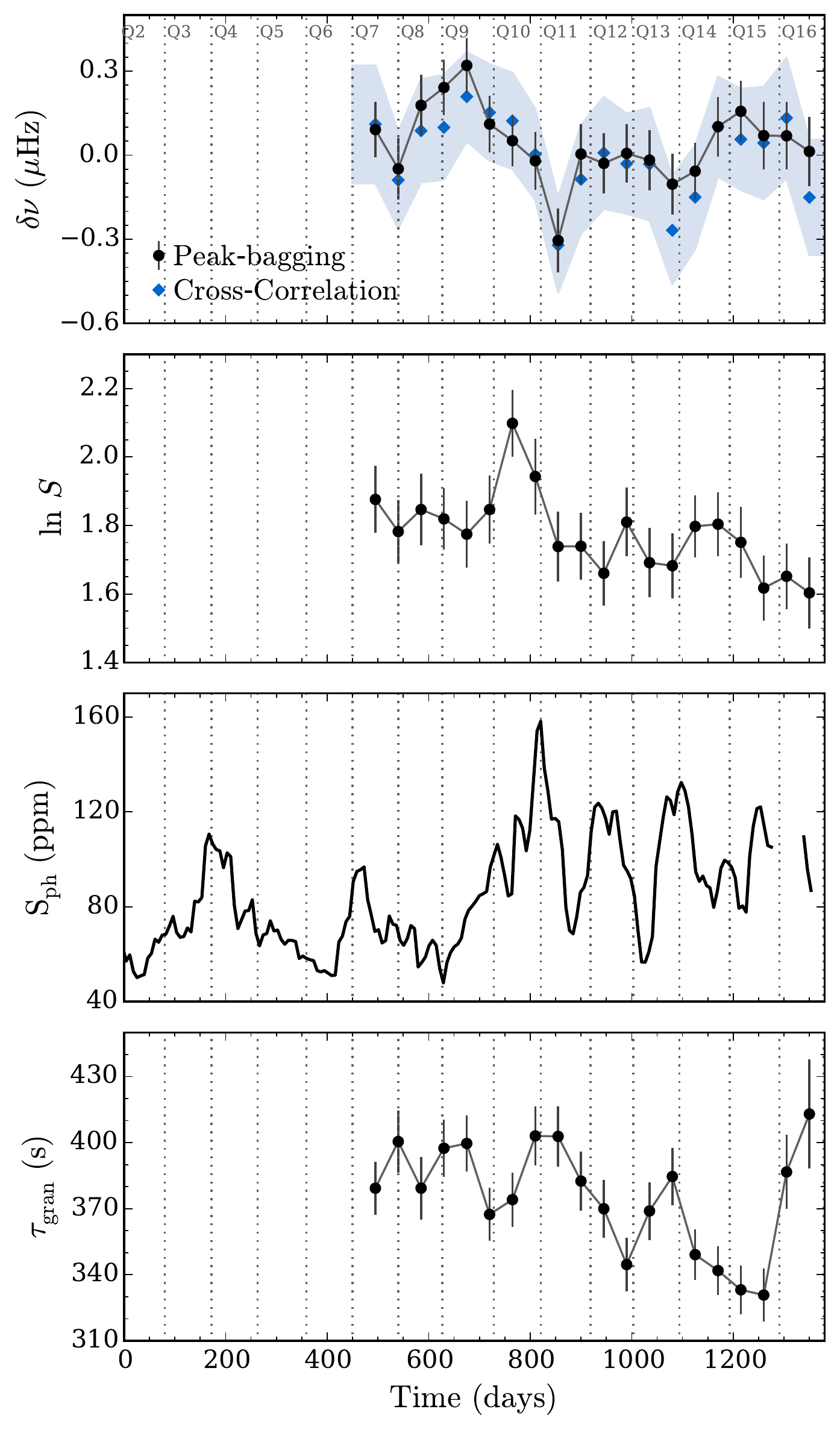}
\caption{Same as in Figure 3, but for KIC 7940546. Results in Table~\ref{tab:7940546}.}\label{fig:7940546}
\end{figure}

\begin{figure}[ht]
\includegraphics[width=\hsize]{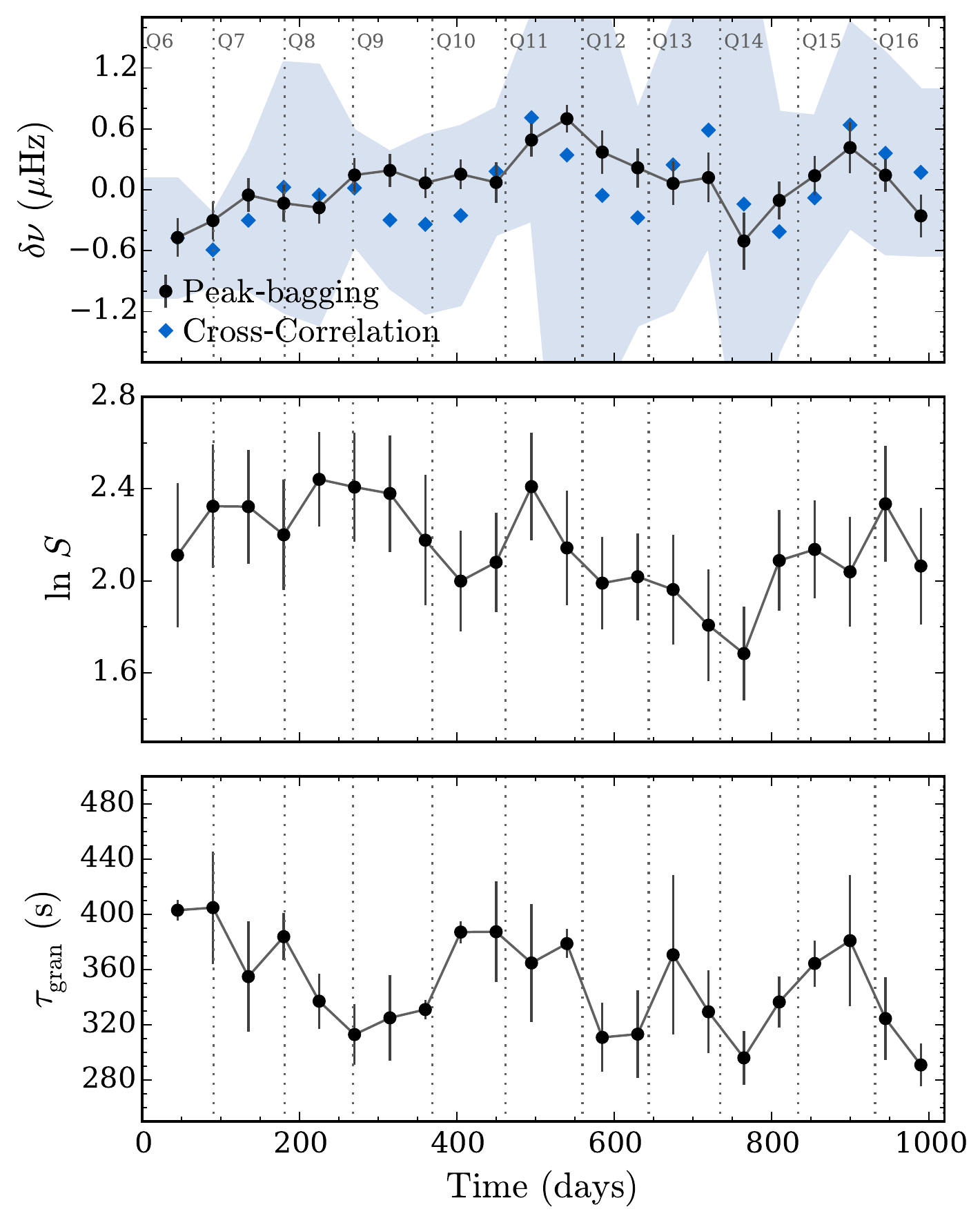}\vspace{-0.2cm}
\caption{Same as in Figure 3, but for KIC 8077137. Results in Table~\ref{tab:8077137}.}\label{fig:8077137}
\end{figure}

\begin{figure}[ht]
\includegraphics[width=\hsize]{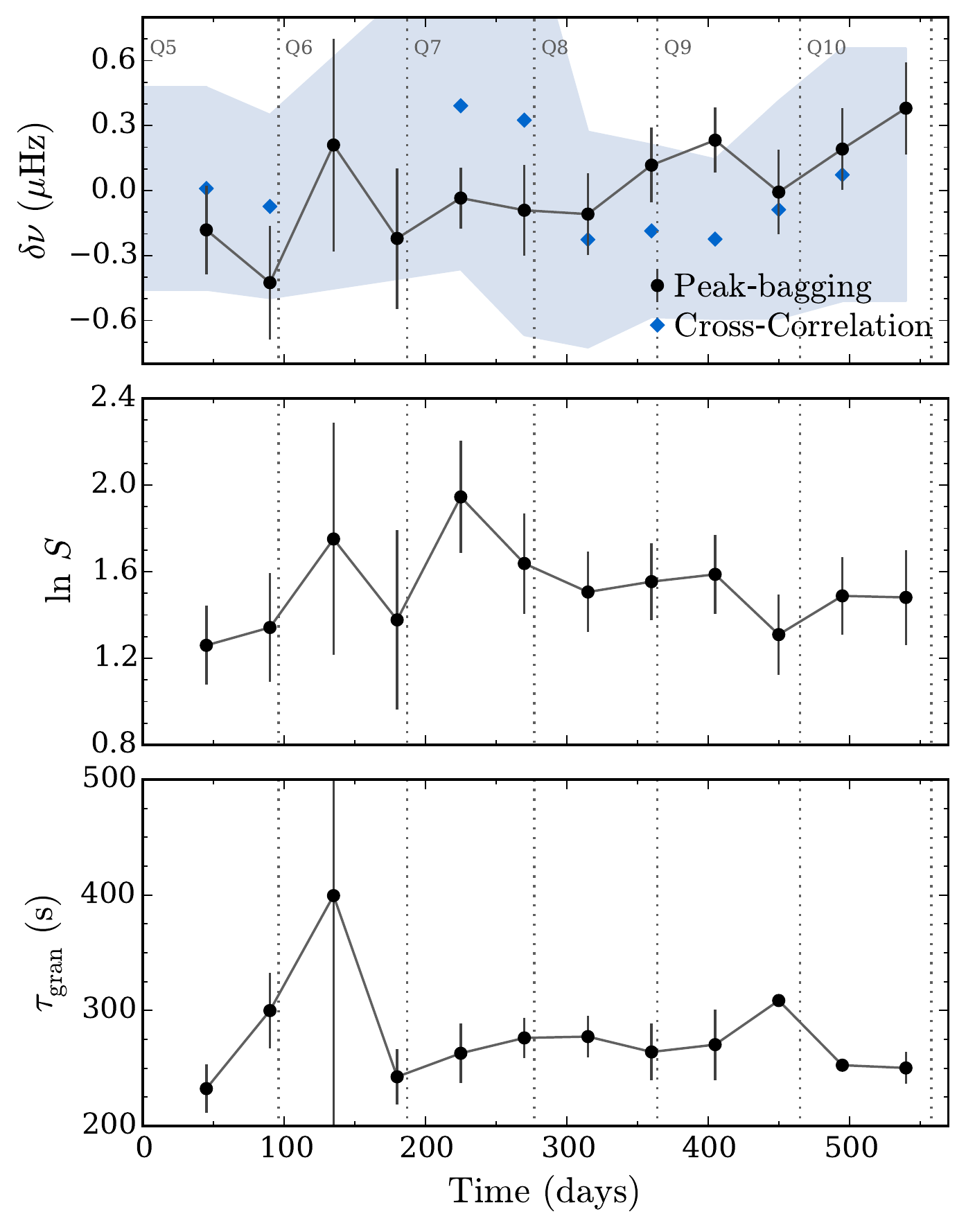}\vspace{-0.2cm}
\caption{Same as in Figure 3, but for KIC 8150065. Results in Table~\ref{tab:8150065}.}\label{fig:8150065}\vspace{-1.5cm}
\end{figure}

\begin{figure}[ht]
\includegraphics[width=\hsize]{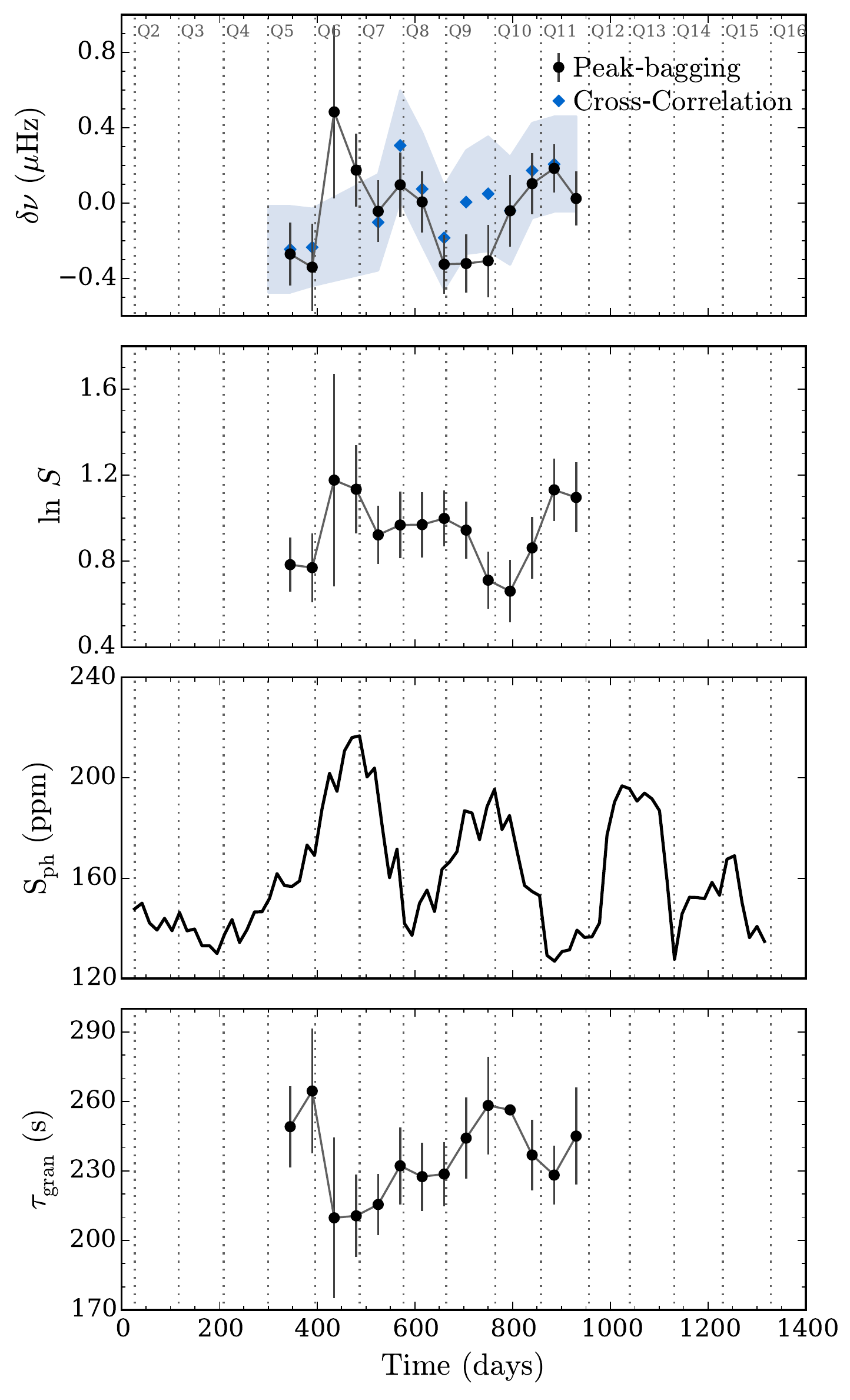}
\caption{Same as in Figure 3, but for KIC 8179536. Results in Table~\ref{tab:8179536}.}\label{fig:8179536}
\end{figure}

\begin{figure}[ht]
\includegraphics[width=\hsize]{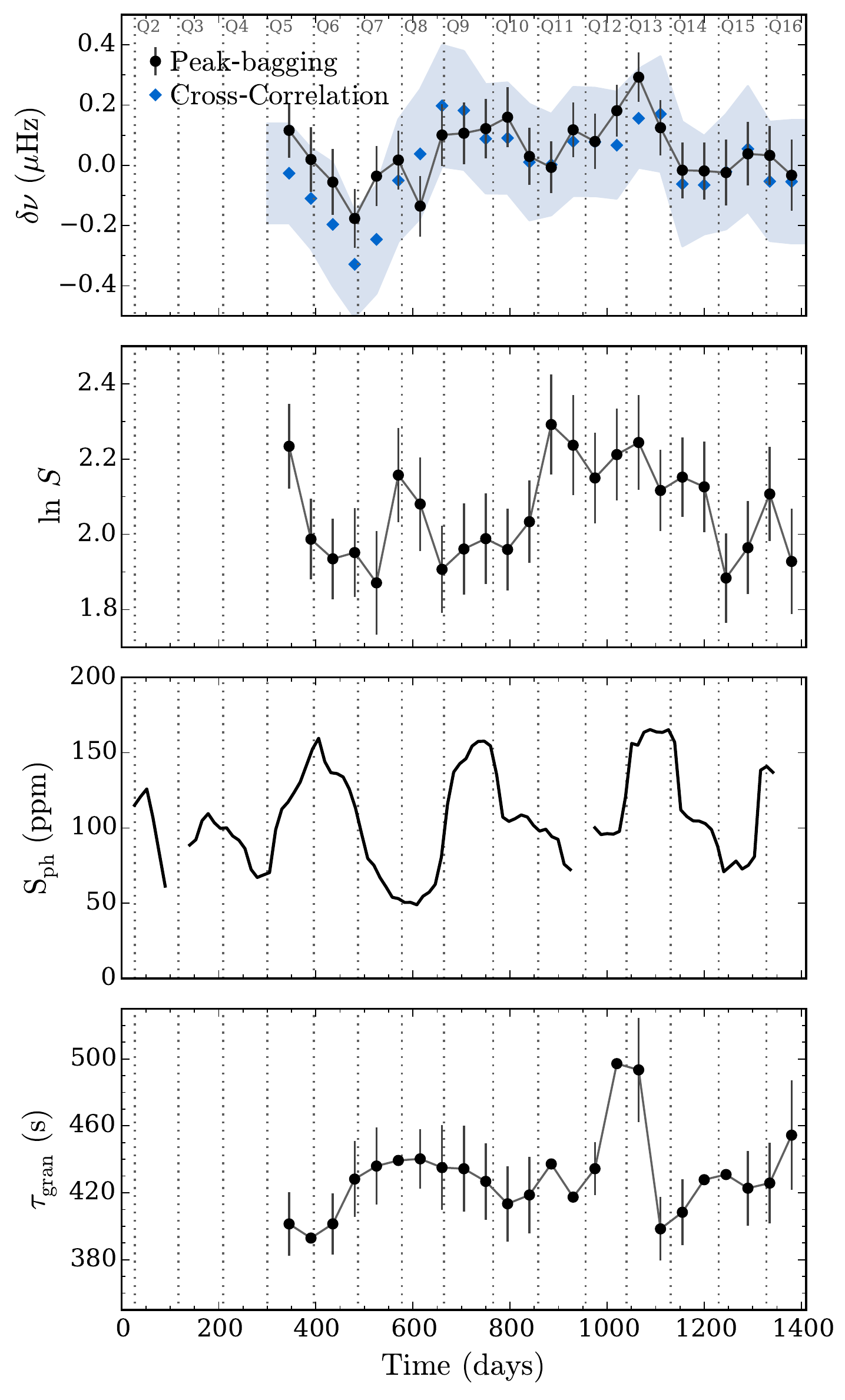}
\caption{Same as in Figure 3, but for KIC 8228742. Results in Table~\ref{tab:8228742}.}\label{fig:8228742}
\end{figure}

\begin{figure}[ht]
\includegraphics[width=\hsize]{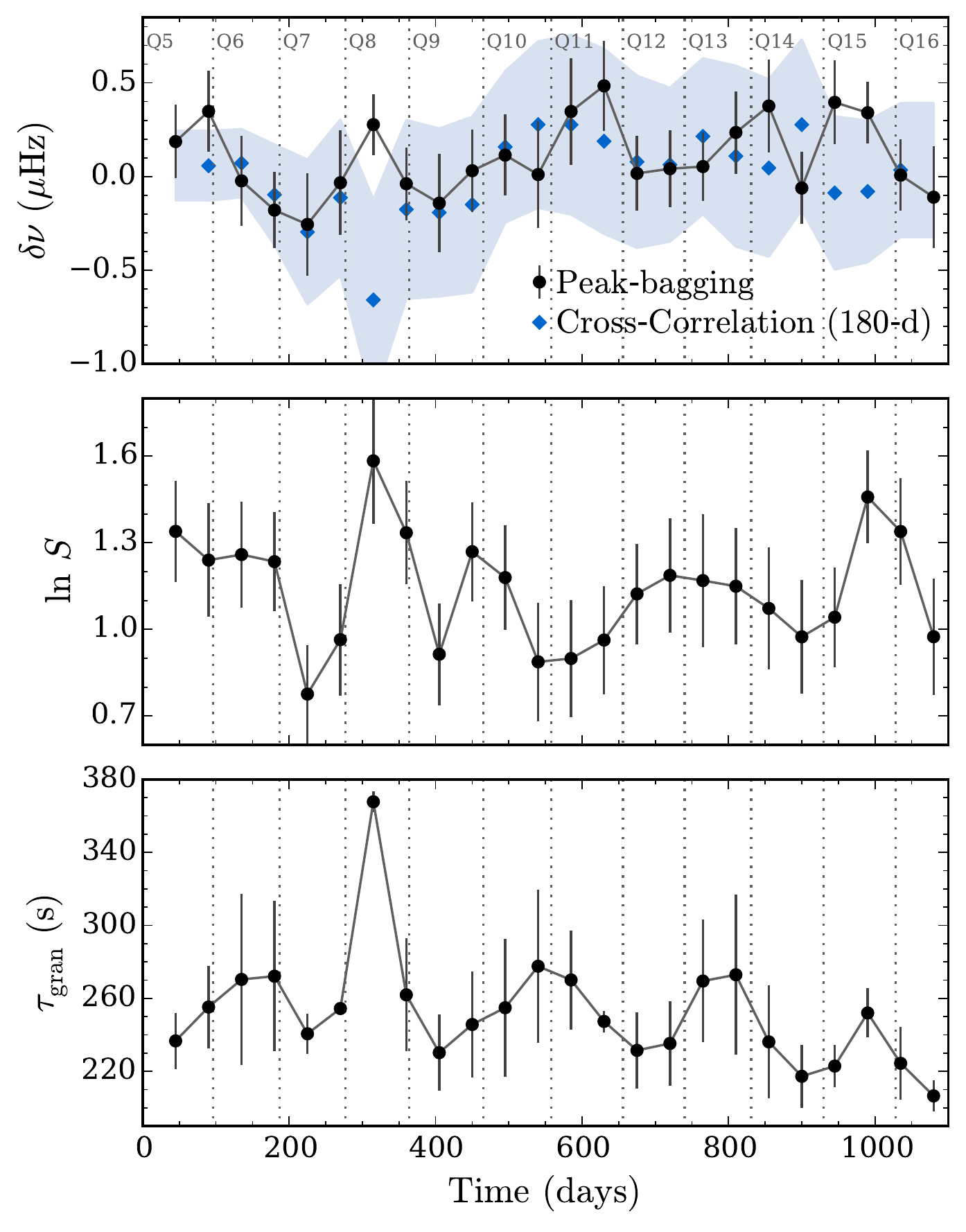}\vspace{-0.2cm}
\caption{Same as in Figure 3, but for KIC 8292840. The frequency shifts from the cross-correlation method were obtained with 180-d sub-series. Results in Table~\ref{tab:8292840}.}\label{fig:8292840}\vspace{-0.3cm}
\end{figure}

\begin{figure}[ht]
\includegraphics[width=\hsize]{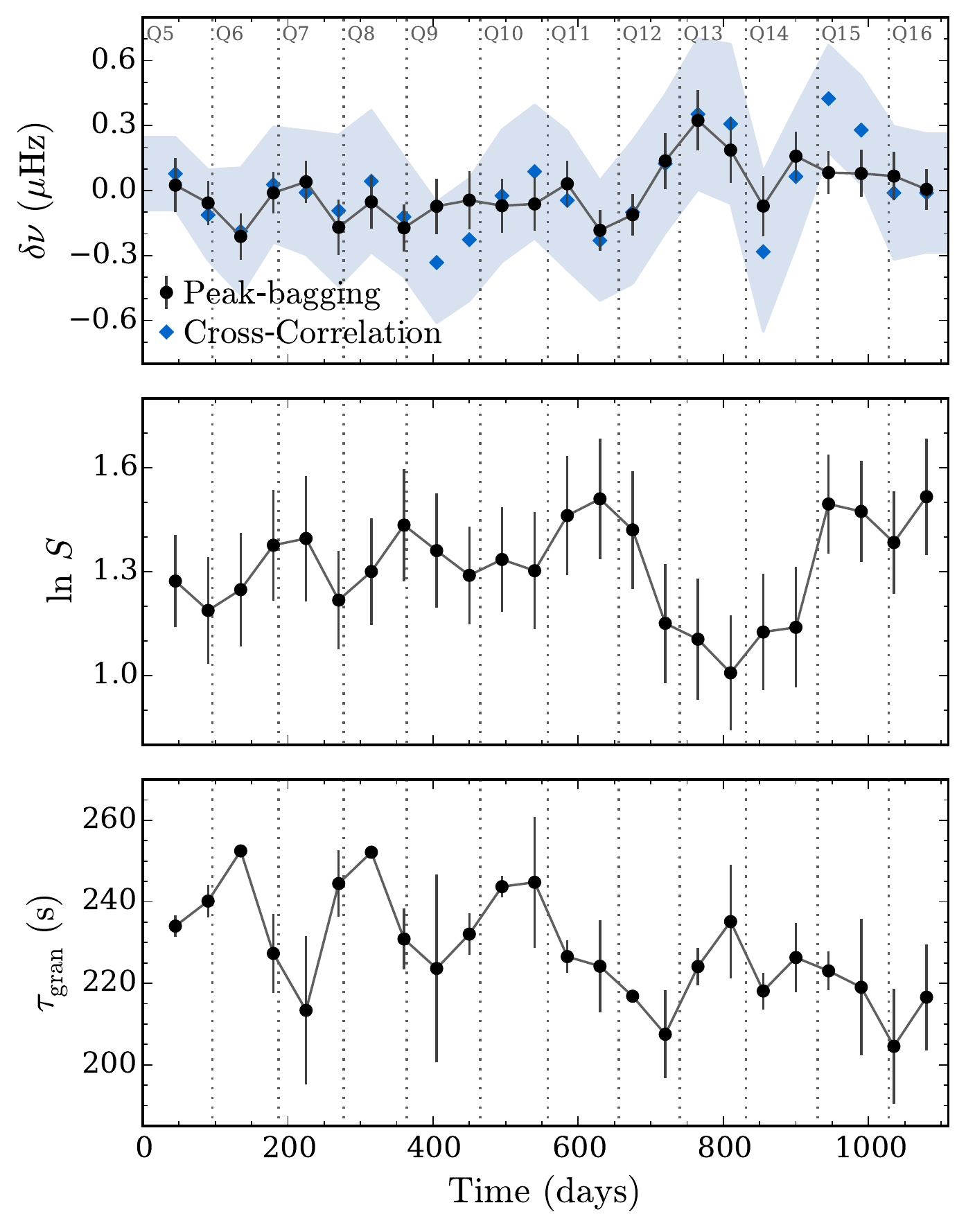}\vspace{-0.2cm}
\caption{Same as in Figure 3, but for KIC 8394589. Results in Table~\ref{tab:8394589}.}\label{fig:8394589}\vspace{-1cm}
\end{figure}

\begin{figure}[ht]
\includegraphics[width=\hsize]{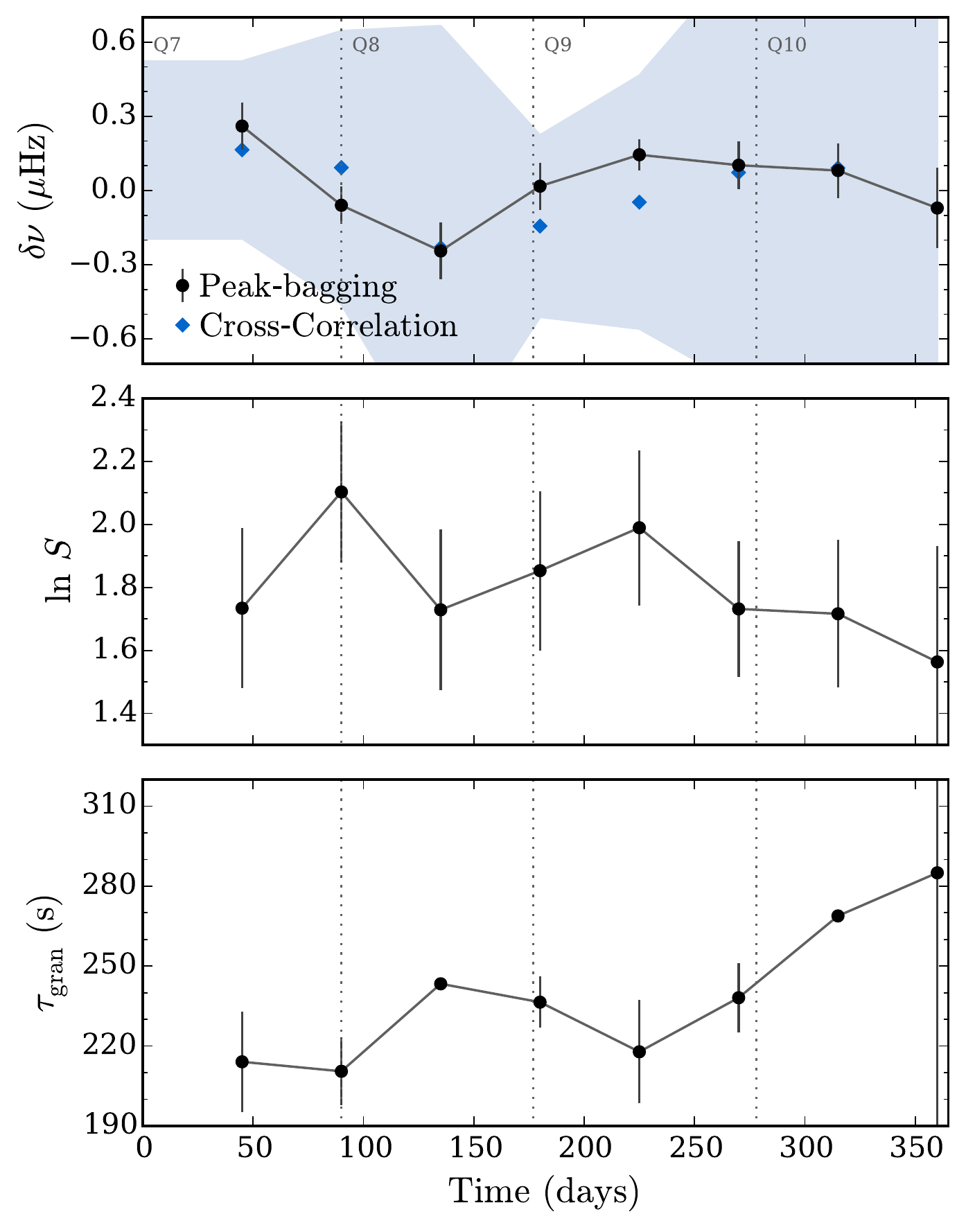}\vspace{-0.2cm}
\caption{Same as in Figure 3, but for KIC 8424992. Results in Table~\ref{tab:8424992}.}\label{fig:8424992}\vspace{-0.3cm}
\end{figure}

\begin{figure}[ht]\vspace{-0.2cm}
\includegraphics[width=\hsize]{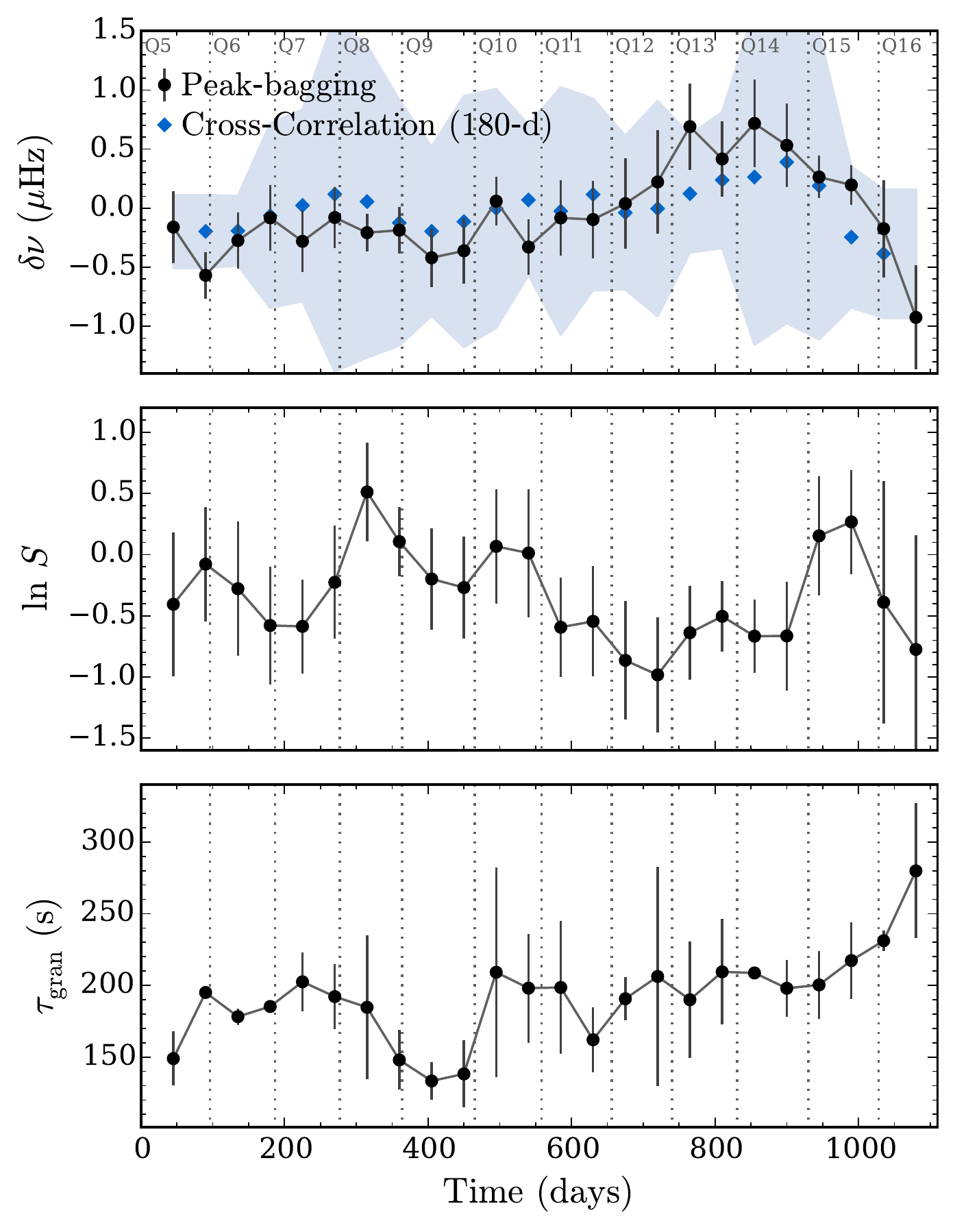}\vspace{-0.2cm}
\caption{Same as in Figure 3, but for KIC 8478994. The frequency shifts from the cross-correlation method were obtained with 180-d sub-series. Results in Table~\ref{tab:8478994}.}\label{fig:8478994}\vspace{-1cm}
\end{figure}

\begin{figure}[ht]
\includegraphics[width=\hsize]{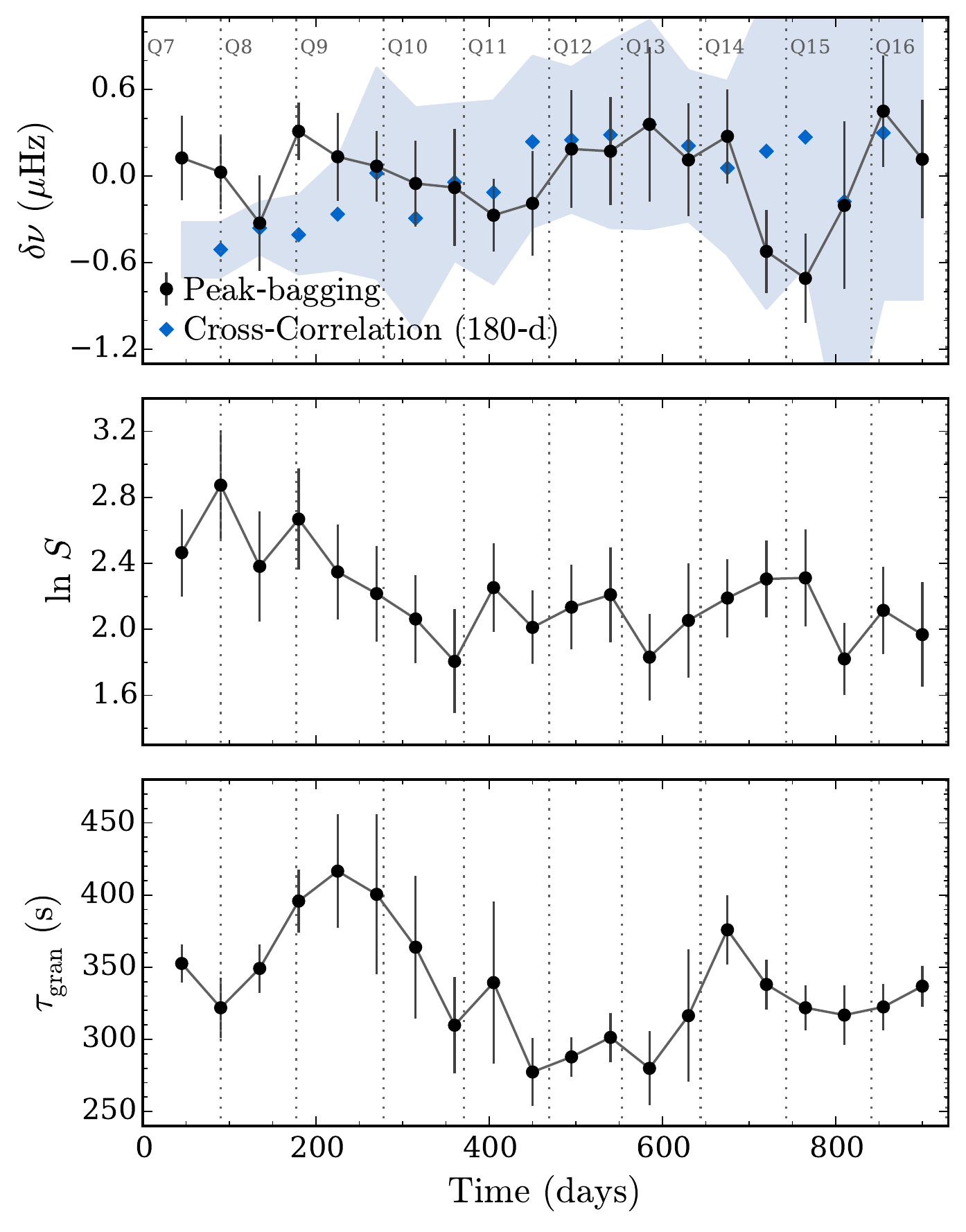}
\caption{Same as in Figure 3, but for KIC 8494142. The frequency shifts from the cross-correlation method were obtained with 180-d sub-series. Results in Table~\ref{tab:8494142}.}\label{fig:8494142}
\end{figure}

\begin{figure}[ht]
\includegraphics[width=\hsize]{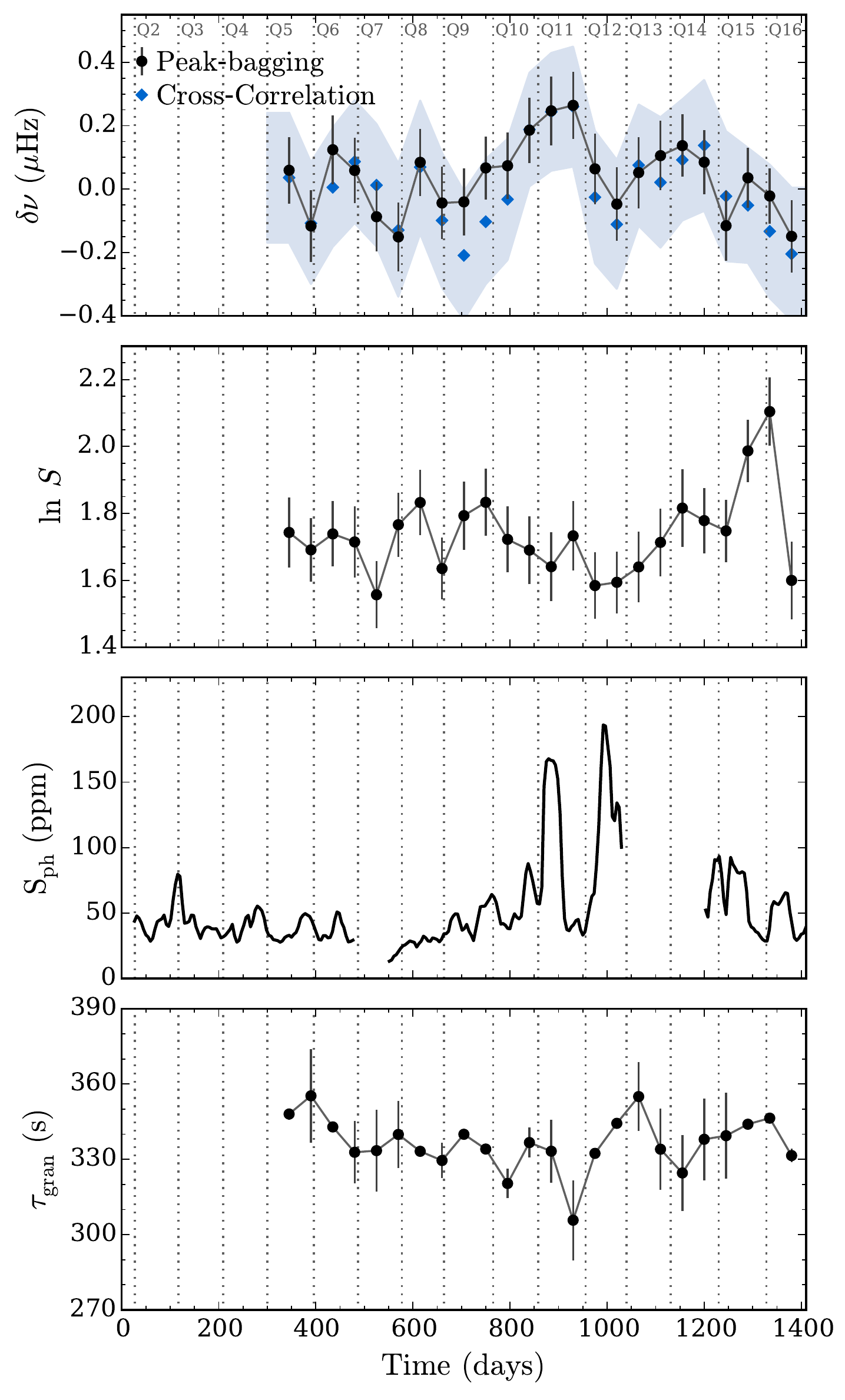}
\caption{Same as in Figure 3, but for KIC 8694723. Results in Table~\ref{tab:8694723}.}\label{fig:8694723}
\end{figure}

\begin{figure}[ht]
\includegraphics[width=\hsize]{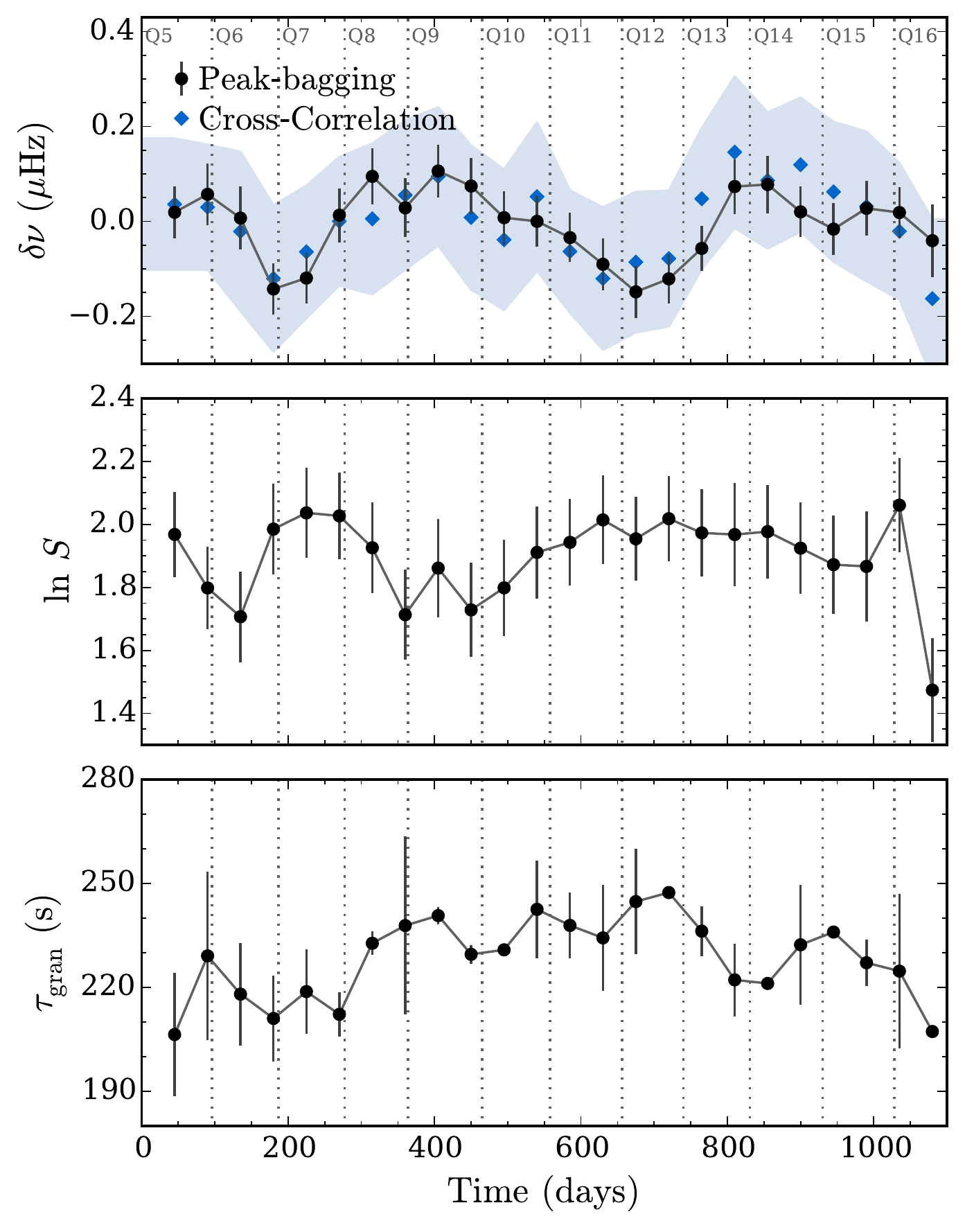}\vspace{-0.2cm}
\caption{Same as in Figure 3, but for KIC 8760414. Results in Table~\ref{tab:8760414}.}\label{fig:8760414}
\end{figure}

\begin{figure}[ht]
\includegraphics[width=\hsize]{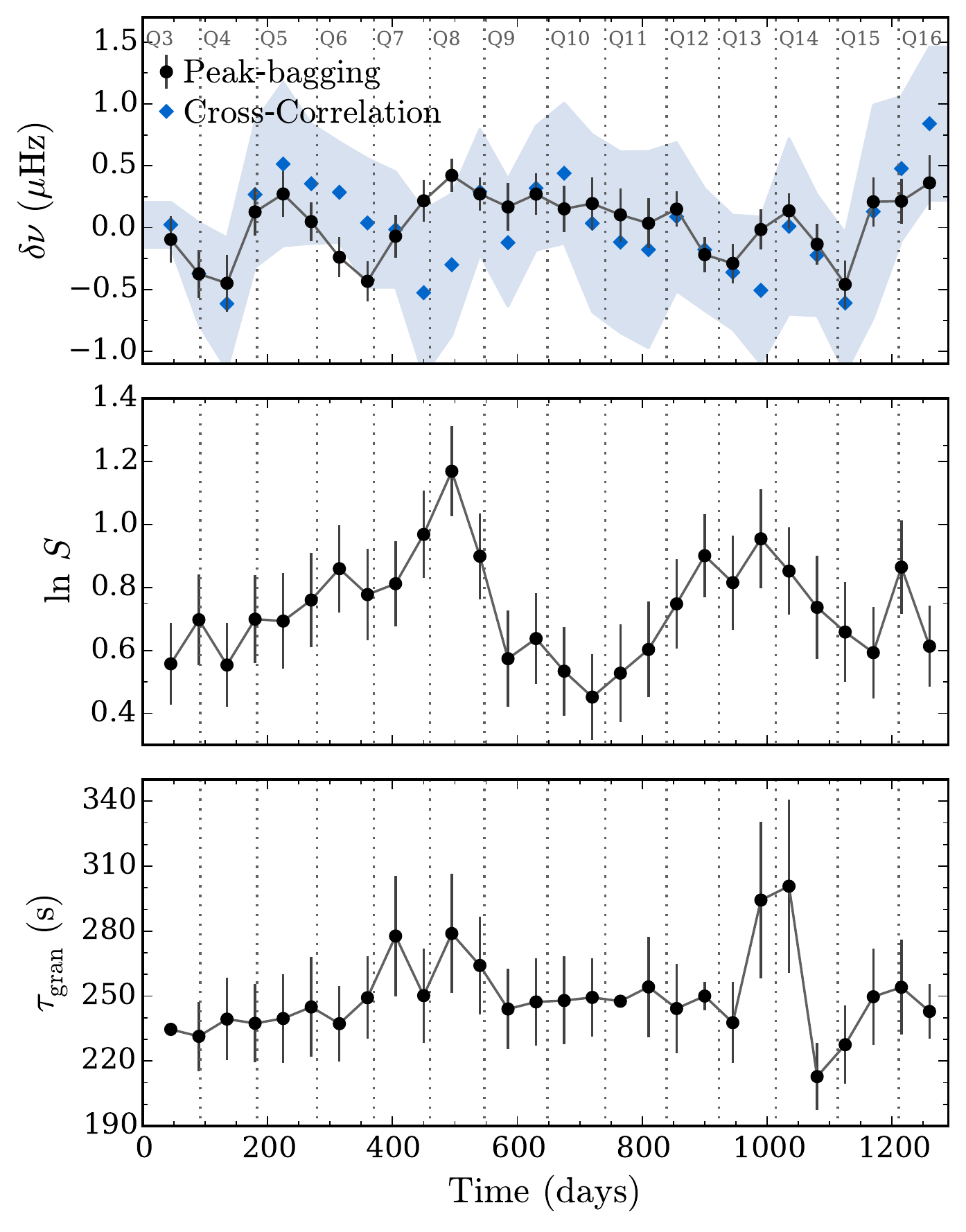}\vspace{-0.2cm}
\caption{Same as in Figure 3, but for KIC 8866102. Results in Table~\ref{tab:8866102}.}\label{fig:8866102}\vspace{-1cm}
\end{figure}

\begin{figure}[ht]
\includegraphics[width=\hsize]{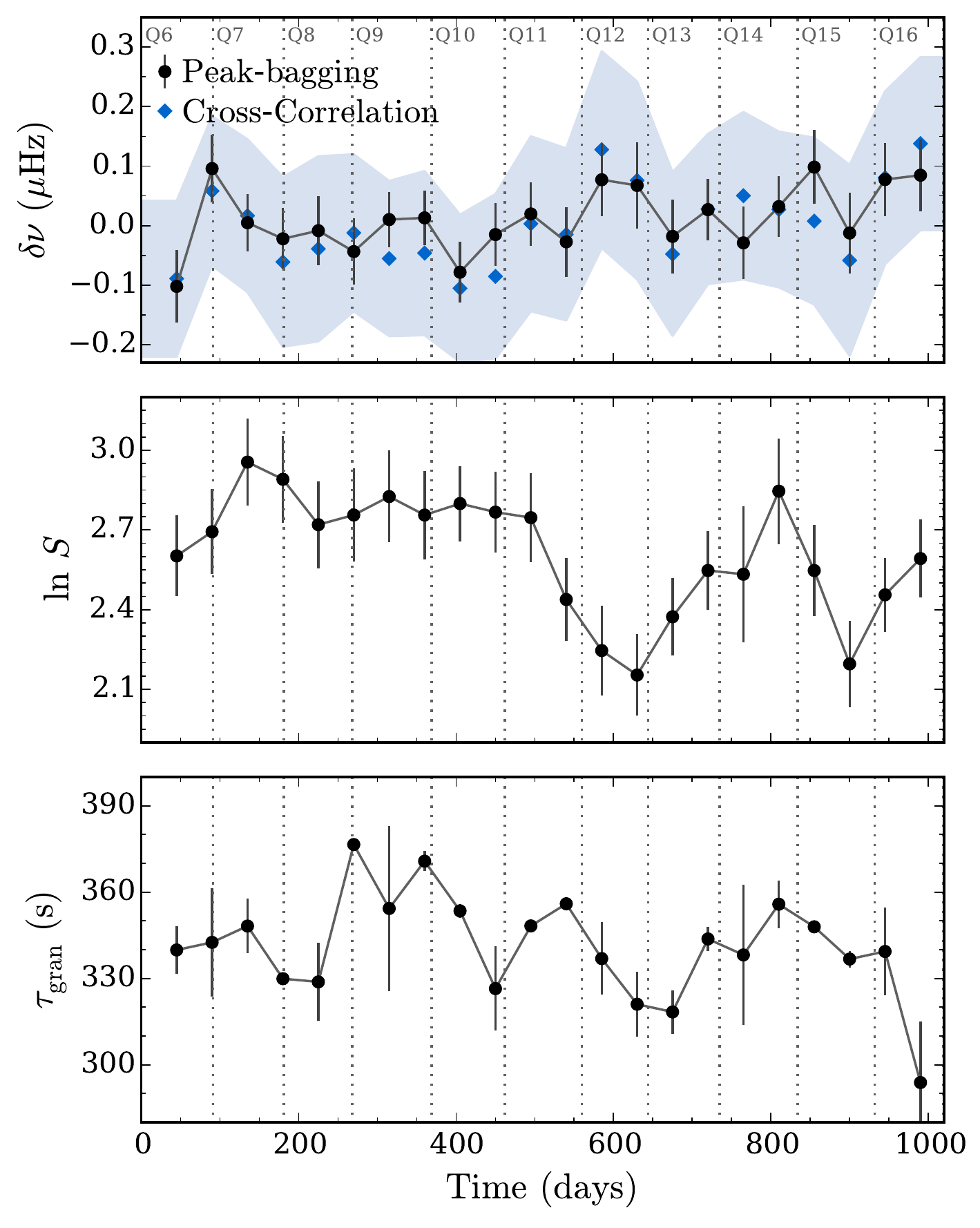}
\caption{Same as in Figure 3, but for KIC 8938364. Results in Table~\ref{tab:8938364}.}\label{fig:8938364}
\end{figure}

\begin{figure}[ht]
\includegraphics[width=\hsize]{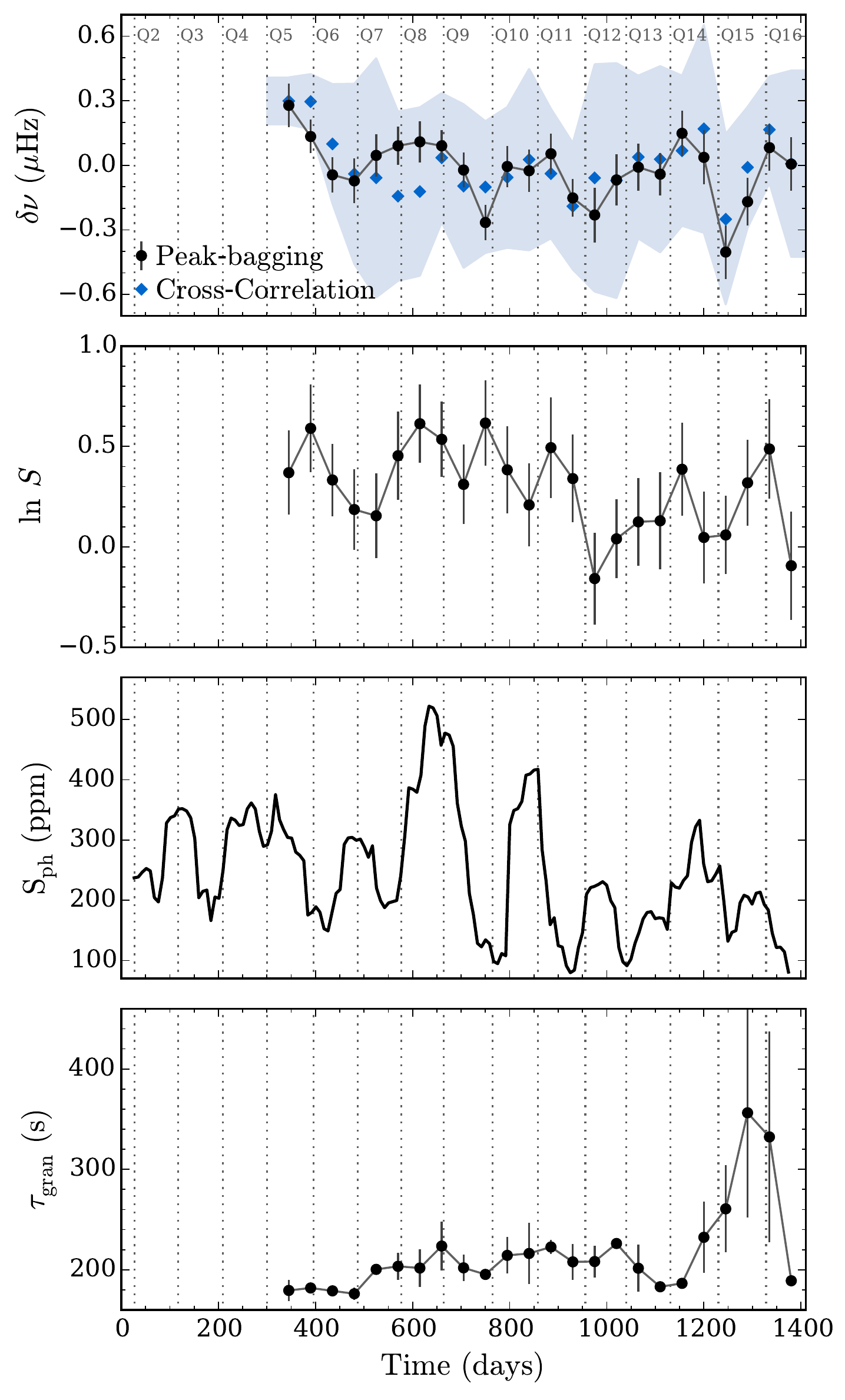}
\caption{Same as in Figure 3, but for KIC 9025370. Results in Table~\ref{tab:9025370}.}\label{fig:9025370}
\end{figure}

\begin{figure}[ht]
\includegraphics[width=\hsize]{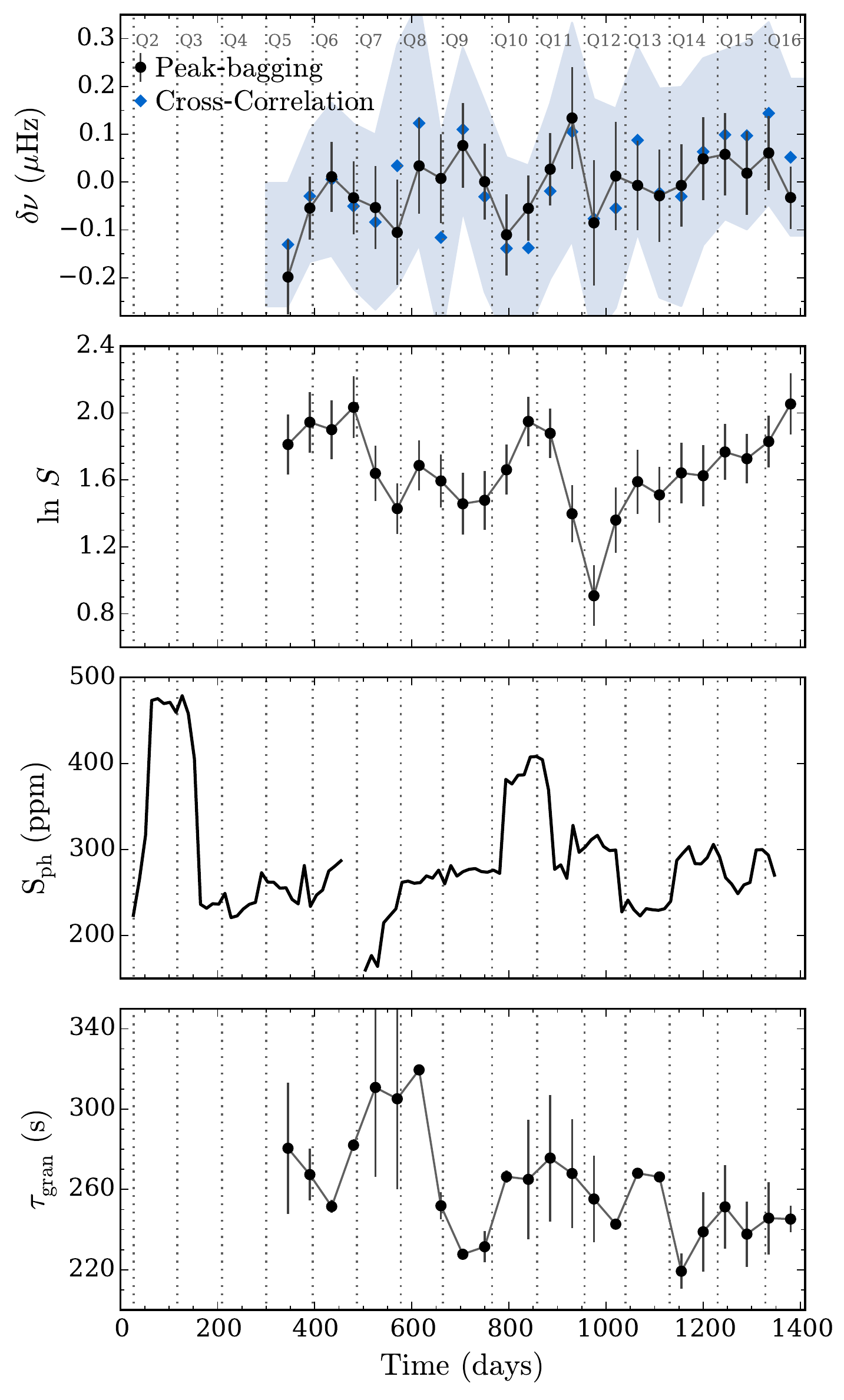}
\caption{Same as in Figure 3, but for KIC 9098294. Results in Table~\ref{tab:9098294}.}\label{fig:9098294}
\end{figure}

\FloatBarrier
\nopagebreak
%!TEX root = peakbagging.tex
\begin{figure}[ht]
\includegraphics[width=\hsize]{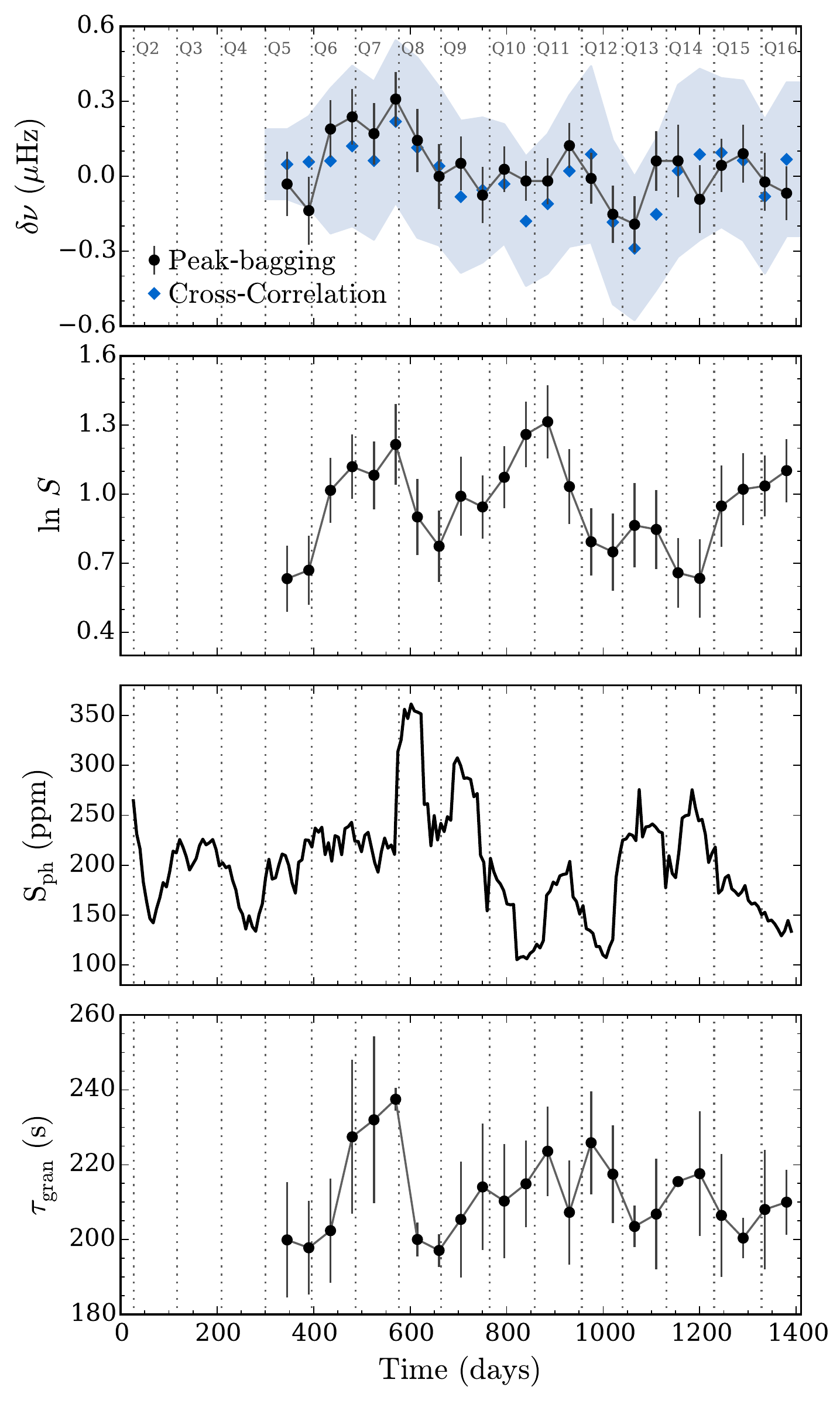}
\caption{Same as in Figure 3, but for KIC 9139151. Results in Table~\ref{tab:9139151}.}\label{fig:9139151}
\end{figure}

\begin{figure}[ht]
\includegraphics[width=\hsize]{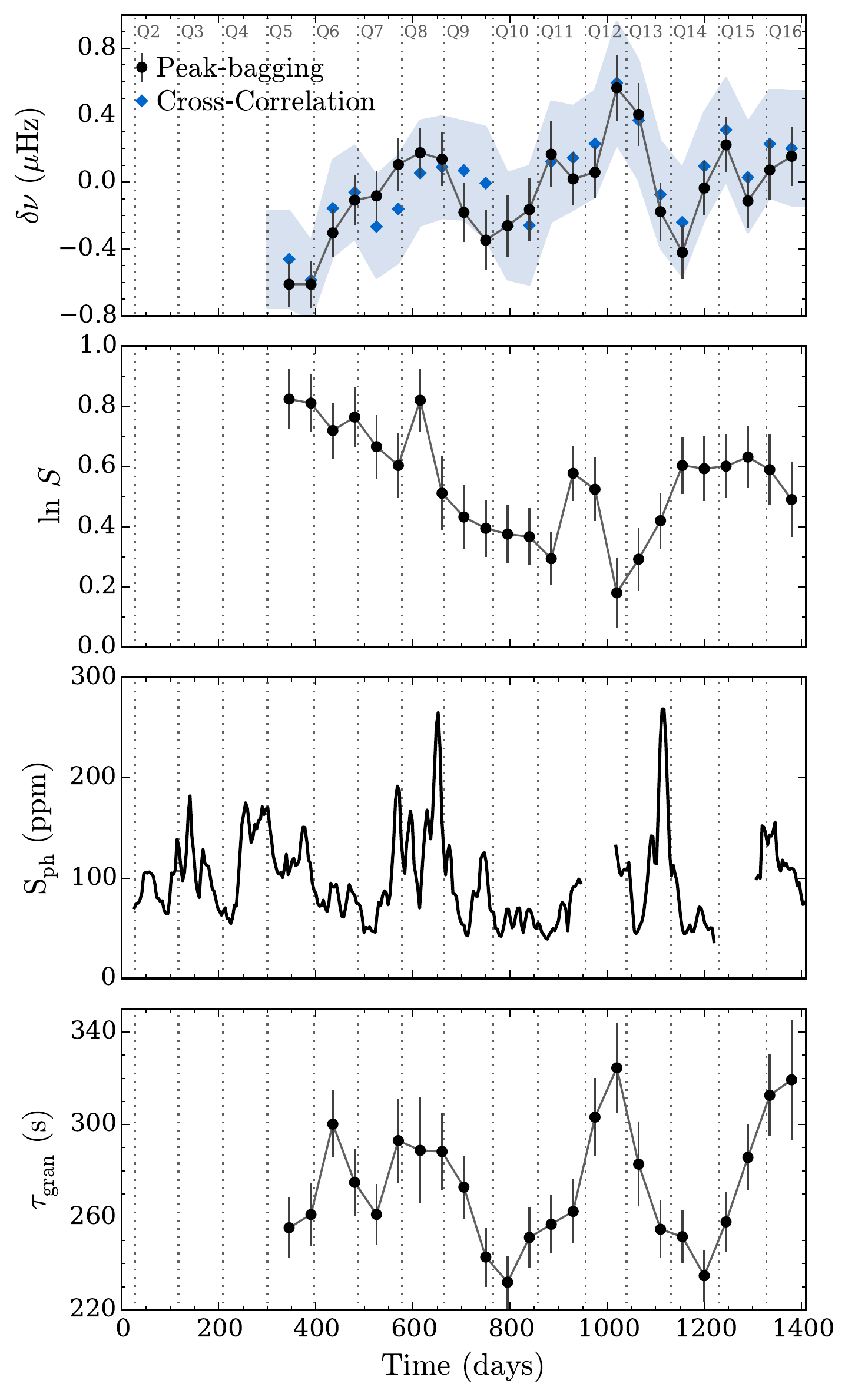}
\caption{Same as in Figure 3, but for KIC 9139163. Results in Table~\ref{tab:9139163}.}\label{fig:9139163}
\end{figure}

\begin{figure}[ht]
\includegraphics[width=\hsize]{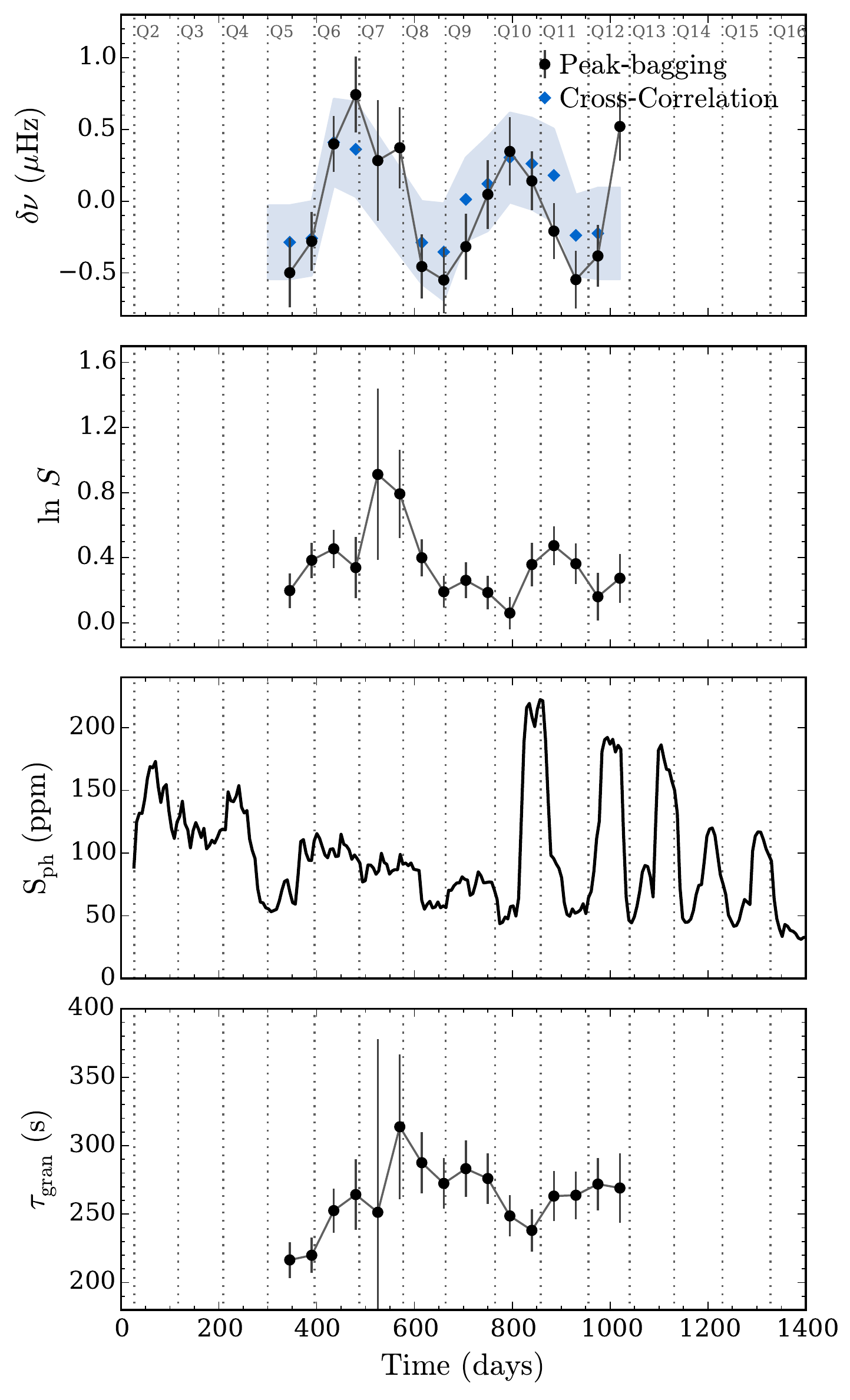}
\caption{Same as in Figure 3, but for KIC 9206432. Results in Table~\ref{tab:9206432}.}\label{fig:9206432}
\end{figure}

\begin{figure}[ht]
\includegraphics[width=\hsize]{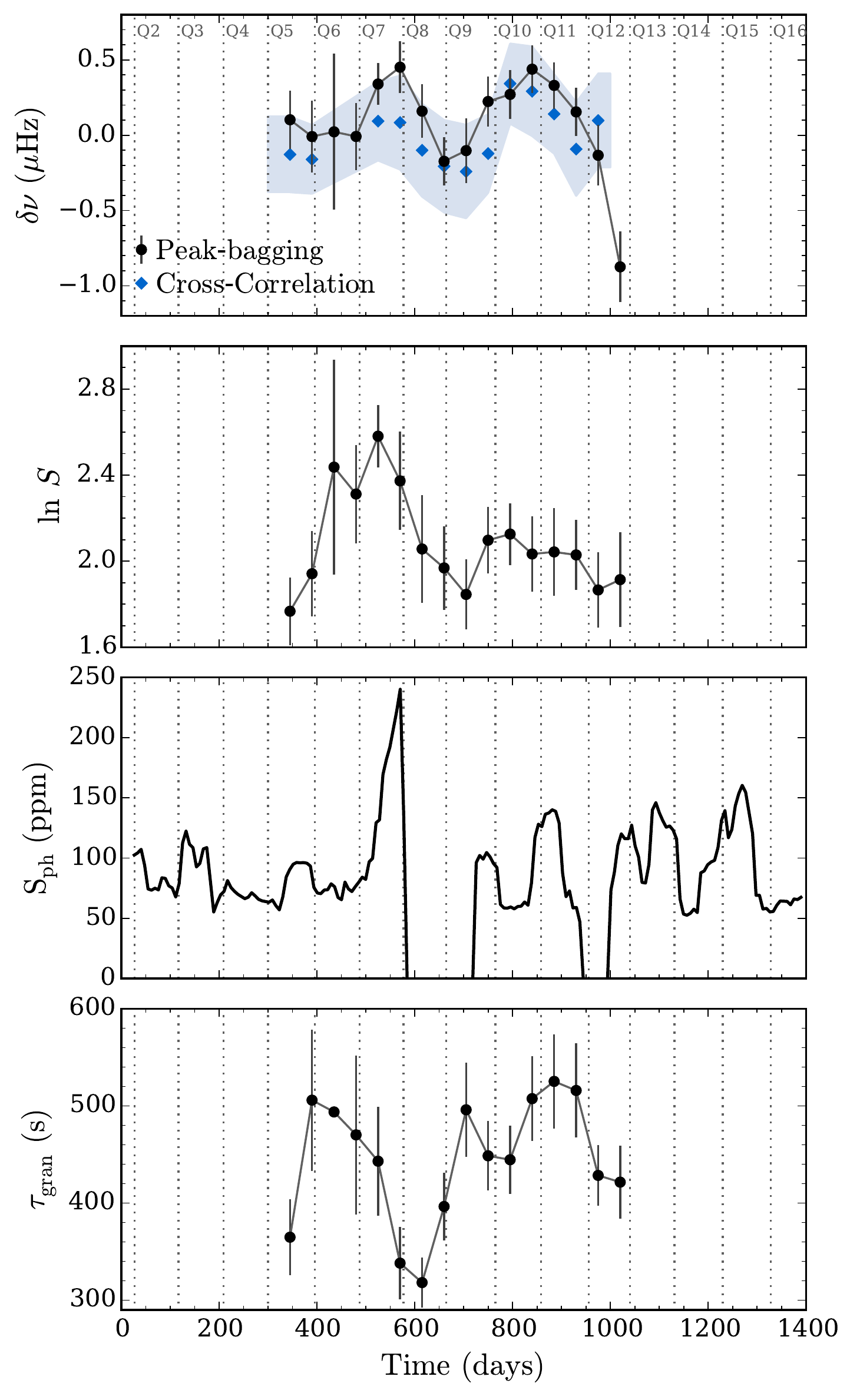}
\caption{Same as in Figure 3, but for KIC 9353712. Results in Table~\ref{tab:9353712}.}\label{fig:9353712}
\end{figure}

\begin{figure}[ht]
\includegraphics[width=\hsize]{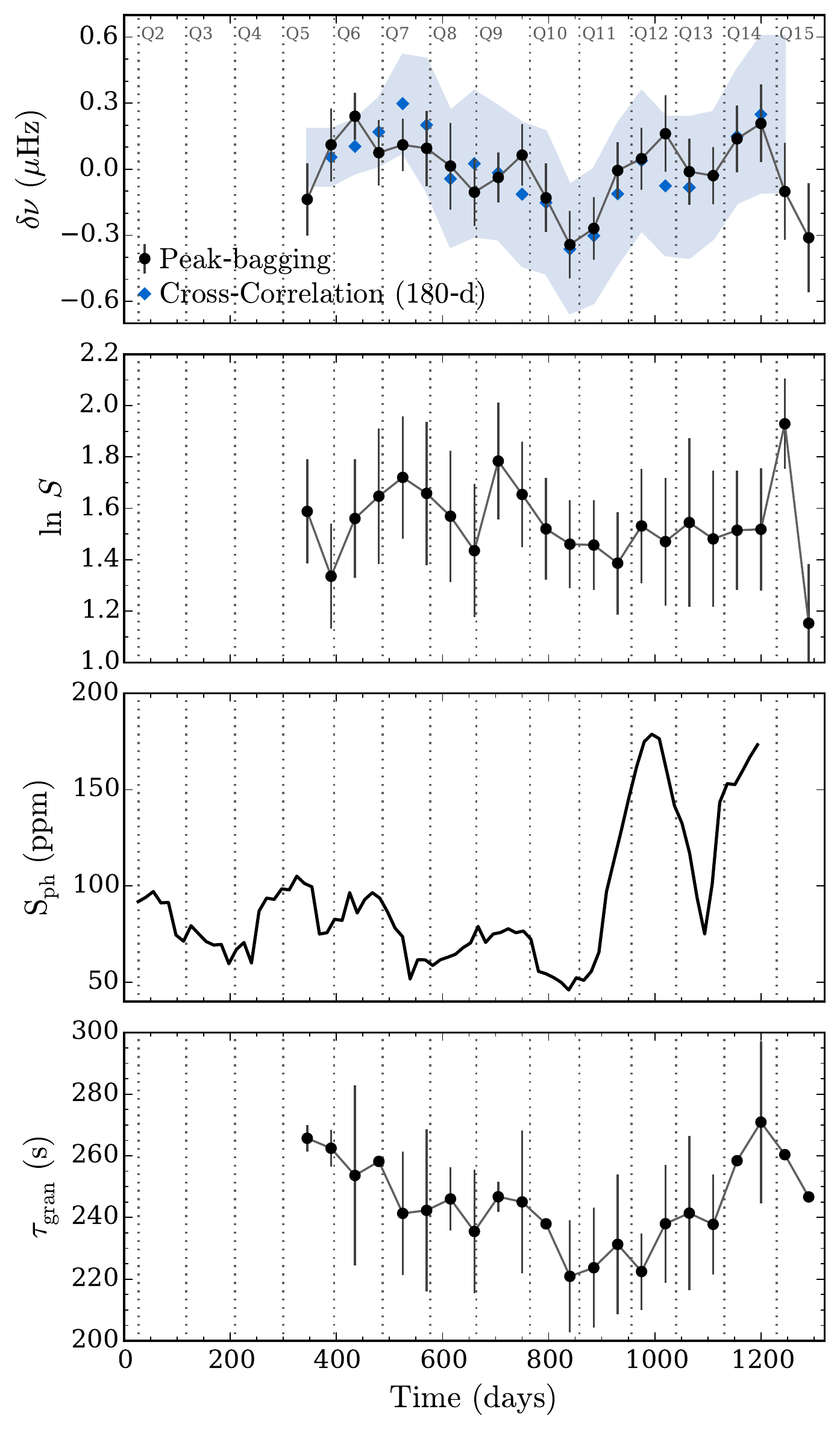}
\caption{Same as in Figure 3, but for KIC 9410862. The frequency shifts from the cross-correlation method were obtained with 180-d sub-series. Results in Table~\ref{tab:9410862}.}\label{fig:9410862}
\end{figure}

\begin{figure}[ht]
\includegraphics[width=\hsize]{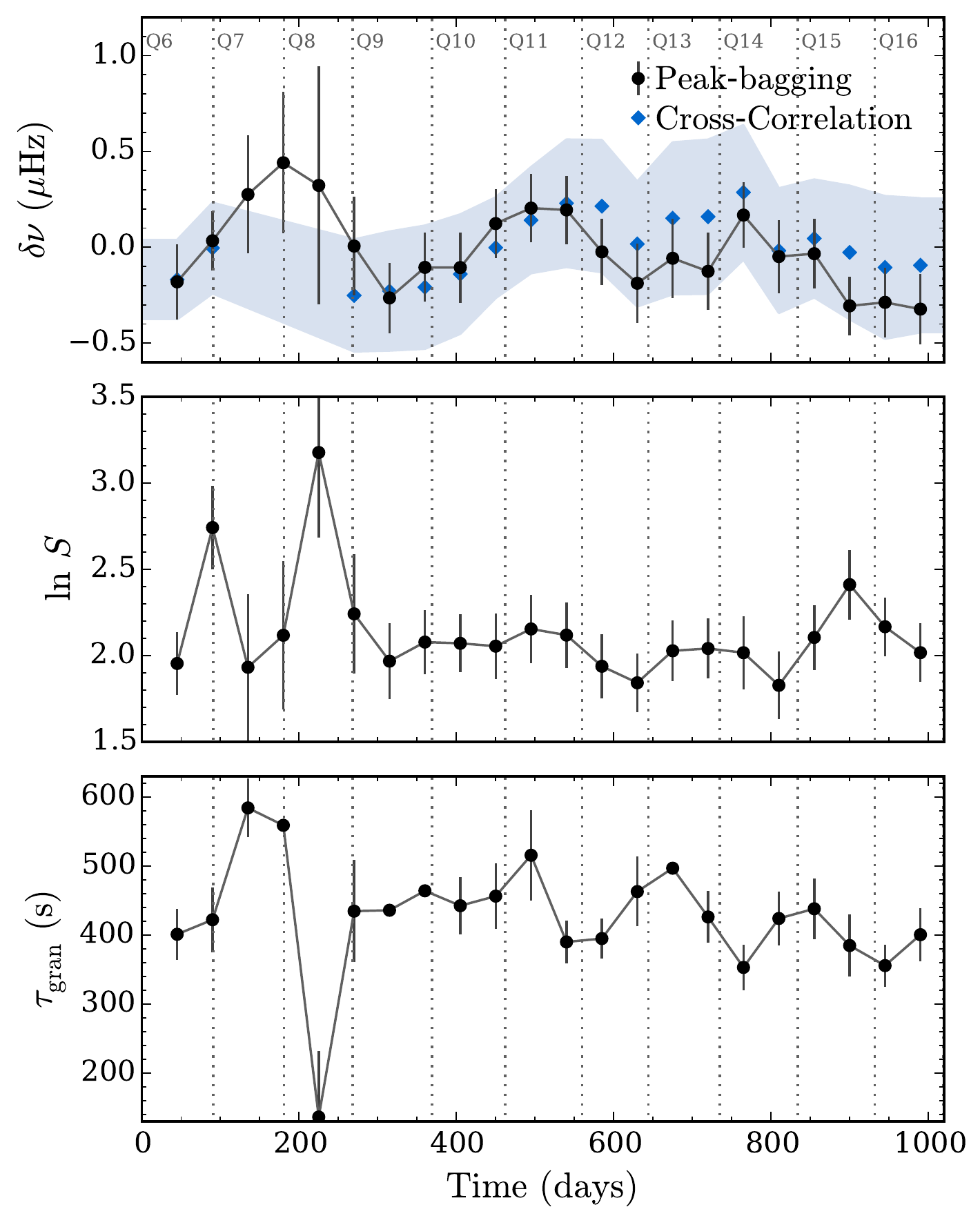}
\caption{Same as in Figure 3, but for KIC 9592705. Results in Table~\ref{tab:9592705}.}\label{fig:9592705}
\end{figure}

\begin{figure}[ht]
\includegraphics[width=\hsize]{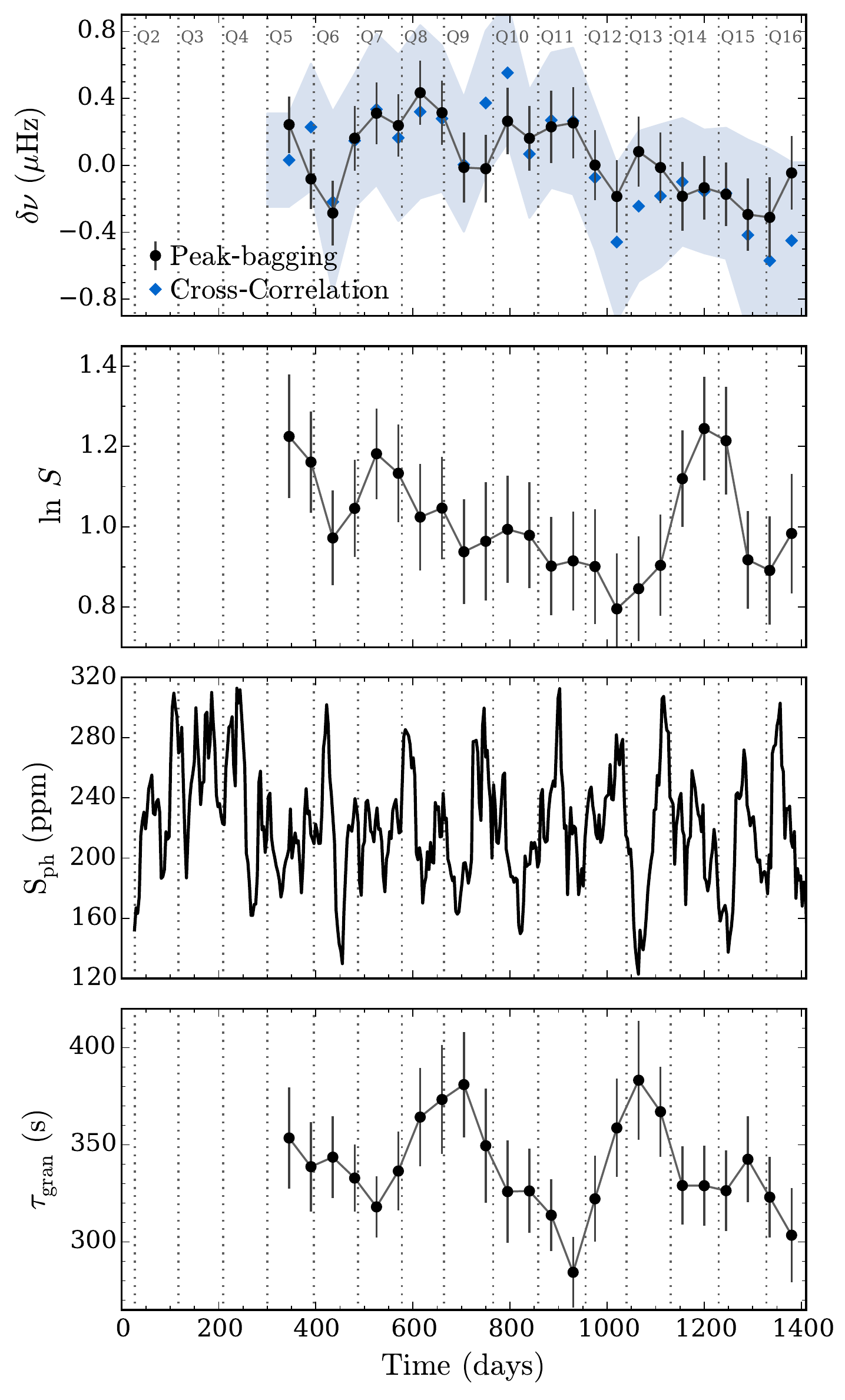}
\caption{Same as in Figure 3, but for KIC 9812850. Results in Table~\ref{tab:9812850}.}\label{fig:9812850}
\end{figure}

\begin{figure}[ht]
\includegraphics[width=\hsize]{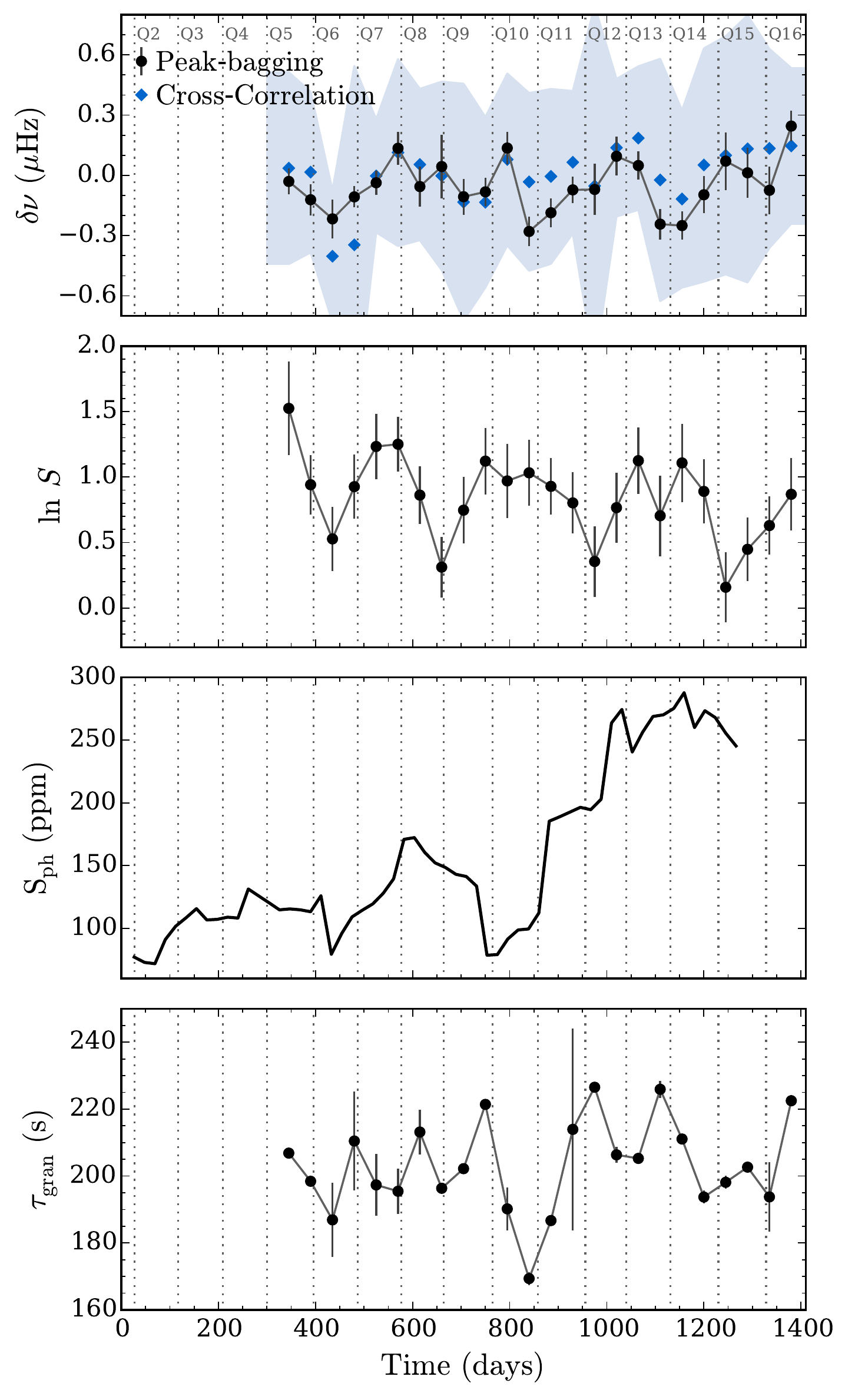}
\caption{Same as in Figure 3, but for KIC 9955598. Results in Table~\ref{tab:9955598}.}\label{fig:9955598}
\end{figure}

\begin{figure}[ht]
\includegraphics[width=\hsize]{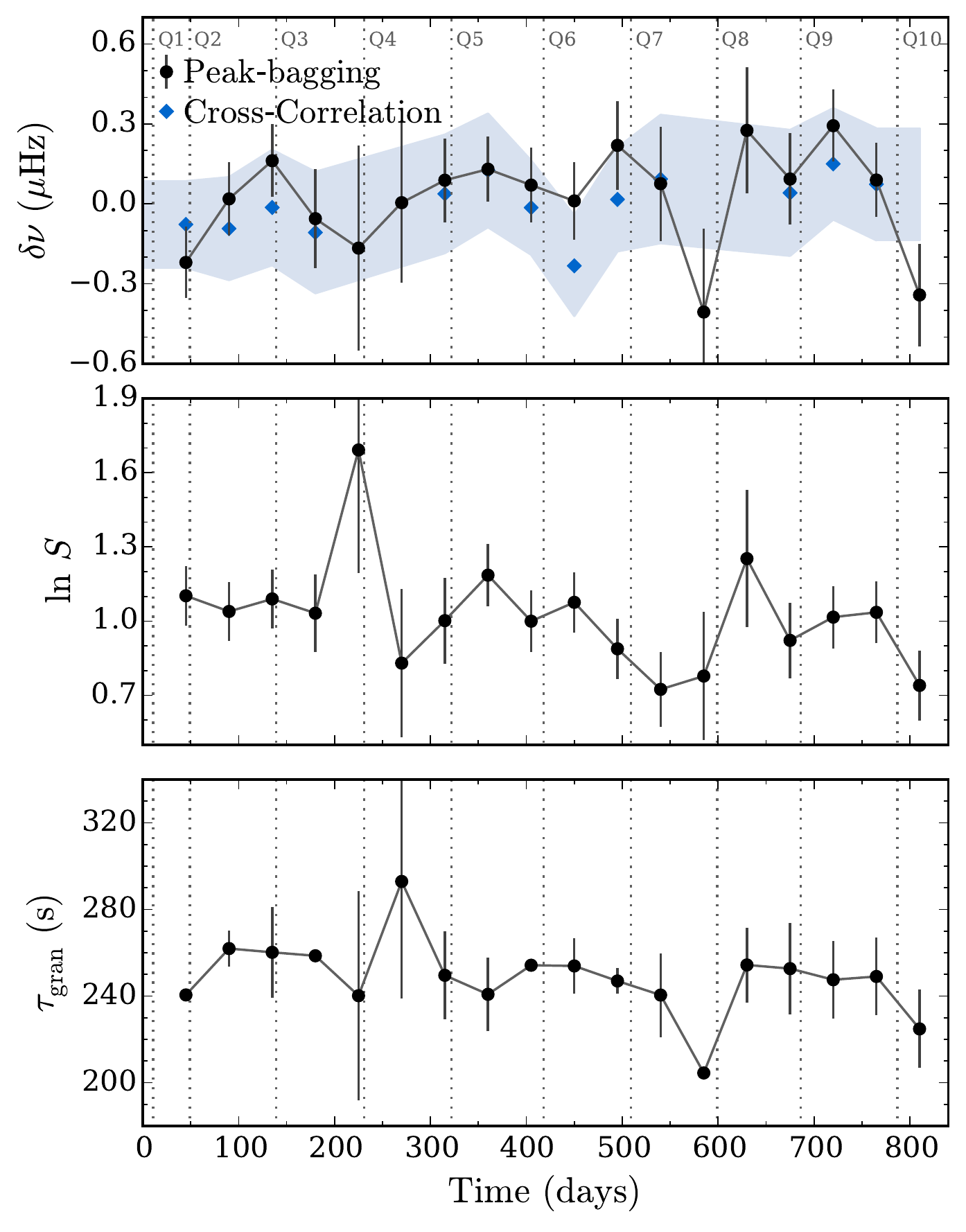}
\caption{Same as in Figure 3, but for KIC 9965715. Results in Table~\ref{tab:9965715}.}\label{fig:9965715}
\end{figure}

\begin{figure}[ht]
\includegraphics[width=\hsize]{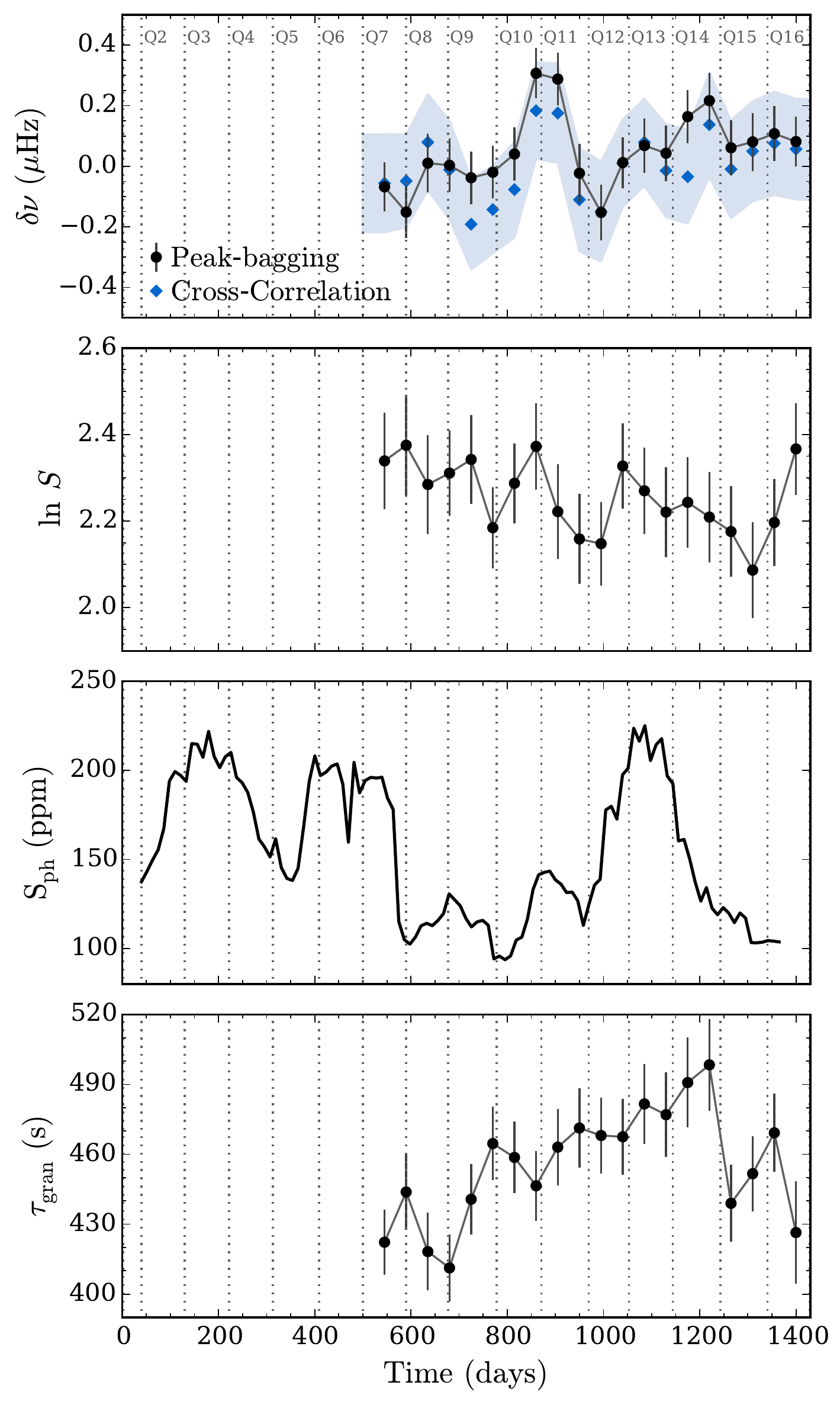}
\caption{Same as in Figure 3, but for KIC 10068307. Results in Table~\ref{tab:10068307}.}\label{fig:10068307}
\end{figure}

\begin{figure}[ht]
\includegraphics[width=\hsize]{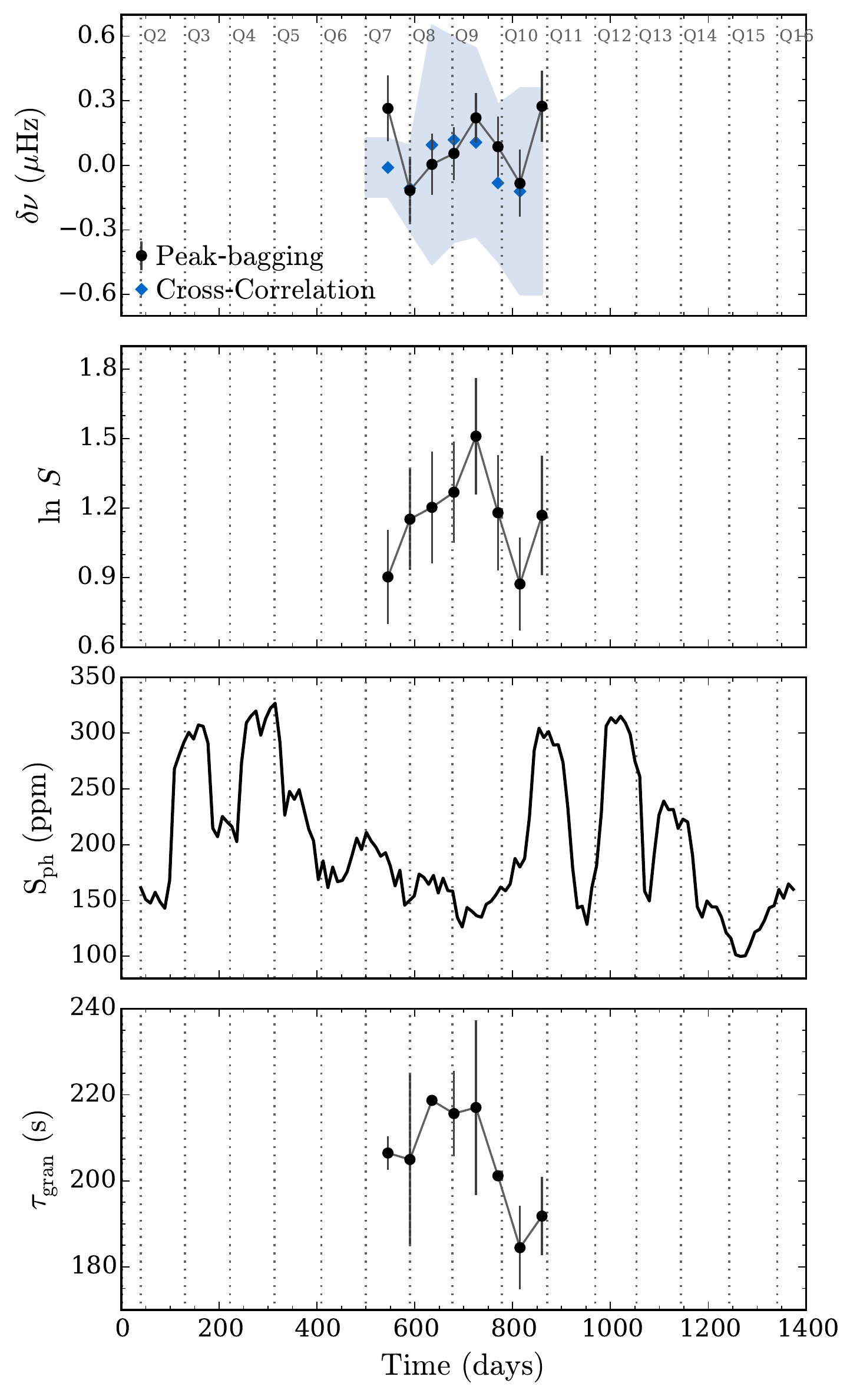}
\caption{Same as in Figure 3, but for KIC 10079226. Results in Table~\ref{tab:10079226}.}\label{fig:10079226}
\end{figure}

\begin{figure}[ht]
\includegraphics[width=\hsize]{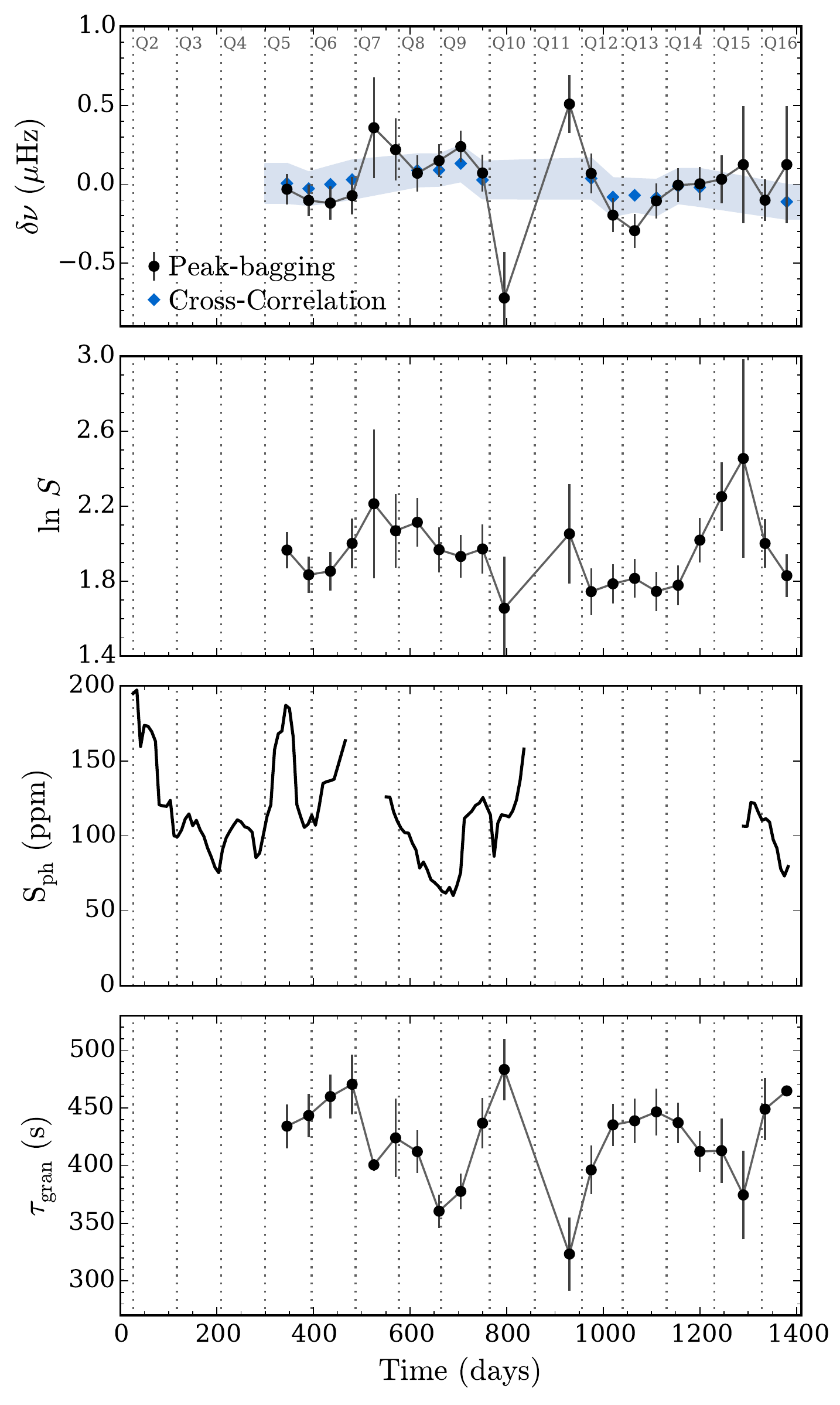}
\caption{Same as in Figure 3, but for KIC 10162436. Results in Table~\ref{tab:10162436}.}\label{fig:10162436}
\end{figure}

\begin{figure}[ht]
\includegraphics[width=\hsize]{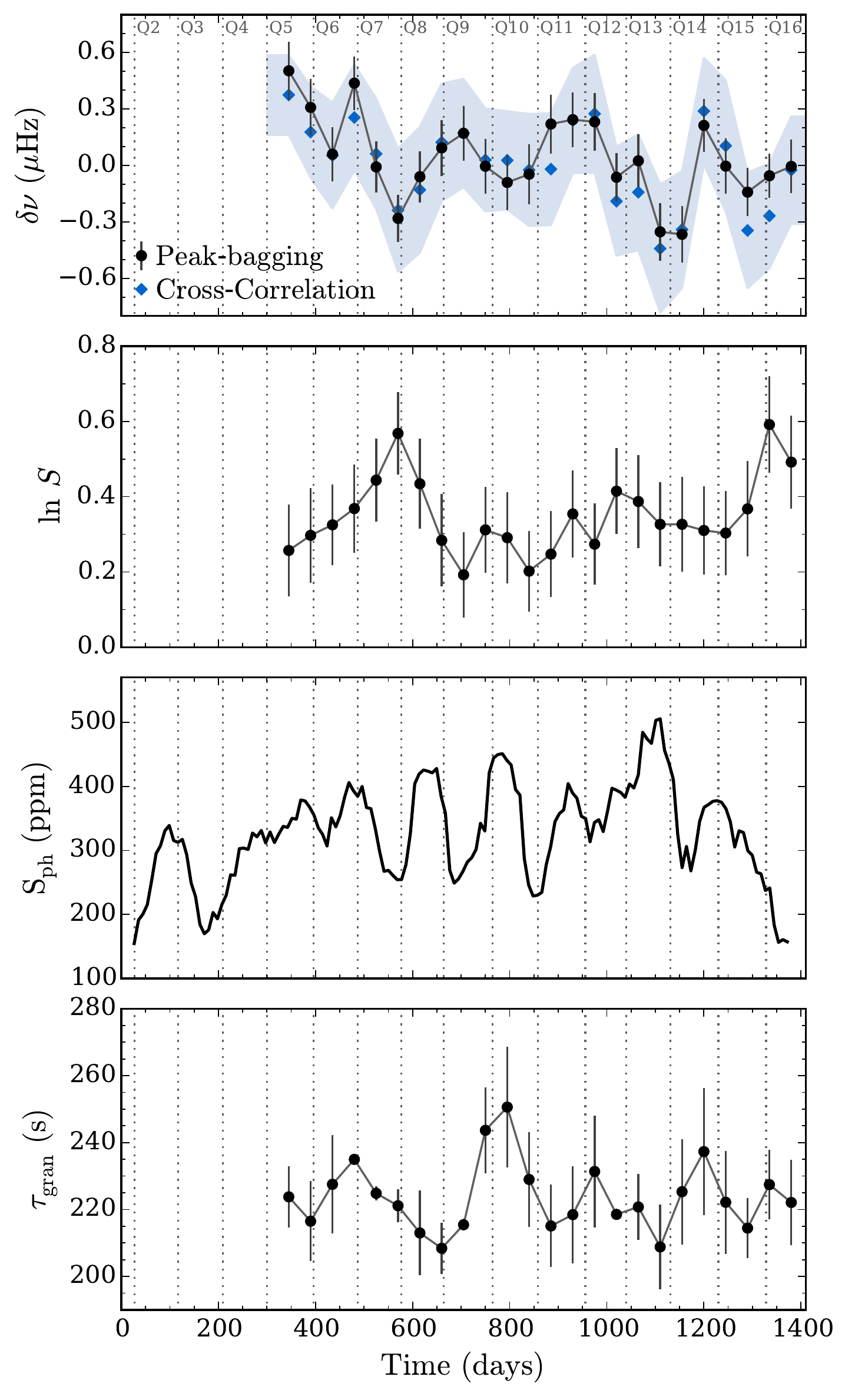}
\caption{Same as in Figure 3, but for KIC 10454113. Results in Table~\ref{tab:10454113}.}\label{fig:10454113}
\end{figure}

\begin{figure}[ht]
\includegraphics[width=\hsize]{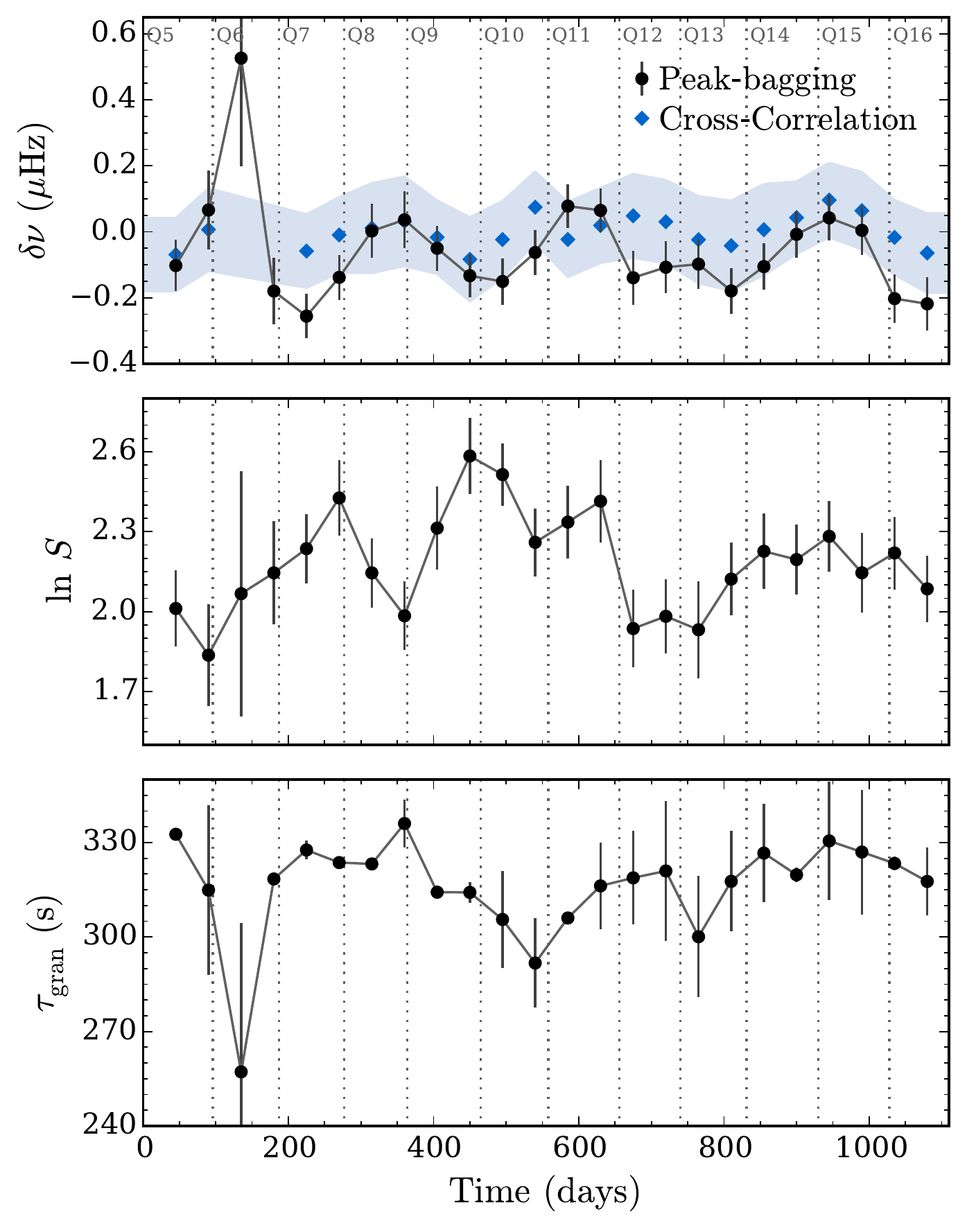}\vspace{-0.2cm}
\caption{Same as in Figure 3, but for KIC 10516096. Results in Table~\ref{tab:10516096}.}\label{fig:10516096}
\end{figure}

\begin{figure}[ht]
\includegraphics[width=\hsize]{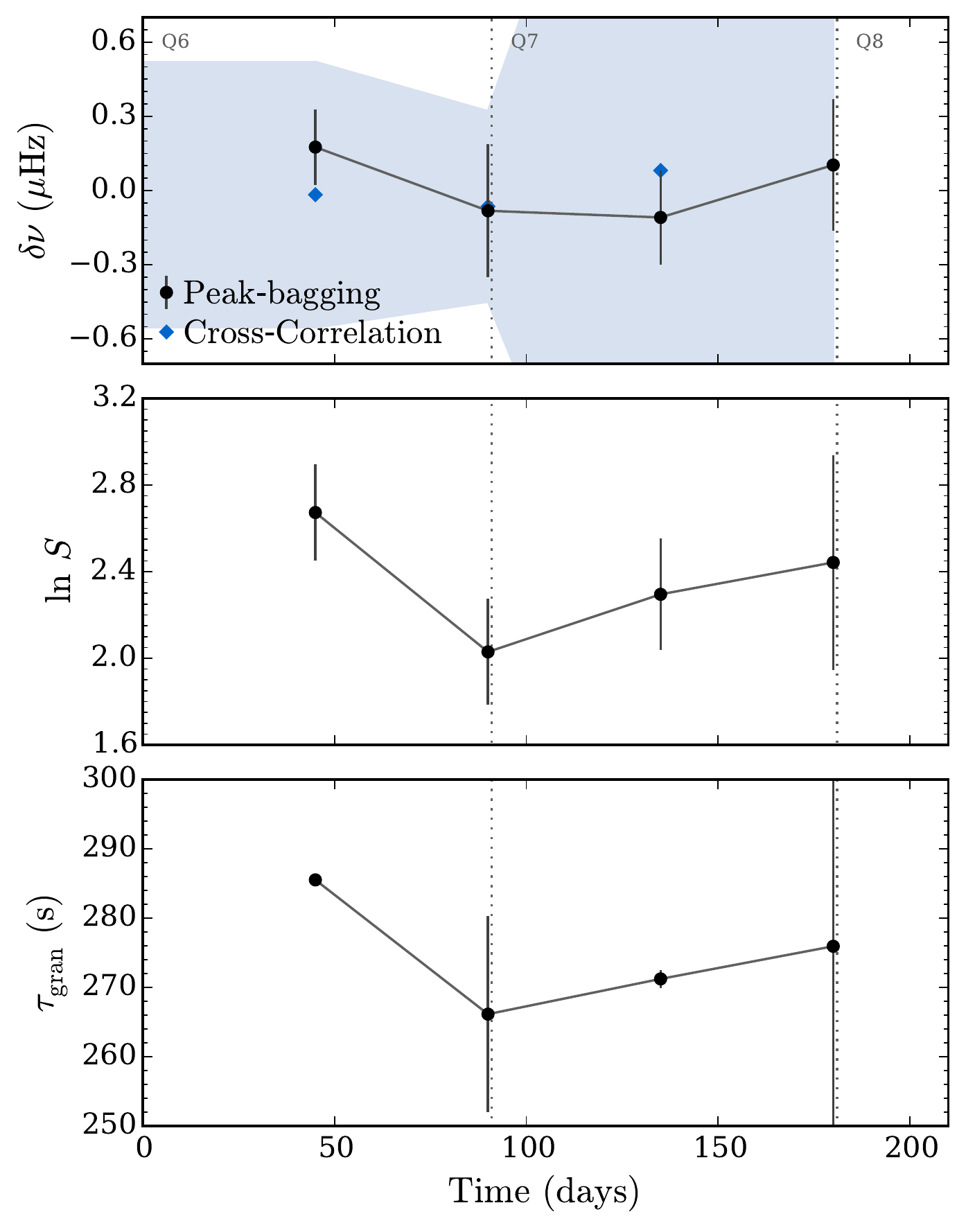}\vspace{-0.2cm}
\caption{Same as in Figure 3, but for KIC 10586004. Results in Table~73.}\label{fig:10586004}\vspace{-1cm}%\ref{tab:10586004}
\end{figure}

\FloatBarrier
\nopagebreak
%!TEX root = peakbagging.tex
\begin{figure}[ht]
\includegraphics[width=\hsize]{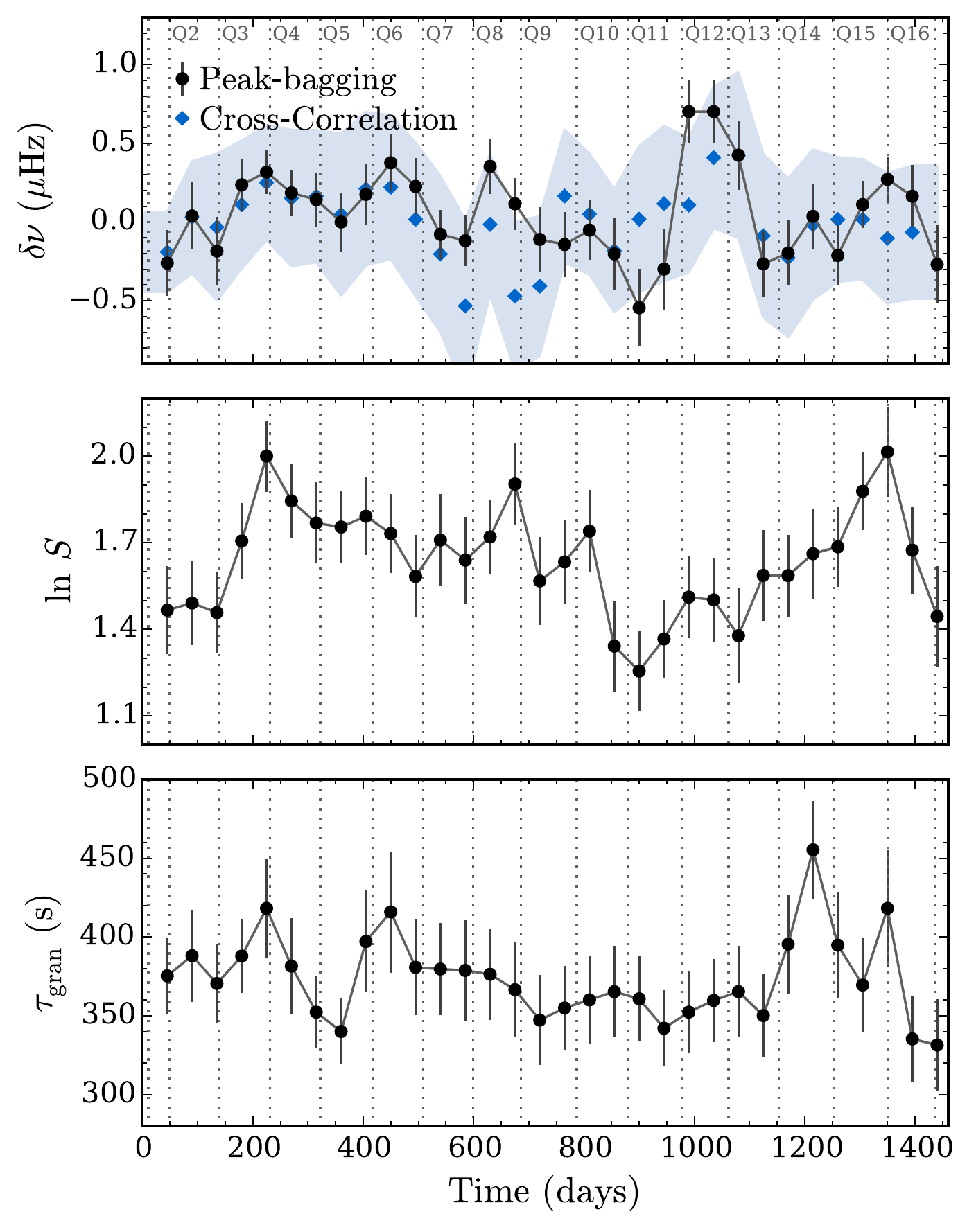}\vspace{-0.3cm}
\caption{Same as in Figure 3, but for KIC 10666592. Results in Table~74.}\label{fig:10666592}\vspace{-0.3cm}%\ref{tab:10666592}
\end{figure}

\begin{figure}[ht]
\includegraphics[width=\hsize]{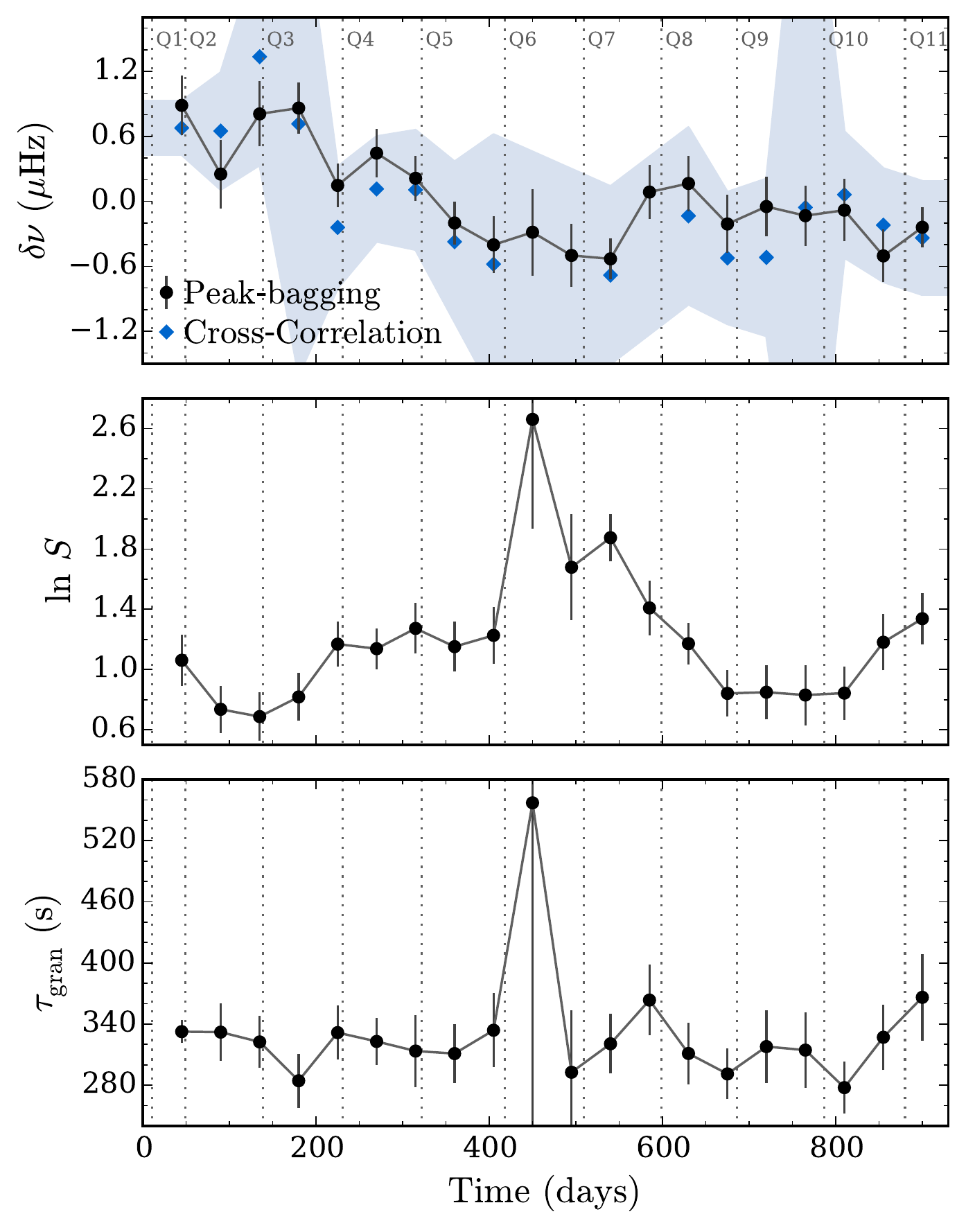}\vspace{-0.3cm}
\caption{Same as in Figure 3, but for KIC 10730618. Results in Table~75.}\label{fig:10730618}\vspace{-1.5cm}%\ref{tab:10730618}
\end{figure}

\begin{figure}[ht]
\includegraphics[width=\hsize]{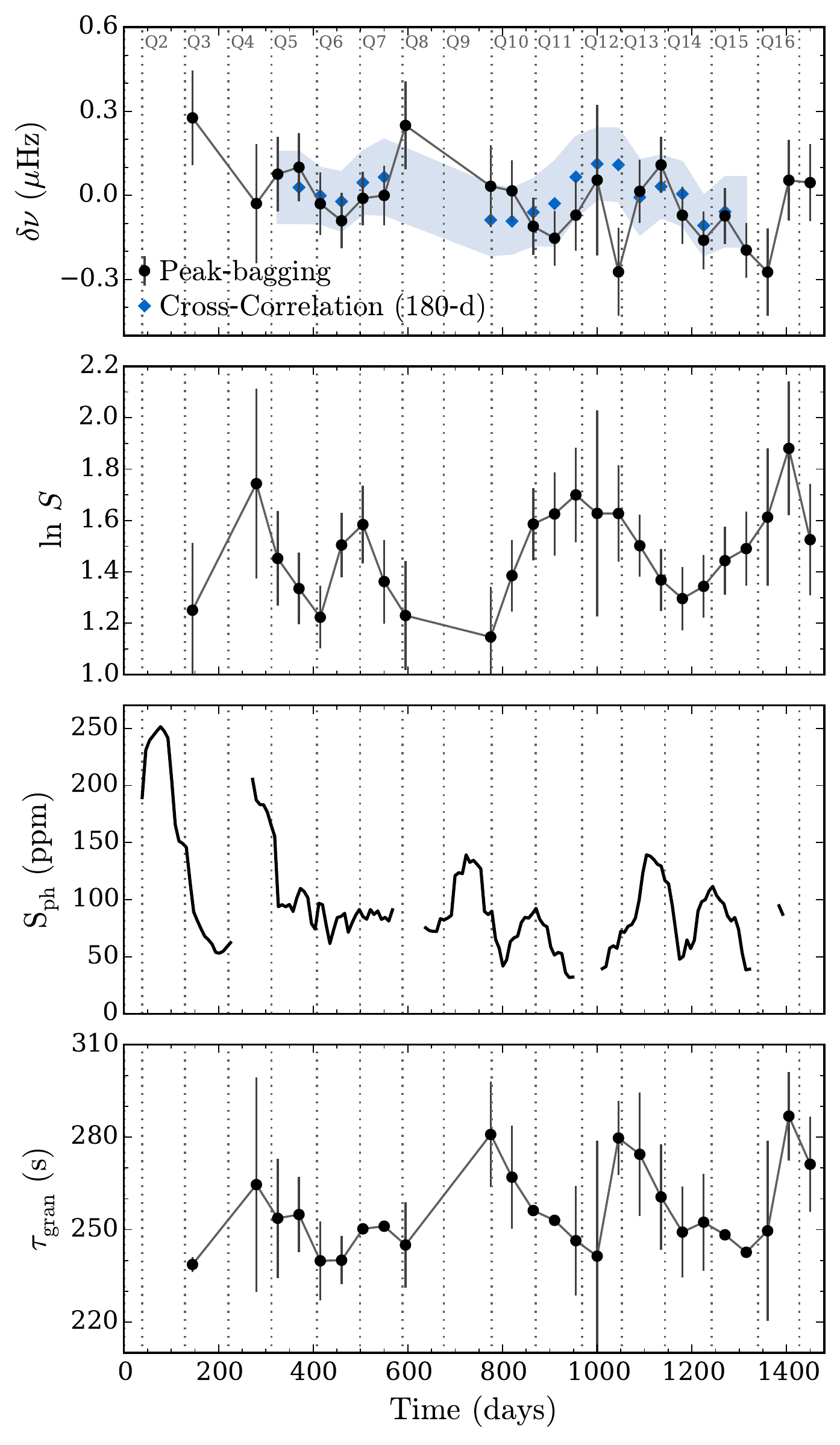}
\caption{Same as in Figure 3, but for KIC 10963065. Results in Table~76.}\label{fig:10963065}%\ref{tab:10963065}
\end{figure}

\begin{figure}[ht]
\includegraphics[width=\hsize]{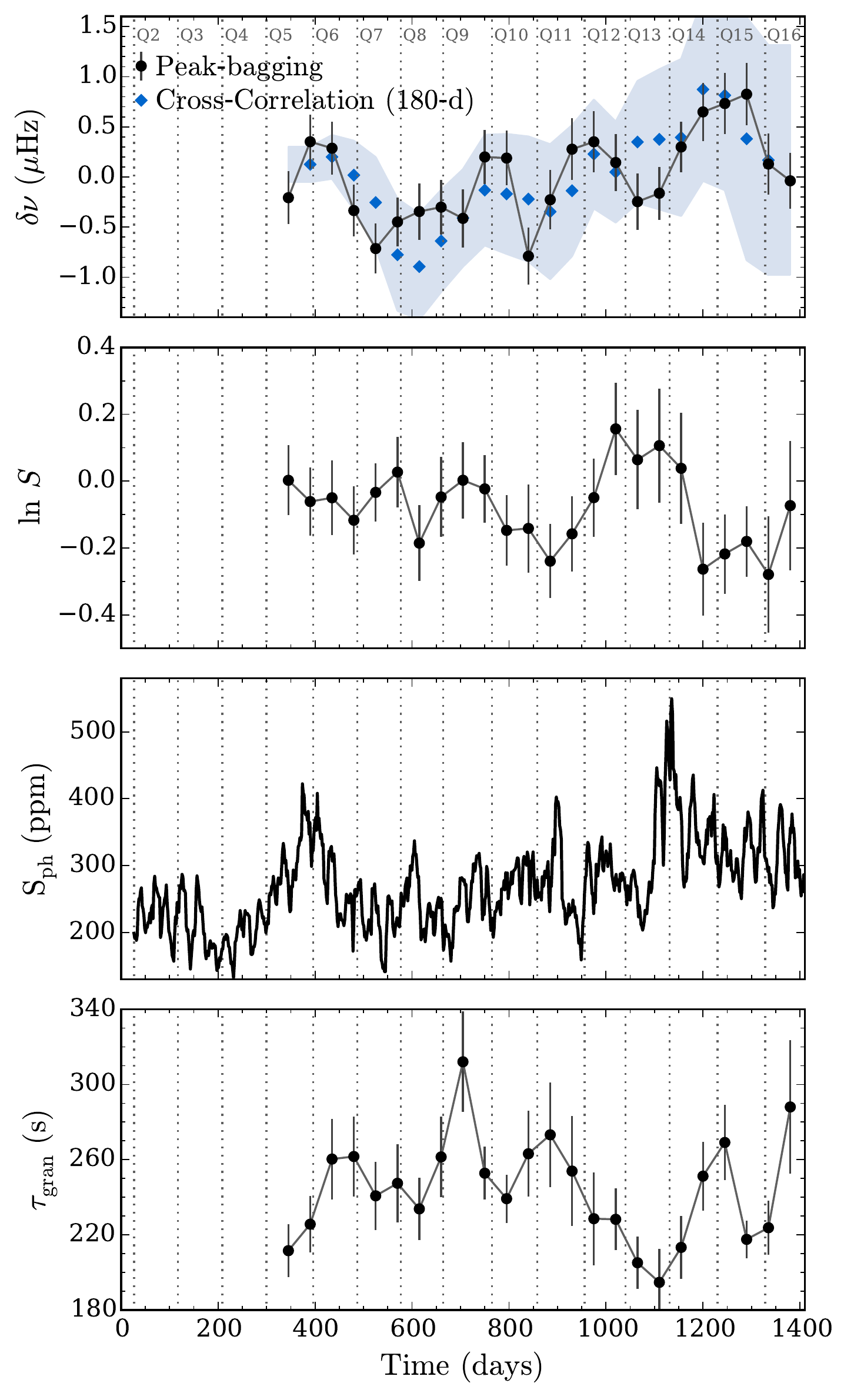}
\caption{Same as in Figure 3, but for KIC 11081729. The frequency shifts from the cross-correlation method were obtained with 180-d sub-series. Results in Table~77.}\label{fig:11081729}%\ref{tab:11081729}
\end{figure}

\begin{figure}[ht]
\includegraphics[width=\hsize]{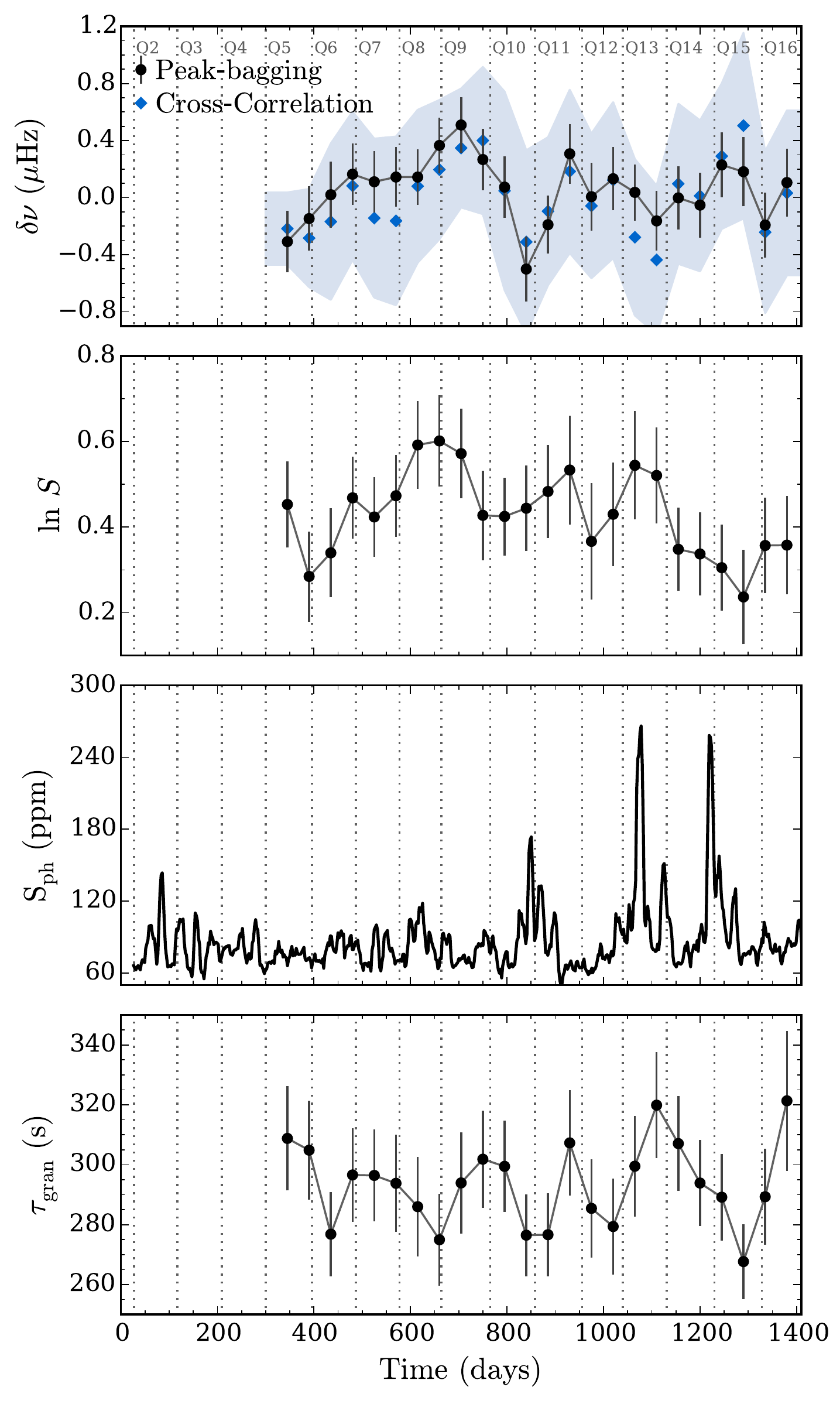}
\caption{Same as in Figure 3, but for KIC 11253226. Results in Table~78.}\label{fig:11253226}%\ref{tab:11253226}
\end{figure}

\begin{figure}[ht]
\includegraphics[width=\hsize]{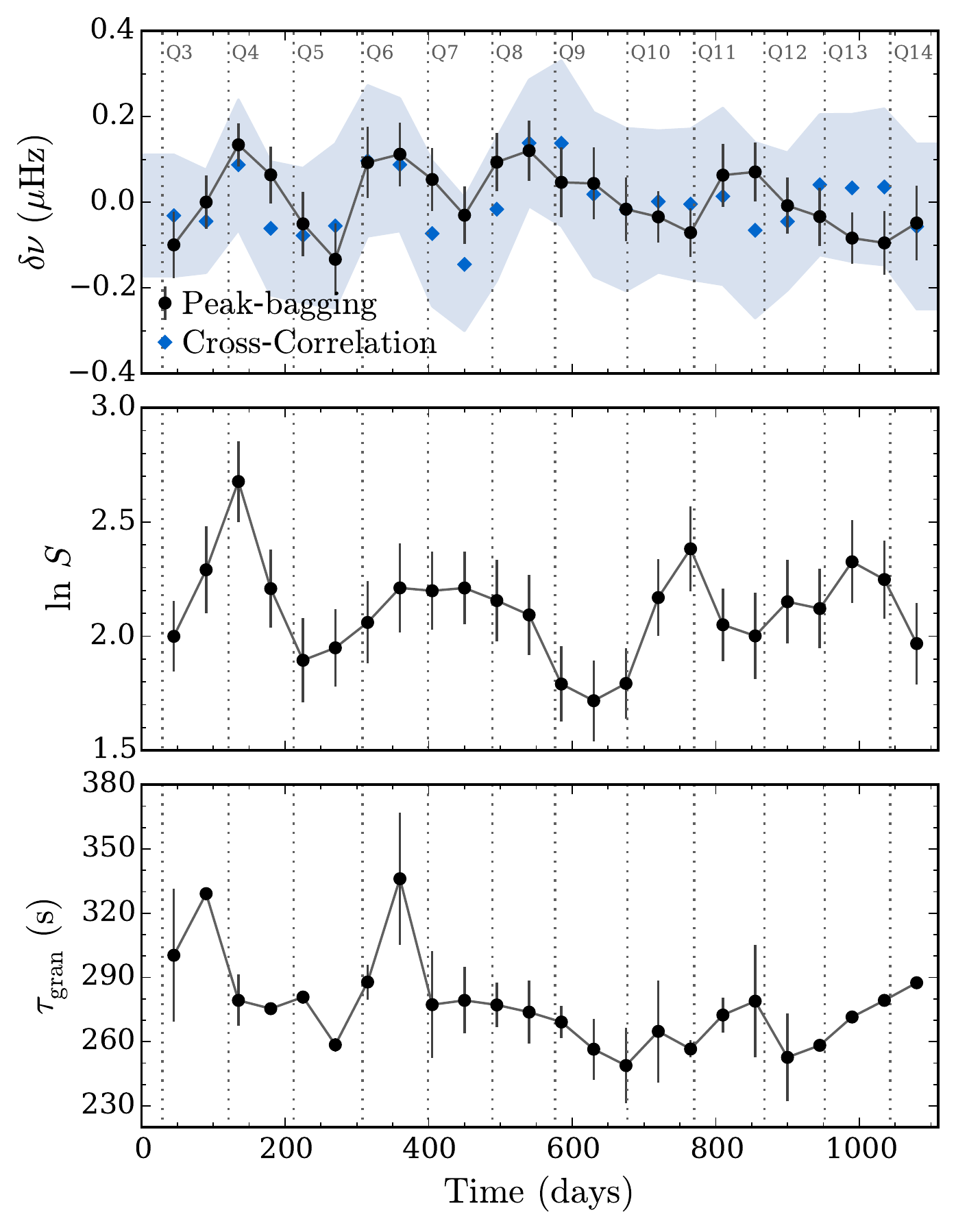}\vspace{-0.2cm}
\caption{Same as in Figure 3, but for KIC 11295426. Results in Table~79.}\label{fig:11295426}%\ref{tab:11295426}
\end{figure}

\begin{figure}[ht]
\includegraphics[width=\hsize]{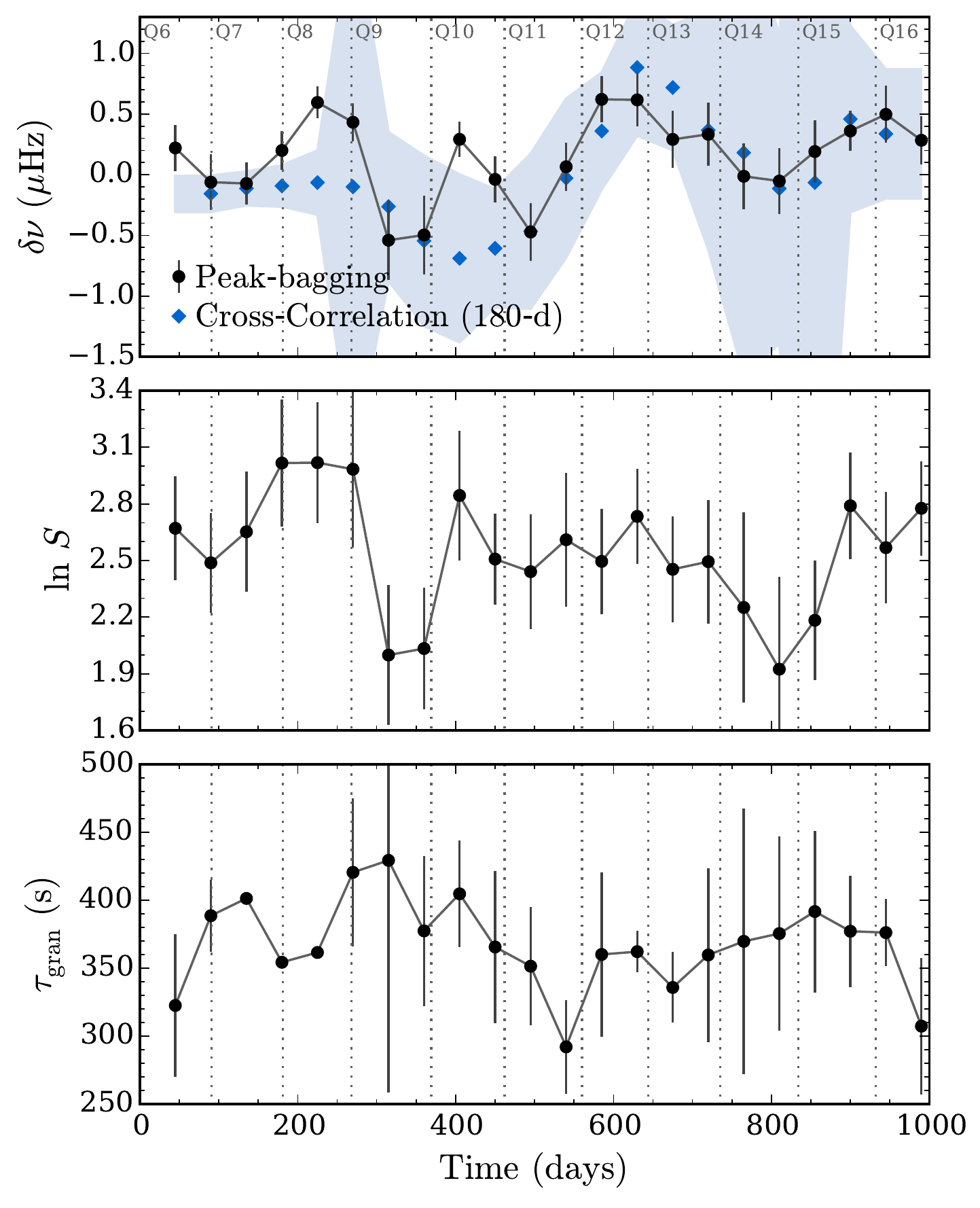}\vspace{-0.2cm}
\caption{Same as in Figure 3, but for KIC 11401755. The frequency shifts from the cross-correlation method were obtained with 180-d sub-series. Results in Table~80.}\label{fig:11401755}\vspace{-1.5cm}%\ref{tab:11401755}
\end{figure}

\begin{figure}[ht]
\includegraphics[width=\hsize]{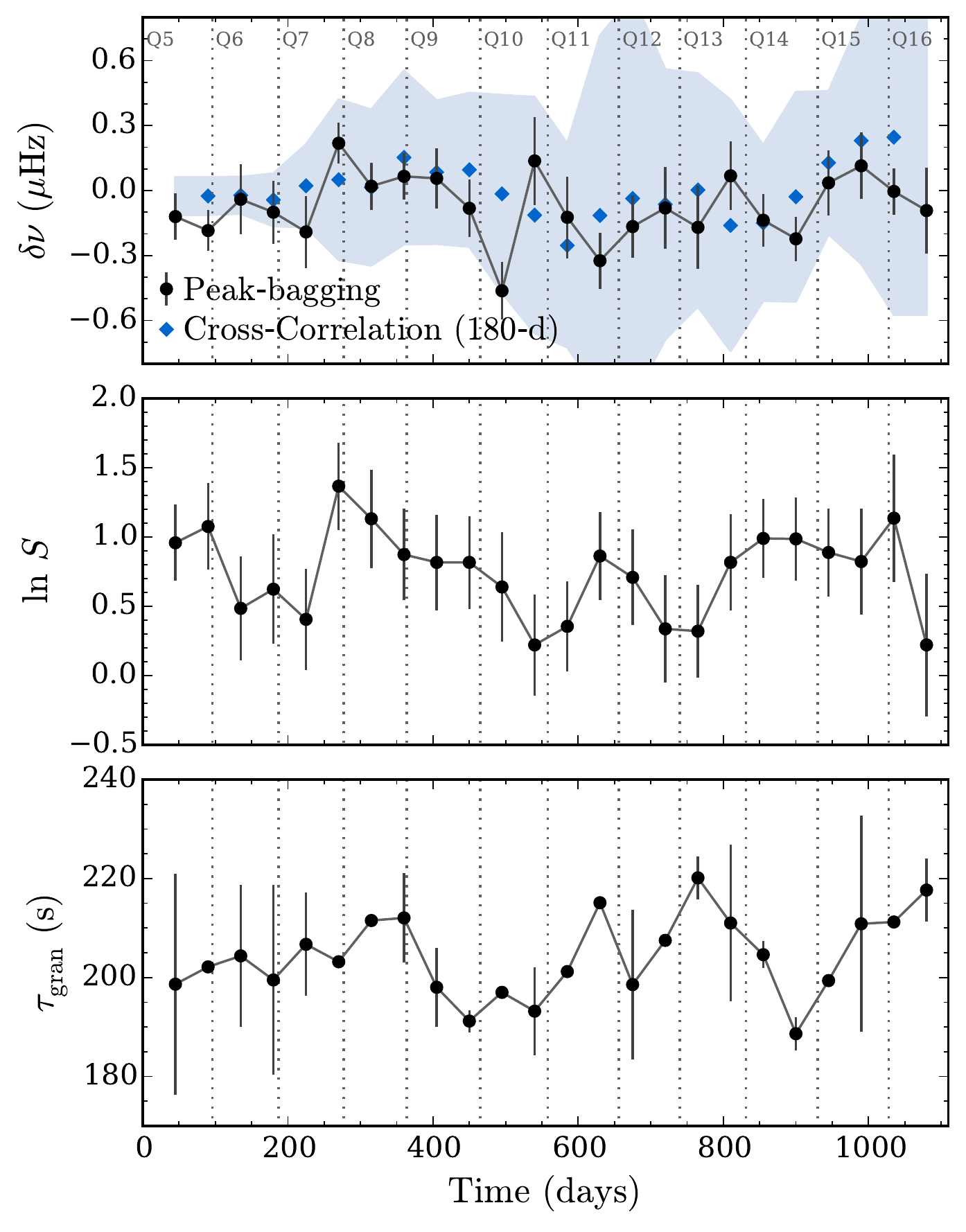}\vspace{-0.2cm}
\caption{Same as in Figure 3, but for KIC 11772920. The frequency shifts from the cross-correlation method were obtained with 180-d sub-series. Results in Table~81.}\label{fig:11772920}\vspace{-0.3cm}%\ref{tab:11772920}
\end{figure}

\begin{figure}[ht]
\includegraphics[width=\hsize]{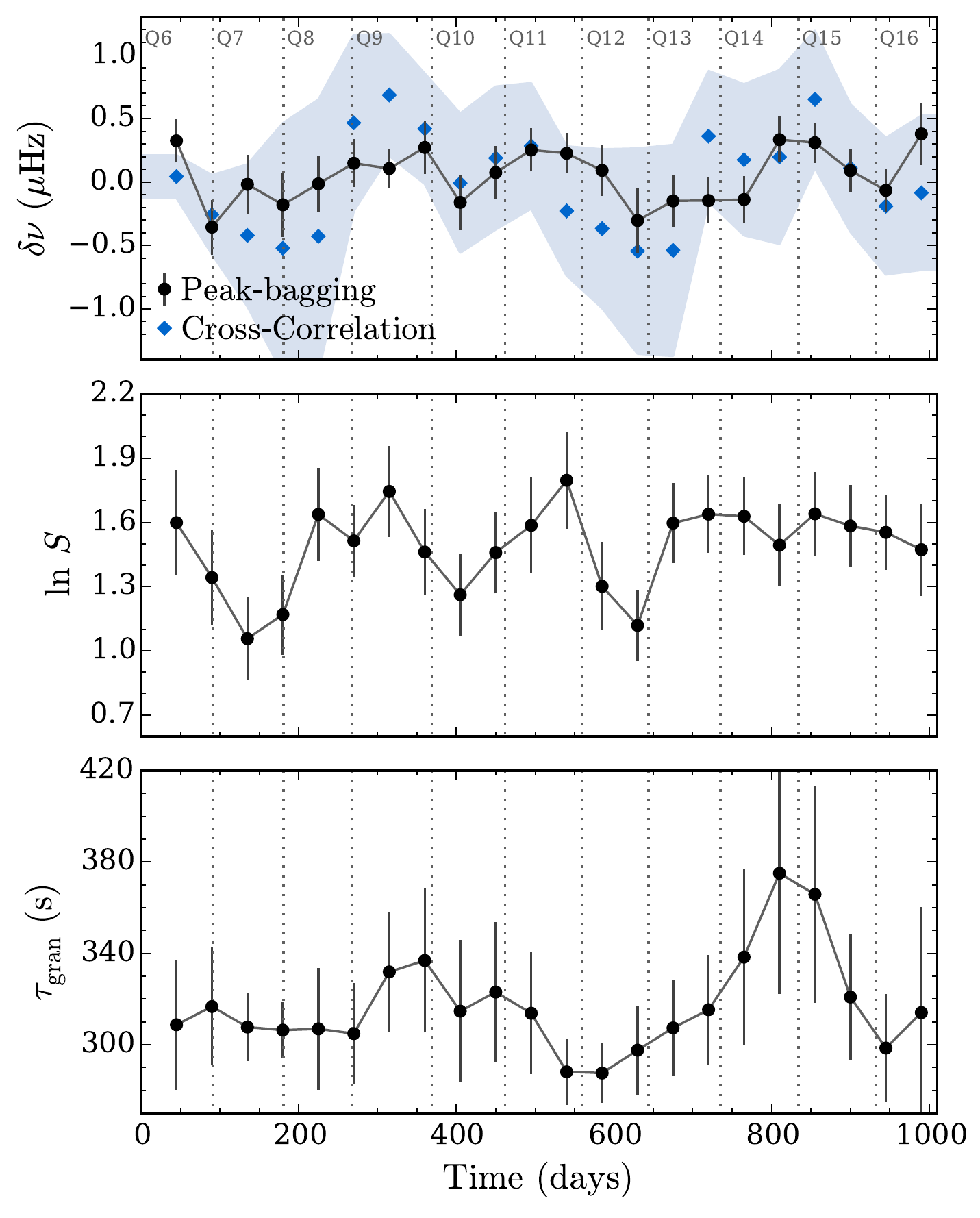}\vspace{-0.2cm}
\caption{Same as in Figure 3, but for KIC 11807274. Results in Table~82.}\label{fig:11807274}\vspace{-1cm}%\ref{tab:11807274}
\end{figure}

\begin{figure}[ht]
\includegraphics[width=\hsize]{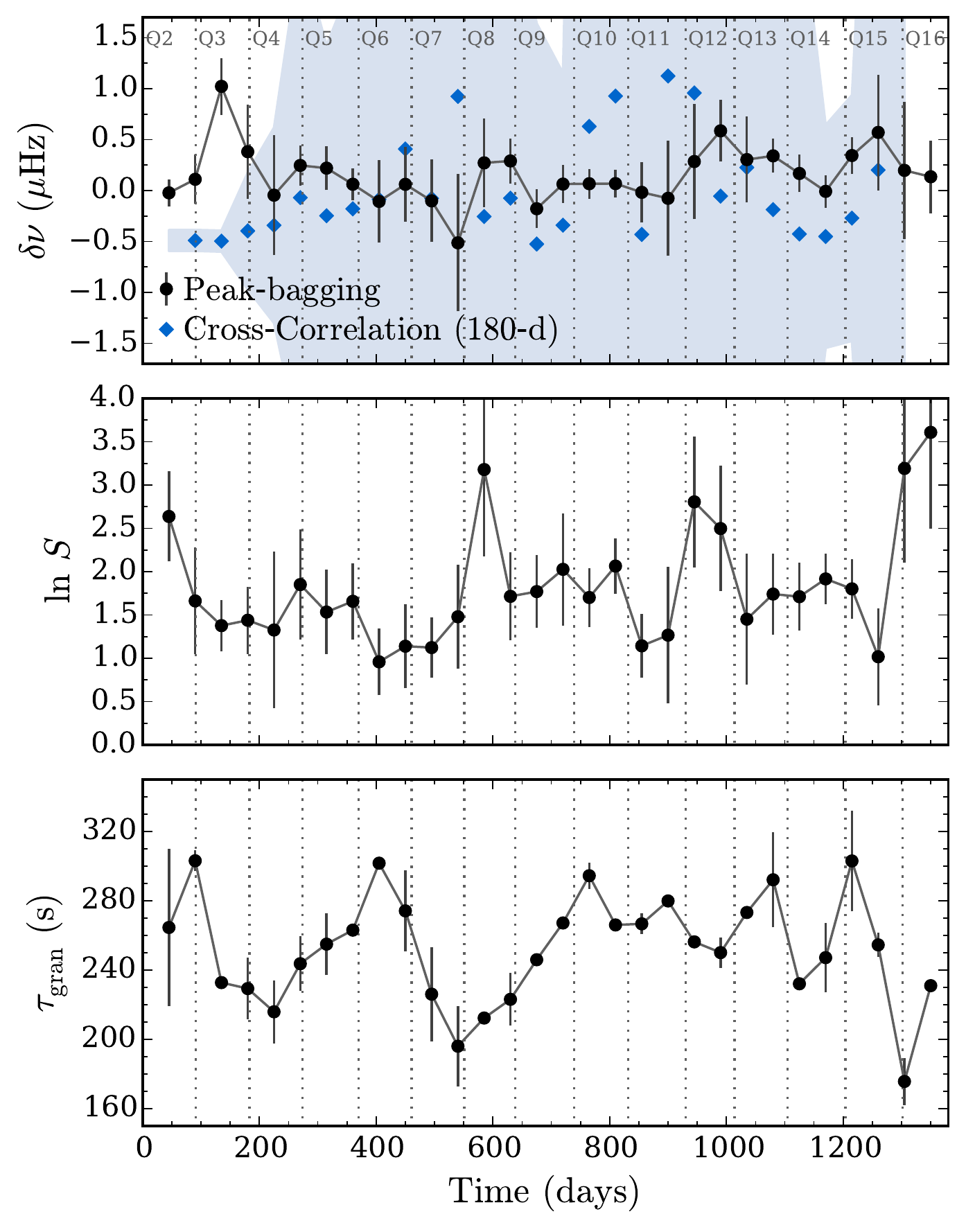}
\caption{Same as in Figure 3, but for KIC 11904151. The frequency shifts from the cross-correlation method were obtained with 180-d sub-series. Results in Table~83.}\label{fig:11904151}%\ref{tab:11904151}
\end{figure}

\begin{figure}[ht]
\includegraphics[width=\hsize]{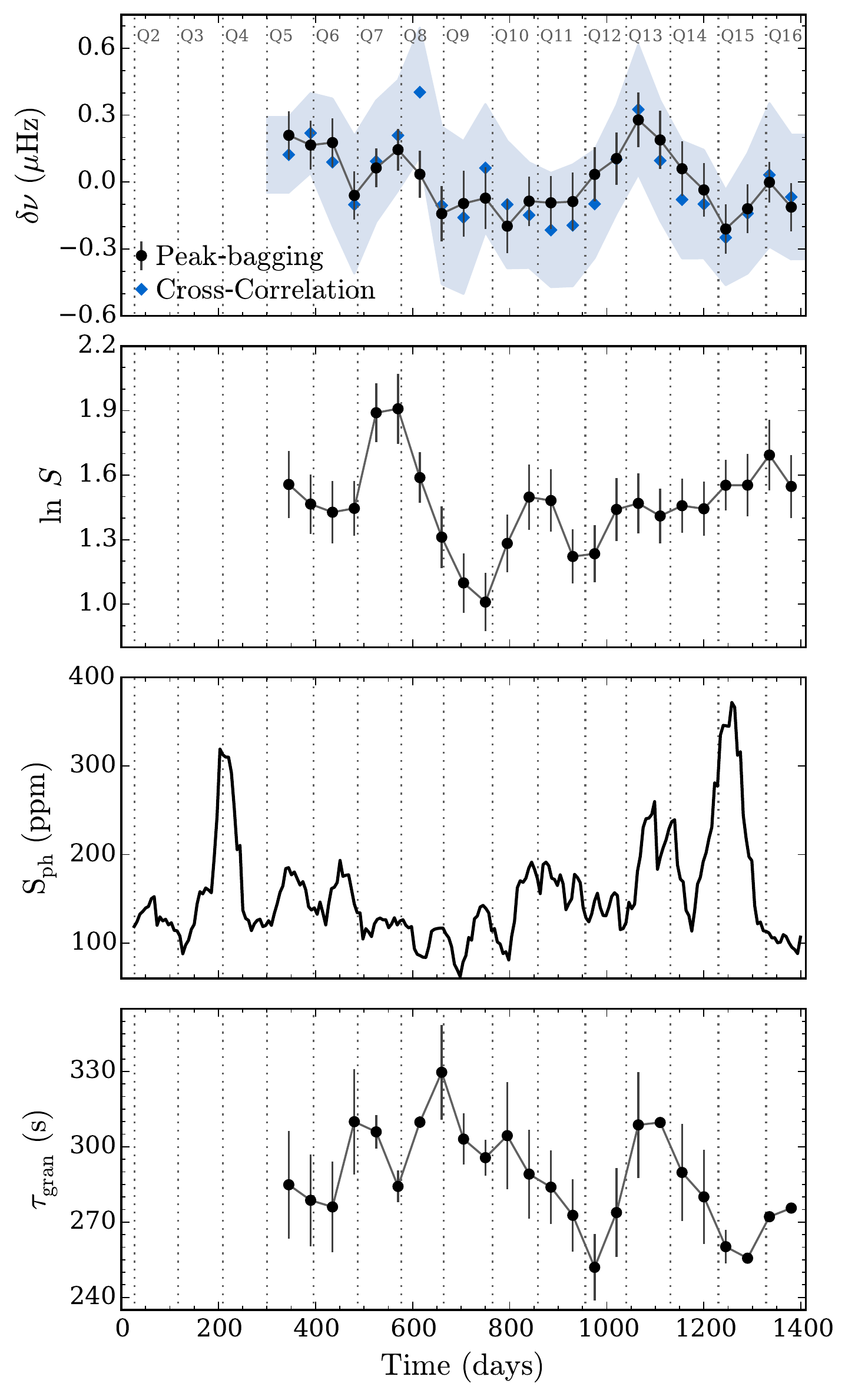}
\caption{Same as in Figure 3, but for KIC 12009504. Results in Table~84.}\label{fig:12009504}%\ref{tab:12009504}
\end{figure}

\begin{figure}[ht]
\includegraphics[width=\hsize]{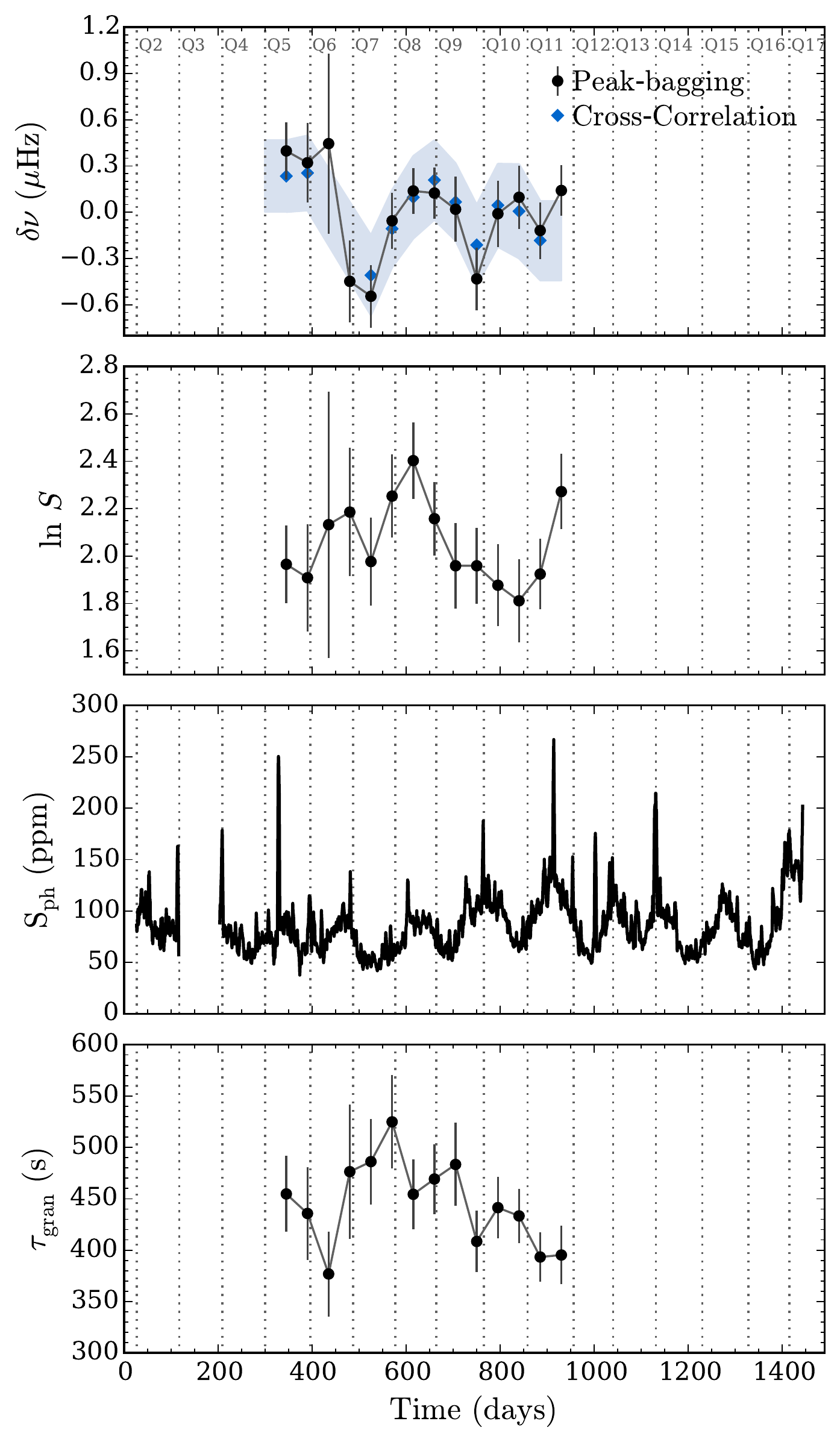}
\caption{Same as in Figure 3, but for KIC 12069127. Results in Table~85.}\label{fig:12069127}%\ref{tab:12069127}
\end{figure}

\begin{figure}[ht]
\includegraphics[width=\hsize]{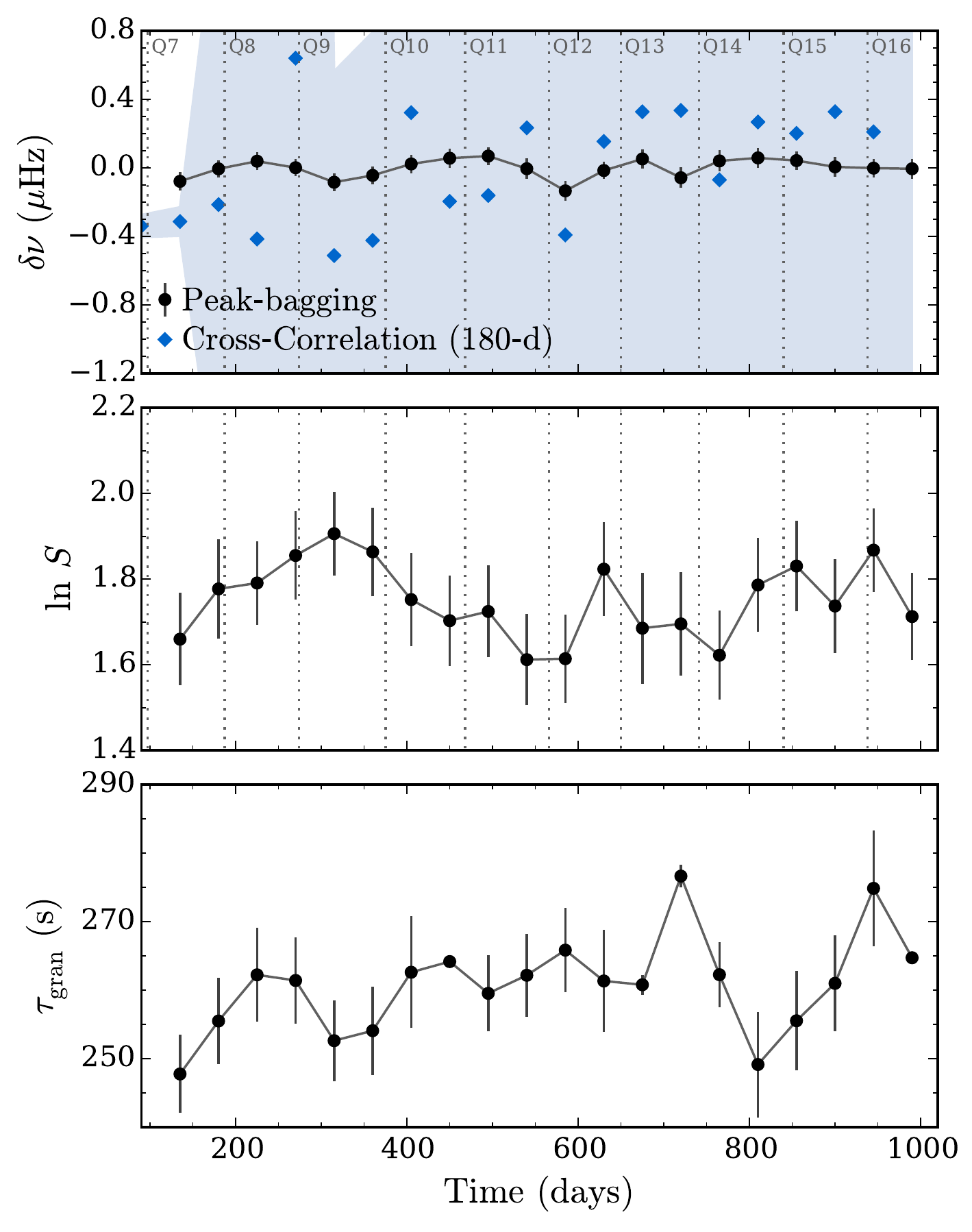}\vspace{-0.2cm}
\caption{Same as in Figure 3, but for KIC 12069424. The frequency shifts from the cross-correlation method were obtained with 180-d sub-series. Results in Table~86.}\label{fig:12069424}\vspace{-0.3cm}%\ref{tab:12969424}
\end{figure}

\begin{figure}[ht]
\includegraphics[width=\hsize]{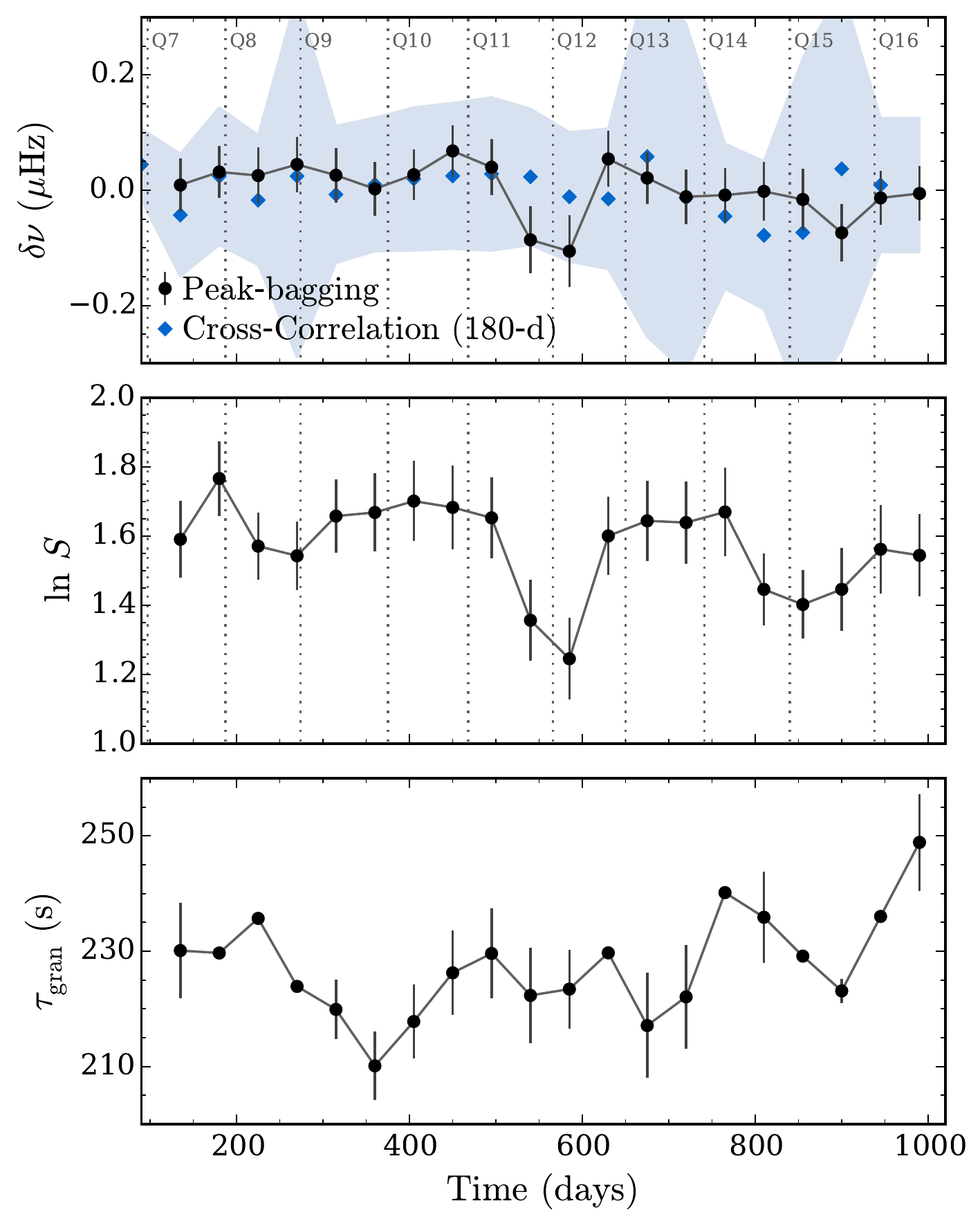}\vspace{-0.2cm}
\caption{Same as in Figure 3, but for KIC 12069449. The frequency shifts from the cross-correlation method were obtained with 180-d sub-series. Results in Table~87.}\label{fig:12069449}\vspace{-1.8cm}%\ref{tab:12969449}
\end{figure}

\begin{figure}[ht]
\includegraphics[width=\hsize]{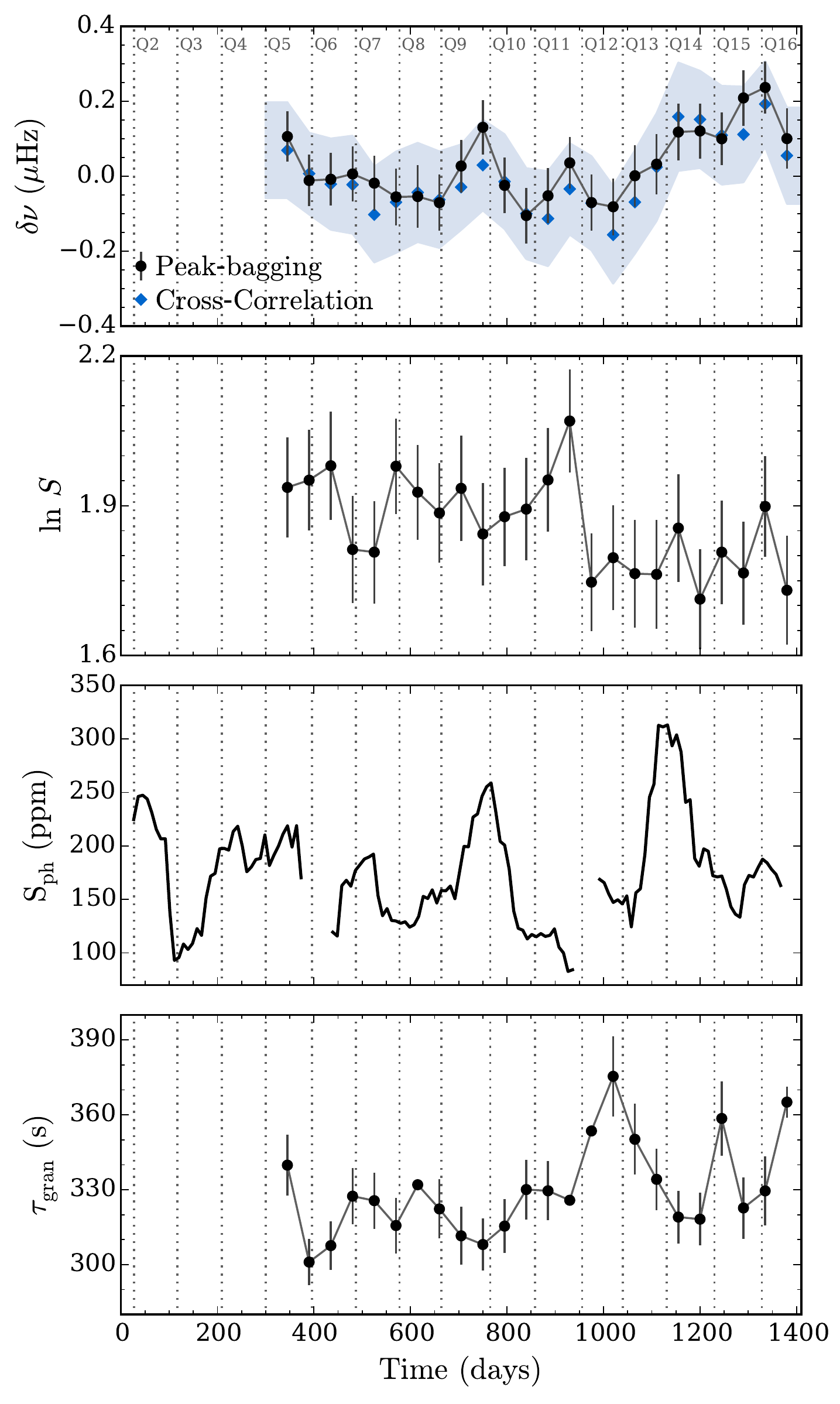}
\caption{Same as in Figure 3, but for KIC 12258514. Results in Table~88.}\label{fig:12258514}%\ref{tab:12258514}
\end{figure}

\begin{figure}[ht]
\includegraphics[width=\hsize]{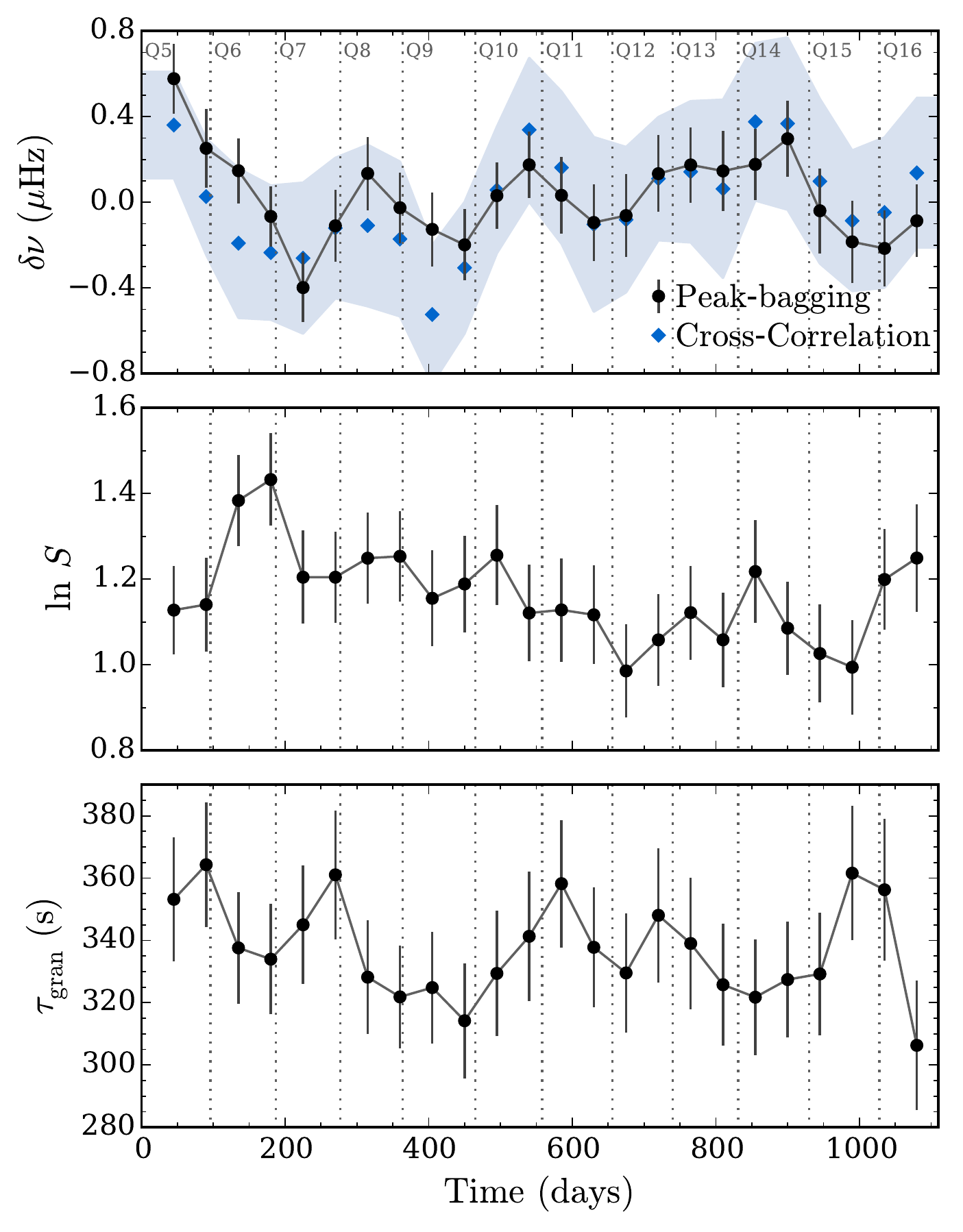}
\caption{Same as in Figure 3, but for KIC 12317678. Results in Table~89.}\label{fig:12317678}%\ref{tab:12317678}
\end{figure}

\FloatBarrier
\begin{table}\centering\fontsize{9.}{7.}\selectfont
\begin{tabular}{ccc|ccccc|c}
\multicolumn{9}{c}{KIC 1435467}\\ \hline\hline
time & duty & $\tau_\text{gran}$ &\multicolumn{5}{c|}{Peak-bagging}&Cross-correlation\\
(d)& cycle & (s)&$\delta\nu_{l=0}$ ($\mu$Hz) & $\delta\nu_{l=1}$ ($\mu$Hz) & $\delta\nu_{l=2}$ ($\mu$Hz) & $\delta\nu$ ($\mu$Hz)& $\ln\,S$ & $\delta\nu\,(\mu$Hz)\\\hline
345 & 0.98 & 336.4 $\pm$ 17.1 & 0.38 $\pm$ 0.24 & -0.30 $\pm$ 0.23 & -0.06 $\pm$ 0.46 & 0.02 $\pm$ 0.17 & 1.04 $\pm$ 0.12 & 0.10 $\pm$ 0.24\\
390 & 0.97 & 298.7 $\pm$ 14.6 & 0.35 $\pm$ 0.25 & -0.42 $\pm$ 0.24 & -0.34 $\pm$ 0.44 & -0.05 $\pm$ 0.17 & 1.05 $\pm$ 0.12 & -0.11 $\pm$ 0.31\\
435 & 0.96 & 294.1 $\pm$ 14.2 & -0.28 $\pm$ 0.27 & -0.28 $\pm$ 0.25 & 0.26 $\pm$ 0.44 & -0.28 $\pm$ 0.18 & 1.00 $\pm$ 0.11 & -0.31 $\pm$ 0.35\\
480 & 0.97 & 329.5 $\pm$ 16.8 & -0.42 $\pm$ 0.25 & 0.12 $\pm$ 0.21 & -0.50 $\pm$ 0.34 & -0.11 $\pm$ 0.16 & 1.18 $\pm$ 0.13 & -0.21 $\pm$ 0.37\\
525 & 0.96 & 339.1 $\pm$ 18.7 & -0.06 $\pm$ 0.22 & 0.34 $\pm$ 0.20 & -0.93 $\pm$ 0.38 & 0.15 $\pm$ 0.15 & 1.19 $\pm$ 0.12 & 0.11 $\pm$ 0.40\\
570 & 0.81 & 335.8 $\pm$ 20.7 & 0.01 $\pm$ 0.24 & 0.17 $\pm$ 0.20 & -0.53 $\pm$ 0.51 & 0.10 $\pm$ 0.15 & 1.10 $\pm$ 0.13 & 0.24 $\pm$ 0.46\\
615 & 0.78 & 316.9 $\pm$ 17.2 & -0.34 $\pm$ 0.29 & -0.04 $\pm$ 0.23 & 0.54 $\pm$ 0.57 & -0.16 $\pm$ 0.18 & 0.89 $\pm$ 0.12 & -0.17 $\pm$ 0.38\\
660 & 0.89 & 323.3 $\pm$ 15.7 & -0.32 $\pm$ 0.24 & -0.05 $\pm$ 0.21 & -1.47 $\pm$ 0.48 & -0.16 $\pm$ 0.16 & 1.03 $\pm$ 0.12 & -0.22 $\pm$ 0.42\\
705 & 0.93 & 317.7 $\pm$ 16.4 & -0.05 $\pm$ 0.27 & -0.04 $\pm$ 0.24 & 0.02 $\pm$ 0.47 & -0.04 $\pm$ 0.18 & 0.88 $\pm$ 0.13 & -0.08 $\pm$ 0.38\\
750 & 0.97 & 290.2 $\pm$ 14.7 & 0.28 $\pm$ 0.29 & 0.15 $\pm$ 0.24 & 0.06 $\pm$ 0.37 & 0.20 $\pm$ 0.18 & 0.93 $\pm$ 0.12 & -0.07 $\pm$ 0.34\\
795 & 0.97 & 310.0 $\pm$ 16.6 & 0.47 $\pm$ 0.28 & 0.17 $\pm$ 0.24 & -0.28 $\pm$ 0.39 & 0.30 $\pm$ 0.18 & 0.93 $\pm$ 0.12 & -0.15 $\pm$ 0.33\\
840 & 0.98 & 338.3 $\pm$ 17.2 & 0.66 $\pm$ 0.26 & 0.31 $\pm$ 0.23 & -0.44 $\pm$ 0.41 & 0.46 $\pm$ 0.17 & 0.97 $\pm$ 0.13 & 0.07 $\pm$ 0.39\\
885 & 0.94 & 335.2 $\pm$ 15.7 & 0.70 $\pm$ 0.24 & 0.25 $\pm$ 0.21 & 0.01 $\pm$ 0.41 & 0.45 $\pm$ 0.16 & 1.05 $\pm$ 0.12 & 0.23 $\pm$ 0.38\\
930 & 0.92 & 331.1 $\pm$ 16.6 & 0.43 $\pm$ 0.28 & 0.47 $\pm$ 0.21 & -0.45 $\pm$ 0.41 & 0.46 $\pm$ 0.17 & 0.93 $\pm$ 0.13 & 0.46 $\pm$ 0.35\\
975 & 0.90 & 327.5 $\pm$ 17.9 & -0.24 $\pm$ 0.29 & 0.47 $\pm$ 0.23 & -1.17 $\pm$ 0.44 & 0.20 $\pm$ 0.18 & 0.87 $\pm$ 0.13 & 0.19 $\pm$ 0.41\\
1020 & 0.90 & 306.1 $\pm$ 15.6 & -0.76 $\pm$ 0.30 & 0.18 $\pm$ 0.27 & -0.65 $\pm$ 0.48 & -0.25 $\pm$ 0.20 & 0.79 $\pm$ 0.11 & -0.31 $\pm$ 0.35\\
1065 & 0.95 & 323.5 $\pm$ 17.1 & -1.17 $\pm$ 0.35 & 0.14 $\pm$ 0.27 & 1.47 $\pm$ 0.57 & -0.35 $\pm$ 0.21 & 0.95 $\pm$ 0.11 & -0.34 $\pm$ 0.39\\
1110 & 0.90 & 328.2 $\pm$ 19.2 & 0.09 $\pm$ 0.31 & -0.03 $\pm$ 0.24 & 1.59 $\pm$ 0.48 & 0.02 $\pm$ 0.19 & 1.01 $\pm$ 0.12 & 0.12 $\pm$ 0.44\\
1155 & 0.89 & 304.4 $\pm$ 16.5 & 0.82 $\pm$ 0.31 & -0.13 $\pm$ 0.25 & 1.20 $\pm$ 0.51 & 0.25 $\pm$ 0.20 & 0.94 $\pm$ 0.13 & 0.12 $\pm$ 0.43\\
1200 & 0.95 & 307.9 $\pm$ 16.3 & -0.13 $\pm$ 0.32 & -0.24 $\pm$ 0.25 & 0.36 $\pm$ 0.52 & -0.20 $\pm$ 0.20 & 0.91 $\pm$ 0.14 & -0.21 $\pm$ 0.38\\
1245 & 0.90 & 313.7 $\pm$ 16.5 & 0.02 $\pm$ 0.28 & 0.03 $\pm$ 0.23 & 0.25 $\pm$ 0.43 & 0.02 $\pm$ 0.18 & 0.93 $\pm$ 0.11 & 0.15 $\pm$ 0.36\\
1290 & 0.89 & 322.4 $\pm$ 17.3 & 0.61 $\pm$ 0.23 & 0.02 $\pm$ 0.23 & -0.49 $\pm$ 0.36 & 0.31 $\pm$ 0.16 & 1.01 $\pm$ 0.11 & 0.20 $\pm$ 0.31\\
1335 & 0.85 & 327.7 $\pm$ 18.1 & 0.51 $\pm$ 0.25 & -0.04 $\pm$ 0.24 & -0.15 $\pm$ 0.41 & 0.22 $\pm$ 0.17 & 0.95 $\pm$ 0.13 & -0.01 $\pm$ 0.33\\
1380 & 0.85 & 329.3 $\pm$ 17.0 & 0.45 $\pm$ 0.34 & 0.12 $\pm$ 0.27 & 0.25 $\pm$ 0.61 & 0.24 $\pm$ 0.21 & 0.75 $\pm$ 0.13 & 0.19 $\pm$ 0.36\\
\end{tabular}
\parbox{2\hsize}{\caption{Results for KIC 1435467 (same as in Table~\ref{tab:fshifts8006161}). {\it First column}: Time of the sub-series. {\it Second column}: Duty-cycle for each sub-series. {\it Third column}: Characteristic timescale of the granulation component. {\it Fourth to Eighth Columns}: Results obtained from the peak-bagging analysis (radial orders used to compute the mean parameters range between $n=17$ and $n=21$): mean frequency shifts (for radial ($\delta\nu_{l=0}$), dipolar ($\delta\nu_{l=1}$), and quadrupolar ($\delta\nu_{l=2}$) modes, and when combining the $l=0$ and $l=1$ modes ($\delta\nu$)) and logarithmic mode heights. {\it Ninth column}: Frequency shifts obtained with the cross-correlation method described in \citet{Kiefer2017}. Results shown in Figure \ref{fig:1435467}.}\label{tab:1435467}}
\end{table}

\begin{table*}[ht]\centering\fontsize{9.}{7.}\selectfont
\begin{tabular}{ccc|ccccc|c}
\multicolumn{9}{c}{KIC 2837475}\\ \hline\hline
time & duty & $\tau_\text{gran}$ &\multicolumn{5}{c|}{Peak-bagging}&Cross-correlation\\
(d)& cycle & (s)&$\delta\nu_{l=0}$ ($\mu$Hz) & $\delta\nu_{l=1}$ ($\mu$Hz) & $\delta\nu_{l=2}$ ($\mu$Hz) & $\delta\nu$ ($\mu$Hz)& $\ln\,S$ & $\delta\nu\,(\mu$Hz)\\\hline
345 & 0.98 & 291.9 $\pm$ 14.9 & 0.00 $\pm$ 0.34 & 0.71 $\pm$ 0.32 & 0.57 $\pm$ 0.60 & 0.37 $\pm$ 0.23 & 0.44 $\pm$ 0.11 & 0.46 $\pm$ 0.26\\
390 & 0.97 & 282.4 $\pm$ 15.8 & 0.00 $\pm$ 0.37 & 0.64 $\pm$ 0.31 & 0.11 $\pm$ 0.59 & 0.38 $\pm$ 0.24 & 0.34 $\pm$ 0.11 & 0.37 $\pm$ 0.43\\
435 & 0.96 & 297.7 $\pm$ 16.9 & 0.29 $\pm$ 0.35 & -0.33 $\pm$ 0.32 & -0.40 $\pm$ 0.58 & -0.05 $\pm$ 0.24 & 0.22 $\pm$ 0.11 & -0.14 $\pm$ 0.76\\
480 & 0.97 & 285.7 $\pm$ 16.9 & 0.55 $\pm$ 0.29 & -0.90 $\pm$ 0.36 & -0.56 $\pm$ 0.64 & -0.02 $\pm$ 0.23 & 0.33 $\pm$ 0.11 & 0.16 $\pm$ 0.68\\
525 & 0.96 & 286.8 $\pm$ 16.9 & -0.14 $\pm$ 0.31 & -0.43 $\pm$ 0.33 & 1.11 $\pm$ 0.50 & -0.27 $\pm$ 0.23 & 0.37 $\pm$ 0.10 & -0.05 $\pm$ 0.62\\
\end{tabular}
\caption{Same as in Table 3, but for KIC 2837475. Radial orders used to compute the mean parameters range between $n=16$ and $n=20$. Results shown in Figure \ref{fig:2837475}.}\label{tab:2837475}
\end{table*}

\begin{table*}[ht]\centering\fontsize{9.}{7.}\selectfont
\begin{tabular}{ccc|ccccc|c}
\multicolumn{9}{c}{KIC 3425851}\\ \hline\hline
time & duty & $\tau_\text{gran}$ &\multicolumn{5}{c|}{Peak-bagging}&Cross-correlation\\
(d)& cycle & (s)&$\delta\nu_{l=0}$ ($\mu$Hz) & $\delta\nu_{l=1}$ ($\mu$Hz) & $\delta\nu_{l=2}$ ($\mu$Hz) & $\delta\nu$ ($\mu$Hz)& $\ln\,S$ & $\delta\nu\,(\mu$Hz)\\\hline
45 & 0.98 & 265.9 $\pm$ 0.2 & 1.23 $\pm$ 0.61 & 0.83 $\pm$ 0.80 & -- & 1.08 $\pm$ 0.48 & 0.82 $\pm$ 0.38 & -0.10 $\pm$ 0.34\\
90 & 0.97 & 249.4 $\pm$ 1.1 & 0.14 $\pm$ 0.81 & -0.48 $\pm$ 0.85 & -- & -0.15 $\pm$ 0.59 & 0.69 $\pm$ 0.29 & -0.04 $\pm$ 0.92\\
135 & 0.96 & 223.3 $\pm$ 1.2 & -0.43 $\pm$ 0.76 & -0.48 $\pm$ 0.80 & -- & -0.46 $\pm$ 0.55 & 0.68 $\pm$ 0.30 & -0.11 $\pm$ 3.97\\
180 & 0.78 & 237.2 $\pm$ 14.8 & -0.69 $\pm$ 0.66 & 0.42 $\pm$ 0.92 & -- & -0.31 $\pm$ 0.54 & 0.82 $\pm$ 0.37 & 0.25 $\pm$ 4.83\\
\end{tabular}
\caption{Same as in Table 3, but for KIC 3425851. Radial orders used to compute the mean parameters range between $n=18$ and $n=22$. Results shown in Figure \ref{fig:3425851}.}\label{tab:3425851}
\end{table*}

\begin{table*}[ht]\centering\fontsize{9.}{7.}\selectfont
\begin{tabular}{ccc|ccccc|c}
\multicolumn{9}{c}{KIC 3427720}\\ \hline\hline
time & duty & $\tau_\text{gran}$ &\multicolumn{5}{c|}{Peak-bagging}&Cross-correlation\\
(d)& cycle & (s)&$\delta\nu_{l=0}$ ($\mu$Hz) & $\delta\nu_{l=1}$ ($\mu$Hz) & $\delta\nu_{l=2}$ ($\mu$Hz) & $\delta\nu$ ($\mu$Hz)& $\ln\,S$ & $\delta\nu\,(\mu$Hz)\\\hline
345 & 0.97 & 313.2 $\pm$ 7.5 & -0.25 $\pm$ 0.14 & -0.23 $\pm$ 0.14 & 0.13 $\pm$ 0.17 & -0.24 $\pm$ 0.10 & 1.01 $\pm$ 0.15 & -0.24 $\pm$ 0.13\\
390 & 0.97 & 318.5 $\pm$ 3.5 & -0.34 $\pm$ 0.15 & 0.03 $\pm$ 0.15 & -0.12 $\pm$ 0.19 & -0.15 $\pm$ 0.10 & 1.02 $\pm$ 0.16 & -0.17 $\pm$ 0.16\\
435 & 0.96 & 237.8 $\pm$ 0.6 & -0.26 $\pm$ 0.16 & 0.00 $\pm$ 0.14 & 0.20 $\pm$ 0.21 & -0.11 $\pm$ 0.10 & 1.16 $\pm$ 0.18 & -0.12 $\pm$ 0.25\\
480 & 0.97 & 229.6 $\pm$ 7.9 & -0.24 $\pm$ 0.15 & -0.08 $\pm$ 0.12 & 0.57 $\pm$ 0.19 & -0.14 $\pm$ 0.09 & 1.25 $\pm$ 0.19 & -0.11 $\pm$ 0.24\\
525 & 0.96 & 252.1 $\pm$ 6.3 & -0.15 $\pm$ 0.17 & -0.27 $\pm$ 0.14 & 0.15 $\pm$ 0.24 & -0.22 $\pm$ 0.11 & 1.06 $\pm$ 0.17 & 0.00 $\pm$ 0.27\\
\end{tabular}
\caption{Same as in Table 3, but for KIC 3427720. Radial orders used to compute the mean parameters range between $n=19$ and $n=23$. Results shown in Figure \ref{fig:3427720}.}\label{tab:3427720}
\end{table*}

\begin{table*}[ht]\centering\fontsize{9.}{7.}\selectfont
\begin{tabular}{ccc|ccccc|c}
\multicolumn{9}{c}{KIC 3456181}\\ \hline\hline
time & duty & $\tau_\text{gran}$ &\multicolumn{5}{c|}{Peak-bagging}&Cross-correlation\\
(d)& cycle & (s)&$\delta\nu_{l=0}$ ($\mu$Hz) & $\delta\nu_{l=1}$ ($\mu$Hz) & $\delta\nu_{l=2}$ ($\mu$Hz) & $\delta\nu$ ($\mu$Hz)& $\ln\,S$ & $\delta\nu\,(\mu$Hz)\\\hline
45 & 0.98 & 447.3 $\pm$ 26.5 & -0.75 $\pm$ 0.35 & -0.05 $\pm$ 0.19 & 1.66 $\pm$ 0.60 & -0.21 $\pm$ 0.17 & 1.88 $\pm$ 0.13 & -0.07 $\pm$ 0.18\\
90 & 0.54 & 521.3 $\pm$ 44.7 & -0.30 $\pm$ 0.37 & 0.05 $\pm$ 0.24 & 0.02 $\pm$ 0.66 & -0.06 $\pm$ 0.20 & 1.96 $\pm$ 0.16 & -0.08 $\pm$ 0.18\\
135 & 0.05 & 666.4 $\pm$ 230.6 & 0.11 $\pm$ 0.60 & 0.10 $\pm$ 0.66 & 0.12 $\pm$ 0.67 & 0.10 $\pm$ 0.45 & 2.35 $\pm$ 0.41 & --\\
180 & 0.42 & 410.2 $\pm$ 37.8 & 0.12 $\pm$ 0.36 & 0.24 $\pm$ 0.22 & -0.91 $\pm$ 0.36 & 0.20 $\pm$ 0.19 & 2.02 $\pm$ 0.19 & --\\
225 & 0.90 & 422.9 $\pm$ 23.4 & 0.21 $\pm$ 0.24 & 0.03 $\pm$ 0.18 & -1.17 $\pm$ 0.32 & 0.09 $\pm$ 0.14 & 1.99 $\pm$ 0.13 & -0.02 $\pm$ 0.19\\
\end{tabular}
\caption{Same as in Table 3, but for KIC 3456181. Radial orders used to compute the mean parameters range between $n=15$ and $n=19$. Results shown in Figure \ref{fig:3456181}.}\label{tab:3456181}
\end{table*}

\begin{table*}[ht]\centering\fontsize{9.}{7.}\selectfont
\begin{tabular}{ccc|ccccc|c}
\multicolumn{9}{c}{KIC 3544595}\\ \hline\hline
time & duty & $\tau_\text{gran}$ &\multicolumn{5}{c|}{Peak-bagging}&Cross-correlation\\
(d)& cycle & (s)&$\delta\nu_{l=0}$ ($\mu$Hz) & $\delta\nu_{l=1}$ ($\mu$Hz) & $\delta\nu_{l=2}$ ($\mu$Hz) & $\delta\nu$ ($\mu$Hz)& $\ln\,S$ & $\delta\nu\,(\mu$Hz)\\\hline
45 & 0.95 & 185.5 $\pm$ 24.1 & -0.05 $\pm$ 0.21 & -0.18 $\pm$ 0.26 & -0.69 $\pm$ 0.68 & -0.10 $\pm$ 0.16 & 0.84 $\pm$ 0.41 & --\\
90 & 0.92 & 177.4 $\pm$ 4.8 & -0.03 $\pm$ 0.23 & 0.06 $\pm$ 0.27 & 0.90 $\pm$ 0.61 & 0.01 $\pm$ 0.17 & 0.69 $\pm$ 0.43 & -0.10 $\pm$ 0.09\\
135 & 0.89 & 171.4 $\pm$ 8.3 & 0.02 $\pm$ 0.19 & 0.28 $\pm$ 0.27 & -0.14 $\pm$ 0.64 & 0.10 $\pm$ 0.16 & 0.42 $\pm$ 0.37 & -0.08 $\pm$ 0.10\\
180 & 0.93 & 170.4 $\pm$ 12.3 & -0.15 $\pm$ 0.19 & -0.10 $\pm$ 0.26 & -0.14 $\pm$ 0.53 & -0.13 $\pm$ 0.16 & 0.66 $\pm$ 0.37 & 0.00 $\pm$ 0.31\\
225 & 0.97 & 186.4 $\pm$ 32.3 & -0.03 $\pm$ 0.26 & -0.11 $\pm$ 0.26 & -0.30 $\pm$ 0.54 & -0.07 $\pm$ 0.18 & 0.62 $\pm$ 0.31 & -0.07 $\pm$ 0.23\\
\end{tabular}
\caption{Same as in Table 3, but for KIC 3544595. {Radial orders used to compute the mean parameters range between $n=20$ and $n=24$.} Note that the frequency shifts from the cross-correlation method (last column) were obtained with 180-d sub-series. Results shown in Figure~\ref{fig:3544595}.}\label{tab:3544595}
\end{table*}

\begin{table*}[ht]\centering\fontsize{9.}{7.}\selectfont
\begin{tabular}{ccc|ccccc|c}
\multicolumn{9}{c}{KIC 3632418}\\ \hline\hline
time & duty & $\tau_\text{gran}$ &\multicolumn{5}{c|}{Peak-bagging}&Cross-correlation\\
(d)& cycle & (s)&$\delta\nu_{l=0}$ ($\mu$Hz) & $\delta\nu_{l=1}$ ($\mu$Hz) & $\delta\nu_{l=2}$ ($\mu$Hz) & $\delta\nu$ ($\mu$Hz)& $\ln\,S$ & $\delta\nu\,(\mu$Hz)\\\hline
345 & 0.98 & 401.0 $\pm$ 15.8 & 0.02 $\pm$ 0.16 & 0.18 $\pm$ 0.15 & -0.34 $\pm$ 0.27 & 0.11 $\pm$ 0.11 & 1.89 $\pm$ 0.10 & -0.02 $\pm$ 0.21\\
390 & 0.97 & 440.9 $\pm$ 16.9 & -0.33 $\pm$ 0.18 & 0.16 $\pm$ 0.15 & 0.05 $\pm$ 0.30 & -0.04 $\pm$ 0.11 & 1.80 $\pm$ 0.09 & -0.01 $\pm$ 0.18\\
435 & 0.96 & 402.6 $\pm$ 6.3 & 0.18 $\pm$ 0.17 & 0.10 $\pm$ 0.14 & 0.31 $\pm$ 0.29 & 0.14 $\pm$ 0.11 & 1.77 $\pm$ 0.09 & 0.12 $\pm$ 0.20\\
480 & 0.97 & 359.4 $\pm$ 14.3 & 0.34 $\pm$ 0.15 & 0.11 $\pm$ 0.15 & 0.21 $\pm$ 0.29 & 0.22 $\pm$ 0.11 & 1.69 $\pm$ 0.10 & 0.17 $\pm$ 0.19\\
525 & 0.96 & 398.3 $\pm$ 15.5 & 0.36 $\pm$ 0.15 & 0.07 $\pm$ 0.14 & 0.05 $\pm$ 0.28 & 0.20 $\pm$ 0.11 & 1.68 $\pm$ 0.11 & 0.16 $\pm$ 0.17\\
\end{tabular}
\caption{Same as in Table 3, but for KIC 3632418. Radial orders used to compute the mean parameters range between $n=17$ and $n=21$. Results shown in Figure \ref{fig:3632418}.}\label{tab:3632418}\vspace{-0.2cm}
\end{table*}

\begin{table*}[ht]\centering\fontsize{9.}{7.}\selectfont
\begin{tabular}{ccc|ccccc|c}
\multicolumn{9}{c}{KIC 3656476}\\ \hline\hline
time & duty & $\tau_\text{gran}$ &\multicolumn{5}{c|}{Peak-bagging}&Cross-correlation\\
(d)& cycle & (s)&$\delta\nu_{l=0}$ ($\mu$Hz) & $\delta\nu_{l=1}$ ($\mu$Hz) & $\delta\nu_{l=2}$ ($\mu$Hz) & $\delta\nu$ ($\mu$Hz)& $\ln\,S$ & $\delta\nu\,(\mu$Hz)\\\hline
345 & 0.98 & 294.5 $\pm$ 2.0 & 0.01 $\pm$ 0.07 & 0.16 $\pm$ 0.07 & 0.01 $\pm$ 0.11 & 0.08 $\pm$ 0.05 & 2.63 $\pm$ 0.16 & 0.01 $\pm$ 0.10\\
390 & 0.54 & 295.3 $\pm$ 7.4 & 0.10 $\pm$ 0.10 & 0.22 $\pm$ 0.09 & -0.18 $\pm$ 0.11 & 0.17 $\pm$ 0.06 & 2.79 $\pm$ 0.21 & -0.00 $\pm$ 0.09\\
435 & 0.05 & 299.9 $\pm$ 46.2 & -0.08 $\pm$ 0.45 & -0.26 $\pm$ 0.57 & 0.49 $\pm$ 0.57 & -0.15 $\pm$ 0.35 & 2.29 $\pm$ 0.55 & --\\
480 & 0.42 & 332.2 $\pm$ 2.3 & -0.03 $\pm$ 0.13 & 0.29 $\pm$ 0.13 & -0.39 $\pm$ 0.14 & 0.13 $\pm$ 0.09 & 2.49 $\pm$ 0.25 & --\\
525 & 0.90 & 320.7 $\pm$ 17.9 & 0.12 $\pm$ 0.09 & 0.11 $\pm$ 0.10 & -0.28 $\pm$ 0.11 & 0.11 $\pm$ 0.07 & 2.17 $\pm$ 0.17 & -0.01 $\pm$ 0.11\\
\end{tabular}
\caption{Same as in Table 3, but for KIC 3656476. Radial orders used to compute the mean parameters range between $n=16$ and $n=20$. Results shown in Figure \ref{fig:3656476}.}\label{tab:3656476}\vspace{-1cm}
\end{table*}

\begin{table*}[ht]\centering\fontsize{9.}{7.}\selectfont
\begin{tabular}{ccc|ccccc|c}
\multicolumn{9}{c}{KIC 3735871}\\ \hline\hline
time & duty & $\tau_\text{gran}$ &\multicolumn{5}{c|}{Peak-bagging}&Cross-correlation\\
(d)& cycle & (s)&$\delta\nu_{l=0}$ ($\mu$Hz) & $\delta\nu_{l=1}$ ($\mu$Hz) & $\delta\nu_{l=2}$ ($\mu$Hz) & $\delta\nu$ ($\mu$Hz)& $\ln\,S$ & $\delta\nu\,(\mu$Hz)\\\hline
345 & 0.98 & 197.1 $\pm$ 17.6 & -0.04 $\pm$ 0.23 & -0.45 $\pm$ 0.27 & 0.05 $\pm$ 0.49 & -0.21 $\pm$ 0.17 & 0.65 $\pm$ 0.21 & -0.18 $\pm$ 0.34\\
390 & 0.97 & 249.0 $\pm$ 16.8 & 0.08 $\pm$ 0.26 & 0.08 $\pm$ 0.24 & -0.13 $\pm$ 0.42 & 0.08 $\pm$ 0.18 & 0.85 $\pm$ 0.25 & -0.08 $\pm$ 0.19\\
435 & 0.96 & 204.1 $\pm$ 3.1 & 0.25 $\pm$ 0.14 & 0.16 $\pm$ 0.21 & 0.64 $\pm$ 0.37 & 0.22 $\pm$ 0.12 & 1.03 $\pm$ 0.22 & 0.07 $\pm$ 0.48\\
480 & 0.97 & 195.7 $\pm$ 6.9 & 0.01 $\pm$ 0.12 & 0.43 $\pm$ 0.18 & -0.31 $\pm$ 0.32 & 0.15 $\pm$ 0.10 & 1.10 $\pm$ 0.23 & 0.20 $\pm$ 0.55\\
525 & 0.96 & 209.2 $\pm$ 11.5 & -0.43 $\pm$ 0.18 & 0.24 $\pm$ 0.21 & -0.58 $\pm$ 0.42 & -0.15 $\pm$ 0.14 & 0.96 $\pm$ 0.23 & -0.16 $\pm$ 0.55\\
\end{tabular}
\caption{Same as in Table 3, but for KIC 3735871. Radial orders used to compute the mean parameters range between $n=20$ and $n=24$. Results shown in Figure \ref{fig:3735871}.}\label{tab:3735871}\vspace{-0.2cm}
\end{table*}

\begin{table*}[ht]\centering\fontsize{9.}{7.}\selectfont
\begin{tabular}{ccc|ccccc|c}
\multicolumn{9}{c}{KIC 4141376}\\ \hline\hline
time & duty & $\tau_\text{gran}$ &\multicolumn{5}{c|}{Peak-bagging}&Cross-correlation\\
(d)& cycle & (s)&$\delta\nu_{l=0}$ ($\mu$Hz) & $\delta\nu_{l=1}$ ($\mu$Hz) & $\delta\nu_{l=2}$ ($\mu$Hz) & $\delta\nu$ ($\mu$Hz)& $\ln\,S$ & $\delta\nu\,(\mu$Hz)\\\hline
45 & 0.98 & 295.1 $\pm$ 17.8 & -0.61 $\pm$ 0.65 & -0.05 $\pm$ 0.50 & -- & -0.26 $\pm$ 0.40 & 1.34 $\pm$ 0.41 & --\\
90 & 0.97 & 166.4 $\pm$ 13.6 & 0.13 $\pm$ 0.82 & 0.14 $\pm$ 0.61 & -- & 0.13 $\pm$ 0.49 & 1.09 $\pm$ 0.53 & -0.29 $\pm$ 0.62\\
135 & 0.96 & 168.7 $\pm$ 17.2 & 0.59 $\pm$ 0.84 & -0.13 $\pm$ 0.74 & -- & 0.18 $\pm$ 0.55 & 0.89 $\pm$ 0.39 & -0.35 $\pm$ 0.11\\
180 & 0.78 & 153.9 $\pm$ 1.2 & -0.18 $\pm$ 0.30 & -0.35 $\pm$ 0.70 & -- & -0.21 $\pm$ 0.28 & 1.30 $\pm$ 0.37 & -0.26 $\pm$ 0.20\\
225 & 0.72 & 555.8 $\pm$ 703.6 & -0.61 $\pm$ 0.40 & -0.09 $\pm$ 0.56 & -- & -0.43 $\pm$ 0.33 & 1.46 $\pm$ 0.38 & -0.09 $\pm$ 0.80\\
\end{tabular}
\caption{Same as in Table 3, but for KIC 4141376. {Radial orders used to compute the mean parameters range between $n=18$ and $n=22$.} Note that the frequency shifts from the cross-correlation method (last column) were obtained with 180-d sub-series. Results shown in Figure~\ref{fig:4141376}.}\label{tab:4141376}\vspace{-1.cm}
\end{table*}

\begin{table*}[ht]\centering\fontsize{9.}{7.}\selectfont
\begin{tabular}{ccc|ccccc|c}
\multicolumn{9}{c}{KIC 4349452}\\ \hline\hline
time & duty & $\tau_\text{gran}$ &\multicolumn{5}{c|}{Peak-bagging}&Cross-correlation\\
(d)& cycle & (s)&$\delta\nu_{l=0}$ ($\mu$Hz) & $\delta\nu_{l=1}$ ($\mu$Hz) & $\delta\nu_{l=2}$ ($\mu$Hz) & $\delta\nu$ ($\mu$Hz)& $\ln\,S$ & $\delta\nu\,(\mu$Hz)\\\hline
45 & 0.97 & 311.0 $\pm$ 21.5 & 0.54 $\pm$ 0.42 & -0.19 $\pm$ 0.55 & 0.06 $\pm$ 0.89 & 0.27 $\pm$ 0.34 & 0.63 $\pm$ 0.34 & --\\
90 & 0.97 & 317.4 $\pm$ 24.9 & 1.04 $\pm$ 0.44 & 0.11 $\pm$ 0.48 & 0.13 $\pm$ 0.83 & 0.62 $\pm$ 0.32 & 0.88 $\pm$ 0.34 & -0.33 $\pm$ 0.11\\
135 & 0.96 & 224.3 $\pm$ 34.0 & -0.48 $\pm$ 0.46 & -0.46 $\pm$ 0.56 & -0.77 $\pm$ 0.86 & -0.47 $\pm$ 0.36 & 0.72 $\pm$ 0.27 & -0.36 $\pm$ 0.11\\
180 & 0.97 & 245.3 $\pm$ 5.4 & -1.42 $\pm$ 0.48 & -0.76 $\pm$ 0.57 & -0.90 $\pm$ 0.90 & -1.15 $\pm$ 0.37 & 0.57 $\pm$ 0.22 & -0.32 $\pm$ 0.18\\
225 & 0.96 & 244.8 $\pm$ 1.4 & -0.27 $\pm$ 0.40 & 0.84 $\pm$ 0.47 & 0.66 $\pm$ 0.87 & 0.20 $\pm$ 0.31 & 0.82 $\pm$ 0.22 & -0.18 $\pm$ 1.02\\
\end{tabular}
\caption{Same as in Table 3, but for KIC 4349452. {Radial orders used to compute the mean parameters range between $n=19$ and $n=23$.} Note that the frequency shifts from the cross-correlation method (last column) were obtained with 180-d sub-series. Results shown in Figure~\ref{fig:4349452}.}\label{tab:4349452}\vspace{-0.2cm}
\end{table*}

\begin{table*}[ht]\centering\fontsize{9.}{7.}\selectfont
\begin{tabular}{ccc|ccccc|c}
\multicolumn{9}{c}{KIC 4914423}\\ \hline\hline
time & duty & $\tau_\text{gran}$ &\multicolumn{5}{c|}{Peak-bagging}&Cross-correlation\\
(d)& cycle & (s)&$\delta\nu_{l=0}$ ($\mu$Hz) & $\delta\nu_{l=1}$ ($\mu$Hz) & $\delta\nu_{l=2}$ ($\mu$Hz) & $\delta\nu$ ($\mu$Hz)& $\ln\,S$ & $\delta\nu\,(\mu$Hz)\\\hline
45 & 0.95 & 318.7 $\pm$ 38.7 & -0.75 $\pm$ 0.87 & 0.38 $\pm$ 0.59 & -- & 0.02 $\pm$ 0.49 & 2.09 $\pm$ 0.39 & --\\
90 & 0.93 & 281.4 $\pm$ 17.4 & -0.96 $\pm$ 0.99 & 0.65 $\pm$ 0.54 & -- & 0.28 $\pm$ 0.48 & 1.98 $\pm$ 0.52 & --\\
135 & 0.92 & 233.3 $\pm$ 23.1 & 0.08 $\pm$ 1.05 & 0.09 $\pm$ 0.89 & -- & 0.08 $\pm$ 0.68 & 1.61 $\pm$ 0.47 & -0.37 $\pm$ 0.31\\
180 & 0.92 & 265.7 $\pm$ 43.3 & -0.31 $\pm$ 0.52 & 0.41 $\pm$ 0.52 & -- & 0.06 $\pm$ 0.37 & 2.32 $\pm$ 0.31 & -0.14 $\pm$ 1.09\\
225 & 0.96 & 293.6 $\pm$ 43.0 & -0.55 $\pm$ 0.76 & 0.35 $\pm$ 0.48 & -- & 0.09 $\pm$ 0.41 & 2.35 $\pm$ 0.31 & 0.10 $\pm$ 1.53\\
\end{tabular}
\caption{Same as in Table 3, but for KIC 4914423. {Radial orders used to compute the mean parameters range between $n=16$ and $n=20$.} Note that the frequency shifts from the cross-correlation method (last column) were obtained with 180-d sub-series. Results shown in Figure~\ref{fig:4914423}.}\label{tab:4914423}\vspace{-1.cm}
\end{table*}

\begin{table*}[ht]\centering\fontsize{9.}{7.}\selectfont
\begin{tabular}{ccc|ccccc|c}
\multicolumn{9}{c}{KIC 4914923}\\ \hline\hline
time & duty & $\tau_\text{gran}$ &\multicolumn{5}{c|}{Peak-bagging}&Cross-correlation\\
(d)& cycle & (s)&$\delta\nu_{l=0}$ ($\mu$Hz) & $\delta\nu_{l=1}$ ($\mu$Hz) & $\delta\nu_{l=2}$ ($\mu$Hz) & $\delta\nu$ ($\mu$Hz)& $\ln\,S$ & $\delta\nu\,(\mu$Hz)\\\hline
345 & 0.98 & 293.3 $\pm$ 16.8 & -0.12 $\pm$ 0.11 & -0.09 $\pm$ 0.09 & -0.13 $\pm$ 0.14 & -0.10 $\pm$ 0.07 & 2.17 $\pm$ 0.17 & -0.10 $\pm$ 0.11\\
390 & 0.54 & 294.8 $\pm$ 24.1 & 0.04 $\pm$ 0.14 & -0.02 $\pm$ 0.11 & -0.08 $\pm$ 0.19 & 0.00 $\pm$ 0.09 & 2.34 $\pm$ 0.25 & -0.05 $\pm$ 0.11\\
435 & 0.05 & 314.3 $\pm$ 1.1 & -0.02 $\pm$ 0.47 & 0.17 $\pm$ 0.44 & 0.47 $\pm$ 0.74 & 0.08 $\pm$ 0.32 & 3.15 $\pm$ 0.54 & --\\
480 & 0.41 & 301.6 $\pm$ 26.4 & 0.05 $\pm$ 0.15 & -0.03 $\pm$ 0.15 & 0.23 $\pm$ 0.27 & 0.01 $\pm$ 0.11 & 2.11 $\pm$ 0.24 & --\\
525 & 0.90 & 294.1 $\pm$ 15.9 & 0.17 $\pm$ 0.11 & -0.01 $\pm$ 0.09 & -0.32 $\pm$ 0.19 & 0.06 $\pm$ 0.07 & 2.00 $\pm$ 0.15 & -0.02 $\pm$ 0.11\\
\end{tabular}
\caption{Same as in Table 3, but for KIC 4914923. Radial orders used to compute the mean parameters range between $n=17$ and $n=21$. Results shown in Figure \ref{fig:4914923}.}\label{tab:4914923}
\end{table*}

\begin{table*}[ht]\centering\fontsize{9.}{7.}\selectfont
\begin{tabular}{ccc|ccccc|c}
\multicolumn{9}{c}{KIC 5184732}\\ \hline\hline
time & duty & $\tau_\text{gran}$ &\multicolumn{5}{c|}{Peak-bagging}&Cross-correlation\\
(d)& cycle & (s)&$\delta\nu_{l=0}$ ($\mu$Hz) & $\delta\nu_{l=1}$ ($\mu$Hz) & $\delta\nu_{l=2}$ ($\mu$Hz) & $\delta\nu$ ($\mu$Hz)& $\ln\,S$ & $\delta\nu\,(\mu$Hz)\\\hline
545 & 0.96 & 278.2 $\pm$ 0.4 & -0.25 $\pm$ 0.09 & -0.05 $\pm$ 0.10 & -0.07 $\pm$ 0.13 & -0.16 $\pm$ 0.07 & 1.75 $\pm$ 0.12 & -0.08 $\pm$ 0.16\\
590 & 0.78 & 261.3 $\pm$ 10.4 & -0.09 $\pm$ 0.08 & 0.09 $\pm$ 0.10 & -0.00 $\pm$ 0.14 & -0.01 $\pm$ 0.06 & 1.86 $\pm$ 0.12 & -0.02 $\pm$ 0.14\\
635 & 0.73 & 264.5 $\pm$ 11.3 & 0.11 $\pm$ 0.08 & 0.10 $\pm$ 0.10 & 0.01 $\pm$ 0.13 & 0.11 $\pm$ 0.06 & 2.10 $\pm$ 0.12 & 0.03 $\pm$ 0.14\\
680 & 0.92 & 276.1 $\pm$ 8.8 & -0.07 $\pm$ 0.08 & -0.08 $\pm$ 0.08 & -0.13 $\pm$ 0.11 & -0.07 $\pm$ 0.06 & 2.05 $\pm$ 0.11 & -0.09 $\pm$ 0.13\\
725 & 0.98 & 270.8 $\pm$ 3.8 & -0.08 $\pm$ 0.08 & -0.15 $\pm$ 0.09 & -0.24 $\pm$ 0.12 & -0.11 $\pm$ 0.06 & 1.94 $\pm$ 0.11 & -0.14 $\pm$ 0.13\\
\end{tabular}
\caption{Same as in Table 3, but for KIC 5184732. Figures in the main text (see Figure \ref{fig:KIC5184732}). Radial orders used to compute the mean parameters range between $n=18$ and $n=22$.}\label{tab:5184732}
\end{table*}

\nopagebreak
\FloatBarrier
\nopagebreak

\begin{table}[ht]\centering\fontsize{9.}{7.}\selectfont
\begin{tabular}{ccc|ccccc|c}
\multicolumn{9}{c}{KIC 5773345}\\ \hline\hline
time & duty & $\tau_\text{gran}$ &\multicolumn{5}{c|}{Peak-bagging}&Cross-correlation\\
(d)& cycle & (s)&$\delta\nu_{l=0}$ ($\mu$Hz) & $\delta\nu_{l=1}$ ($\mu$Hz) & $\delta\nu_{l=2}$ ($\mu$Hz) & $\delta\nu$ ($\mu$Hz)& $\ln\,S$ & $\delta\nu\,(\mu$Hz)\\\hline
45 & 0.98 & 404.1 $\pm$ 18.7 & 0.08 $\pm$ 0.20 & -0.10 $\pm$ 0.16 & -1.02 $\pm$ 0.35 & -0.02 $\pm$ 0.13 & 1.72 $\pm$ 0.11 & -0.09 $\pm$ 0.23\\
90 & 0.97 & 360.0 $\pm$ 16.2 & -0.16 $\pm$ 0.23 & 0.12 $\pm$ 0.18 & -0.51 $\pm$ 0.41 & 0.01 $\pm$ 0.14 & 1.63 $\pm$ 0.11 & 0.04 $\pm$ 0.21\\
135 & 0.96 & 352.0 $\pm$ 15.5 & 0.12 $\pm$ 0.19 & 0.22 $\pm$ 0.16 & -0.61 $\pm$ 0.30 & 0.18 $\pm$ 0.12 & 1.60 $\pm$ 0.11 & 0.13 $\pm$ 0.20\\
180 & 0.78 & 387.0 $\pm$ 20.5 & 0.08 $\pm$ 0.19 & -0.11 $\pm$ 0.17 & -0.00 $\pm$ 0.33 & -0.03 $\pm$ 0.12 & 1.59 $\pm$ 0.10 & -0.02 $\pm$ 0.22\\
225 & 0.71 & 412.1 $\pm$ 23.3 & -0.34 $\pm$ 0.24 & -0.43 $\pm$ 0.17 & 0.88 $\pm$ 0.35 & -0.40 $\pm$ 0.14 & 1.85 $\pm$ 0.12 & -0.19 $\pm$ 0.22\\
270 & 0.42 & 412.7 $\pm$ 29.6 & -0.26 $\pm$ 0.29 & -0.24 $\pm$ 0.22 & 0.59 $\pm$ 0.41 & -0.25 $\pm$ 0.18 & 1.87 $\pm$ 0.17 & --\\
360 & 0.41 & 419.4 $\pm$ 27.4 & 0.41 $\pm$ 0.24 & 0.51 $\pm$ 0.22 & -0.04 $\pm$ 0.45 & 0.47 $\pm$ 0.16 & 1.74 $\pm$ 0.17 & --\\
405 & 0.90 & 414.6 $\pm$ 20.0 & -0.01 $\pm$ 0.18 & 0.24 $\pm$ 0.18 & 0.09 $\pm$ 0.36 & 0.12 $\pm$ 0.13 & 1.68 $\pm$ 0.11 & 0.26 $\pm$ 0.22\\
450 & 0.97 & 411.1 $\pm$ 20.4 & -0.41 $\pm$ 0.21 & -0.11 $\pm$ 0.18 & 0.22 $\pm$ 0.36 & -0.23 $\pm$ 0.13 & 1.70 $\pm$ 0.12 & -0.13 $\pm$ 0.20\\
495 & 0.94 & 441.6 $\pm$ 23.2 & 0.25 $\pm$ 0.18 & -0.15 $\pm$ 0.16 & -0.04 $\pm$ 0.34 & 0.02 $\pm$ 0.12 & 1.82 $\pm$ 0.12 & -0.01 $\pm$ 0.23\\
540 & 0.67 & 410.0 $\pm$ 32.4 & 0.75 $\pm$ 0.18 & -0.24 $\pm$ 0.18 & 0.29 $\pm$ 0.44 & 0.26 $\pm$ 0.13 & 1.86 $\pm$ 0.15 & --\\
\end{tabular}
\parbox{2\hsize}{\caption{Same as in Table 3, but for KIC 5773345. Radial orders used to compute the mean parameters range between $n=15$ and $n=19$. Results shown in Figure \ref{fig:5773345}.}\label{tab:5773345}}
\end{table}

\begin{table}[ht]\centering\fontsize{9.}{7.}\selectfont
\begin{tabular}{ccc|ccccc|c}
\multicolumn{9}{c}{KIC 5866724}\\ \hline\hline
time & duty & $\tau_\text{gran}$ &\multicolumn{5}{c|}{Peak-bagging}&Cross-correlation\\
(d)& cycle & (s)&$\delta\nu_{l=0}$ ($\mu$Hz) & $\delta\nu_{l=1}$ ($\mu$Hz) & $\delta\nu_{l=2}$ ($\mu$Hz) & $\delta\nu$ ($\mu$Hz)& $\ln\,S$ & $\delta\nu\,(\mu$Hz)\\\hline
45 & 0.95 & 236.5 $\pm$ 1.0 & -0.20 $\pm$ 0.41 & 0.33 $\pm$ 0.32 & 0.30 $\pm$ 0.78 & 0.12 $\pm$ 0.25 & 1.48 $\pm$ 0.37 & --\\
90 & 0.93 & 245.6 $\pm$ 0.0 & 0.05 $\pm$ 0.40 & 0.21 $\pm$ 0.34 & -0.01 $\pm$ 0.83 & 0.14 $\pm$ 0.26 & 0.98 $\pm$ 0.34 & 0.18 $\pm$ 0.18\\
135 & 0.92 & 240.5 $\pm$ 10.0 & 0.23 $\pm$ 0.36 & 0.51 $\pm$ 0.56 & -0.06 $\pm$ 0.84 & 0.31 $\pm$ 0.30 & 1.04 $\pm$ 0.37 & 0.20 $\pm$ 0.18\\
180 & 0.92 & 240.2 $\pm$ 15.1 & 1.05 $\pm$ 0.44 & 0.35 $\pm$ 0.57 & 0.39 $\pm$ 0.81 & 0.79 $\pm$ 0.35 & 0.91 $\pm$ 0.27 & 0.21 $\pm$ 0.25\\
225 & 0.96 & 244.8 $\pm$ 26.5 & 0.68 $\pm$ 0.29 & 0.32 $\pm$ 0.28 & 1.12 $\pm$ 0.76 & 0.49 $\pm$ 0.20 & 1.29 $\pm$ 0.30 & 0.19 $\pm$ 0.42\\
\end{tabular}
\parbox{2\hsize}{\caption{Same as in Table 3, but for KIC 5866724. {Radial orders used to compute the mean parameters range between $n=19$ and $n=23$.} Note that the frequency shifts from the cross-correlation method (last column) were obtained with 180-d sub-series. Results shown in Figure~\ref{fig:5866724}.}\label{tab:5866724}}
\end{table}

\begin{table*}[ht]\centering\fontsize{9.}{7.}\selectfont
\begin{tabular}{ccc|ccccc|c}
\multicolumn{9}{c}{KIC 5950854}\\ \hline\hline
time & duty & $\tau_\text{gran}$ &\multicolumn{5}{c|}{Peak-bagging}&Cross-correlation\\
(d)& cycle & (s)&$\delta\nu_{l=0}$ ($\mu$Hz) & $\delta\nu_{l=1}$ ($\mu$Hz) & $\delta\nu_{l=2}$ ($\mu$Hz) & $\delta\nu$ ($\mu$Hz)& $\ln\,S$ & $\delta\nu\,(\mu$Hz)\\\hline
45 & 0.97 & 291.6 $\pm$ 6.5 & 0.73 $\pm$ 0.13 & -0.19 $\pm$ 0.12 & 0.13 $\pm$ 0.37 & 0.22 $\pm$ 0.09 & 2.19 $\pm$ 0.24 & -0.13 $\pm$ 0.43\\
90 & 0.54 & 291.7 $\pm$ 20.8 & 0.70 $\pm$ 0.25 & -0.16 $\pm$ 0.24 & 0.01 $\pm$ 0.66 & 0.24 $\pm$ 0.17 & 1.95 $\pm$ 0.35 & -0.11 $\pm$ 0.22\\
135 & 0.05 & 263.3 $\pm$ 139.9 & -0.66 $\pm$ 0.71 & -0.58 $\pm$ 0.57 & 0.28 $\pm$ 0.94 & -0.61 $\pm$ 0.45 & 2.52 $\pm$ 0.63 & --\\
180 & 0.33 & 226.4 $\pm$ 1.5 & -0.60 $\pm$ 0.61 & 0.31 $\pm$ 0.57 & -0.29 $\pm$ 0.76 & -0.12 $\pm$ 0.42 & 1.06 $\pm$ 0.32 & --\\
225 & 0.58 & 263.2 $\pm$ 10.8 & 1.05 $\pm$ 0.16 & 0.18 $\pm$ 0.20 & 0.65 $\pm$ 0.46 & 0.70 $\pm$ 0.12 & 1.85 $\pm$ 0.21 & 0.37 $\pm$ 0.97\\
\end{tabular}
\caption{Same as in Table 3, but for KIC 5950854. Radial orders used to compute the mean parameters range between $n=17$ and $n=21$. Results shown in Figure \ref{fig:5950854}.}\label{tab:5950854}
\end{table*}

\begin{table*}[ht]\centering\fontsize{9.}{7.}\selectfont
\begin{tabular}{ccc|ccccc|c}
\multicolumn{9}{c}{KIC 6106415}\\ \hline\hline
time & duty & $\tau_\text{gran}$ &\multicolumn{5}{c|}{Peak-bagging}&Cross-correlation\\
(d)& cycle & (s)&$\delta\nu_{l=0}$ ($\mu$Hz) & $\delta\nu_{l=1}$ ($\mu$Hz) & $\delta\nu_{l=2}$ ($\mu$Hz) & $\delta\nu$ ($\mu$Hz)& $\ln\,S$ & $\delta\nu\,(\mu$Hz)\\\hline
45 & 0.97 & 205.5 $\pm$ 5.7 & -0.08 $\pm$ 0.09 & 0.17 $\pm$ 0.11 & 0.10 $\pm$ 0.13 & 0.03 $\pm$ 0.07 & 1.48 $\pm$ 0.09 & 0.00 $\pm$ 0.10\\
90 & 0.97 & 220.0 $\pm$ 6.9 & -0.06 $\pm$ 0.08 & 0.11 $\pm$ 0.10 & 0.06 $\pm$ 0.13 & 0.01 $\pm$ 0.07 & 1.48 $\pm$ 0.09 & 0.06 $\pm$ 0.08\\
135 & 0.96 & 237.9 $\pm$ 8.3 & 0.02 $\pm$ 0.09 & 0.15 $\pm$ 0.11 & -0.07 $\pm$ 0.13 & 0.08 $\pm$ 0.07 & 1.40 $\pm$ 0.10 & 0.09 $\pm$ 0.09\\
180 & 0.77 & 234.1 $\pm$ 8.6 & 0.04 $\pm$ 0.10 & 0.08 $\pm$ 0.11 & 0.25 $\pm$ 0.15 & 0.06 $\pm$ 0.08 & 1.27 $\pm$ 0.09 & 0.03 $\pm$ 0.09\\
225 & 0.71 & 237.1 $\pm$ 9.3 & 0.07 $\pm$ 0.11 & -0.12 $\pm$ 0.12 & 0.26 $\pm$ 0.16 & -0.01 $\pm$ 0.08 & 1.20 $\pm$ 0.09 & -0.06 $\pm$ 0.11\\
\end{tabular}
\caption{Same as in Table 3, but for KIC 6106415. Radial orders used to compute the mean parameters range between $n=19$ and $n=23$. Results shown in Figure \ref{fig:6106415}.}\label{tab:6106415}
\end{table*}

\begin{table*}[ht]\centering\fontsize{9.}{7.}\selectfont
\begin{tabular}{ccc|ccccc|c}
\multicolumn{9}{c}{KIC 6116048}\\ \hline\hline
time & duty & $\tau_\text{gran}$ &\multicolumn{5}{c|}{Peak-bagging}&Cross-correlation\\
(d)& cycle & (s)&$\delta\nu_{l=0}$ ($\mu$Hz) & $\delta\nu_{l=1}$ ($\mu$Hz) & $\delta\nu_{l=2}$ ($\mu$Hz) & $\delta\nu$ ($\mu$Hz)& $\ln\,S$ & $\delta\nu\,(\mu$Hz)\\\hline
345 & 0.98 & 262.3 $\pm$ 12.8 & 0.04 $\pm$ 0.10 & 0.26 $\pm$ 0.10 & -0.01 $\pm$ 0.15 & 0.15 $\pm$ 0.07 & 1.62 $\pm$ 0.11 & 0.23 $\pm$ 0.14\\
390 & 0.97 & 254.3 $\pm$ 12.3 & 0.02 $\pm$ 0.10 & -0.05 $\pm$ 0.10 & 0.11 $\pm$ 0.14 & -0.02 $\pm$ 0.07 & 1.65 $\pm$ 0.10 & -0.03 $\pm$ 0.13\\
435 & 0.96 & 248.6 $\pm$ 11.6 & -0.05 $\pm$ 0.11 & 0.05 $\pm$ 0.10 & 0.12 $\pm$ 0.14 & 0.00 $\pm$ 0.07 & 1.70 $\pm$ 0.11 & -0.00 $\pm$ 0.14\\
480 & 0.97 & 269.2 $\pm$ 13.9 & 0.04 $\pm$ 0.10 & 0.28 $\pm$ 0.10 & 0.21 $\pm$ 0.16 & 0.16 $\pm$ 0.07 & 1.72 $\pm$ 0.11 & 0.22 $\pm$ 0.12\\
525 & 0.96 & 256.1 $\pm$ 1.6 & 0.19 $\pm$ 0.10 & 0.15 $\pm$ 0.11 & 0.46 $\pm$ 0.15 & 0.17 $\pm$ 0.07 & 1.60 $\pm$ 0.11 & 0.17 $\pm$ 0.14\\
\end{tabular}
\caption{Same as in Table 3, but for KIC 6116048. Radial orders used to compute the mean parameters range between $n=17$ and $n=21$. Results shown in Figure \ref{fig:6116048}.}\label{tab:6116048}
\end{table*}

\begin{table*}[ht]\centering\fontsize{9.}{7.}\selectfont
\begin{tabular}{ccc|ccccc|c}
\multicolumn{9}{c}{KIC 6225718}\\ \hline\hline
time & duty & $\tau_\text{gran}$ &\multicolumn{5}{c|}{Peak-bagging}&Cross-correlation\\
(d)& cycle & (s)&$\delta\nu_{l=0}$ ($\mu$Hz) & $\delta\nu_{l=1}$ ($\mu$Hz) & $\delta\nu_{l=2}$ ($\mu$Hz) & $\delta\nu$ ($\mu$Hz)& $\ln\,S$ & $\delta\nu\,(\mu$Hz)\\\hline
45 & 0.98 & 218.0 $\pm$ 6.8 & -0.01 $\pm$ 0.12 & 0.01 $\pm$ 0.12 & 0.38 $\pm$ 0.18 & 0.00 $\pm$ 0.09 & 1.05 $\pm$ 0.09 & -0.00 $\pm$ 0.19\\
90 & 0.97 & 234.1 $\pm$ 10.9 & 0.07 $\pm$ 0.12 & -0.06 $\pm$ 0.12 & 0.23 $\pm$ 0.18 & 0.00 $\pm$ 0.09 & 1.03 $\pm$ 0.09 & -0.04 $\pm$ 0.16\\
135 & 0.96 & 242.7 $\pm$ 0.5 & -0.06 $\pm$ 0.12 & 0.14 $\pm$ 0.12 & -0.10 $\pm$ 0.17 & 0.04 $\pm$ 0.08 & 1.09 $\pm$ 0.09 & 0.07 $\pm$ 0.15\\
180 & 0.78 & 230.4 $\pm$ 8.3 & 0.05 $\pm$ 0.12 & 0.39 $\pm$ 0.12 & -0.02 $\pm$ 0.19 & 0.21 $\pm$ 0.08 & 1.06 $\pm$ 0.09 & 0.27 $\pm$ 0.16\\
225 & 0.72 & 230.0 $\pm$ 0.0 & 0.07 $\pm$ 0.12 & 0.07 $\pm$ 0.12 & 0.01 $\pm$ 0.19 & 0.07 $\pm$ 0.09 & 1.06 $\pm$ 0.09 & 0.08 $\pm$ 0.19\\
\end{tabular}
\caption{Same as in Table 3, but for KIC 6225718. Radial orders used to compute the mean parameters range between $n=18$ and $n=22$. Results shown in Figure \ref{fig:6225718}.}\label{tab:6225718}\vspace{-1cm}
\end{table*}

\begin{table*}[ht]\centering\fontsize{9.}{7.}\selectfont
\begin{tabular}{ccc|ccccc|c}
\multicolumn{9}{c}{KIC 6278762}\\ \hline\hline
time & duty & $\tau_\text{gran}$ &\multicolumn{5}{c|}{Peak-bagging}&Cross-correlation\\
(d)& cycle & (s)&$\delta\nu_{l=0}$ ($\mu$Hz) & $\delta\nu_{l=1}$ ($\mu$Hz) & $\delta\nu_{l=2}$ ($\mu$Hz) & $\delta\nu$ ($\mu$Hz)& $\ln\,S$ & $\delta\nu\,(\mu$Hz)\\\hline
45 & 0.91 & 185.3 $\pm$ 21.5 & -0.55 $\pm$ 0.15 & -0.39 $\pm$ 0.15 & -- & -0.47 $\pm$ 0.11 & 0.06 $\pm$ 0.27 & -0.12 $\pm$ 0.09\\
90 & 0.85 & 211.9 $\pm$ 35.2 & -0.56 $\pm$ 0.16 & 0.24 $\pm$ 0.21 & -- & -0.27 $\pm$ 0.13 & 0.20 $\pm$ 0.27 & -0.05 $\pm$ 0.12\\
135 & 0.84 & 197.5 $\pm$ 23.6 & -0.07 $\pm$ 0.14 & 0.30 $\pm$ 0.15 & -- & 0.10 $\pm$ 0.10 & 0.34 $\pm$ 0.29 & 0.17 $\pm$ 0.79\\
180 & 0.82 & 165.7 $\pm$ 17.2 & -0.36 $\pm$ 0.16 & -0.25 $\pm$ 0.15 & -- & -0.30 $\pm$ 0.11 & 0.13 $\pm$ 0.27 & --\\
\end{tabular}
\caption{Same as in Table 3, but for KIC 6278762. Radial orders used to compute the mean parameters range between $n=21$ and $n=25$. Results shown in Figure \ref{fig:6278762}.}\label{tab:6278762}
\end{table*}

\begin{table*}[ht]\centering\fontsize{9.}{7.}\selectfont
\begin{tabular}{ccc|ccccc|c}
\multicolumn{9}{c}{KIC 6508366}\\ \hline\hline
time & duty & $\tau_\text{gran}$ &\multicolumn{5}{c|}{Peak-bagging}&Cross-correlation\\
(d)& cycle & (s)&$\delta\nu_{l=0}$ ($\mu$Hz) & $\delta\nu_{l=1}$ ($\mu$Hz) & $\delta\nu_{l=2}$ ($\mu$Hz) & $\delta\nu$ ($\mu$Hz)& $\ln\,S$ & $\delta\nu\,(\mu$Hz)\\\hline
345 & 0.98 & 412.2 $\pm$ 21.6 & 0.16 $\pm$ 0.27 & -0.36 $\pm$ 0.24 & -0.60 $\pm$ 0.43 & -0.12 $\pm$ 0.18 & 1.39 $\pm$ 0.11 & -0.26 $\pm$ 0.32\\
390 & 0.97 & 408.2 $\pm$ 21.1 & 0.21 $\pm$ 0.28 & -0.02 $\pm$ 0.24 & -0.12 $\pm$ 0.48 & 0.08 $\pm$ 0.18 & 1.29 $\pm$ 0.11 & -0.18 $\pm$ 0.34\\
435 & 0.96 & 405.0 $\pm$ 20.3 & -0.41 $\pm$ 0.33 & -0.13 $\pm$ 0.30 & -0.36 $\pm$ 0.52 & -0.26 $\pm$ 0.22 & 1.11 $\pm$ 0.11 & -0.12 $\pm$ 0.42\\
480 & 0.97 & 377.8 $\pm$ 17.6 & -0.05 $\pm$ 0.29 & 0.30 $\pm$ 0.31 & -0.67 $\pm$ 0.48 & 0.11 $\pm$ 0.21 & 1.13 $\pm$ 0.12 & 0.16 $\pm$ 0.46\\
525 & 0.96 & 383.9 $\pm$ 18.3 & -0.28 $\pm$ 0.25 & 0.47 $\pm$ 0.26 & 0.14 $\pm$ 0.39 & 0.08 $\pm$ 0.18 & 1.39 $\pm$ 0.10 & -0.00 $\pm$ 0.43\\
\end{tabular}
\caption{Same as in Table 3, but for KIC 6508366. Radial orders used to compute the mean parameters range between $n=15$ and $n=19$. Results shown in Figure \ref{fig:6508366}.}\label{tab:6508366}\vspace{-0.1cm}
\end{table*}

\begin{table*}[ht]\centering\fontsize{9.}{7.}\selectfont
\begin{tabular}{ccc|ccccc|c}
\multicolumn{9}{c}{KIC 6521045}\\ \hline\hline
time & duty & $\tau_\text{gran}$ &\multicolumn{5}{c|}{Peak-bagging}&Cross-correlation\\
(d)& cycle & (s)&$\delta\nu_{l=0}$ ($\mu$Hz) & $\delta\nu_{l=1}$ ($\mu$Hz) & $\delta\nu_{l=2}$ ($\mu$Hz) & $\delta\nu$ ($\mu$Hz)& $\ln\,S$ & $\delta\nu\,(\mu$Hz)\\\hline
45 & 0.95 & 334.8 $\pm$ 30.7 & -0.13 $\pm$ 0.09 & 0.01 $\pm$ 0.17 & -0.26 $\pm$ 0.29 & -0.08 $\pm$ 0.08 & 2.64 $\pm$ 0.19 & --\\
90 & 0.93 & 298.8 $\pm$ 18.5 & -0.05 $\pm$ 0.13 & -0.27 $\pm$ 0.20 & -0.18 $\pm$ 0.27 & -0.09 $\pm$ 0.11 & 2.57 $\pm$ 0.19 & -0.02 $\pm$ 0.14\\
135 & 0.92 & 323.0 $\pm$ 1.2 & 0.04 $\pm$ 0.26 & -0.10 $\pm$ 0.23 & -0.03 $\pm$ 0.46 & 0.01 $\pm$ 0.18 & 2.17 $\pm$ 0.20 & 0.02 $\pm$ 0.14\\
180 & 0.92 & 364.2 $\pm$ 31.2 & 0.18 $\pm$ 0.23 & 0.29 $\pm$ 0.21 & -0.63 $\pm$ 0.44 & 0.24 $\pm$ 0.16 & 2.22 $\pm$ 0.21 & 0.04 $\pm$ 0.15\\
225 & 0.96 & 359.9 $\pm$ 35.1 & 0.08 $\pm$ 0.14 & 0.05 $\pm$ 0.11 & -0.18 $\pm$ 0.22 & 0.06 $\pm$ 0.09 & 2.98 $\pm$ 0.23 & -0.01 $\pm$ 0.19\\
\end{tabular}
\caption{Same as in Table 3, but for KIC 6521045. {Radial orders used to compute the mean parameters range between $n=16$ and $n=20$.} Note that the frequency shifts from the cross-correlation method (last column) were obtained with 180-d sub-series. Results shown in Figure~\ref{fig:6521045}.}\label{tab:6521045}\vspace{-2cm}
\end{table*}

\begin{table*}[ht]\centering\fontsize{9.}{7.}\selectfont
\begin{tabular}{ccc|ccccc|c}
\multicolumn{9}{c}{KIC 6603624}\\ \hline\hline
time & duty & $\tau_\text{gran}$ &\multicolumn{5}{c|}{Peak-bagging}&Cross-correlation\\
(d)& cycle & (s)&$\delta\nu_{l=0}$ ($\mu$Hz) & $\delta\nu_{l=1}$ ($\mu$Hz) & $\delta\nu_{l=2}$ ($\mu$Hz) & $\delta\nu$ ($\mu$Hz)& $\ln\,S$ & $\delta\nu\,(\mu$Hz)\\\hline
45 & 0.98 & 239.9 $\pm$ 0.4 & -0.07 $\pm$ 0.07 & -0.03 $\pm$ 0.07 & 0.05 $\pm$ 0.06 & -0.05 $\pm$ 0.05 & 2.30 $\pm$ 0.16 & -0.04 $\pm$ 0.12\\
90 & 0.97 & 240.3 $\pm$ 11.4 & -0.05 $\pm$ 0.07 & 0.03 $\pm$ 0.06 & 0.04 $\pm$ 0.07 & -0.01 $\pm$ 0.05 & 2.37 $\pm$ 0.15 & -0.02 $\pm$ 0.10\\
135 & 0.96 & 241.9 $\pm$ 1.5 & 0.05 $\pm$ 0.06 & 0.17 $\pm$ 0.06 & -0.01 $\pm$ 0.06 & 0.12 $\pm$ 0.04 & 2.45 $\pm$ 0.14 & 0.09 $\pm$ 0.12\\
180 & 0.97 & 255.5 $\pm$ 1.5 & 0.06 $\pm$ 0.07 & 0.10 $\pm$ 0.05 & -0.09 $\pm$ 0.07 & 0.09 $\pm$ 0.04 & 2.48 $\pm$ 0.14 & 0.05 $\pm$ 0.11\\
225 & 0.96 & 252.9 $\pm$ 9.9 & 0.05 $\pm$ 0.05 & -0.02 $\pm$ 0.05 & -0.12 $\pm$ 0.06 & 0.02 $\pm$ 0.04 & 2.49 $\pm$ 0.14 & 0.01 $\pm$ 0.33\\
\end{tabular}
\caption{Same as in Table 3, but for KIC 6603624. Radial orders used to compute the mean parameters range between $n=18$ and $n=22$. Results shown in Figure \ref{fig:6603624}.}\label{tab:6603624}\vspace{-0.1cm}
\end{table*}

\begin{table*}[ht]\centering\fontsize{9.}{7.}\selectfont
\begin{tabular}{ccc|ccccc|c}
\multicolumn{9}{c}{KIC 6679371}\\ \hline\hline
time & duty & $\tau_\text{gran}$ &\multicolumn{5}{c|}{Peak-bagging}&Cross-correlation\\
(d)& cycle & (s)&$\delta\nu_{l=0}$ ($\mu$Hz) & $\delta\nu_{l=1}$ ($\mu$Hz) & $\delta\nu_{l=2}$ ($\mu$Hz) & $\delta\nu$ ($\mu$Hz)& $\ln\,S$ & $\delta\nu\,(\mu$Hz)\\\hline
345 & 0.98 & 385.8 $\pm$ 19.2 & 0.07 $\pm$ 0.21 & -0.08 $\pm$ 0.24 & 0.63 $\pm$ 0.60 & 0.01 $\pm$ 0.16 & 1.68 $\pm$ 0.10 & 0.04 $\pm$ 0.26\\
390 & 0.97 & 409.4 $\pm$ 18.9 & 0.31 $\pm$ 0.21 & 0.00 $\pm$ 0.22 & -0.90 $\pm$ 0.50 & 0.17 $\pm$ 0.15 & 1.62 $\pm$ 0.11 & 0.13 $\pm$ 0.26\\
435 & 0.96 & 392.8 $\pm$ 18.4 & 0.70 $\pm$ 0.21 & -0.05 $\pm$ 0.22 & -0.79 $\pm$ 0.52 & 0.34 $\pm$ 0.15 & 1.55 $\pm$ 0.11 & 0.11 $\pm$ 0.31\\
480 & 0.97 & 354.8 $\pm$ 18.2 & 0.31 $\pm$ 0.28 & -0.02 $\pm$ 0.28 & 0.74 $\pm$ 0.46 & 0.15 $\pm$ 0.20 & 1.28 $\pm$ 0.11 & 0.02 $\pm$ 0.34\\
525 & 0.96 & 343.4 $\pm$ 17.2 & -0.14 $\pm$ 0.28 & -0.56 $\pm$ 0.28 & 0.35 $\pm$ 0.59 & -0.35 $\pm$ 0.20 & 1.10 $\pm$ 0.11 & -0.13 $\pm$ 0.35\\
\end{tabular}
\caption{Same as in Table 3, but for KIC 6679371. Radial orders used to compute the mean parameters range between $n=15$ and $n=19$. Results shown in Figure \ref{fig:6679371}.}\label{tab:6679371}\vspace{-2cm}
\end{table*}

\begin{table*}[ht]\centering\fontsize{9.}{7.}\selectfont
\begin{tabular}{ccc|ccccc|c}
\multicolumn{9}{c}{KIC 6933899}\\ \hline\hline
time & duty & $\tau_\text{gran}$ &\multicolumn{5}{c|}{Peak-bagging}&Cross-correlation\\
(d)& cycle & (s)&$\delta\nu_{l=0}$ ($\mu$Hz) & $\delta\nu_{l=1}$ ($\mu$Hz) & $\delta\nu_{l=2}$ ($\mu$Hz) & $\delta\nu$ ($\mu$Hz)& $\ln\,S$ & $\delta\nu\,(\mu$Hz)\\\hline
45 & 0.98 & 369.9 $\pm$ 1.7 & 0.06 $\pm$ 0.08 & 0.01 $\pm$ 0.08 & -0.11 $\pm$ 0.12 & 0.04 $\pm$ 0.06 & 2.72 $\pm$ 0.13 & -0.04 $\pm$ 0.13\\
90 & 0.97 & 386.4 $\pm$ 20.3 & 0.07 $\pm$ 0.09 & -0.07 $\pm$ 0.08 & -0.32 $\pm$ 0.13 & -0.00 $\pm$ 0.06 & 2.50 $\pm$ 0.14 & -0.10 $\pm$ 0.11\\
135 & 0.96 & 380.4 $\pm$ 20.6 & 0.05 $\pm$ 0.09 & 0.01 $\pm$ 0.09 & -0.28 $\pm$ 0.12 & 0.03 $\pm$ 0.07 & 2.44 $\pm$ 0.12 & -0.05 $\pm$ 0.11\\
180 & 0.97 & 438.4 $\pm$ 24.8 & 0.03 $\pm$ 0.11 & -0.10 $\pm$ 0.10 & -0.02 $\pm$ 0.16 & -0.04 $\pm$ 0.07 & 2.24 $\pm$ 0.13 & -0.05 $\pm$ 0.14\\
225 & 0.96 & 491.8 $\pm$ 30.7 & 0.11 $\pm$ 0.10 & -0.12 $\pm$ 0.10 & 0.17 $\pm$ 0.15 & -0.01 $\pm$ 0.07 & 2.41 $\pm$ 0.14 & -0.08 $\pm$ 0.11\\
\end{tabular}
\caption{Same as in Table 3, but for KIC 6933899. Radial orders used to compute the mean parameters range between $n=16$ and $n=20$. Results shown in Figure \ref{fig:6933899}.}\label{tab:6933899}\vspace{-0.1cm}
\end{table*}

\begin{table*}[ht]\centering\fontsize{9.}{7.}\selectfont
\begin{tabular}{ccc|ccccc|c}
\multicolumn{9}{c}{KIC 7103006}\\ \hline\hline
time & duty & $\tau_\text{gran}$ &\multicolumn{5}{c|}{Peak-bagging}&Cross-correlation\\
(d)& cycle & (s)&$\delta\nu_{l=0}$ ($\mu$Hz) & $\delta\nu_{l=1}$ ($\mu$Hz) & $\delta\nu_{l=2}$ ($\mu$Hz) & $\delta\nu$ ($\mu$Hz)& $\ln\,S$ & $\delta\nu\,(\mu$Hz)\\\hline
345 & 0.98 & 393.8 $\pm$ 22.4 & 0.54 $\pm$ 0.21 & 0.21 $\pm$ 0.20 & -0.59 $\pm$ 0.44 & 0.36 $\pm$ 0.15 & 1.22 $\pm$ 0.13 & 0.13 $\pm$ 0.27\\
390 & 0.97 & 394.8 $\pm$ 23.8 & 0.21 $\pm$ 0.25 & -0.01 $\pm$ 0.20 & 0.09 $\pm$ 0.50 & 0.08 $\pm$ 0.16 & 1.07 $\pm$ 0.13 & 0.08 $\pm$ 0.29\\
435 & 0.96 & 382.5 $\pm$ 22.3 & 0.32 $\pm$ 0.22 & -0.32 $\pm$ 0.24 & -0.71 $\pm$ 0.47 & 0.02 $\pm$ 0.16 & 1.13 $\pm$ 0.13 & 0.03 $\pm$ 0.32\\
480 & 0.97 & 367.9 $\pm$ 19.8 & 0.43 $\pm$ 0.23 & -0.28 $\pm$ 0.21 & -0.96 $\pm$ 0.44 & 0.05 $\pm$ 0.16 & 1.27 $\pm$ 0.13 & -0.05 $\pm$ 0.32\\
525 & 0.96 & 362.9 $\pm$ 20.4 & 0.03 $\pm$ 0.25 & -0.17 $\pm$ 0.20 & -0.20 $\pm$ 0.49 & -0.09 $\pm$ 0.15 & 1.11 $\pm$ 0.14 & -0.12 $\pm$ 0.30\\
\end{tabular}
\caption{Same as in Table 3, but for KIC 7103006. Radial orders used to compute the mean parameters range between $n=16$ and $n=20$. Results shown in Figure \ref{fig:7103006}.}\label{tab:7103006}\vspace{-2cm}
\end{table*}

\begin{table*}[ht]\centering\fontsize{9.}{7.}\selectfont
\begin{tabular}{ccc|ccccc|c}
\multicolumn{9}{c}{KIC 7106245}\\ \hline\hline
time & duty & $\tau_\text{gran}$ &\multicolumn{5}{c|}{Peak-bagging}&Cross-correlation\\
(d)& cycle & (s)&$\delta\nu_{l=0}$ ($\mu$Hz) & $\delta\nu_{l=1}$ ($\mu$Hz) & $\delta\nu_{l=2}$ ($\mu$Hz) & $\delta\nu$ ($\mu$Hz)& $\ln\,S$ & $\delta\nu\,(\mu$Hz)\\\hline
45 & 0.98 & 199.2 $\pm$ 0.8 & 0.30 $\pm$ 0.31 & 0.17 $\pm$ 0.23 & -0.43 $\pm$ 0.41 & 0.21 $\pm$ 0.18 & 1.34 $\pm$ 0.27 & --\\
90 & 0.97 & 209.8 $\pm$ 16.9 & -0.02 $\pm$ 0.21 & 0.19 $\pm$ 0.27 & -0.00 $\pm$ 0.65 & 0.06 $\pm$ 0.17 & 1.53 $\pm$ 0.33 & 0.19 $\pm$ 0.11\\
135 & 0.96 & 230.5 $\pm$ 13.5 & -0.16 $\pm$ 0.18 & 0.25 $\pm$ 0.20 & 0.35 $\pm$ 0.47 & 0.03 $\pm$ 0.13 & 1.72 $\pm$ 0.27 & 0.10 $\pm$ 0.12\\
180 & 0.97 & 220.8 $\pm$ 15.5 & -0.56 $\pm$ 0.20 & -0.11 $\pm$ 0.17 & 0.32 $\pm$ 0.47 & -0.30 $\pm$ 0.13 & 1.64 $\pm$ 0.27 & 0.02 $\pm$ 0.17\\
225 & 0.96 & 220.5 $\pm$ 9.3 & -0.57 $\pm$ 0.28 & -0.15 $\pm$ 0.18 & 0.16 $\pm$ 0.65 & -0.27 $\pm$ 0.15 & 1.41 $\pm$ 0.34 & -0.13 $\pm$ 0.23\\
\end{tabular}
\caption{Same as in Table 3, but for KIC 7106245. {Radial orders used to compute the mean parameters range between $n=17$ and $n=21$.} Note that the frequency shifts from the cross-correlation method (last column) were obtained with 180-d sub-series. Results shown in Figure~\ref{fig:7106245}.}\label{tab:7106245}\vspace{-0.1cm}
\end{table*}

\begin{table*}[ht]\centering\fontsize{9.}{7.}\selectfont
\begin{tabular}{ccc|ccccc|c}
\multicolumn{9}{c}{KIC 7206837}\\ \hline\hline
time & duty & $\tau_\text{gran}$ &\multicolumn{5}{c|}{Peak-bagging}&Cross-correlation\\
(d)& cycle & (s)&$\delta\nu_{l=0}$ ($\mu$Hz) & $\delta\nu_{l=1}$ ($\mu$Hz) & $\delta\nu_{l=2}$ ($\mu$Hz) & $\delta\nu$ ($\mu$Hz)& $\ln\,S$ & $\delta\nu\,(\mu$Hz)\\\hline
345 & 0.98 & 275.3 $\pm$ 19.9 & -0.02 $\pm$ 0.23 & 0.15 $\pm$ 0.23 & 0.33 $\pm$ 0.45 & 0.07 $\pm$ 0.16 & 1.04 $\pm$ 0.16 & 0.03 $\pm$ 0.20\\
390 & 0.97 & 277.8 $\pm$ 22.6 & -0.23 $\pm$ 0.26 & 0.02 $\pm$ 0.28 & -0.46 $\pm$ 0.54 & -0.11 $\pm$ 0.19 & 0.80 $\pm$ 0.15 & -0.10 $\pm$ 0.26\\
435 & 0.96 & 297.4 $\pm$ 24.5 & 0.06 $\pm$ 0.25 & -0.73 $\pm$ 0.34 & -0.88 $\pm$ 0.62 & -0.22 $\pm$ 0.20 & 0.93 $\pm$ 0.14 & -0.27 $\pm$ 0.44\\
480 & 0.97 & 298.6 $\pm$ 17.3 & -0.12 $\pm$ 0.22 & -0.90 $\pm$ 0.31 & 0.03 $\pm$ 0.53 & -0.38 $\pm$ 0.18 & 0.99 $\pm$ 0.14 & -0.40 $\pm$ 0.44\\
525 & 0.96 & 271.4 $\pm$ 20.4 & -0.57 $\pm$ 0.25 & 0.07 $\pm$ 0.28 & -0.55 $\pm$ 0.54 & -0.28 $\pm$ 0.19 & 1.00 $\pm$ 0.15 & -0.22 $\pm$ 0.41\\
\end{tabular}
\caption{Same as in Table 3, but for KIC 7206837. Radial orders used to compute the mean parameters range between $n=18$ and $n=22$. Results shown in Figure \ref{fig:7206837}.}\label{tab:7206837}\vspace{-2cm}
\end{table*}

\begin{table*}[ht]\centering\fontsize{9.}{7.}\selectfont
\begin{tabular}{ccc|ccccc|c}
\multicolumn{9}{c}{KIC 7296438}\\ \hline\hline
time & duty & $\tau_\text{gran}$ &\multicolumn{5}{c|}{Peak-bagging}&Cross-correlation\\
(d)& cycle & (s)&$\delta\nu_{l=0}$ ($\mu$Hz) & $\delta\nu_{l=1}$ ($\mu$Hz) & $\delta\nu_{l=2}$ ($\mu$Hz) & $\delta\nu$ ($\mu$Hz)& $\ln\,S$ & $\delta\nu\,(\mu$Hz)\\\hline
545 & 0.97 & 308.1 $\pm$ 7.9 & 0.11 $\pm$ 0.09 & 0.13 $\pm$ 0.08 & -0.11 $\pm$ 0.15 & 0.12 $\pm$ 0.06 & 2.43 $\pm$ 0.17 & 0.00 $\pm$ 0.13\\
590 & 0.78 & 289.1 $\pm$ 13.7 & 0.04 $\pm$ 0.10 & 0.11 $\pm$ 0.09 & -0.22 $\pm$ 0.17 & 0.08 $\pm$ 0.07 & 2.51 $\pm$ 0.16 & 0.02 $\pm$ 0.12\\
635 & 0.73 & 274.3 $\pm$ 21.7 & 0.11 $\pm$ 0.13 & -0.01 $\pm$ 0.13 & 0.16 $\pm$ 0.18 & 0.05 $\pm$ 0.09 & 2.18 $\pm$ 0.16 & -0.04 $\pm$ 0.19\\
680 & 0.92 & 278.8 $\pm$ 19.3 & -0.10 $\pm$ 0.12 & -0.03 $\pm$ 0.10 & -0.06 $\pm$ 0.16 & -0.06 $\pm$ 0.07 & 2.33 $\pm$ 0.16 & -0.07 $\pm$ 0.15\\
725 & 0.98 & 286.8 $\pm$ 15.2 & -0.03 $\pm$ 0.09 & 0.04 $\pm$ 0.10 & -0.26 $\pm$ 0.19 & 0.00 $\pm$ 0.07 & 2.23 $\pm$ 0.16 & -0.03 $\pm$ 0.15\\
770 & 0.97 & 291.3 $\pm$ 2.8 & 0.01 $\pm$ 0.09 & 0.02 $\pm$ 0.12 & 0.12 $\pm$ 0.18 & 0.02 $\pm$ 0.07 & 2.12 $\pm$ 0.15 & 0.09 $\pm$ 0.15\\
815 & 0.97 & 297.2 $\pm$ 13.7 & 0.12 $\pm$ 0.09 & 0.09 $\pm$ 0.09 & -0.08 $\pm$ 0.17 & 0.10 $\pm$ 0.06 & 2.16 $\pm$ 0.15 & 0.11 $\pm$ 0.15\\
860 & 0.97 & 277.1 $\pm$ 15.9 & -0.06 $\pm$ 0.08 & 0.02 $\pm$ 0.10 & -0.14 $\pm$ 0.17 & -0.03 $\pm$ 0.06 & 2.24 $\pm$ 0.14 & 0.02 $\pm$ 0.15\\
905 & 0.94 & 282.0 $\pm$ 15.4 & -0.16 $\pm$ 0.11 & -0.06 $\pm$ 0.11 & 0.01 $\pm$ 0.22 & -0.11 $\pm$ 0.08 & 2.17 $\pm$ 0.14 & -0.11 $\pm$ 0.16\\
950 & 0.66 & 289.7 $\pm$ 7.9 & 0.18 $\pm$ 0.14 & -0.23 $\pm$ 0.12 & -0.04 $\pm$ 0.23 & -0.05 $\pm$ 0.09 & 2.26 $\pm$ 0.17 & --\\
\end{tabular}
\caption{Same as in Table 3, but for KIC 7296438. Radial orders used to compute the mean parameters range between $n=17$ and $n=21$. Results shown in Figure \ref{fig:7296438}.}\label{tab:7296438}
\end{table*}

\begin{table*}[ht]\centering\fontsize{9.}{7.}\selectfont
\begin{tabular}{ccc|ccccc|c}
\multicolumn{9}{c}{KIC 7510397}\\ \hline\hline
time & duty & $\tau_\text{gran}$ &\multicolumn{5}{c|}{Peak-bagging}&Cross-correlation\\
(d)& cycle & (s)&$\delta\nu_{l=0}$ ($\mu$Hz) & $\delta\nu_{l=1}$ ($\mu$Hz) & $\delta\nu_{l=2}$ ($\mu$Hz) & $\delta\nu$ ($\mu$Hz)& $\ln\,S$ & $\delta\nu\,(\mu$Hz)\\\hline
45 & 0.96 & 425.3 $\pm$ 22.1 & 0.11 $\pm$ 0.14 & 0.16 $\pm$ 0.12 & 0.17 $\pm$ 0.15 & 0.14 $\pm$ 0.09 & 1.18 $\pm$ 0.12 & 0.04 $\pm$ 0.19\\
90 & 0.77 & 429.9 $\pm$ 24.0 & 0.03 $\pm$ 0.15 & 0.10 $\pm$ 0.13 & 0.01 $\pm$ 0.18 & 0.07 $\pm$ 0.10 & 1.19 $\pm$ 0.11 & -0.02 $\pm$ 0.19\\
135 & 0.73 & 361.8 $\pm$ 19.1 & 0.06 $\pm$ 0.14 & -0.07 $\pm$ 0.12 & 0.03 $\pm$ 0.18 & -0.02 $\pm$ 0.09 & 1.43 $\pm$ 0.12 & -0.07 $\pm$ 0.18\\
180 & 0.92 & 396.3 $\pm$ 18.7 & 0.13 $\pm$ 0.13 & -0.04 $\pm$ 0.11 & -0.05 $\pm$ 0.16 & 0.03 $\pm$ 0.08 & 1.41 $\pm$ 0.11 & -0.03 $\pm$ 0.16\\
225 & 0.98 & 416.9 $\pm$ 19.7 & 0.24 $\pm$ 0.12 & -0.09 $\pm$ 0.11 & 0.13 $\pm$ 0.15 & 0.07 $\pm$ 0.08 & 1.44 $\pm$ 0.11 & 0.06 $\pm$ 0.16\\
\end{tabular}
\caption{Same as in Table 3, but for KIC 7510397. Radial orders used to compute the mean parameters range between $n=16$ and $n=20$. Results shown in Figure \ref{fig:7510397}.}\label{tab:7510397}
\end{table*}

\begin{table*}[ht]\centering\fontsize{9.}{7.}\selectfont
\begin{tabular}{ccc|ccccc|c}
\multicolumn{9}{c}{KIC 7670943}\\ \hline\hline
time & duty & $\tau_\text{gran}$ &\multicolumn{5}{c|}{Peak-bagging}&Cross-correlation\\
(d)& cycle & (s)&$\delta\nu_{l=0}$ ($\mu$Hz) & $\delta\nu_{l=1}$ ($\mu$Hz) & $\delta\nu_{l=2}$ ($\mu$Hz) & $\delta\nu$ ($\mu$Hz)& $\ln\,S$ & $\delta\nu\,(\mu$Hz)\\\hline
45 & 0.98 & 276.0 $\pm$ 26.3 & 0.35 $\pm$ 0.39 & -0.36 $\pm$ 0.45 & 0.28 $\pm$ 1.59 & 0.04 $\pm$ 0.29 & 1.13 $\pm$ 0.20 & --\\
90 & 0.97 & 285.4 $\pm$ 34.4 & -0.19 $\pm$ 0.50 & -0.62 $\pm$ 0.62 & 0.18 $\pm$ 1.60 & -0.36 $\pm$ 0.39 & 1.02 $\pm$ 0.22 & -0.25 $\pm$ 0.13\\
135 & 0.96 & 247.6 $\pm$ 30.7 & 0.24 $\pm$ 0.56 & -0.77 $\pm$ 0.66 & -0.04 $\pm$ 1.75 & -0.19 $\pm$ 0.43 & 0.84 $\pm$ 0.22 & -0.28 $\pm$ 0.29\\
180 & 0.78 & 244.8 $\pm$ 34.2 & 0.16 $\pm$ 0.44 & -0.66 $\pm$ 0.56 & -0.06 $\pm$ 1.62 & -0.15 $\pm$ 0.35 & 1.52 $\pm$ 0.33 & -0.21 $\pm$ 0.30\\
225 & 0.72 & 235.4 $\pm$ 25.2 & 1.42 $\pm$ 0.32 & 1.47 $\pm$ 0.39 & -1.18 $\pm$ 1.57 & 1.44 $\pm$ 0.25 & 1.64 $\pm$ 0.20 & 0.18 $\pm$ 0.65\\
\end{tabular}
\caption{Same as in Table 3, but for KIC 7670943. {Radial orders used to compute the mean parameters range between $n=18$ and $n=22$.} Note that the frequency shifts from the cross-correlation method (last column) were obtained with 180-d sub-series. Results shown in Figure~\ref{fig:7670943}.}\label{tab:7670943}
\end{table*}

\begin{table*}[ht]\centering\fontsize{9.}{7.}\selectfont
\begin{tabular}{ccc|ccccc|c}
\multicolumn{9}{c}{KIC 7680114}\\ \hline\hline
time & duty & $\tau_\text{gran}$ &\multicolumn{5}{c|}{Peak-bagging}&Cross-correlation\\
(d)& cycle & (s)&$\delta\nu_{l=0}$ ($\mu$Hz) & $\delta\nu_{l=1}$ ($\mu$Hz) & $\delta\nu_{l=2}$ ($\mu$Hz) & $\delta\nu$ ($\mu$Hz)& $\ln\,S$ & $\delta\nu\,(\mu$Hz)\\\hline
345 & 0.98 & 357.1 $\pm$ 28.5 & 0.11 $\pm$ 0.10 & 0.19 $\pm$ 0.09 & -0.14 $\pm$ 0.15 & 0.16 $\pm$ 0.07 & 2.44 $\pm$ 0.17 & -0.02 $\pm$ 0.13\\
390 & 0.54 & 363.2 $\pm$ 2.7 & -0.03 $\pm$ 0.15 & 0.29 $\pm$ 0.10 & -0.24 $\pm$ 0.17 & 0.19 $\pm$ 0.08 & 2.55 $\pm$ 0.24 & 0.05 $\pm$ 0.11\\
435 & 0.05 & 320.0 $\pm$ 78.5 & -0.21 $\pm$ 0.44 & -0.26 $\pm$ 0.41 & 0.33 $\pm$ 0.71 & -0.24 $\pm$ 0.30 & 3.67 $\pm$ 0.62 & --\\
480 & 0.33 & 295.9 $\pm$ 35.2 & -0.08 $\pm$ 0.22 & -0.22 $\pm$ 0.22 & -0.81 $\pm$ 0.28 & -0.15 $\pm$ 0.15 & 2.06 $\pm$ 0.27 & --\\
525 & 0.58 & 293.0 $\pm$ 23.2 & 0.01 $\pm$ 0.14 & 0.04 $\pm$ 0.15 & -0.60 $\pm$ 0.19 & 0.03 $\pm$ 0.10 & 2.16 $\pm$ 0.15 & -0.08 $\pm$ 0.17\\
\end{tabular}
\caption{Same as in Table 3, but for KIC 7680114. Radial orders used to compute the mean parameters range between $n=16$ and $n=20$. Results shown in Figure \ref{fig:7680114}.}\label{tab:7680114}\vspace{-1cm}
\end{table*}

\FloatBarrier
\nopagebreak
\begin{table}[ht]\centering\fontsize{9.}{7.}\selectfont
\begin{tabular}{ccc|ccccc|c}
\multicolumn{9}{c}{KIC 7771282}\\ \hline\hline
time & duty & $\tau_\text{gran}$ &\multicolumn{5}{c|}{Peak-bagging}&Cross-correlation\\
(d)& cycle & (s)&$\delta\nu_{l=0}$ ($\mu$Hz) & $\delta\nu_{l=1}$ ($\mu$Hz) & $\delta\nu_{l=2}$ ($\mu$Hz) & $\delta\nu$ ($\mu$Hz)& $\ln\,S$ & $\delta\nu\,(\mu$Hz)\\\hline
345 & 0.97 & 289.3 $\pm$ 19.9 & 0.38 $\pm$ 0.29 & -0.07 $\pm$ 0.31 & -1.39 $\pm$ 0.49 & 0.17 $\pm$ 0.21 & 1.56 $\pm$ 0.17 & 0.16 $\pm$ 0.21\\
390 & 0.54 & 300.0 $\pm$ 42.4 & 1.11 $\pm$ 0.33 & 0.07 $\pm$ 0.38 & -0.82 $\pm$ 0.72 & 0.66 $\pm$ 0.25 & 1.65 $\pm$ 0.28 & 0.17 $\pm$ 0.25\\
435 & 0.05 & 200.8 $\pm$ 16.8 & 0.34 $\pm$ 0.84 & 0.98 $\pm$ 0.76 & 0.21 $\pm$ 0.77 & 0.69 $\pm$ 0.56 & 1.87 $\pm$ 0.60 & --\\
480 & 0.42 & 290.6 $\pm$ 29.3 & 0.14 $\pm$ 0.34 & 1.40 $\pm$ 0.48 & 1.41 $\pm$ 0.65 & 0.55 $\pm$ 0.28 & 1.57 $\pm$ 0.27 & --\\
525 & 0.90 & 300.6 $\pm$ 32.8 & 0.23 $\pm$ 0.24 & 0.42 $\pm$ 0.29 & 1.15 $\pm$ 0.64 & 0.30 $\pm$ 0.19 & 1.72 $\pm$ 0.18 & 0.36 $\pm$ 0.39\\
\end{tabular}
\parbox{2\hsize}{\caption{Same as in Table 3, but for KIC 7771282. Radial orders used to compute the mean parameters range between $n=17$ and $n=21$. Results shown in Figure \ref{fig:7771282}.}\label{tab:7771282}}
\end{table}

\begin{table}[ht]\centering\fontsize{9.}{7.}\selectfont
\begin{tabular}{ccc|ccccc|c}
\multicolumn{9}{c}{KIC 7871531}\\ \hline\hline
time & duty & $\tau_\text{gran}$ &\multicolumn{5}{c|}{Peak-bagging}&Cross-correlation\\
(d)& cycle & (s)&$\delta\nu_{l=0}$ ($\mu$Hz) & $\delta\nu_{l=1}$ ($\mu$Hz) & $\delta\nu_{l=2}$ ($\mu$Hz) & $\delta\nu$ ($\mu$Hz)& $\ln\,S$ & $\delta\nu\,(\mu$Hz)\\\hline
345 & 0.98 & 186.3 $\pm$ 2.9 & 0.14 $\pm$ 0.18 & 0.00 $\pm$ 0.14 & 0.35 $\pm$ 0.29 & 0.05 $\pm$ 0.11 & 0.82 $\pm$ 0.26 & -0.11 $\pm$ 0.10\\
390 & 0.97 & 196.4 $\pm$ 6.6 & 0.00 $\pm$ 0.13 & -0.16 $\pm$ 0.14 & 0.42 $\pm$ 0.27 & -0.07 $\pm$ 0.10 & 0.73 $\pm$ 0.22 & -0.10 $\pm$ 0.25\\
435 & 0.96 & 208.8 $\pm$ 16.7 & 0.05 $\pm$ 0.11 & 0.22 $\pm$ 0.16 & 0.67 $\pm$ 0.27 & 0.11 $\pm$ 0.09 & 0.72 $\pm$ 0.21 & -0.02 $\pm$ 0.60\\
480 & 0.97 & 221.8 $\pm$ 18.8 & -0.20 $\pm$ 0.14 & 0.10 $\pm$ 0.14 & 0.60 $\pm$ 0.32 & -0.05 $\pm$ 0.10 & 0.60 $\pm$ 0.24 & -0.15 $\pm$ 0.42\\
525 & 0.96 & 205.9 $\pm$ 0.0 & -0.30 $\pm$ 0.14 & 0.03 $\pm$ 0.15 & -0.32 $\pm$ 0.39 & -0.15 $\pm$ 0.10 & 0.56 $\pm$ 0.25 & -0.19 $\pm$ 0.50\\
\end{tabular}
\parbox{2\hsize}{\caption{Same as in Table 3, but for KIC 7871531. Radial orders used to compute the mean parameters range between $n=19$ and $n=23$. Results shown in Figure \ref{fig:7871531}.}\label{tab:7871531}}
\end{table}

\begin{table*}[ht]\centering\fontsize{9.}{7.}\selectfont
\begin{tabular}{ccc|ccccc|c}
\multicolumn{9}{c}{KIC 7940546}\\ \hline\hline
time & duty & $\tau_\text{gran}$ &\multicolumn{5}{c|}{Peak-bagging}&Cross-correlation\\
(d)& cycle & (s)&$\delta\nu_{l=0}$ ($\mu$Hz) & $\delta\nu_{l=1}$ ($\mu$Hz) & $\delta\nu_{l=2}$ ($\mu$Hz) & $\delta\nu$ ($\mu$Hz)& $\ln\,S$ & $\delta\nu\,(\mu$Hz)\\\hline
495 & 0.97 & 379.3 $\pm$ 12.1 & 0.11 $\pm$ 0.14 & 0.07 $\pm$ 0.14 & 0.54 $\pm$ 0.28 & 0.09 $\pm$ 0.10 & 1.88 $\pm$ 0.10 & 0.11 $\pm$ 0.21\\
540 & 0.77 & 400.5 $\pm$ 14.0 & -0.01 $\pm$ 0.16 & -0.08 $\pm$ 0.15 & 0.36 $\pm$ 0.29 & -0.05 $\pm$ 0.11 & 1.78 $\pm$ 0.09 & -0.09 $\pm$ 0.17\\
585 & 0.73 & 379.3 $\pm$ 14.3 & 0.09 $\pm$ 0.17 & 0.24 $\pm$ 0.14 & -0.04 $\pm$ 0.31 & 0.18 $\pm$ 0.11 & 1.85 $\pm$ 0.10 & 0.09 $\pm$ 0.18\\
630 & 0.92 & 397.4 $\pm$ 12.9 & 0.28 $\pm$ 0.14 & 0.20 $\pm$ 0.14 & -0.67 $\pm$ 0.26 & 0.24 $\pm$ 0.10 & 1.82 $\pm$ 0.09 & 0.10 $\pm$ 0.19\\
675 & 0.98 & 399.6 $\pm$ 12.6 & 0.56 $\pm$ 0.14 & 0.10 $\pm$ 0.14 & -0.44 $\pm$ 0.26 & 0.32 $\pm$ 0.10 & 1.77 $\pm$ 0.10 & 0.21 $\pm$ 0.16\\
\end{tabular}
\caption{Same as in Table 3, but for KIC 7940546. Radial orders used to compute the mean parameters range between $n=15$ and $n=19$. Results shown in Figure \ref{fig:7940546}.}\label{tab:7940546}
\end{table*}

\begin{table*}[ht]\centering\fontsize{9.}{7.}\selectfont
\begin{tabular}{ccc|ccccc|c}
\multicolumn{9}{c}{KIC 7970740}\\ \hline\hline
time & duty & $\tau_\text{gran}$ &\multicolumn{5}{c|}{Peak-bagging}&Cross-correlation\\
(d)& cycle & (s)&$\delta\nu_{l=0}$ ($\mu$Hz) & $\delta\nu_{l=1}$ ($\mu$Hz) & $\delta\nu_{l=2}$ ($\mu$Hz) & $\delta\nu$ ($\mu$Hz)& $\ln\,S$ & $\delta\nu\,(\mu$Hz)\\\hline
445 & 0.97 & 219.1 $\pm$ 4.7 & 0.06 $\pm$ 0.09 & -0.11 $\pm$ 0.13 & -0.21 $\pm$ 0.21 & 0.00 $\pm$ 0.08 & -0.14 $\pm$ 0.15 & 0.02 $\pm$ 0.12\\
490 & 0.97 & 215.5 $\pm$ 11.1 & -0.02 $\pm$ 0.09 & -0.15 $\pm$ 0.11 & 0.17 $\pm$ 0.24 & -0.07 $\pm$ 0.07 & -0.15 $\pm$ 0.17 & -0.05 $\pm$ 0.13\\
535 & 0.96 & 209.6 $\pm$ 0.3 & -0.01 $\pm$ 0.08 & 0.01 $\pm$ 0.08 & -0.09 $\pm$ 0.14 & -0.00 $\pm$ 0.06 & 0.12 $\pm$ 0.17 & -0.03 $\pm$ 0.17\\
580 & 0.77 & 238.2 $\pm$ 14.6 & -0.11 $\pm$ 0.09 & -0.02 $\pm$ 0.10 & -0.08 $\pm$ 0.14 & -0.06 $\pm$ 0.07 & 0.25 $\pm$ 0.15 & -0.11 $\pm$ 0.16\\
625 & 0.72 & 226.7 $\pm$ 13.5 & -0.24 $\pm$ 0.10 & 0.03 $\pm$ 0.15 & -0.09 $\pm$ 0.17 & -0.15 $\pm$ 0.08 & -0.08 $\pm$ 0.15 & -0.01 $\pm$ 0.24\\
\end{tabular}
\caption{Same as in Table 3, but for KIC 7970740. Figures in the main text (see Figure \ref{fig:KIC7970740}). Radial orders used to compute the mean parameters range between $n=20$ and $n=24$.}\label{tab:7970740}
\end{table*}

\begin{table*}[ht]\centering\fontsize{9.}{7.}\selectfont
\begin{tabular}{ccc|ccccc|c}
\multicolumn{9}{c}{KIC 8077137}\\ \hline\hline
time & duty & $\tau_\text{gran}$ &\multicolumn{5}{c|}{Peak-bagging}&Cross-correlation\\
(d)& cycle & (s)&$\delta\nu_{l=0}$ ($\mu$Hz) & $\delta\nu_{l=1}$ ($\mu$Hz) & $\delta\nu_{l=2}$ ($\mu$Hz) & $\delta\nu$ ($\mu$Hz)& $\ln\,S$ & $\delta\nu\,(\mu$Hz)\\\hline
45 & 0.98 & 403.0 $\pm$ 7.3 & -1.24 $\pm$ 0.33 & -0.09 $\pm$ 0.23 & -0.26 $\pm$ 0.74 & -0.47 $\pm$ 0.19 & 2.11 $\pm$ 0.31 & -0.48 $\pm$ 0.59\\
90 & 0.97 & 404.9 $\pm$ 40.7 & -0.07 $\pm$ 0.27 & -0.51 $\pm$ 0.25 & 1.07 $\pm$ 0.68 & -0.30 $\pm$ 0.19 & 2.32 $\pm$ 0.27 & -0.59 $\pm$ 0.36\\
135 & 0.96 & 354.9 $\pm$ 39.8 & 0.11 $\pm$ 0.22 & -0.27 $\pm$ 0.26 & 1.04 $\pm$ 0.70 & -0.05 $\pm$ 0.17 & 2.32 $\pm$ 0.25 & -0.30 $\pm$ 0.69\\
180 & 0.78 & 383.9 $\pm$ 17.1 & 0.23 $\pm$ 0.27 & -0.44 $\pm$ 0.25 & 0.07 $\pm$ 0.68 & -0.13 $\pm$ 0.18 & 2.20 $\pm$ 0.24 & 0.02 $\pm$ 1.23\\
225 & 0.72 & 337.1 $\pm$ 20.0 & -0.09 $\pm$ 0.21 & -0.29 $\pm$ 0.24 & -0.84 $\pm$ 0.51 & -0.18 $\pm$ 0.16 & 2.44 $\pm$ 0.21 & -0.05 $\pm$ 1.28\\
\end{tabular}
\caption{Same as in Table 3, but for KIC 8077137. Radial orders used to compute the mean parameters range between $n=16$ and $n=20$. Results shown in Figure \ref{fig:8077137}.}\label{tab:8077137}
\end{table*}

\begin{table*}[ht]\centering\fontsize{9.}{7.}\selectfont
\begin{tabular}{ccc|ccccc|c}
\multicolumn{9}{c}{KIC 8150065}\\ \hline\hline
time & duty & $\tau_\text{gran}$ &\multicolumn{5}{c|}{Peak-bagging}&Cross-correlation\\
(d)& cycle & (s)&$\delta\nu_{l=0}$ ($\mu$Hz) & $\delta\nu_{l=1}$ ($\mu$Hz) & $\delta\nu_{l=2}$ ($\mu$Hz) & $\delta\nu$ ($\mu$Hz)& $\ln\,S$ & $\delta\nu\,(\mu$Hz)\\\hline
45 & 0.97 & 232.4 $\pm$ 21.0 & -0.22 $\pm$ 0.38 & -0.17 $\pm$ 0.24 & -1.43 $\pm$ 0.40 & -0.18 $\pm$ 0.20 & 1.26 $\pm$ 0.18 & 0.01 $\pm$ 0.47\\
90 & 0.54 & 299.9 $\pm$ 32.7 & -0.74 $\pm$ 0.45 & -0.26 $\pm$ 0.33 & -1.27 $\pm$ 0.61 & -0.43 $\pm$ 0.26 & 1.34 $\pm$ 0.25 & -0.07 $\pm$ 0.42\\
135 & 0.05 & 399.5 $\pm$ 460.4 & 0.26 $\pm$ 0.68 & 0.16 $\pm$ 0.72 & -0.32 $\pm$ 0.80 & 0.21 $\pm$ 0.49 & 1.75 $\pm$ 0.54 & --\\
180 & 0.33 & 242.6 $\pm$ 23.9 & -0.43 $\pm$ 0.52 & -0.08 $\pm$ 0.42 & 0.36 $\pm$ 0.68 & -0.22 $\pm$ 0.33 & 1.38 $\pm$ 0.42 & --\\
225 & 0.58 & 263.0 $\pm$ 25.9 & -0.19 $\pm$ 0.16 & 0.20 $\pm$ 0.23 & 0.34 $\pm$ 0.35 & -0.06 $\pm$ 0.13 & 1.94 $\pm$ 0.26 & 0.39 $\pm$ 0.75\\
\end{tabular}
\caption{Same as in Table 3, but for KIC 8150065. Radial orders used to compute the mean parameters range between $n=17$ and $n=21$. Results shown in Figure \ref{fig:8150065}.}\label{tab:8150065}
\end{table*}

\begin{table*}[ht]\centering\fontsize{9.}{7.}\selectfont
\begin{tabular}{ccc|ccccc|c}
\multicolumn{9}{c}{KIC 8179536}\\ \hline\hline
time & duty & $\tau_\text{gran}$ &\multicolumn{5}{c|}{Peak-bagging}&Cross-correlation\\
(d)& cycle & (s)&$\delta\nu_{l=0}$ ($\mu$Hz) & $\delta\nu_{l=1}$ ($\mu$Hz) & $\delta\nu_{l=2}$ ($\mu$Hz) & $\delta\nu$ ($\mu$Hz)& $\ln\,S$ & $\delta\nu\,(\mu$Hz)\\\hline
345 & 0.98 & 249.1 $\pm$ 17.5 & -0.21 $\pm$ 0.21 & -0.37 $\pm$ 0.27 & -0.12 $\pm$ 0.47 & -0.27 $\pm$ 0.17 & 0.78 $\pm$ 0.13 & -0.25 $\pm$ 0.23\\
390 & 0.54 & 264.5 $\pm$ 27.0 & -0.14 $\pm$ 0.30 & -0.64 $\pm$ 0.37 & 0.35 $\pm$ 0.54 & -0.34 $\pm$ 0.23 & 0.77 $\pm$ 0.16 & -0.24 $\pm$ 0.21\\
435 & 0.05 & 209.7 $\pm$ 34.6 & 0.68 $\pm$ 0.62 & 0.23 $\pm$ 0.69 & 0.10 $\pm$ 0.65 & 0.48 $\pm$ 0.46 & 1.18 $\pm$ 0.49 & --\\
480 & 0.42 & 210.6 $\pm$ 17.9 & 0.09 $\pm$ 0.25 & 0.30 $\pm$ 0.30 & -0.05 $\pm$ 0.53 & 0.17 $\pm$ 0.19 & 1.13 $\pm$ 0.21 & --\\
525 & 0.90 & 215.5 $\pm$ 13.2 & 0.02 $\pm$ 0.22 & -0.13 $\pm$ 0.25 & 0.31 $\pm$ 0.40 & -0.04 $\pm$ 0.16 & 0.92 $\pm$ 0.13 & -0.10 $\pm$ 0.26\\
\end{tabular}%
\caption{Same as in Table 3, but for KIC 8292840. {Radial orders used to compute the mean parameters range between $n=19$ and $n=23$.} Note that the frequency shifts from the cross-correlation method (last column) were obtained with 180-d sub-series. Results shown in Figure~\ref{fig:8292840}.}\label{tab:8292840}\vspace{-0.2cm}
\end{table*}

\begin{table*}[ht]\centering\fontsize{9.}{7.}\selectfont
%
\caption{Same as in Table 3, but for KIC 8379927. Figures in the main text (see Figure \ref{fig:KIC8379927}). Radial orders used to compute the mean parameters range between $n=18$ and $n=22$.}\label{tab:8379927}\vspace{-1.5cm}
\end{table*}

\begin{table*}[ht]\centering\fontsize{9.}{7.}\selectfont
%
\caption{Same as in Table 3, but for KIC 8478994. {Radial orders used to compute the mean parameters range between $n=24$ and $n=28$.} Note that the frequency shifts from the cross-correlation method (last column) were obtained with 180-d sub-series. Results shown in Figure~\ref{fig:8478994}.}\label{tab:8478994}
\end{table*}

\begin{table*}[ht]\centering\fontsize{9.}{7.}\selectfont
%
\caption{Same as in Table 3, but for KIC 8494142. {Radial orders used to compute the mean parameters range between $n=17$ and $n=21$.} Note that the frequency shifts from the cross-correlation method (last column) were obtained with 180-d sub-series. Results shown in Figure~\ref{fig:8494142}.}\label{tab:8494142}\vspace{-1cm}
\end{table*}

\begin{table*}[ht]\centering\fontsize{9.}{7.}\selectfont
%
\caption{Same as in Table 3, but for KIC 9025370. Radial orders used to compute the mean parameters range between $n=19$ and $n=23$. Results shown in Figure \ref{fig:9025370}.}\label{tab:9025370}\vspace{-1.5cm}
\end{table*}

\FloatBarrier
\nopagebreak

\begin{table}[ht]\centering\fontsize{9.}{7.}\selectfont
%
\parbox{2\hsize}{\caption{Same as in Table 3, but for KIC 9098294. Radial orders used to compute the mean parameters range between $n=17$ and $n=21$. Results shown in Figure \ref{fig:9098294}.}\label{tab:9098294}}\vspace{-0.2cm}
\end{table}

\begin{table}[ht]\centering\fontsize{9.}{7.}\selectfont
%
\parbox{2\hsize}{\caption{Same as in Table 3, but for KIC 9139151. Radial orders used to compute the mean parameters range between $n=19$ and $n=23$. Results shown in Figure \ref{fig:9139151}.}\label{tab:9139151}}\vspace{-1.5cm}
\end{table}

\begin{table*}[ht]\centering\fontsize{9.}{7.}\selectfont
%
\caption{Same as in Table 3, but for KIC 9414417. Figures in the main text (see Figure \ref{fig:KIC9414417}). Radial orders used to compute the mean parameters range between $n=16$ and $n=20$.}\label{tab:9414417}
\end{table*}

\begin{table*}[ht]\centering\fontsize{9.}{7.}\selectfont
%
\caption{Same as in Table 3, but for KIC 10516096. Radial orders used to compute the mean parameters range between $n=16$ and $n=20$. Results shown in Figure \ref{fig:10516096}.}\label{tab:10516096}\vspace{-1.5cm}
\end{table*}

\FloatBarrier
\nopagebreak

\begin{table}[ht]\centering\fontsize{9.}{7.}\selectfont
%
\parbox{2\hsize}{\caption{Same as in Table 3, but for KIC 10586004. Radial orders used to compute the mean parameters range between $n=15$ and $n=19$. Results shown in Figure \ref{fig:10586004}.}\label{tab:10586004}}
\end{table}

\begin{table}[ht]\centering\fontsize{9.}{7.}\selectfont
%
\parbox{2\hsize}{\caption{Same as in Table 3, but for KIC 10644253. Figures in the main text (see Figure \ref{fig:KIC10644253}). Radial orders used to compute the mean parameters range between $n=19$ and $n=23$.}\label{tab:10644253}}
\end{table}

\begin{table*}[ht]\centering\fontsize{9.}{7.}\selectfont
%
\caption{Same as in Table 3, but for KIC 11904151. {Radial orders used to compute the mean parameters range between $n=17$ and $n=21$.} Note that the frequency shifts from the cross-correlation method (last column) were obtained with 180-d sub-series. Results shown in Figure~\ref{fig:11904151}.}\label{tab:11904151}\vspace{-2cm}
\end{table*}

\begin{table*}[ht]\centering\fontsize{9.}{7.}\selectfont
%
\caption{Same as in Table 3, but for KIC 12317678. Radial orders used to compute the mean parameters range between $n=15$ and $n=19$. Results shown in Figure \ref{fig:12317678}.}\label{tab:12317678}\vspace{-1.5cm}
\end{table*}

\end{document}